\documentclass[11pt,a4paper]{article}
\pdfoutput=1
\usepackage[applemac]{inputenc}
\usepackage{amsmath, amsthm}
\usepackage{jheppub}
\usepackage[center]{caption}
\usepackage{subcaption}
\usepackage{multirow}
\usepackage{booktabs}
\usepackage{amsmath}
\usepackage{amsfonts}
\usepackage{amssymb}
\usepackage{graphicx}
\usepackage{indentfirst}
\usepackage{slashed}
\usepackage{mathrsfs}

\newcommand{\refE}[1]{eq.~(\ref{#1})}
\newcommand{\refF}[1]{fig.~\ref{#1}}

\newcommand{\refS}[1]{section~\ref{#1}}

\newcommand{\abs}[1]{\left\lvert#1\right\rvert}

\newcommand{\Li}[0]{\text{Li}}

\newcommand{\Disc}[0]{\operatorname{Disc}}
\newcommand{\Cut}[0]{\operatorname{Cut}}
\renewcommand\Re{\operatorname{\mathfrak{Re}}}

\newcommand{\zbar}{\bar{z}}
\newcommand{\cut}{\text{Cut}}

\newcommand{\bea}{\begin{eqnarray}}
\newcommand{\eea}{\end{eqnarray}}
\newcommand{\bean}{\begin{eqnarray*}}
\newcommand{\eean}{\end{eqnarray*}}

\def\half{\frac{1}{2}}

\def\abs#1{\left| #1\right|}

\def\Li{{\rm Li}}

\def\th{{\theta}}

\def\eps{\epsilon}

\def\bz{{\bar{z}}}
\def\bZ{{\bar{Z}}}

\def\ord{{\cal O}}
\def\cS{{\cal S}}
\def\cP{{\cal P}}

\def\Label#1{\label{#1}
  \smash{\hbox to0pt{\raise1ex\hbox{\tiny[#1]}\hss}}}

\def\beq{\begin{equation}}
\def\eeq{\end{equation}}
\def\bsp#1\esp{\begin{split}#1\end{split}}

\newcommand{\cA}{\mathcal{A}}
\newcommand{\cH}{\mathcal{H}}

\def \sh{{\,\amalg\hskip -2.9pt\amalg\,}}

\renewcommand{\ln}{\log}

\preprint{IPhT-t14/004
\rightline{PPP/14/06}
\rightline{DCPT/14/12}
\rightline{Edinburgh 2014/02}}

\title{From multiple unitarity cuts to the coproduct of Feynman integrals}

\author[a,b]{Samuel Abreu,}
\author[b]{Ruth Britto,}
\author[c]{Claude Duhr,}
\author[a]{and Einan Gardi}

\affiliation[a]{Higgs Centre for Theoretical Physics, School of Physics and Astronomy, \\
The University of Edinburgh,  
Edinburgh EH9 3JZ, Scotland, UK}
\affiliation[b]{Institut de Physique Th\'eorique, Orme des Merisiers, CEA-Saclay, 91191 Gif-sur-Yvette cedex, France}
\affiliation[c]{Institute for Particle Physics Phenomenology, University of Durham,\\
Durham, DH1 3LE, United Kingdom}

\emailAdd{abreu.samuel@ed.ac.uk}
\emailAdd{ruth.britto@cea.fr}
\emailAdd{claude.duhr@durham.ac.uk}
\emailAdd{einan.gardi@ed.ac.uk}

\abstract{We develop techniques for computing and analyzing multiple unitarity cuts of Feynman integrals, and reconstructing the integral from these cuts.
We study the relations among unitarity cuts of a Feynman integral computed via diagrammatic cutting rules, the discontinuity across the corresponding branch cut, and the coproduct of the integral.
For single unitarity cuts, these relations are familiar. Here we show that they can be generalized to sequences of unitarity cuts in different channels.
 Using concrete one- and two-loop scalar integral examples we demonstrate that it is possible to reconstruct a Feynman integral from either single or double unitarity cuts. Our results offer insight into the analytic structure of Feynman integrals as well as a new approach to computing them.
}


\begin{document}
\maketitle

\section{Introduction}
\label{sec:intro}

The precise determination of physical observables in quantum field theory involves computing multiloop Feynman integrals.  The difficulty of these integrals has led to their extensive study and the development of various specialized integration techniques.  

One approach to computing Feynman integrals has been to analyze the discontinuities across their branch cuts.  Like the integrals themselves, their discontinuities can be computed by diagrammatic rules \cite{Cutkosky:1960sp,tHooft:1973pz,Veltman:1994wz}. According to these rules, a given cut diagram is separated into two parts, with the intermediate particles at the interface of the two parts restricted to their mass shells, resulting in a cut Feynman integral.  This on-shell restriction can simplify the integration, and its result, considerably. The discontinuity with respect to a particular Mandelstam invariant $s_i$ is then recovered by the \emph{unitarity cut}, which is the sum over all diagrams that are cut in this channel.
The original uncut integral may then be reconstructed directly from one of its discontinuities by a dispersion relation \cite{Landau:1959fi,Cutkosky:1960sp,tHooft:1973pz,Remiddi:1981hn,Veltman:1994wz}.  Alternatively, modern unitarity methods \cite{Bern:1994zx,Bern:1994cg,Britto:2004nc,Britto:2005ha,Buchbinder:2005wp,Forde:2007mi,Britto:2007tt,Mastrolia:2009dr,Kosower:2011ty} make use of discontinuities to constrain an integral through its expansion in a basis of Feynman integrals.
The goal of this paper is to extend the study of discontinuities of Feynman integrals and their relation with cut diagrams, in the light of modern mathematical tools.

A large class of Feynman integrals can be expressed in terms of transcendental functions called multiple polylogarithms, which are defined by certain iterated integrals and include classical polylogarithms as a special case.  
Multiple polylogarithms, and iterated integrals in general, carry a lot of unexpected algebraic structure. In particular, they form a Hopf algebra~\cite{Goncharov:2005sla,GoncharovMixedTate}, which is a natural tool to capture discontinuities.  By now, there is considerable evidence that the coproduct (and the symbol) of a Feynman integral of transcendental weight~$n$, with massless propagators, satisfies a condition known as the first entry condition~\cite{Gaiotto:2011dt}:  the terms in the coproduct of transcendental weight $(1,n-1)$ can be written in the form 
\beq\label{eq:intro}
\sum_{i}\log(-s_i) \otimes f_{s_i}\,,
\eeq
where the sum ranges over all Mandelstam invariants $s_i$, and $f_{s_i}$ is the discontinuity of the integral with respect to the variable $s_i$. The deeper structure of the coproduct contains information about sequential discontinuities.

While it is known that the function $f_{s_i}$ is related to the unitarity cut on the corresponding channel~\cite{Cutkosky:1960sp,tHooft:1973pz,Veltman:1994wz}, here we generalize this relation and show that  \emph{multiple unitarity cuts} correspond to certain sequential discontinuities, which are in turn related to entries of the coproduct.  
We find that cuts and discontinuities taken in physical channels  corresponding to Mandelstam invariants are encoded into specific terms in the coproduct characterized by a set of simple logarithms that have the same set of discontinuities. It is noteworthy, however, that the arguments of these logarithms are in general not simple Mandelstam invariants, but they are related to the latter through algebraic relations specific to each Feynman integral. In this way we obtain a correspondence among three a priori unrelated concepts in the case of Feynman integrals: the computation of cut integrals, the mathematical operation of computing the discontinuity across a branch cut and the coproduct of the integral.

The aforementioned correspondence among cuts, discontinuities, and coproducts are the main result of this paper. In order to formulate it precisely, however, we need to define all the operations rigorously. 
We begin by extending the diagrammatic cutting rules of refs.~\cite{tHooft:1973pz,Veltman:1994wz}, which have only been formulated for single unitarity cuts so far, to allow for sequential unitarity cuts in multiple channels. 
Similarly to the case of a single unitarity cut, we restrict the computation to real kinematics. Beyond single unitarity cuts, the results depend crucially on the phase-space boundaries imposed by the kinematic region where each cut diagram is computed, and not only on the set of cut propagators. We conjecture relations between sequential discontinuities, entries of the coproduct and multiple unitarity cuts. To support these conjectures, we analyze them in the context of several non-trivial examples at one loop,  as well as the three-point ladder at two loops.

Because the techniques for computing cut diagrams are less well developed compared to standard Feynman diagrams, we proceed to establish such techniques.   Working in real kinematics allows us to use explicit real phase-space parametrizations. Furthermore, we show that at the multi-loop level one can efficiently use lower-order information by identifying subdiagrams whose external legs are 
cut propagators of the original integral. 
The calculations are checked in several different ways. First we verify that
the unitarity cut in the $s_i$ channel does indeed correspond to the expected function
 $f_{s_i}$ in eq.~\eqref{eq:intro}. Further checks are provided by the nontrivial cancellation of infrared singularities.
Indeed, even if the original Feynman integral is finite in $D=4$ dimensions, it is often necessary to regularize the corresponding cut integrals; we use dimensional regularization. Although individual cut diagrams can be infrared divergent, their sum is finite, through a mechanism similar to the cancellation of infrared divergences in a total cross section. Having established the techniques, we apply them to the calculation of multiple cuts in several one- and two-loop examples. These in turn are used to verify the relations discussed above between multiple unitarity cuts, sequential discontinuities and the coproduct.

Finally, we discuss the possibility to reconstruct the full uncut integral from the knowledge of its cuts, either through dispersion relations or through algebraic manipulations based on the Hopf algebra structure. Using the three-mass triangle and the two-loop three-point ladder as examples, we show that single and double dispersive integrals can be recast into the form of iterated integrals which greatly simplifies their evaluation. Alternatively, we demonstrate how to reconstruct first the symbol and then the full function from the knowledge of its (multiple) cuts, by means of simple algebraic manipulations on the symbol tensor. We thus see the computation of multiple unitarity cuts as a promising approach to computing Feynman integrals in the future.

The paper is organized as follows.
In section 2, we give a brief review of multiple polylogarithms and their Hopf algebra, and we discuss the class of \emph{pure transcendental} functions that we expect to be able to analyze.

In section 3, we present definitions of the three types of discontinuities that we consider:
\emph{Disc} is the difference in value as a function crosses its branch cut; \emph{Cut} is the value obtained by cutting diagrams into parts; and $\delta$ is a function identified algebraically inside the coproduct.  Each of these discontinuities is defined not just for a single cut, but for sequences of unitarity cuts in different Mandelstam invariants or related variables.   We conjecture a precise relation between Cut and Disc, and one between Disc and $\delta$.  By combining the two relations, we claim that diagrammatic cuts correspond to functions within the coproduct. We close this section by presenting several consequences of the correspondence in the case of ladder-type three-point functions.

In section 4, we give examples of our relations at one-loop and present our technique for evaluating cut integrals.  We discuss in detail the examples of the {three-mass triangle}, the {four-mass box} and the {two-mass-hard box}, and we show that the conjectured relations holds for these examples.

Sections 5 and 6 contain the main example of this paper, namely the two-loop three-point ladder integral with massless propagators.  We discuss techniques used in multi-loop cut calculations. In section 5, we compute single unitarity cuts and show that the relations conjectured in section 3 are satisfied.  In section 6, we compute sequences of two unitarity cuts, and we show that the conjectured relations still hold.

In section 7, we discuss the reconstruction of Feynman integrals from their cuts, based on the examples of the one-loop triangle and the ladder diagram. We show how dispersion relations can be recast into simple iterated integrals which allows to easily evaluate these integrals from the knowledge of their cuts. We then show how their symbols can be reconstructed from even limited knowledge of their cuts by using the integrability condition, and how to obtain the full functions from the knowledge of their symbols.  In section 8, we close with discussion of outstanding issues and suggestions for future study.  Appendix A summarizes our key conventions for evaluating Feynman diagrams and cut diagrams. Appendix B collects results of one-loop diagrams, cut and uncut, that are used throughout the paper. In appendix C we give explicit results for single unitarity cuts of the two-loop ladder. Finally, in appendix D we summarize the calculation for two sets of double cuts of the two-loop ladder, and give explicit expressions for their result.

\section{The Hopf algebra of multiple polylogarithms}
\label{sec:Hopf}

Feynman integrals in dimensional regularization usually evaluate to transcendental functions whose branch cuts are related to the physical discontinuities of $S$-matrix elements. Although it is known that generic Feynman integrals can involve elliptic functions~\cite{Caffo:1998du,MullerStach:2011ru,CaronHuot:2012ab,Adams:2013nia,Bloch:2013tra,Remiddi:2013joa}, large classes of Feynman integrals can be expressed through the classical logarithm and polylogarithm functions,
\beq
\log z = \int_1^z\frac{dt}{t} {\rm~~and~~} \textrm{Li}_n(z) = \int_0^z\frac{dt}{t}\,\textrm{Li}_{n-1}(t) {\rm~~with~~} \textrm{Li}_1(z) = -\log(1-z)\,,
\eeq
and generalizations thereof (see, e.g., refs.~\cite{Remiddi:1999ew,Gehrmann:2000zt,Kalmykov:2000qe,Kalmykov:2007dk,Bonciani:2010ms,Ablinger:2011te,Ablinger:2013cf}, and references therein). In the following we will concentrate exclusively on integrals that can be expressed entirely through polylogarithmic functions. Of special interest in this context are the so-called \emph{multiple polylogarithms}, and in the rest of this section we will review some of their mathematical properties.

\subsection{Multiple polylogarithms}
Multiple polylogarithms are defined by the iterated integral~\cite{Goncharov:1998kja,GoncharovMixedTate}
 \beq\label{eq:Mult_PolyLog_def}
 G(a_1,\ldots,a_n;z)=\,\int_0^z\,{d t\over t-a_1}\,G(a_2,\ldots,a_n;t)\,,\\
\eeq
with $a_i, z\in \mathbb{C}$. In the special case where all the $a_i$'s are zero, we define, using the obvious vector notation $\vec a_n=(\underbrace{a,\dots,a}_{n})$,
\beq
G(\vec 0_n;z) = {1\over n!}\,\ln^n z\,.
\eeq
The number $n$ of integrations in eq.~\eqref{eq:Mult_PolyLog_def}, or equivalently the number of $a_i$'s, is called the \emph{weight} of the multiple polylogarithm.
In the following we denote by $\overline\cH$ the $\mathbb{Q}$-vector space spanned by all multiple polylogarithms.
In addition, $\overline\cH$ can be turned into an algebra. Indeed, iterated integrals form a \emph{shuffle algebra}, 
\beq
G(\vec a_1;z)\,G(\vec a_2;z) = \sum_{\vec a\,\in\,\vec a_1\sh \vec a_2}G(\vec a;z)\,,
\eeq
where $\vec a_1\sh \vec a_2$ denotes the set of all shuffles of $\vec a_1$ and $\vec a_2$, i.e., the set of all permutations of their union that preserve the relative orderings inside $\vec a_1$ and $\vec a_2$. It is obvious that the shuffle product preserves the weight, and hence the product of two multiple polylogarithms of weight $n_1$ and $n_2$ is a linear combination of multiple polylogarithms of weight $n_1+n_2$. We can formalize this statement by saying that the algebra of multiple polylogarithms is \emph{graded} by the weight,
\beq
\overline\cH = \bigoplus_{n=0}^\infty\overline\cH_n{\rm~~with~~} \overline\cH_{n_1}\cdot\overline\cH_{n_2}\subset\overline\cH_{n_1+n_2}\,,
\eeq
where $\overline\cH_n$ is the $\mathbb{Q}$-vector space spanned by all multiple polylogarithms of weight $n$, and we define $\overline\cH_0=\mathbb{Q}$.

Multiple polylogarithms can be endowed with more algebraic structures. If we look at the quotient space $\cH = \overline\cH/(\pi\,\overline\cH)$ (the algebra $\overline\cH$ modulo $\pi$), then $\cH$ is a Hopf algebra~\cite{Goncharov:2005sla,GoncharovMixedTate}. In particular, $\cH$ can be equipped with a \emph{coproduct} $\Delta:\cH\to\cH\otimes\cH$, which is coassociative,
\beq
(\textrm{id}\otimes\Delta)\,\Delta = (\Delta\otimes\textrm{id})\,\Delta\,,
\eeq
respects the multiplication,
\beq
\Delta(a\cdot b) = \Delta(a)\cdot\Delta(b)\,,
\eeq
and respects the weight,
\beq
\cH_n  \stackrel{\Delta}{\longrightarrow} \bigoplus_{k=0}^n\cH_k\otimes\cH_{n-k}\,.
\eeq
The coproduct of the ordinary logarithm and the classical polylogarithms are 
\beq
\Delta(\log z) = 1\otimes\log z+\log z\otimes 1 {\rm~~and~~}
\Delta(\textrm{Li}_n(z)) = 1\otimes\textrm{Li}_n(z) + \sum_{k=0}^{n-1}\textrm{Li}_{n-k}(z)\otimes\frac{\log^kz}{k!}\,.
\eeq
For the definition of the coproduct of general multiple polylogarithms we refer to refs.~\cite{Goncharov:2005sla,GoncharovMixedTate}. 

The coassociativity of the coproduct implies that it can be iterated in a unique way. If $(n_1,\ldots,n_k)$ is a partition of $n$, we define
\beq
\Delta_{n_1,\ldots,n_k} : \cH_{n}\to\cH_{n_1}\otimes\ldots\otimes\cH_{n_k}\,.
\eeq
Note that the maximal iteration of the coproduct, corresponding to the partition $(1,\ldots,1)$, agrees with the symbol of a transcendental function $F$~\cite{ChenSymbol,Goncharov:2010jf,Goncharov:2009tja,Brown:2009qja,Duhr:2011zq}
\beq
\cS(F) \equiv \Delta_{1,\ldots,1}(F)\in\cH_1\otimes\ldots\otimes\cH_1\,.
\eeq
Since every element of $\cH_1$ is a logarithm, the `$\log$' sign is usually dropped when talking about the symbol of a function.
Note that not every element in $\cH_1\otimes\ldots\otimes\cH_1$ corresponds to the symbol of a function in $\cH$. Instead, one can show that if we take an element
\beq
s = \sum_{i_1,\ldots,i_n}c_{i_1,\ldots,i_n}\,\log x_{i_1}\otimes\ldots\otimes \log x_{i_n} \in\cH_1\otimes\ldots\otimes\cH_1\,,
\eeq
then there is a function $F\in\cH_n$ such that $\cS(F) = s$ if and only if $s$ satisfies the \emph{integrability condition}
\beq\label{eq:integrability}
\sum_{i_1,\ldots,i_n}c_{i_1,\ldots,i_n}\,d\log x_{i_k}\wedge d\log x_{i_{k+1}}\,\log x_{i_1}\otimes\ldots\otimes\log x_{k-1}\otimes\log x_{k+2}\otimes\ldots\otimes \log x_{i_n} = 0\,,
\eeq
where $\wedge$ denotes the usual wedge product on differential forms.

While $\cH$ is a Hopf algebra, we are practically interested in the full algebra $\overline\cH$ where we have kept all factors of $\pi$. Based on similar ideas in the context of motivic multiple zeta values~\cite{Brown:2011ik}, it was argued in ref.~\cite{Duhr:2012fh} that we can reintroduce $\pi$ into the construction by considering the trivial comodule $\overline\cH=\mathbb{Q}[i\pi]\otimes\cH$. The coproduct is then lifted to a comodule map $\Delta:\overline\cH\to \overline\cH\otimes \cH$ which acts on $i\pi$ according to $\Delta(i\pi)=i\pi\otimes1$. In the following we will, by abuse of language, refer to the comodule as the Hopf algebra $\overline\cH$ of multiple polylogarithms.  

Let us conclude this review of multiple polylogarithms and their Hopf algebra structure by discussing how differentiation and taking discontinuities (see section~\ref{sec:disccutdelta} for precise definition of discontinuity in this work) interact with the coproduct. In ref.~\cite{Duhr:2012fh} it was argued that the following identities hold:
\beq\label{eq:disc_coproduct}
\Delta\,\frac{\partial}{\partial z} = \left(\textrm{id}\otimes\frac{\partial}{\partial z}\right)\,\Delta
{\rm~~and~~}
\Delta\,\textrm{Disc} = (\textrm{Disc}\otimes\textrm{id})\,\Delta\,.
\eeq
In other words, differentiation only acts in the last entry of the coproduct, while taking discontinuities only acts in the first entry.
Since the discontinuities are proportional to $i\pi$, which appears only in isolation in the first entry of the coproduct, it follows from the last relation that for an element $f_n$ of weight $n$ in $\overline\cH$,
\bea
\Disc f_n \cong \mu\left[(\Disc \otimes \textrm{id})\,\Delta_{1,n-1} f_n\right]\,
\label{eq:disc-with-mu}
\eea
where $\mu:\overline\cH\otimes\overline\cH\to\overline\cH$ denotes the multiplication in $\overline\cH$, i.e.\ we simply multiply the two factors in the coproduct, and $\cong$ denotes equivalence modulo $\pi^2$, because the weight $(n-1)$ part of the coproduct in the right-hand side is only defined modulo $\pi$.

\subsection{Pure Feynman integrals}
In the rest of this paper we will be concerned with connected Feynman integrals in dimensional regularization. Close to $D=4-2\eps$ dimensions, an $L$-loop Feynman integral $F^{(L)}$ then defines a Laurent series,
\beq
F^{(L)}(\eps) = \sum_{k=-2L}^\infty F_k^{(L)}\,\eps^k\,.
\label{laurentExp}
\eeq
In the following we will concentrate on situations where the coefficients of the Laurent series can be written exclusively in terms of multiple polylogarithms and rational functions, and a well-known conjecture states that the weight of the transcendental functions (and numbers) that enter the coefficient $F_k^{(L)}$ of an $L$ loop integral is less than or equal to $2L-k$. If all the polylogarithms in $F_k^{(L)}$ have the same weight, the integral is said to have \emph{uniform (transcendental) weight}. In addition, we say that an integral is \emph{pure} if the coefficients $F_k^{(L)}$ do not contain rational or algebraic functions of the external kinematical variables.

It is clear that pure integrals are the natural objects to study when trying to link Hopf algebraic ideas for multiple polylogarithms to Feynman integrals. For this reason we will only be concerned with pure integrals in the rest of this paper. 
However, the question naturally arises of how restrictive this assumption is. 
In ref.~\cite{ArkaniHamed:2010gh} it was noted that if a Feynman integral has unit leading singularity~\cite{Cachazo:2008vp}, i.e., if all the residues of the integrand, obtained by integrating over compact complex contours around the poles of the integrand, are equal to one, then the corresponding integral is pure. Furthermore, it is well known that Feynman integrals satisfy integration-by-parts identities~\cite{Chetyrkin:1981qh}, which, loosely speaking, allow one to express a loop integral with a given propagator structure in terms of a minimal set of so-called master integrals. In ref.~\cite{Henn:2013pwa} it was conjectured that it is always possible to choose the master integrals to be pure integrals, and the conjecture was shown to hold in several nontrivial cases~\cite{Henn:2013tua,Henn:2013woa,Henn:2013nsa,Argeri:2014qva}. Hence, if this conjecture is true, it should always be possible to restrict the computation of the master integrals to pure integrals, which justifies the restriction to this particular class of integrals.

Another restriction on the class of Feynman integrals considered in this paper is that we consider all propagators to be massless. In this case, it is known that the branch points of the integral, seen as a function of the invariants $s_{ij} = (p_i + p_j)^2$, where $p_i$ are the external momenta (which can be massive or massless), are the points where one of the invariants is zero or infinite~\cite{Landau:1959fi}. It follows then from eq.~\eqref{eq:disc_coproduct} that the first entry of the coproduct of a Feynman integral can only have discontinuities in these precise locations. In particular, this implies the so-called \emph{first entry condition}, i.e., the statement that the first entries of the symbol of a Feynman integral with massless propagators can only be (logarithms of) Mandelstam invariants~\cite{Gaiotto:2011dt}. This observation, combined with the fact that Feynman integrals can be given a dispersive representation, provides the motivation for the rest of this paper, namely the study of the discontinuities of a pure Feynman integral with massless propagators through the lens of the Hopf algebraic language reviewed at the beginning of this section.

\subsection{The symbol alphabet}
\label{sec:symbolAlphabet}

The most natural kinematic variables for a given integral might be more complicated functions of the momentum invariants. 
Indeed, it is known that the Laurent expansion coefficients in \refE{laurentExp} are \emph{periods} (defined, loosely speaking, as integrals of rational functions), which implies that the arguments of the polylogarithmic functions are expected to be algebraic functions of the external scales \cite{Bogner:2007mn}. In practice it is more convenient to find a parametrization of the kinematics such that the arguments of all polylogarithmic functions are rational. More precisely, if we have a Feynman integral depending on $n$ independent scales $s_i$ (e.g. Mandelstam invariants), we want to find $n-1$ independent variables $z_i$ such that
\beq\label{eq:s_to_xi}
s_i/s_n = f_i(z_1,\ldots,z_{n-1})\,,
\eeq
where the $f_i$ are rational functions such that all the arguments of the polylogarithms are rational functions of the $z_i$ variables. While no general algorithm is known that allows one to find the parametrization~\eqref{eq:s_to_xi}, such a parametrization exists for all known examples of Feynman integrals that can be expressed in terms of multiple polylogarithms. We will therefore from now on assume that such a parametrization exists. The inverse relations to (\ref{eq:s_to_xi}), expressing $z_i$ in terms of the Mandelstam invariants $\{s_j\}$, are algebraic functions, often involving square roots of polynomials of the invariants $\{s_j\}$. Concrete examples will be given in Section~\ref{sec:OneLoop}.

If a parametrization of the type~\eqref{eq:s_to_xi} has been determined for a given Feynman integral, it is easy to see that the entries of the symbol of this integral will be rational functions of the $z_i$. Moreover, due to the additivity of the symbol, we can assume that the entries of the symbol are polynomials with integer coefficients\footnote{We allow the polynomials to be constants.} in the variables $z_i$, which without loss of generality we may assume to be irreducible over $\mathbb{Z}$. In other words, once a rational parametrization~\eqref{eq:s_to_xi} has been determined, we can assign to every Feynman integral a set $\cA\subset \mathbb{Z}[z_1,\ldots,z_{n-1}]$ of irreducible polynomials. In the following we call the set $\cA$ the \emph{symbol alphabet} of the integral, and its elements, which we generically denote by $x_i$, will be called the \emph{letters} of the alphabet. Some comments are in order: First, we note that the symbol alphabet $\cA$ is not unique, but it is tightly connected to the choice of the rational parametrization~\eqref{eq:s_to_xi}. A different choice for the rational functions $f_i$ may result in a different symbol alphabet $\cA$. Second, we emphasize that although the parametrization~\eqref{eq:s_to_xi} only involves the external scales, its form is in general dependent on the loop order and/or the order in the expansion in the dimensional regulator $\eps$ and the topology of the integral under consideration. Third, it is easy to see that once a symbol alphabet $\cA$ is fixed, the symbol of a polylogarithmic function of weight $k$ takes values in $\mathbb{Q}\otimes_{\mathbb{Z}}\mathbb{Z}[\cA]^{\otimes k}$, the $k$-fold tensor product (with rational coefficients) of the free abelian group of rational functions whose generators are the polynomials in the set $\cA$. Finally, we note that it is expected that the arguments of the polylogarithms take values in a subset of the free abelian group $\mathbb{Z}[\cA]$, and an explicit (conjectural) construction of this subset was presented in ref.~\cite{Duhr:2011zq}.

In practical applications it is often advantageous to know the symbol alphabet underlying a specific problem a priori. For example, if the alphabet is determined, it is possible to write ans\"atze for the symbols and/or the function spaces for Feynman integrals or amplitudes, which can then be fixed using additional physical information (e.g., behaviour in certain limits)~\cite{Dixon:2011pw,Dixon:2011nj,Drummond:2012bg,Chavez:2012kn,Dixon:2012yy,Drummond:2013nda,Dixon:2013eka,Dixon:2014voa}. Unfortunately, as already mentioned, no general algorithm to determine a rational parametrization~\eqref{eq:s_to_xi}, and thus the letters $x_i\in\cA$ is known. One possible way to determine the alphabet is to analyze the differential equations satisfied by Feynman integrals~\cite{Kotikov:1990kg,Kotikov:1991hm,Kotikov:1991pm,Gehrmann:1999as,Henn:2013pwa}, where the letters $x_i$ appear as the singularities of the differential equations.\footnote{We note, however, that also in that case a rational parametrization has to be determined by independent means.} In the rest of this paper we argue that another way of determining the letters $x_i$ consists in analyzing (iterated) unitarity cuts of Feynman integrals. Indeed, as we will argue in the next section, cut integrals are tightly connected to the entries in the coproduct (and hence the symbol) and the discontinuities of a Feynman integral, but they are sometimes easier to compute because the transcendental weight is reduced. The precise connection between (iterated) unitarity cuts, discontinuities and coproducts of Feynman integrals is the subject of the next sections.

\section{Three definitions of discontinuities}
\label{sec:disccutdelta}

In this section we present our definitions and conventions for the discontinuities of Feynman integrals with respect to external momentum invariants, also called \emph{cut channels}.  There are three operations giving systematically related results: a discontinuity across a branch cut of the Feynman integral, which we denote by Disc and define in section~\ref{sec:disc_def} below; unitarity cuts computed via Cutkosky rules and the diagrammatic rules of refs.~\cite{tHooft:1973pz,Veltman:1994wz}, which we extend here to multiple cuts and denote by Cut (section~\ref{sec:cut_def}); and a corresponding $\delta$ operation on the coproduct of the integral
(section~\ref{sec:delta_def}).
Discontinuities taken with respect to one invariant are familiar, but sequential discontinuities must be constructed with care in order to derive equivalent results from the three operations. 
In this section, we present these concepts in general terms and then close with a simple example.  More concrete illustrations, which further support the conjectured connections among the three operations, appear in the following sections. 

Let $F$ be a pure Feynman integral,
and let $s$ and $s_i$ denote Mandelstam invariants (squared sums of external momenta), labeled by $i$ in the case where we consider several of them. 
These invariants come with an $i\varepsilon$ prescription inherited from the Feynman rules for propagators.  In the case of planar integrals, such as the examples we will consider in the following sections, the integral is originally calculated in the Euclidean region, where all Mandelstam invariants of consecutive legs are negative, so that branch cuts are avoided.
 It may then be analytically continued to any other kinematic region by the prescription $s_i \to s_i + i \varepsilon$.  

\subsection{Disc: Discontinuity across branch cuts\label{sec:disc_def}}

The operator $\Disc_s F$ gives the direct value of the discontinuity of $F$ as the variable $s$ crosses the real axis.  If there is no branch cut in the kinematic region being considered, or if $F$ does not depend on $s$, then the value is zero.  Concretely,
\bea
\Disc_s \left[F(s\pm i 0)\right]= \lim_{\varepsilon \to 0}\left[F(s\pm i \varepsilon)-F(s \mp i \varepsilon)\right],
\label{eq:def-disc}
\eea
where the $i\varepsilon$ prescription must be inserted correctly in order to obtain the appropriate sign of the discontinuity.  
For example, $\Disc_s \log(s+i0) = 2\pi i\,\theta(-s)$.
We will discuss the sign in more detail at the end of this section, when we relate $\Disc$ to the other definitions of discontinuities.

The sequential discontinuity operator $\Disc_{s_1,\ldots,s_k}$ is defined recursively:
\bea
\Disc_{s_1,\ldots,s_k} F &\equiv& \Disc_{s_k} \left( \Disc_{s_1,\ldots,s_{k-1}} F \right).
\label{eq:disc-seq}
\eea

Note that $\Disc$ may be computed in any kinematic region after careful analytic continuation, but if it is to be related to the value of  $\Cut$, it should be computed in the same region as the corresponding multiple cut.
In particular, sequential $\Disc$ will be computed in different regions at each step. We will sometimes write
\bea
\Disc_{s_1,\ldots,s_k;R} F
\eea 
to make explicit the region $R$ in which $\Disc_{s_1,\ldots,s_k}$ is to be computed, after having analytically continued $F$ to this same region.

\subsection{Cut:  Cut integral \label{sec:cut_def}}

The operator $\Cut_s$ gives the sum of \emph{cut} Feynman integrals, in which some propagators in the integrand of $F$ are replaced by Dirac delta functions.  These propagators themselves are called \emph{cut propagators}.  The sum is taken over all combinations of cut propagators that separate the diagram into two  parts, in which the momentum flowing through the cut propagators from one part to the other corresponds to the Mandelstam invariant $s$.   Furthermore, each cut is associated with a consistent direction of energy flow between the two parts of the diagram, in each of the cut propagators.
In this work, we follow the conventions for cutting rules established in refs.~\cite{tHooft:1973pz,Veltman:1994wz}, and extend them for sequential cuts.

\paragraph{First cut.} Let us first review the cutting rules of refs.~\cite{tHooft:1973pz,Veltman:1994wz}.  We start by enumerating all possible partitions of the vertices of a Feynman diagram into two sets, colored black ($b$) and white ($w$).  Each such colored diagram is then evaluated according to the following rules:
\begin{itemize}
\item Black vertices, and propagators joining two black vertices, are computed according to the usual Feynman rules.
\item White vertices, and propagators joining two white vertices, are complex-conjugated with respect to the usual Feynman rules.  
\item Propagators joining a black and a white vertex are \emph{cut} with an on-shell delta function, a factor of $2 \pi$ to capture the complex residue correctly, {\it and} a theta function restricting energy to flow in the direction $b \to w$.
\end{itemize}
For a massless scalar theory, the rules for the first cut may be depicted as:
\begin{align}
\raisebox{-0.1mm}{\includegraphics[keepaspectratio=true, height=0.2cm]{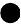}}=i
\qquad \qquad \qquad
\raisebox{-0.1mm}{\includegraphics[keepaspectratio=true, height=0.2cm]{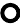}}=-i
\end{align}
\begin{align}
\raisebox{-0.1mm}{\includegraphics[keepaspectratio=true, height=0.55cm]{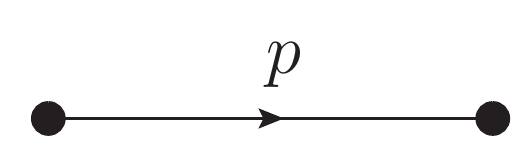}}=\frac{i}{p^2+i\varepsilon}
\qquad \qquad
\raisebox{-0.1mm}{\includegraphics[keepaspectratio=true, height=0.55cm]{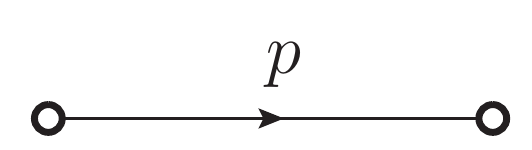}}=\frac{-i}{p^2-i\varepsilon}
\end{align}
\begin{align}
\raisebox{-6.7mm}{\includegraphics[keepaspectratio=true, height=1.6cm]{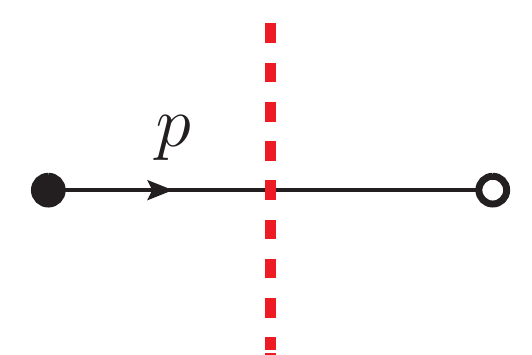}}
= 2\pi\, \delta\left(p^2\right)\theta\left(p_0\right)
\end{align}
The dashed line indicating a cut propagator is given for reference and does not add any further information. 
We write $\Cut_s$ to denote the sum of all diagrams belonging to the same momentum channel, i.e., in each of these diagrams, if $p$ is the sum of all momenta through cut propagators flowing in the direction from black to white, then $p^2=s$.  Note that cut diagrams in a given momentum channel will appear in pairs that are black/white color reversals --- but of each pair, only one of the two can be consistent with the energies of the fixed external momenta, giving a potentially nonzero result.

We note that $\Cut_s F(x_1,\ldots,x_k)$ is a function of the variables $x_i$ mentioned above, which we recall can be complicated algebraic functions of the Mandelstam invariants. Finding the correct $x_i$ in which to express a given Feynman integral is a nontrivial problem. Since cut Feynman integrals depend on the same variables as uncut diagrams but are simpler functions, the $x_i$ can be more easily identified by computing cuts, as illustrated in \refE{3mass_triangle_cut_result}.

\paragraph{Sequential cuts.} 
The diagrammatic cutting rules of refs.~\cite{tHooft:1973pz,Veltman:1994wz} reviewed so far allow us to consistently define cut integrals corresponding to a single unitarity cut. The aim of this paper is however the study of sequences of unitarity cuts. The cutting rules of refs.~\cite{tHooft:1973pz,Veltman:1994wz} are insufficient in that case, as they only allow us to partition a diagram in two parts, corresponding to connected areas of black and white vertices. We now present an extension of the cutting rules to sequences of unitarity cuts on different channels. 
At this stage, we only state the rules, whose consistency is then backed up by the results we find in the remainder of this paper.

\begin{figure}
\begin{center}
\includegraphics[keepaspectratio=true, height=9cm]{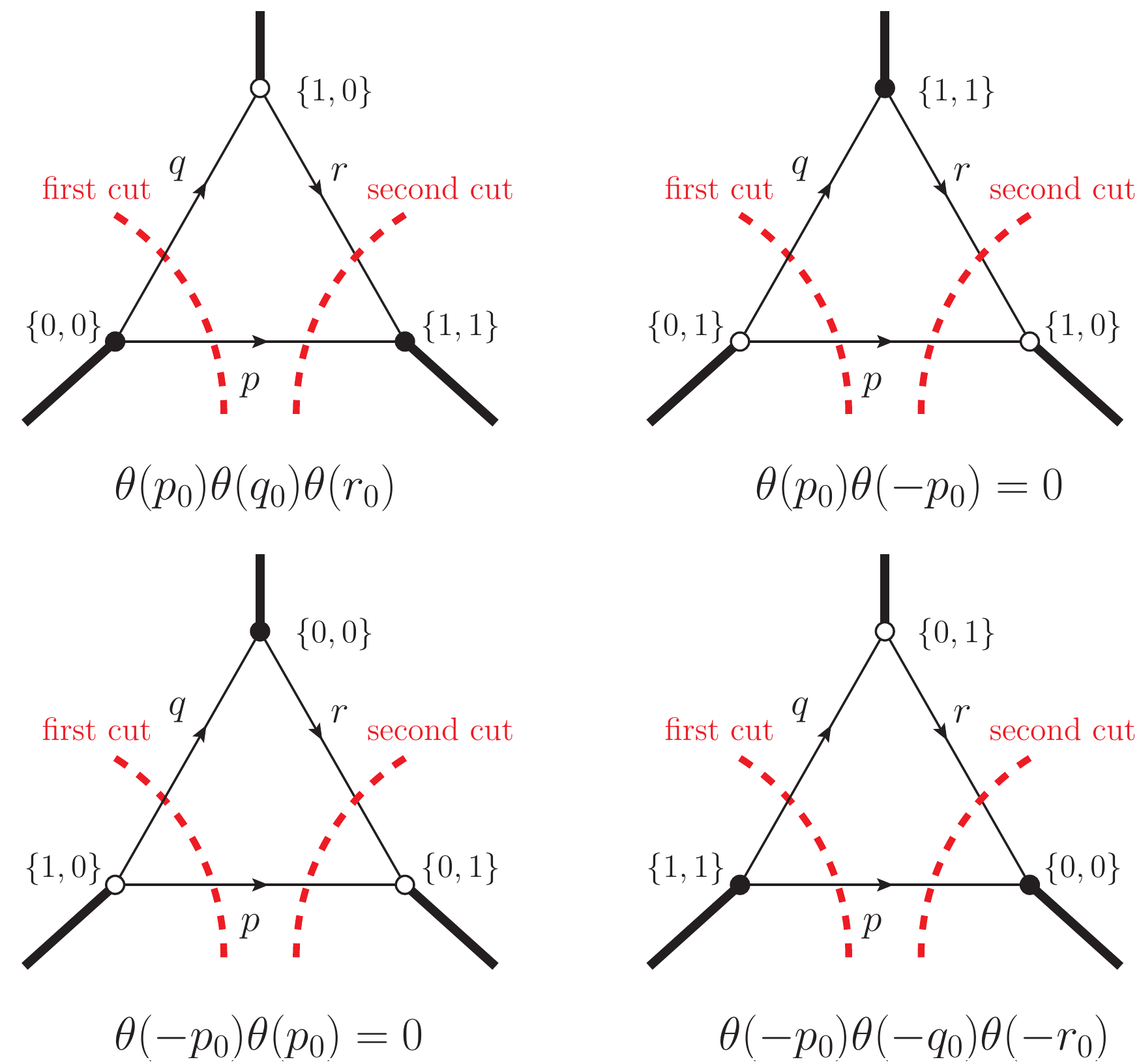}
\end{center}
\caption{Sequential cuts of a triangle diagram, whose vertices $v$ are labelled by all possible color sequences $(c_1(v),c_2(v))$ encoding the cuts.  Energy flows from $0$ to $1$ for each cut, giving the restrictions listed below each diagram. }
\label{fig:cutcolors1}
\end{figure}

In a sequence of diagrammatic cuts, energy-flow conditions are overlaid, and complex conjugation of vertices and propagators is applied sequentially.  We continue to use black and white vertex coloring to show complex conjugation.  We illustrate an example in \refF{fig:cutcolors1}, which will be discussed below.

Consider a multiple-channel cut, $\Cut_{s_1, \ldots, s_k}I$.  It is represented by the sum of all diagrams with a color-partition of vertices for each of the cut invariants $s_i=p_i^2$.  Assign a sequence of colors $(c_1(v),\ldots,c_k(v))$ to each vertex $v$  of the diagram, where each $c_i$ takes the value 0 or 1.  For a given $i$, the colors $c_i$ partition the vertices into two sets, such that the total momentum flowing from vertices labeled 0 to vertices labeled 1 is equal to $p_i$.  A vertex $v$ is finally colored according to $c(v) \equiv \sum_{i=1}^k c_i(v)$ modulo 2, with black for $c(v)=0$ and white for $c(v)=1$.  The rules for evaluating a diagram are as follows.
\begin{itemize}
\item Black vertices are computed according to the usual Feynman rules; white vertices are computed according to complex-conjugated Feynman rules.
\item A propagator joining vertices $u$ and $v$ is uncut if $c_i(u)=c_i(v)$ for all $i$.  Then, if the vertices are black, i.e. $c(u)=c(v)=0$, then the propagator is computed according to the usual Feynman rules, and if the vertices are white, i.e. $c(u)=c(v)=1$, then the propagator is computed according to complex-conjugated Feynman rules.
\item A propagator joining vertices $u$ and $v$ is cut if $c_i(u) \neq c_i(v)$ for any $i$.  There is a theta function restricting the direction of energy flow from 0 to 1 for each $i$ for which $c_i(u) \neq c_i(v)$.  If different cuts impose conflicting energy flows, then the product of the theta functions is zero and the diagram gives no contribution.
\item 
We exclude crossed cuts, as they do not correspond to the types of discontinuities captured by Disc and $\delta$.\footnote{A similar restriction was proposed in refs.~\cite{steinmann1,steinmann2,stapp1}.}  In other words, each new cut must be located within a region of identically-colored vertices with respect to the previous cuts.  
In terms of the color labels, this is equivalent to requiring that for any two values of $i,j$, exactly three of the four possible distinct color sequences $(c_i(v),c_j(v))$ are present in the diagram.
\item
Likewise, we exclude sequential cuts in which the channels are not all distinct. This restriction is made only because we have not found a general relation between such cuts and Disc or $\delta$.  In principle, there is no obstacle to computing these cut diagrams.
\item
We restrict ourselves to the use of real kinematics, both for internal and external momenta. This implies, in particular, that diagrams with on-shell massless three-point vertices must vanish in dimensional regularization. The consistency of this choice will be verified in the examples considered in subsequent sections.

\end{itemize}
For massless scalar theory, the rules for sequential cut diagrams may then be depicted thus:
\begin{align}
\raisebox{-0.1mm}{\includegraphics[keepaspectratio=true, height=0.2cm]{./diagrams/blackVertex.pdf}}=i
\qquad \qquad \qquad
\raisebox{-0.1mm}{\includegraphics[keepaspectratio=true, height=0.2cm]{./diagrams/whiteVertex.pdf}}=-i
\end{align}
\begin{align}
\raisebox{-0.1mm}{\includegraphics[keepaspectratio=true, height=0.55cm]{./diagrams/propagator.pdf}}=\frac{i}{p^2+i\varepsilon}
\qquad \qquad
\raisebox{-0.1mm}{\includegraphics[keepaspectratio=true, height=0.55cm]{./diagrams/whitePropagator.pdf}}=\frac{-i}{p^2-i\varepsilon}
\end{align}
\begin{align}\nonumber
\raisebox{-4.7mm}{\includegraphics[keepaspectratio=true, height=1.2cm]{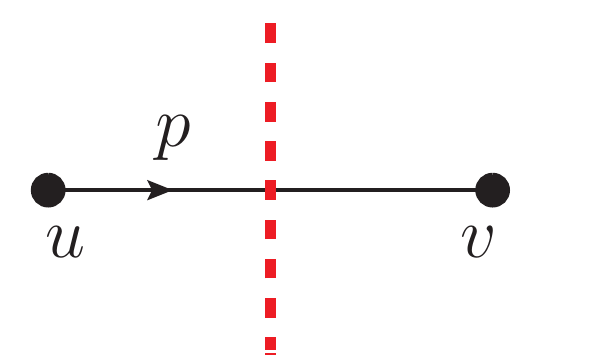}}
= 
\raisebox{-4.7mm}{\includegraphics[keepaspectratio=true, height=1.2cm]{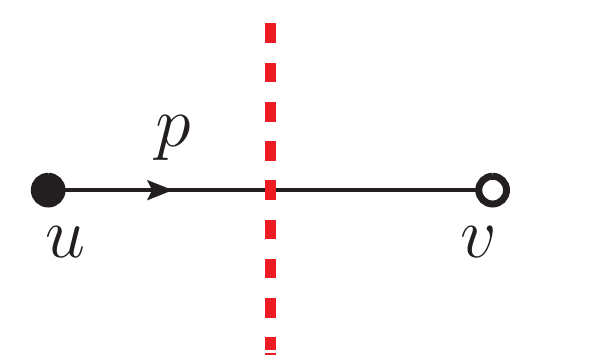}}
=&
\raisebox{-4.7mm}{\includegraphics[keepaspectratio=true, height=1.2cm]{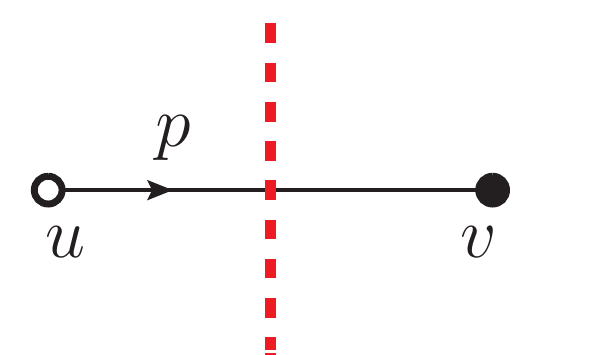}}
= 
\raisebox{-4.7mm}{\includegraphics[keepaspectratio=true, height=1.2cm]{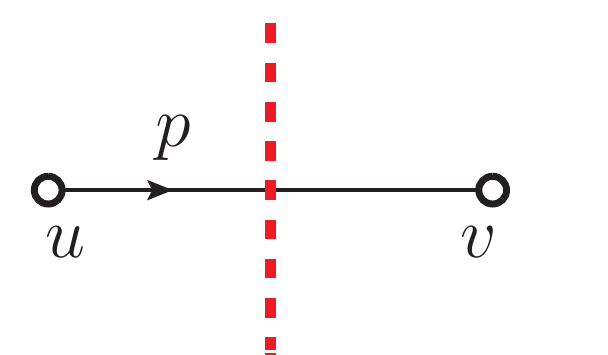}}\\
=&
2\pi\, \delta\left(p^2\right)
\prod_{i:c_i(u)\neq c_i(v)}
\theta\left([c_i(v)-c_i(u)]p_0\right)
\end{align}

Let us make some comments about the diagrammatic cutting rules for multiple cuts we just introduced. First,
we note that these rules are consistent with the corresponding rules for single unitarity cuts presented at the beginning of this section.
Second, using these rules, it is clear that sequential cuts are independent of the order of cuts.
Indeed, none of our rules depends on the order in which the cuts are listed.
Finally, the dashed line 
is an incomplete shorthand merely indicating the location of the delta functions, but not specifying the direction of energy flow, for which one needs to refer to  the color indices.
Our diagrams might also include multiple cut lines on individual propagators, such as
\begin{align}
\raisebox{-6.7mm}{\includegraphics[keepaspectratio=true, height=1.4cm]{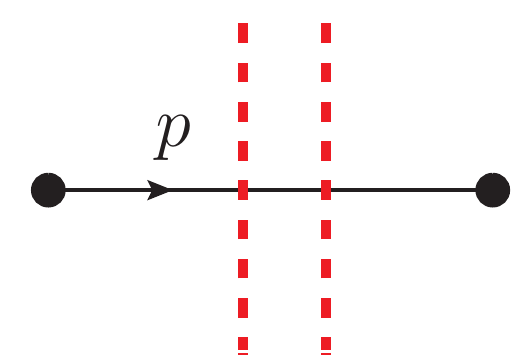}}.
\end{align}

We also introduce notation allowing us to consider individual diagrams contributing to a particular cut, and possibly restricted to a particular kinematic region.   When no region is specified, for the planar examples given in this paper, it is assumed that the cut invariants are taken to be positive while all other consecutive Mandelstam invariants are negative.  We write
\begin{equation}
\cut_{s,[e_1\cdots e_w],R}D
\end{equation}
to denote a diagram $D$ cut in the channel $s$, in which exactly the propagators $e_1\cdots e_w$ are cut, and computed in the kinematic region $R$. Rules of complex conjugation and energy flow will be apparent in the context of such a diagram.

\paragraph{Examples of sequential cuts.}  We briefly illustrate the diagrammatics of sequential cuts.  Consider taking two cuts of a triangle integral.  At one-loop order, a cut in a given channel is associated to a unique pair of propagators.  We list the four possible color partitions $\{c_1(v),\ldots,c_k(v)\}$ in \refF{fig:cutcolors1}.
The  first graph is evaluated according to the rules above, giving
\bean
e^{{\gamma_E} \epsilon}
\int \frac{d^{D}k}{\pi^{D/2}} i^2(-i)(2\pi)^3 \,\delta(p^2)\delta(q^2)\delta(r^2)\,\theta(p_0)\theta(q_0)\theta(r_0).
\eean
The second and third graphs evaluate to zero, since the color partitions give conflicting restrictions for the energy flow on the propagator labeled $p$.  The fourth graph is similar to the first, but with energy flow located on the support of $\theta(-p_0)\theta(-q_0)\theta(-r_0)$.  Just as for a single unitarity cut, in which only one of the two colorings is compatible with a given assigment of external momenta, there can be at most one nonzero diagram for a given topology of sequential cuts subject to fixed external momenta.  In the examples calculated in the following sections of this paper, we will thus omit writing the sequences of colors $(c_1(v),\ldots,c_k(v))$.  We may also omit writing the theta functions for energy flow in the cut integrals.

\begin{figure}
\begin{center}
\includegraphics[keepaspectratio=true, height=4cm]{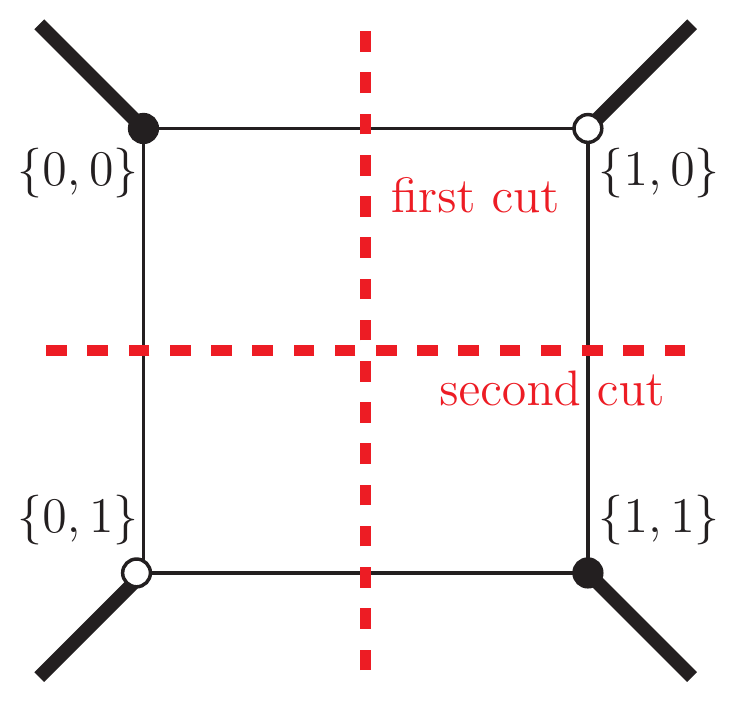}
\end{center}
\caption{An example of crossed cuts, which we do not allow.}
\label{fig:crossedcutsbox}
\end{figure}

We include an example of crossed cuts, which we do not allow, in \refF{fig:crossedcutsbox}.  Notice that there are four distinct color sequences in the diagram, while we only allow three for any given pair of cuts.

\subsection{$\delta$: Entries of the coproduct\label{sec:delta_def}}

If $F$ is of transcendental weight $n$ and has all its symbol entries drawn from the alphabet $\cA$, then we can write without loss of generality
\bea
\Delta_{\underbrace{\scriptstyle 1,1,\ldots,1}_{k\textrm{ times}},n-k}F &=&
\sum_{(x_{i_1},\ldots,x_{i_k})\in\cA^k} \log x_{i_1} \otimes \cdots \otimes \log x_{i_k} \otimes 
g_{x_{i_1},\ldots,x_{i_k}}\,,
\eea 
and we define
\bea\label{eq:deltaDef}
\delta_{x_{j_1},\ldots,x_{j_k}}F &\cong& 
\sum_{(x_{i_1},\ldots,x_{i_k})\in\cA^k}
\delta_{i_1j_1}\,\ldots\,\delta_{i_kj_k}\,
g_{x_{i_1},\ldots,x_{i_k}},
\eea
where the congruence symbol indicates that  $\delta_{x_{j_1},\ldots,x_{j_k}}F$ can be defined only modulo $\pi $.  If the integral contains overall numerical factors of $\pi$, they should be factored out before performing this operation.  

The definition of $\delta_{x_{j_1},\ldots,x_{j_k}}F$ is motivated by the relation~\refE{eq:disc-with-mu} between discontinuities and coproducts. In particular, if $\delta_x F \cong g_x$, then $\Disc_x F \cong \mu[(\Disc_x \otimes\, \textrm{id})(\log x \otimes g_x)] = \pm 2\pi i\, g_x$.  The sign is determined by the $i \varepsilon$ prescription for $x$ in $F$, and the precise form of the relation between the $\Disc$ and $\delta$ operations will be discussed in more detail in the following subsection.

\subsection{Relations among Disc, Cut, and $\delta$}

\paragraph{Cut diagrams and discontinuities.}
The rules for evaluating cut diagrams are designed to compute their discontinuities.  The fact that such a relation exists at all follows from the largest time equation.  For the first cut, the derivation may be found in refs.~\cite{tHooft:1973pz,Veltman:1994wz}.  The original relation is  
\bea
F+F^* = -\sum_s \Cut_s F,
\label{eq:cutting}
\eea
 where the sum runs over all momentum channels.   In terms of diagrams with colored vertices, the left-hand side is the all-black diagram plus the all-white diagram.  The right-hand side is -1 times the sum of all diagrams with mixed colors.
We can isolate a single momentum channel $s$ by 
analytic continuation into a kinematic region where among all the invariants, only $s$ is on its branch cut.  Specifically, for planar integrals such as the examples given in this paper, we take $s>0$ while all other invariants of consecutive momenta are negative.  There, 
the left-hand side of \refE{eq:cutting} can be recast\footnote{The apparent difference in relative sign between \refE{eq:def-disc} and \refE{eq:cutting} is due to an explicit overall factor of $i$ in every diagram, due to the Fourier transform from position to momentum space.  Note therefore that \refE{eq:def-disc} should \emph{not} be interpreted as the imaginary part of the function, and is in fact typically real-valued.
}
as $\Disc_s F$, while the right-hand side collapses to a single term:
\bea
\Disc_s F =-\Cut_s F.
\label{eq:oldcutting}
\eea

For sequential cuts, we claim that $\Cut_{s_1,\ldots,s_k} F$ captures discontinuities through the relation
\bea
 \Cut_{s_1,\ldots,s_k} F = (-1)^k \Disc_{s_1,\ldots,s_k}F.
\label{eq:cutequalsdisc}
\eea
We recall that no two of the invariants $s_1,\ldots,s_k$ should be identical, nor may any pair of them cross each other in the sense given in the cutting rules above.
Eq.~(\ref{eq:cutequalsdisc}) is one of the main results of this paper: it generalizes the well-established relation between single unitarity cuts and the discontinuity of a Feynman integral across the branch cut associated with a given Mandelstam invariant \cite{Cutkosky:1960sp,tHooft:1973pz,Veltman:1994wz}.

Eq.~(\ref{eq:cutequalsdisc}) holds {\em in a specific kinematic region}:
Let $R_j$ denote the kinematic region in which $F$ has branch cuts in the invariants  $s_1,\ldots,s_j$, but not in any of the other invariants on which $F$ depends.  
  The left-hand side of \refE{eq:cutequalsdisc} is evaluated in the region $R_k$.\footnote{We sometimes find it convenient to evaluate the cut integral in a different kinematic region and then perform an analytic continuation back to the region $R_k$.}  
 On the right-hand side, we proceed step by step according to the definition in \refE{eq:disc-seq}, and each $\Disc_{s_i}$ is evaluated after analytic continuation to 
 the region $R_i$. 

In the case of planar integrals, $R_j$ is the  region in which the  $s_i$ are positive for $i=1,\ldots,j$ while all other invariants are negative.  For integrals that are finite in integer dimensions, at least one invariant must remain negative in order to distinguish $R_j$ from the original Euclidean region $R_0$, in terms of the scales of \refE{eq:s_to_xi}.

While sequential cuts are independent of the order in which the channels are listed, the correspondences to $\Disc$ are derived in sequence, so that the right-hand side of \refE{eq:cutequalsdisc} takes a different form when channels on the left-hand side are permuted.  Thus, \refE{eq:cutequalsdisc} implies relations among the $\Disc_{s_1,\ldots,s_k} F$, which will in turn imply nontrivial relations among the $\delta_{x_1,\ldots,x_k} F$.

\paragraph{Coproduct and discontinuities.}

Recall from \refE{eq:disc-with-mu} that for an element $f$ of weight $n$ of the Hopf algebra,
\bea
\Disc f \cong \mu\left[(\Disc \otimes \textrm{id}) (\Delta_{1,n-1}f)\right]\,,
\label{temp0943}
\eea
To be precise, $f$ should not include overall factors of $\pi$. If it does, these are stripped out before performing the operation the operation on the right-hand side, and then reinstated.
It follows from this relation  that the discontinuity of any element of the Hopf algebra is captured by the operation $\delta$ as defined in \refE{eq:deltaDef}. 
To apply the relation, we must take great care with the sequential analytic continuation of the discontinuities and the locations of the branch cuts.  
Since the  first entries of $\Delta_{1,n-1}f$ are of weight 1, the $\Disc$ operation on the right-hand side is computing discontinuities of ordinary logarithms.  Let us specialize to the discontinuity computed in a specific channel $s$.  We can expand the coproduct in terms of the full symbol alphabet by writing
\bea
(\Disc_s \otimes \textrm{id}) (\Delta_{1,n-1}f) \cong (\Disc_s \otimes \textrm{id})
\sum_{x \in {\cal A}} \left( \log(\pm x) \otimes \delta_x f \right)
\label{eq:firstdiscdelta}
\eea
This relation applies generally, so it is valid not just for the original Feynman integral $F$ but also for its sequential discontinuities. We recall, as noted below \refE{temp0943}, that overall factors of $\pi$ in $f$ are handled separately: this is particularly noteworthy if $f$ is a sequential discontinuity where powers of $\pi$ will have been generated from previous discontinuities.
In the relation with $\Cut$, $\Disc$ is computed in a specific region. We require that the sign in the argument of the logarithm in \refE{eq:firstdiscdelta} be chosen so that the argument is positive and the expression is thus real-valued in the kinematic region for which $f$ is away from its branch cut in $s$.  Then, in taking $\Disc_s$, the coproduct will be analytically continued to the region in which there is a branch cut in $s$.  In this new region, the arguments of the logarithms may become negative, and if the letter $x$ depends on the invariant $s$, then there will be a nonzero contribution to $\Disc$.

Sequential discontinuities of a Feynman integral $F$ are computed by the sequential use of \refE{eq:firstdiscdelta}. We thus claim they are captured by $\delta$ in the relation 
\bea
\Disc_{s_1,\ldots,s_k} F \cong  \Theta\, \sum_{(x_1,\ldots,x_k) \in {\cal A}^k } \left( \prod_{i=1}^k a_i(s_i,x_i) \right)  \delta_{x_1,\ldots,x_k} F,
\label{eq:discequalsdelta}
\eea
where the sum runs over all ordered sequences $(x_1,\ldots,x_k)$ of $k$ letters. We recall that the congruence symbol in \refE{eq:discequalsdelta} indicates that despite the fact that the discontinuity function $\Disc_{s_1,\ldots, s_k}F$ is unique, the right-hand side only captures terms whose coproduct is nonvanishing, and it therefore holds modulo $(2\pi i)^{k+1}$. Furthermore,
since the coproduct is the same in all kinematic regions, we have inserted the schematic factor $\Theta$ to express the restriction to the region where the left-hand side is to be compared with $\Cut$.
Finally, the factors $a_i(s_i,x_i)$ are related to the discontinuity of a real-valued logarithm after analytic continuation from one kinematic region ($R_{i-1}$) to another ($R_i$).  Specifically,
\bea
a_i(s_i,x_i) = \Disc_{s_i;R_i} \big[\!\big[ \log(\pm x_i) \big]\!\big]\!_{R_{i-1}},
\eea
where the double-bracket means that the sign of the argument of the logarithm should be chosen so that the argument is positive in the region $R_{i-1}$, or equivalently,
\bea
\big[\!\big[ \log(\pm x_i) \big]\!\big]\!_{R_{i-1}} = \left.\log(x_i)\right|_{R_{i-1} \cap \{x_i>0\}}
+ \left.\log(-x_i)\right|_{R_{i-1} \cap \{x_i<0\}}
\eea

In the simplest cases, each $a_i(s_i, x_i)$ will simply take one of the values $\pm 2\pi i$ or $0$.  In more complicated cases, one might find a further division into nonempty subregions of phase space.

We note that although we focus on cut integrals, the mathematical relation between $\Disc$ and $\delta$ applies in a more general context.  The essential requirement is that the function has no cut in $s_i$ in region $R_{i-1}$, but does in region $R_i$.

In this paper, we give evidence for the validity of \refE{eq:cutequalsdisc} and \refE{eq:discequalsdelta} by matching cut diagrams and coproduct entries directly, as well as by computing discontinuities in some cases.

\subsection{Example of the relations}
\label{sec:ex-relations}

We close this section with a simple example of the proposed relations \refE{eq:cutequalsdisc} and \refE{eq:discequalsdelta}. 

Consider a three-point planar Feynman integral in $D=4$ dimension.  After normalization to unit leading singularity, it will be a dimensionless function of two ratios of Mandelstam invariants,
\bea\label{eq:FRatios}
F\left(\frac{p_2^2}{p_1^2},\frac{p_3^2}{p_1^2} \right).
\eea
Define variables $z, \bz$ such that
\bea
\frac{p_2^2}{p_1^2}=z\bz, \qquad \frac{p_3^2}{p_1^2}=(1-z)(1-\bz), \qquad z>\bz.
\label{eq:zzbar-first}
\eea
Suppose that we know that the symbol alphabet can be taken to be
\bea
{\cal A}_\triangle = \{z, \bz, 1-z, 1-\bz \}.
\eea
This is, in fact, the alphabet of the three-point ladder in $D=4$ dimensions with massless propagators and any number of rungs \cite{Usyukina:1993ch,Drummond:2012bg}, and thus illustrates the parametrization \refE{eq:s_to_xi}.   We will see in the examples of the following sections how the symbol alphabet may be deduced from explicit cut computations.

The integral $F$ is originally defined in the Euclidean region where all $p_i^2<0$.  In terms of real-valued $z, \bz$, there are three separate components of the Euclidean region \cite{Chavez:2012kn}.  For concreteness, we choose the component $R_{0,<}=\{\bz<z<0\},$ but the relations work equally well starting from either of the other components.

Let us take the first cut in the channel $s_1=p_2^2$.  We analytically continue $F$ to the region $R_1$ of the first cut, where $p_2^2>0$ and $p_1^2, p_3^2<0$.  In terms of $z$ and $\bz$, $R_1=\{\bz<0<z<1\}$.
For each letter $x_1\in{\cal A}_\triangle$, the logarithms $\log x_1$   in the definition of $a_1(p_2^2,x_1)$ are written with positive arguments in the region $R_{0,<}$.  For example, in $a_1(p^2_2,z)$ we compute the discontinuity of the analytic continuation of $\log(-z)$ rather than $\log(z)$.
Since according to the usual Feynman rules the invariants have a positive imaginary part, $p_2^2+i\varepsilon$, we can deduce the corresponding imaginary parts in $z+i\varepsilon$ and $\zbar-i\varepsilon$ for the symbol alphabet, and we get:
\begin{align}\bsp\label{ap2R2}
a_1(p^2_2,z)&=\Disc_{p^2_2;R_1}\ln(-z-i0)=-2\pi i\, ,\\
a_1(p^2_2,\zbar)&=\Disc_{p^2_2;R_1}\ln(-\zbar+i0)=0\, ,\\
a_1(p^2_2,1-z)&=\Disc_{p^2_2;R_1}\ln(1-z-i0)=0\, ,\\
a_1(p^2_2,1-\zbar)&=\Disc_{p^2_2;R_1}\ln(1-\zbar+i0)=0\, .
\esp\end{align}
The discontinuities $\Disc$ have been computed directly according to the definition \refE{eq:def-disc}.
According to \refE{eq:cutequalsdisc} and \refE{eq:discequalsdelta}, our relations among the three kinds of discontinuites are then given by
\begin{equation}
\Cut_{p_2^2}F=-\Disc_{p_2^2}F \cong (2\pi i)\,\Theta\, \delta_z F\,.
\label{eq:example-1}
\end{equation}

Let us take the second cut in the channel $s_2=p_3^2$.  We analytically continue $F$ (for the cut) and $\Disc_{s_1}F$ (for the discontinuity) to the region $R_2$ where $p_2^2, p_3^2>0$ and $p_1^2<0$.  In terms of $z$ and $\bz$, $R_2=\{\bz<0, z>1\}$.  
The $a_1$'s are the same as above.  To compute the $a_2(p^2_3,x_2)$'s, we write the logarithms of the alphabet, $x_2\in{\cal A}_\triangle$, with positive arguments in the region $R_1$.
According to our cutting rules, the imaginary part of $p_3^2$ was conjugated in the process of applying the first cut, so we deduce the signs of the imaginary parts in $z-i\varepsilon$ and $\zbar-i\varepsilon$ from $p_3^2-i\varepsilon$.
\begin{align}\bsp\label{ap2R2}
a_2(p^2_3,z)&=\Disc_{p^2_3;R_2}\ln(z-i0)=0\, ,\\
a_2(p^2_3,\zbar)&=\Disc_{p^2_3;R_2}\ln(-\zbar+i0)=0\, ,\\
a_2(p^2_3,1-z)&=\Disc_{p^2_3;R_2}\ln(1-z+i0)=2\pi i\, ,\\
a_2(p^2_3,1-\zbar)&=\Disc_{p^2_3;R_2}\ln(1-\zbar+i0)=0\, .
\esp\end{align}
The only surviving term is $a_1(p_2^2,z)a_2(p_3^2,1-z)=-(2\pi i)^2$, and the multiple cut and iterated discontinuity are then given by:
\begin{equation}
\Cut_{p_2^2,p_3^2}F=\Disc_{p_2^2,p_3^2}F \cong -(2\pi i)^2\,\Theta\,\delta_{z,1-z}F\,.
\label{eq:example-2}
\end{equation}

Finally, we consider cutting in all three channels with the operation $\Cut_{p_2^2,p_3^2,p_1^2}$. The region in which we would hope to detect this triple cut has all $p_i^2>0$. Because $F$ is a function of ratios of the Mandelstam invariants, \refE{eq:FRatios}, this region is indistinguishable from the branch cut-free Euclidean region in $D=4$.   Therefore the specified region of validity for the relation \refE{eq:cutequalsdisc} is empty.

For completeness and for future reference, we close with the full list of relations for single and double cuts of this class of integrals.
\begin{subequations}
\bea
&& \Cut_{p_1^2}F= -\Disc_{p_1^2}F
\cong - (2\pi i)\,\Theta\, [\delta_z + \delta_{1-z}]F\,,
\label{C1}
\\
&&
\Cut_{p_2^2}F=-\Disc_{p_2^2}F 
\cong (2\pi i)\,\Theta\, \delta_z F\,,
\label{C2}
\\
&&
\Cut_{p_3^2}F=-\Disc_{p_3^2}F
\cong   (2\pi i) \,\Theta\, \delta_{1-z} F\,,
\label{C3}
\eea
\label{eq:tablecut1}
\end{subequations}
\begin{subequations}
\bea
&& \Cut_{p_1^2,p_2^2}F= \Disc_{p_1^2,p_2^2} F
\cong  (2\pi i)^2  \,\Theta\, [ \delta_{z,\bz} +  \delta_{1-z,\bz}]F\,,
\label{C12}
\\
&&
\Cut_{p_2^2,p_1^2}F=\Disc_{p_2^2,p_1^2} F 
\cong  (2\pi i)^2 \,\Theta\,[\delta_{z,\bz}+\delta_{z,1-z}]F\,,
\label{C21}
\\
&&
\Cut_{p_1^2,p_3^2}F=\Disc_{p_1^2,p_3^2}F
\cong   (2\pi i)^2 \,\Theta\, [  \delta_{z,1-z}+ \delta_{1-z,1-z}]F\,,
\label{C13}
\\
&&
 \Cut_{p_3^2,p_1^2}F=\Disc_{p_3^2,p_1^2}F
 \cong  (2\pi i)^2 \,\Theta\, [\delta_{1-z,\bz}+ \delta_{1-z,1-z}]F\,,
\label{C31}
\\
&&
\Cut_{p_2^2,p_3^2}F=\Disc_{p_2^2,p_3^2}F
\cong
-(2\pi i)^2\,\Theta\,\delta_{z,1-z}F\,,  
\label{C23}
\\
&&
\Cut_{p_3^2,p_2^2}F = \Disc_{p_3^2,p_2^2}F
\cong
-(2\pi i)^2\,\Theta\,\delta_{1-z,\bz}F\,.  
\label{C32} 
\eea
 \label{eq:tableeqns}
\end{subequations}
We now see concretely that because cuts act simultaneously in the various channels, there are nontrivial relations among entries of the coproduct.  For example, $\Cut_{p_2^2,p_3^2}F=\Cut_{p_3^2,p_2^2}F$ implies that $\delta_{z,1-z}F \cong \delta_{1-z,\bz}F$.   We discuss this point further in \refS{sec:integrable}.

\def\zz{z}
\def\utwo{u_2}
\def\uthree{u_3}
\def\cT{\mathcal{T}}

\section{One-loop examples\label{sec:OneLoop}}

In this section, we present three simple examples of discontinuities of one-loop integrals to demonstrate the relations discussed in the previous section, with explicit computations of cuts, showing the natural appearance of the symbol alphabet.  We first consider the three-mass triangle in some depth, which is an illuminating introduction to the two-loop ladder example in the following section, as their kinematic analyses have many common features.  The second example is the four-mass box, whose functional form is closely related to the triangle although the cut diagrams are quite different.  Finally, we discuss the infrared-divergent ``two-mass-hard'' box, which demonstrates the validity of consistent dimensional regularization.

\subsection{Three-mass triangle}
\label{sec_threeMassTriangle}
\begin{figure}[]
\begin{subfigure}[]{0.32\linewidth}
\centering
\includegraphics[keepaspectratio=true, height=4cm]{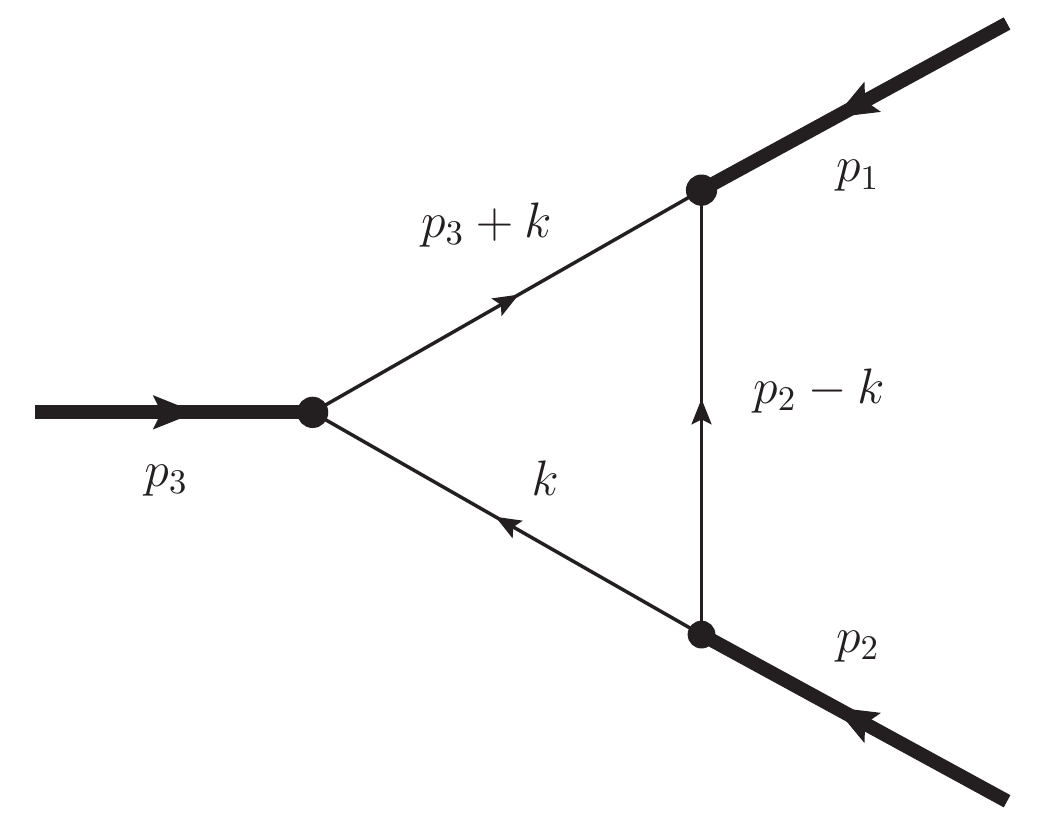}
\caption{$T(p_1^2,p_2^2,p_3^2)$}
\label{fig:triangleUncut} 
\end{subfigure}
\begin{subfigure}[]{0.33\linewidth}
\centering
\includegraphics[keepaspectratio=true, height=4cm]{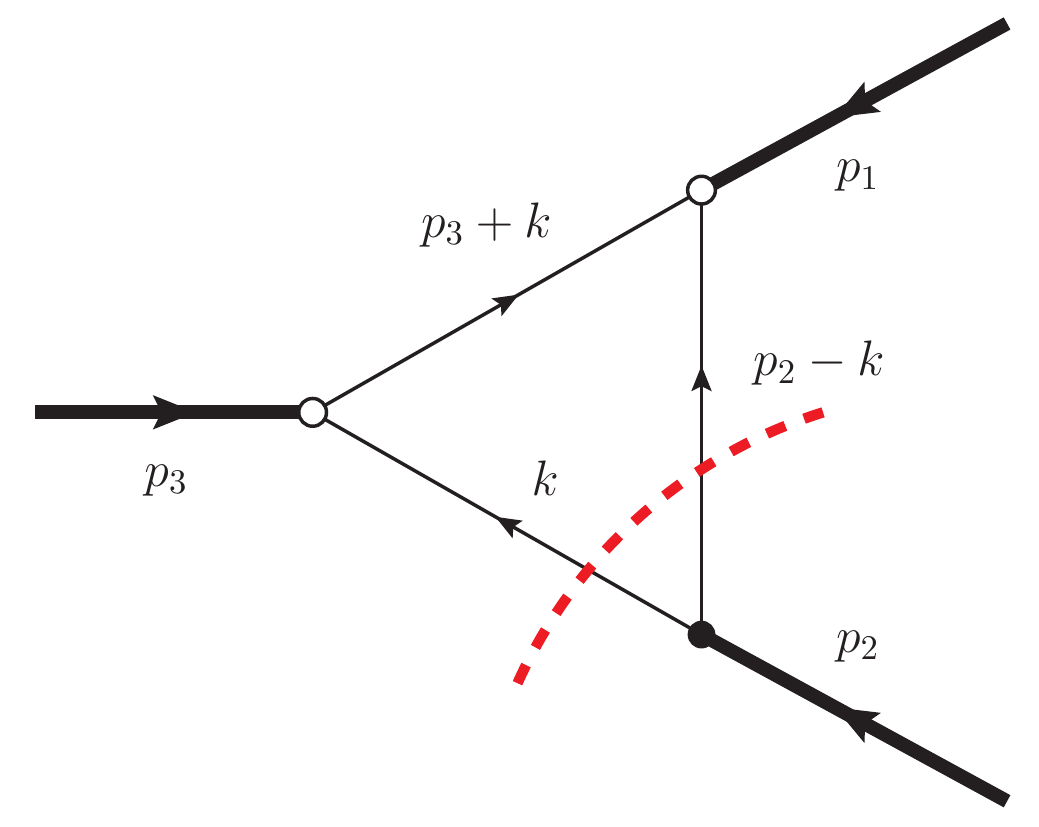}
\caption{$\textrm{Cut}_{p_2^2}T(p_1^2,p_2^2,p_3^2)$}
\label{fig:triangleCut2} 
\end{subfigure} 
\begin{subfigure}[]{0.33\linewidth}
\centering
\includegraphics[keepaspectratio=true, height=4cm]{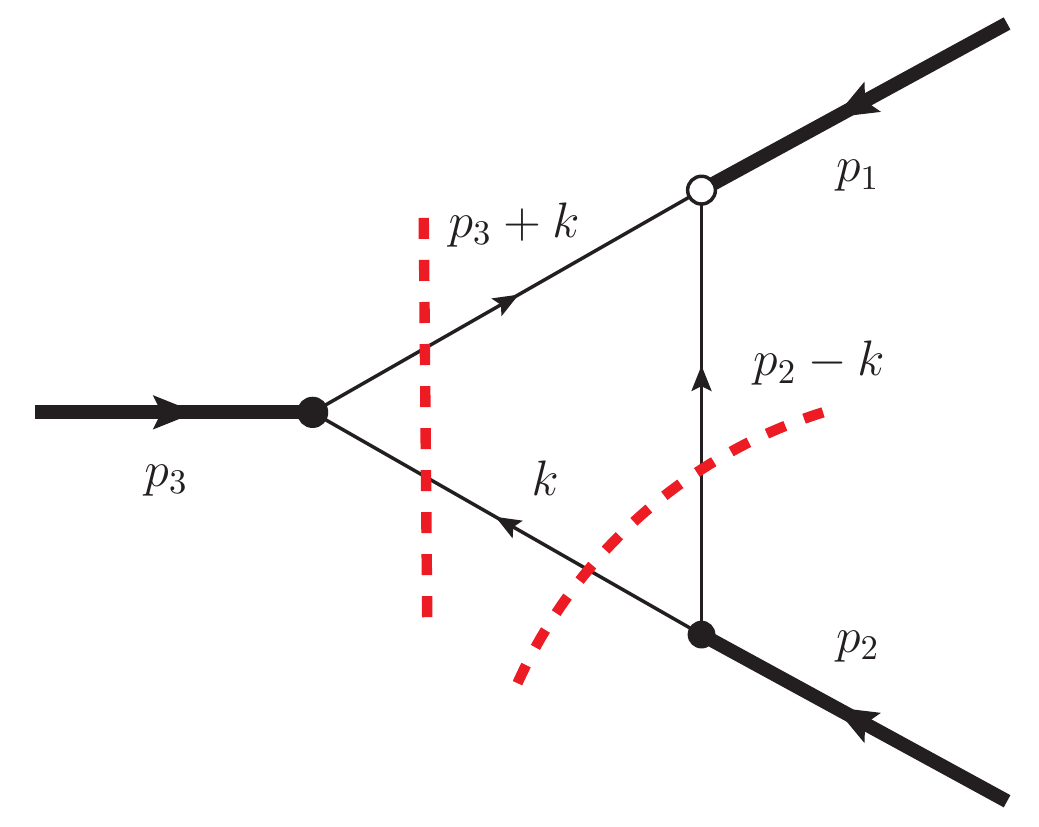}
\caption{$\left(\textrm{Cut}_{p_3^2}\circ\textrm{Cut}_{p_2^2}\right)T(p_1^2,p_2^2,p_3^2)$}
\label{fig:triangleCut32} 
\end{subfigure} 
\caption{The triangle integral, with loop momentum defined as in the text; and with cuts in the $p_2^2$ and $p_3^2$ channels.}
\label{fig:triangle}
\end{figure}

\paragraph{The triangle in $D=4$ dimensions.}
We begin by analyzing the three-mass triangle integral with massless propagators. According to our conventions, which are summarized in appendix \ref{app:conventions}, the three-mass triangle integral in $D=4-2\eps$ dimensions is defined by
\beq
T(p_1^2,p_2^2,p_3^2) 
\equiv -  e^{\gamma_E \eps} 
\int \frac{d^{D}k}{\pi^{D/2}}~ \frac{1}{k^2\, (p_2-k)^2\, (p_3+k)^2}\,,
\label{eq:deftriangleintegral}
\eeq
where $\gamma_E=-\Gamma'(1)$ denotes the Euler-Mascheroni constant. 
As the focus of the paper will be the computation of cut diagrams, it is of utmost importance to keep track of all imaginary parts. 
We follow the conventions for massless scalar theory listed in the preceding section.
In particular, until cuts are introduced, all vertices (denoted by a black dot, see fig.~\ref{fig:triangle}) are proportional to $i$, and all propagators have an explicit factor of $i$ in the numerator and follow the usual Feynman $+i\varepsilon$ prescription. These factors lead to the explicit minus sign in \refE{eq:deftriangleintegral}.  Note that we do not include a factor of $i^{-1}$ per loop into the definition of the integration measure.

Many different expressions are known for the three-mass triangle integral, both in arbitrary dimensions~\cite{Bern:1993kr, Anastasiou:1999ui} as well as an expansion around four space-time dimensions in dimensional regularization~\cite{Usyukina:1992jd,Davydychev:2000na,Birthwright:2004kk,Chavez:2012kn}. Note that the three-mass triangle integral is finite in four dimensions, and we therefore put $\eps=0$ and only analyze the structure of the integral in exactly four dimensions.
We follow \cite{Chavez:2012kn} in writing the result of the integral in the form\beq
T(p_1^2,p_2^2,p_3^2) 
=
-\frac{i}{p_1^2}
\frac{2}{\zz-\bar\zz} \cP_2(\zz) + \ord(\eps),
\label{eq:triangle}
\eeq
where
\bea
\cP_2(z)
&=&  \Li_2(z)-\Li_2(\bz) + \frac{1}{2} \ln(z\bz) \ln\left(\frac{1-z}{1-\bz}\right) \, ,
\label{eq:p2asmpl}
\eea
and
\bea
&& \zz = \half \left(1+\utwo-\uthree+\sqrt{\lambda}\right)\,,
\qquad 
\bar\zz = \half \left(1+\utwo-\uthree-\sqrt{\lambda}\right)\,,
\label{eq:z}
\\
&& \lambda \equiv \lambda(1,u_2,u_3)\,,
\label{eq:deflambda}
\\
&&
  \lambda(a,b,c)=a^2+b^2+c^2-2ab-2ac-2bc\,,
  \eea
  where $\lambda(a,b,c)$ is the well known K\"all\'en function and we have defined dimensionless ratios of Mandelstam invariants,
\bea
  u_i = \frac{p_i^2}{p_1^2}, \qquad i=2,3\,.
\label{eq:ui}
 \eea
Some comments are in order: we see that the three-mass triangle is of homogeneous transcendental weight two, i.e., it is only a function of dilogarithms and products of ordinary logarithms. It is, however, \emph{not} a pure function in the sense of the definition in section~\ref{sec:Hopf}, but it is multiplied by an algebraic function of the three external scales $p_i^2$ (or equivalently, a rational function of $z$, $\bz$ and $p_1^2$), which is the leading singularity. In the following we are only interested in the transcendental contribution, and we therefore define, for arbitrary values of the dimensional regulator $\epsilon$,
\beq
T(p_1^2,p_2^2,p_3^2) \equiv -\frac{i}{p_1^2}
\frac{2}{\zz-\bar\zz} \cT(p_1^2,p_2^2,p_3^2)\,,
\label{eq:defcalT}
\eeq
such that $\cT(p_1^2,p_2^2,p_3^2) = \cP_2(z) + \ord(\eps)$ is a pure function at every order in the $\eps$ expansion.
It is clear from \refE{eq:p2asmpl} that the symbol alphabet of the triangle in $D=4$ may be taken to be
\bea
\cA_\triangle = \{z,\bz,1-z,1-\bz\}\, .
\label{eq:alph-tri}
\eea

The variables $z, \bz$ correspond exactly to the ones introduced in \refE{eq:zzbar-first} for general planar three-point integrals,
so that  $z, \bz$, satisfy the relations
\bea
\zz \bar\zz = \utwo, \qquad (1-\zz)(1-\bar\zz)=\uthree.
\label{eq:zalpha}
\eea
Note that eq.~\eqref{eq:zalpha} corresponds to the rational parametrization~\eqref{eq:s_to_xi} in the case of the triangle integral. Without advance knowledge of the integrated expression, it seems difficult to guess the parametrization a priori.  However, we will see in \refE{3mass_triangle_cut_result} how these variables arise naturally in the calculation of cut diagrams. 

We note that, for positive values of $\lambda$, we always have $\zz > \bar\zz$.
Since \refE{eq:zalpha} is symmetric in $z$ and $\bz$, there is a second solution in which $\bz>z$, which could be interpreted as taking the negative branch of the square root in \refE{eq:z}.  In most of our calculations, we will indeed restrict ourselves to the region where $\zz > \bar\zz$, for concreteness.
In the regions where all $p_i$ have the same sign, there is a portion of kinematic phase space in which $\lambda$ is negative, so that $(\zz,\bz)$  take complex values. We remark that it is known that new letters appear in the three-mass triangle beyond the constant term in the $\eps$-expansion in dimensional regularization \cite{Chavez:2012kn}.

Let us now consider the discontinuities of the triangle integral.  The first-entry condition for Feynman integrals discussed in section~\ref{sec:Hopf} implies that the symbol of the three-mass triangle can be written in  a form with only the combinations $\utwo=\zz \bar\zz$ and $\uthree=(1-\zz)(1-\bar\zz)$ appearing in the leftmost entry.  
The coproduct of the one-loop three mass triangle can be computed explicitly from \refE{eq:defcalT}, with the result
\beq\bsp
\Delta&\left[\cT(p_1^2,p_2^2,p_3^2)\right] \\ 
&=\cP_2(z)\otimes1 + 1\otimes\cP_2(z) + 
\frac{1}{2}\ln(\zz\bar\zz)\otimes \ln\frac{1-\zz}{1-\bar\zz} +\frac{1}{2} \ln[(1-\zz)(1-\bar\zz)] \otimes \ln\frac{\bar\zz}{\zz} +\ord(\eps)  \\
&= \cP_2(z)\otimes1 + 1\otimes\cP_2(z) +\frac{1}{2} \ln\left(-p_2^2\right)\otimes \ln\frac{1-\zz}{1-\bar\zz} + \frac{1}{2}\ln\left(-p_3^2\right)\otimes \ln\frac{\bar\zz}{\zz} 
 \\
&~~~ + \frac{1}{2}\ln (-p_1^2) \otimes \ln \frac{1-1/\bar\zz}{1-1/\zz} +\ord(\eps),
\label{eq:trianglecoproduct}
\esp\eeq
where in the second equality we have made the first entry condition explicit. Our aim is to interpret the coproduct of the one-loop three-mass triangle in terms of cut diagrams, through the relations of \refS{sec:disccutdelta}.
In the rest of this section we present, as a warm-up, the explicit computation of the unitarity cut of the one-loop three-mass triangle.

\paragraph{Unitarity cuts of the one-loop three-mass triangle.}
Cuts of the triangle are evaluated in kinematic regions in which the cut invariants are positive and the uncut invariants are negative.  
The correspondence between signs of Mandelstam invariants and values of $z,\bz$ is given in Table \ref{table:regionstriangle}.

\begin{table}[!t]
\centering
\begin{tabular}{l c l c c}
\hline\hline
Name && Region of the $p_i^2$ &&  Region of $z, \bz$ \\
\hline
$R_\triangle^1$ && $p_1^2>0$,\,\, $p_2^2,p_3^2<0$ && \multirow{2}*{$\bz < 0$,~~ $1<z$} \\\cmidrule{1-3}
$R_\triangle^{2,3}$ && $p_1^2<0$,\,\, $p_2^2,p_3^2>0$ && \\
\hline
$R_\triangle^2$ && $p_2^2>0$,\,\, $p_1^2,p_3^2<0$ && \multirow{2}*{$\bz<0<z<1$} \\\cmidrule{1-3}
$R_\triangle^{1,3}$ && $p_2^2<0$,\,\, $p_1^2,p_3^2>0$ && \\
\hline
$R_\triangle^3$ && $p_3^2>0$,\,\, $p_1^2,p_2^2<0$ && \multirow{2}*{$0<\bz<1<z$} \\\cmidrule{1-3}
$R_\triangle^{1,2}$ && $p_3^2<0$,\,\, $p_1^2,p_2^2>0$ && \\
\hline
  $R_\triangle^*$ && $p_1^2, p_2^2, p_3^2 >0$, ~and $\lambda<0$ && $z^*=\bz$
\\
\hline\hline
\end{tabular}
\caption{
Some kinematic regions of 3-point integrals, classified according to the signs of the Mandelstam invariants and the sign of $\lambda$, as defined in \refE{eq:deflambda}. In the first six rows, $\lambda>0$, so that $z$ and  $\bz$ are real-valued, and we take $z > \bz$ without loss of generality.}
\label{table:regionstriangle}
\end{table}

In the following we review the cut integral calculation. This triangle is fully symmetric, so it is enough to demonstrate one single-channel cut and one double-channel cut.  For concreteness, we choose the cuts illustrated in fig.~\ref{fig:triangleCut2}.
Although it is not necessary in this example,  we now work in $D=4-2\eps$ dimensions, 
as a warmup to the two-loop integral where the $D$-dimensional formalism will be important at the level of cuts.
We will work in the region which we denote by $R_\triangle^*$, where all the invariants are positive and $\lambda<0$ (and thus $\bz=\zz^*$),
because having $\zz$ and $\bz$ complex simplifies the calculation. 
The cut integral we want to compute reads
\beq
\label{3mass_triangle_cut}
\textrm{Cut}_{p_2^2,R_\triangle^*} T = - (2\pi)^2\,e^{\gamma_E \eps}
\int \frac{d^{D}k}{\pi^{D/2}}\, \frac{ \delta^+(k^2) \,\delta^+\left((p_2-k)^2\right)}{ (p_3+k)^2}\,,
\eeq
with $\delta^+(k^2) = \delta(k^2)\,\theta(k_0)$.
Without loss of generality we can select our frame and parametrize the loop momentum as follows:
\begin{align}\bsp
&p_2 = \sqrt{p_2^2}(1,0,{\bf 0}_{D-2}), \qquad p_3 = \sqrt{p_3^2}(\alpha,\sqrt{\alpha^2-1},{\bf 0}_{D-2}),\\
&k =  (k_0,|k| \cos \th,|k|\sin\th ~{\bf 1}_{D-2}),\label{eq:parametrization_triangle}
\esp\end{align}
where $\th \in [0,\pi]$ and $|k|>0$, and ${\bf 1}_{D-2}$ ranges over unit vectors in the dimensions transverse to $p_2$ and $p_3$.
Momentum conservation fixes the value of $\alpha$ in terms of the momentum invariants to be 
\bean
\alpha = \frac{p_1^2-p_2^2-p_3^2}{2 \sqrt{p_2^2}\sqrt{p_3^2}}
.
\eean
 With this frame and parametrization, the cut integration measure becomes
\bea
  d^{D} k~\delta^+(k^2) &=& d k_0 ~\theta(k_0)~d|k|~d\cos\th~\delta(k_0^2-|k|^2)~
  \frac{2 \pi^{1-\eps}}{\Gamma(1-\eps)} |k|^{2-2\eps} (\sin\th)^{-2\eps}\label{kMeasure}\,.
\eea
The $D$-dimensional cut triangle integral, with energy flow conditions suited for the $p_2$ channel, is
\bea
\textrm{Cut}_{p_2^2,R_\triangle^*} T &=&
-{8\pi e^{\gamma_E \eps}}
\int_{0}^{\infty} 
d k_0 ~ \int_0^\infty d|k|~ \int_{-1}^{1} d\cos\th~\delta(k_0^2-|k|^2)~
  \frac{ |k|^{2-2\eps} (\sin\th)^{-2\eps}}{\Gamma(1-\eps)} 
  \nonumber
\\ &\times& 
 \frac{
  \delta(p_2^2-2 k_0 \sqrt{p_2^2})}{p_3^2 +2\sqrt{p_3^2}( k_0 \alpha
  -|k|\cos\th\sqrt{\alpha^2-1})} \\
&=&- \frac{ 2^{1+2\eps} \pi e^{\gamma_E \eps}}{\Gamma(1-\eps)}
\int_{-1}^1 
d\cos\th~
 (\sin\th)^{-2\eps}
 \frac{ \left({p_2^2}\right)^{-\eps}}{p_3^2+ p_1^2 - p_2^2 - \cos\th \sqrt{\lambda(p_1^2,p_2^2,p_3^2)}}.\nonumber
\eea
Performing the change of variables,
\bea
\cos \th = 2 x - 1, \qquad x \in [0,1],
\eea
and turning to the dimensionless variables~(\ref{eq:ui}) and (\ref{eq:z}),  the cut integral becomes 
\bea
\textrm{Cut}_{p_2^2,R_\triangle^*} T &=&
-\frac{2\pi (p_1^2)^{-1-\eps} u_2^{-\eps}   e^{\gamma_E \eps}}{\Gamma(1-\eps)}
\int_{0}^1 
 d x~
  x^{-\eps} (1-x)^{-\eps}
 \frac{ 1}{ 1-\bz-x \sqrt{\lambda} }\nonumber\\
&=& -\frac{2\pi e^{\gamma_E \eps} \Gamma(1-\eps)}{\Gamma(2-2\eps)} (p_1^2)^{-1-\eps} \frac{u_2^{-\eps}}{1-\bz}
{}_2F_1\left(1,1-\eps;2-2\eps;\frac{\sqrt{\lambda}}{1-\bz}\right)
\nonumber
 \\
&=&
\frac{2\pi}{p_1^2 (z-\bz)} 
\ln\frac{1-\zz}{1-\bz}
+ \ord(\eps)\,.
\label{3mass_triangle_cut_result}
\eea
 It is now trivial to analytically continue to the region $R_\triangle^2$ in which $p_2^2>0$ and $p_1^2, p_3^2<0$.  As anticipated, the variables $\zz$ and $\zbar$ appear naturally in the calculation of the cut integral. We comment that while these variables are useful for expressing the uncut integral, it is harder to identify them through a direct computation of the latter. We see this as one of the advantages of a cut-based computation.
The results for the cuts on different channels can be obtained in a similar way and are collected in appendix~\ref{app_oneLoopRes}. Taking the $\epsilon=0$ limit of eqs.~(\ref{cutP1TriDDim}), (\ref{cutP2TriDDim}) and (\ref{cutP3TriDDim}), we see that a minimal and complete choice for the symbol alphabet of the three-mass triangle in $D=4$ is the set $\cA_\triangle=\left\{z,\bz,1-z,1-\bz\right\}$, in accordance with \refE{eq:alph-tri} derived from the known integrated expression.

Let us now consider a pair of cuts on the $p_2^2$ and $p_3^2$ channels, consistent with energy flow from leg three to leg two (see fig.~\ref{fig:triangleCut32}).
We must work in a region where $p_2^2,p_3^2>0$; we choose $R_\triangle^{2,3}$.  The cut integral is
\beq
\textrm{Cut}_{p_2^2,p_3^2}T =
i(2 \pi )^3\,e^{ \gamma_E \eps}\,
\int \frac{d^{D}k}{\pi^{D/2}}\, \delta^{+}(k^2)\, \delta^{+}\left((p_2-k)^2\right)~ \delta^{+}\left((p_3+k)^2\right)\,.
\eeq
Using the parametrization~\eqref{eq:parametrization_triangle}, we find
\beq\bsp
\textrm{Cut}&_{p_2^2,p_3^2}T =
{16\pi^2i e^{ \gamma_E \eps}}\int_{0}^{\infty} 
d k_0 ~ \int_0^\infty d|k|~ \int_{-1}^{1} d\cos\th~\delta(k_0^2-|k|^2)~
  \frac{|k|^{2-2\eps} (\sin\th)^{-2\eps}}{\Gamma(1-\eps)}   
\\ &\times  
  \delta\left(p_2^2 - 2 k_0 \sqrt{p_2^2}\right) ~\delta\left(p_3^2 + 2\sqrt{p_3^2}( k_0 \alpha
  -|k|\cos\th\sqrt{\alpha^2-1})\right) 
  \theta\left(-\sqrt{p_3^2}\alpha-k_0\right)
  \\
&= \frac{2^{1+2\eps} \pi^2 i  e^{ \gamma_E \eps}(p_2^2)^{-\eps}}{\Gamma(1-\eps)}
\int_{-1}^1 d\cos\th (\sin\th)^{-2\eps}
 \,\delta\left( p_3^2 + \sqrt{p_3^2}\sqrt{p_2^2}(\alpha-\cos\th\sqrt{\alpha^2-1}) \right) \\
&= \frac{4\pi^2 i  e^{ \gamma_E \eps}}{\Gamma(1-\eps)}
 \left( -p_1^2 \right)^{-1-\eps}
 \left( u_2 u_3 \right)^{-\eps} (z-\bz)^{-1+2\eps}
\\
&=
-\frac{4\pi^2 i}{p_1^2 (z-\bz)} + \ord(\eps)\, .
 \label{eq:cuttrip3p2}
 \esp\eeq

\paragraph{Summary and discussion.}

We now interpret the results for the cuts of the triangle integral we just computed in terms of the coproduct.  The relations \refE{eq:cutequalsdisc} and \refE{eq:discequalsdelta} have already been applied to the triangle in \refS{sec:ex-relations}.  For a single cut, we predicted in \refE{eq:example-1} that
\begin{equation}
\Cut_{p_2^2} \cT =-\Disc_{p_2^2} \cT \cong (2\pi i) \, \Theta\,  \delta_z \cT\,. \
\label{eq:example-1-sec4}
\end{equation}
After proper accounting of the prefactors, we see now from the cut result \refE{3mass_triangle_cut_result}, from a direct calculation of the discontinuity of \refE{eq:p2asmpl} using the definition \refE{eq:def-disc}, and from the coproduct \refE{eq:trianglecoproduct}, that these three quantities do agree and are each equal to
\bean
i\pi \ln\frac{1-\zz}{1-\bz}\, .
\eean
For concreteness, we note that in this example $\Theta=\theta(-p_1^2)\theta(p_2^2)\theta(-p_3^2)=\theta(-\bz)\theta(\zz)\theta(1-\zz)$.
 
Proceeding to a sequence of two discontinuities,\footnote{For one-loop integrals, a three-propagator cut has previously been interpreted as a discontinuity of a diagrammatic unitarity cut.  In ref. \cite{Ball:1991bs}, 
it was used in a double dispersion relation to verify the region of integration in phase space for semileptonic $D$ decay.  
    More recently, in ref. \cite{Mastrolia:2006ki}, a similar interpretation was given, in the spirit of the Feynman Tree Theorem \cite{Feynman:1963ax,Feynman:tree1,Feynman:tree2}, capitalizing on progress in  unitarity methods for one-loop amplitudes.} 
    we predicted in \refE{eq:example-2} that 
\begin{equation}
\Cut_{p_2^2,p_3^2}\cT=\Disc_{p_2^2,p_3^2}\cT \cong -(2\pi i)^2\,\Theta\,\delta_{z,1-z}\cT\,.
\label{eq:example-2-sec4}
\end{equation}
We take the cut result from \refE{eq:cuttrip3p2} and the coproduct entry again from \refE{eq:trianglecoproduct}.  The direct discontinuity $\Disc_{p_3^2}(\Disc_{p_2^2}\cT)$ can be computed after having obtained the explicit result from the first discontinuity.  Again, we find agreement, with the common value
\bean
2\pi^2\,.
\eean
Here, we have $\Theta=\theta(-p_1^2)\theta(p_2^2)\theta(p_3^2)=\theta(-\bz)\theta(\zz-1)$. In later examples we will not give explicit expressions for the $\Theta$, as they are easily deduced from the discussion.

\subsection{Four-mass box}

\begin{figure}[]
\begin{subfigure}[]{0.32\linewidth}
\centering
\includegraphics[keepaspectratio=true, height=3cm]{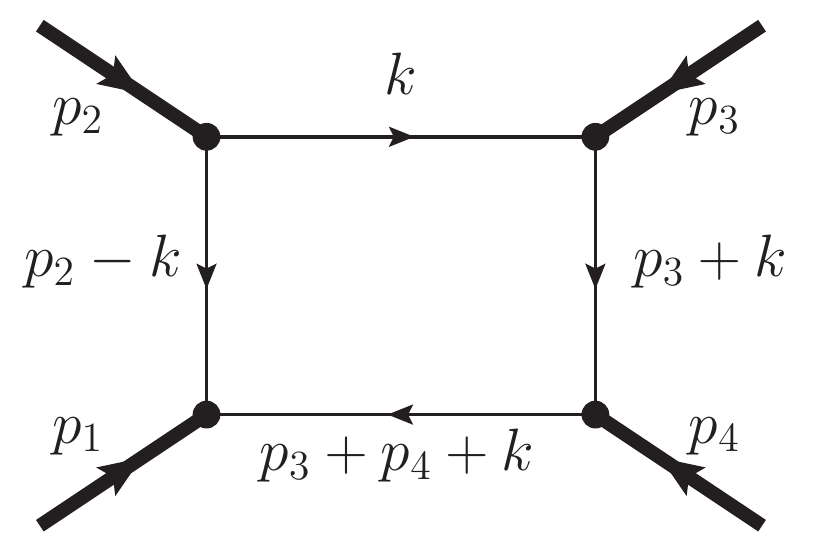}
\caption{$B^{4m}$}
\label{fig:box4mUncut} 
\end{subfigure}
\begin{subfigure}[]{0.33\linewidth}
\centering
\includegraphics[keepaspectratio=true, height=3cm]{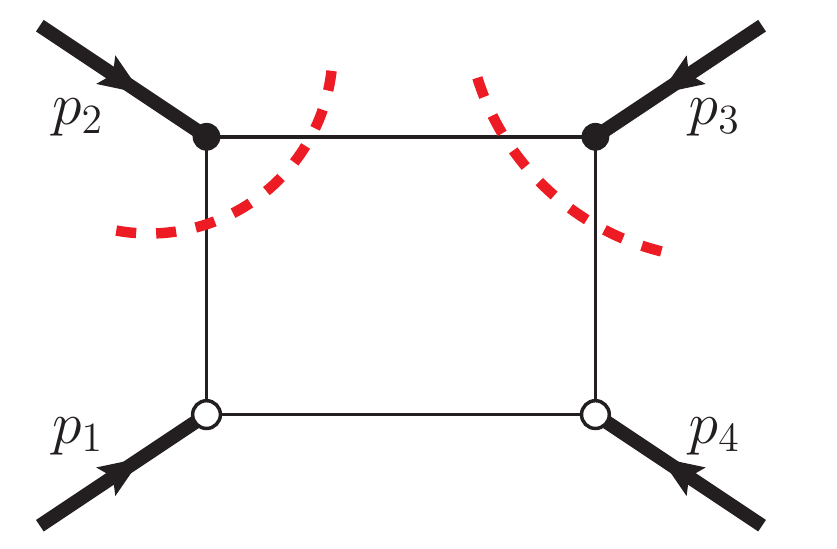}
\caption{$\textrm{Cut}_{p_2^2,p_3^2}B^{4m}$}
\label{fig:box4mCut32} 
\end{subfigure} 
\begin{subfigure}[]{0.33\linewidth}
\centering
\includegraphics[keepaspectratio=true, height=3cm]{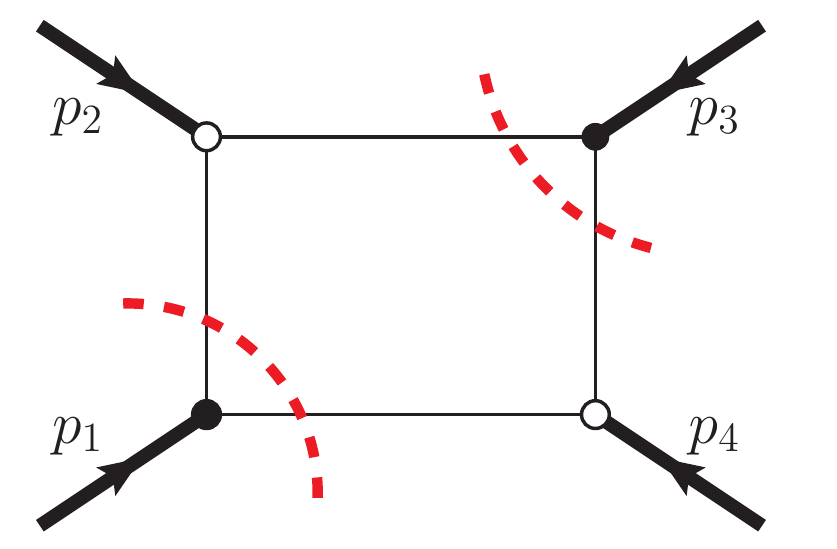}
\caption{$\textrm{Cut}_{p_1^2,p_3^2}B^{4m}$}
\label{fig:box4mCut13} 
\end{subfigure} 
\caption{The four-mass box integral, with pairs of unitarity cuts.}
\label{fig:4mbox}
\end{figure}

The four-mass box is also finite in four dimensions, and may in fact be expressed by the same function as the three-mass triangle \cite{Usyukina:1992jd}.  If we label the momenta at the four corners by $p_1, p_2, p_3, p_4$, as in fig.~\ref{fig:box4mUncut}, and define $s=(p_1+p_2)^2$ and $t=(p_2+p_3)^2$, then the box in the Euclidean region is given by
\bean
B^{4m}(p_1^2,p_2^2,p_3^2,p_4^2,s,t) &\equiv& \frac{1 }{\pi^{2}}
\int d^{4}k~ \frac{1}{k^2 (p_2-k)^2 (p_3+k)^2 (p_3+p_4+k)^2}
\\
&=& - T(p_1^2 p_3^2, p_2^2 p_4^2, s t) \\
&=&  \frac{i}{st}\frac{2}{Z-\bZ} \cP_2(Z),
\eean
where we have introduced variables $Z, \bZ$ defined as follows:
\bean
Z = \half\left(1+ U-V
+ \sqrt{\lambda\left(1,U,V\right)} \right) {\rm~~and~~} \bZ = \half\left(1+ U-V
- \sqrt{\lambda\left(1,U,V\right)} \right).
\eean
Hence
\bean
Z \bZ = U=\frac{p_2^2 p_4^2}{s t} {\rm~~and~~} (1-Z)(1-\bZ)=V=\frac{p_1^2 p_3^2}{s t}.
\eean
The $(1,1)$ component of the coproduct is 
\bea
\Delta_{1,1}B^{4m}=\frac{i}{s t}\frac{1}{Z-\bZ} 
 \left[ \ln\left(Z\bZ\right)\otimes \ln\frac{1-Z}{1-\bZ} + \ln\left((1-Z)(1-\bZ)\right) \otimes \ln\frac{\bZ}{Z} \right]. 
\eea

Let us now show how these variables appear naturally in the calculation of cuts. For the box topology, there are two different types of single unitarity cuts: cuts that isolate one vertex (in one of the $p_i^2$), and cuts that isolate a pair of vertices (in either the $s$ or the $t$ channels). We will look at one example of each configuration: the cut in the $p_3^2$ channel and the cut in the $s$ channel.

Following the cutting rules we established in the previous section, and working in $D=4$, we have
\begin{align}\bsp
\cut_{p_3^2}B^{4m}=&4\int d^4k\frac{\delta^+\left(k^2\right)\delta^+\left((k-p_3)^2\right)}{(k+p_2)^2(k+p_1+p_2)^2}\\
=&4\int_0^\infty da_1 \int_0^\infty da_2\,\delta\left(1-\sum_{i\in S}a_i\right)\int d^4k\frac{\delta^+\left(k^2\right)\delta^+\left((k-p_3)^2\right)}{\left((a_1+a_2)k^2+2 k\cdot\eta+\beta^2\right)^2}\, ,
\esp\end{align}
where in the last line we combined the two uncut propagators using Feynman parameters, and we
have defined the four-vector $\eta$ and the invariant $\beta^2$ as
\begin{equation*}
\eta=a_1p_2+a_2(p_1+p_2)\, ,\qquad\qquad\qquad\beta^2=a_1p_2^2+a_2 s\, .
\end{equation*}
Following the Cheng-Wu theorem, since the denominator of the integrand is homogeneous of degree 2 in $(a_1,a_2)$, we may take $S$ to be any nonempty subset of  $\{1,2\}$.
As far as the integration of the cut loop momentum is concerned, we are in a situation similar to the one of the three-mass triangle, and we thus use a similar parametrization of the momenta,
\begin{align*}\bsp
&p_3 = \sqrt{p_3^2}(1,0,{\bf 0}_{2}), \qquad \eta = \sqrt{\eta^2}(\alpha,\sqrt{\alpha^2-1},{\bf 0}_{2}),\\
&k =  (k_0,|k| \cos \th,|k|\sin\th ~{\bf 1}_{2}),
\esp\end{align*}
where now $\alpha$ and $\eta^2$ are functions of the Feynman parameters. Using \refE{kMeasure} for $D=4$, the integration over the cut loop momentum can be done easily and we get
\begin{align}\bsp
\cut_{p_3^2}B^{4m}&=2\pi\int_0^\infty da_1 \int_0^\infty da_2\frac{\delta\left(1-\sum_{i\in S}a_i\right)}{p_3^2\eta^2+\beta^4+2\beta^2p_3\cdot\eta}\\
&=2\pi\int_0^\infty da_1 \int_0^\infty da_2\frac{\delta\left(1-\sum_{i\in S}a_i\right)}{a_1^2p_2^2t+a_2^2p_4^2 s+a_1a_2\left(st-p_1^2p_3^2+p_2^2p_4^2\right)}\,.
\esp\end{align}
Choosing $S=\{2\}$ so that $a_2=1$, and changing variables to $y=a_1 \frac{p_2^2}{s}$,
\begin{align}\bsp
\cut_{p_3^2}B^{4m}&=\frac{2\pi}{s t}\int_0^\infty dy\frac{1}{(Z+y)(\bar{Z}+y)}\\
&=-\frac{2\pi}{s t}\frac{1}{Z-\bar{Z}}\log\frac{\bZ}{Z}\label{p3CutBox}\,.
\esp\end{align}
As for the three-mass triangle, $Z$ and $\bar{Z}$ appear naturally in the calculation of the cut of the four-mass box diagram, and it is fair to say that this calculation is simpler than the evaluation of the uncut diagram \cite{Usyukina:1992jd}.

The cut in the $s$-channel can be computed following exactly the same steps. We only quote the result,
\begin{equation}
\cut_{s}B^{4m}=\frac{2\pi}{s t}\frac{1}{Z-\bar{Z}}\left(\ln\frac{1-Z}{1-\bar{Z}}+\ln\frac{\bZ}{Z}\right)\label{sCutBox}\,.
\end{equation}

\paragraph{Summary and discussion.} As expected, we find that both cuts verify the relations we anticipated
\begin{equation}
\textrm{Cut}_{p_3^2} B^{4m}=-\textrm{Disc}_{p_3^2} B^{4m}  \cong 
(2\pi i)\,\Theta\,\delta_{1-Z}B^{4m},
\end{equation}
and
\begin{equation}
\textrm{Cut}_{s} B^{4m} =-\textrm{Disc}_{s} B^{4m} \cong
-(2\pi i)\,\Theta\,\left(\delta_{Z}B^{4m}+\delta_{1-Z}B^{4m}\right).
\end{equation}

Multiple cuts also reproduce the expected relation \refE{eq:cutequalsdisc} with $\Disc$. Indeed, since the functional form is the same as for the three-mass triangle, most of the multiple cuts can be analyzed in exactly the same way.  Because the transcendental weight is two, we are limited to a sequence of two discontinuities.  This limitation is consistent with the one encountered when considering $\Cut$, as any real-valued cut of all four propagators of the diagram vanishes.

Now consider the relations between $\Disc$ and $\delta$ from \refE{eq:discequalsdelta}, for a sequence of two channels. 
In view of the  permutation symmetry, we can say without loss of generality that the first cut is in the channel $p_3^2$.  For the second cut channel, we only need to 
distinguish two types:  $p_1^2$, or any of the others.  Suppose we choose $p_2^2$.  Then, the analysis of discontinuities from direct analytic continuation and from the coproduct is exactly the same as in the triangle example.
 The corresponding cut integral, with three delta functions and one of the original propagators, is shown in fig.~\ref{fig:box4mCut32} and produces the leading singularity.
 
 The truly new kind of multiple cut to consider is the discontinuity of $\textrm{Disc}_{p_3^2}B^{4m}$ in the $p_1^2$ channel, shown in fig.~\ref{fig:box4mCut13}.  In a region where $p_1^2, p_3^2>0$, all other invariants are negative, and $\lambda$ is real-valued, we must have $Z/\bZ>0$.  So, either by considering the discontinuity directly, or from the coproduct,  we find
 \bea
\textrm{Disc}_{p_1^2}\textrm{Disc}_{p_3^2}B^{4m}=0.
\label{eq:box-ls-24}
\eea
Recalling the similarity of the functional form of this box to the triangle example, this calculation is analogous to trying to cut the triangle twice in the same channel.  We note that for the box one can set up a cut integral on both the $p_1^2$ and the $p_3^2$ channels: cutting on these two channels amounts to replacing all four propagators by delta functions. This is the familiar ``quadruple cut'' \cite{Britto:2004nc}, which is evaluated at its complex-valued solutions.  
Here, however, we are considering iterated unitarity cuts, where real parametrization of the loop momentum is essential.  Given that there is no real solution to the simultaneous on shell conditions of the four delta functions, we conclude that the cut integral vanishes, in agreement with  eq.~(\ref{eq:box-ls-24}).

\subsection{Two-mass-hard box}
\label{sec_twoMassHard}

We close this section with the example of the two-mass-hard box, namely a box integral (having massless propagators) with two adjacent external massless legs $p_1^2=p_2^2=0$, and two massive ones. This example illustrates several features different from the previous examples: because of the  massless external legs we cannot work directly in four dimensions and the entire analysis is performed using dimensional regularization. A further feature is that the symbol alphabet will now consist exclusively of linear functions of the invariants (this is in contrast to the previous cases analysed where these were algebraic functions involving a square root). As we will see, despite this apparent simplicity, understanding the relations between Cut, Disc and $\delta$ requires some care regarding to the kinematic regions. 
We comment that the two-mass-hard box analysed here will also be needed for our two-loop calculations that follow, where it appears as a subdiagram in some cuts.  

As explained above, because of the infrared divergences of the two-mass-hard box integral, we employ dimensional regularization.   The coproduct structure requires that we work order by order in the regularization parameter.  We use the result of ref.~\cite{Bern:1993kr}, with an additional factor of $i e^{\gamma_E \eps}$ inserted to match our conventions.  In the Euclidean region, the box is given by
\beq\bsp
\label{B2mh_result}
B^{2mh}&(p_3^2,p_4^2,s,t) \equiv \frac{ e^{\gamma_E \eps} }{\pi^{2-\eps}}
\int d^{4-2\eps}k~ \frac{1}{k^2 (k+p_2)^2 (k+p_2+p_3)^2 (k-p_1)^2}
\\ 
&= \frac{i}{st \eps^2} +\frac{i}{st\eps}
\left[
\ln(-p_3^2) + \ln(-p_4^2)-\ln(-s)-2\ln(-t)
\right] \\
& -\frac{i}{12  st} \bigg[
\pi^2 - 6 \ln^2(-p_3^2) - 12 \ln(-p_3^2) \ln(-p_4^2) - 6 \ln^2(-p_4^2)
+ 12 \ln(-p_3^2)\ln(-s)
  \\ &
 + 12\ln(-p_4^2)\ln(-s) -6\ln^2(-s)
+ 24 \ln(-p_3^2)\ln(-t) 
+24\ln(-p_4^2)\ln(-t)\\
&
-24\ln(-s)\ln(-t) -24 \ln^2(-t) - 24 \Li_2\left(1-\frac{t}{p_3^2}\right)- 24 \Li_2\left(1-\frac{t}{p_4^2}\right)
\bigg] + \ord(\eps)\,,
\esp\eeq
where $s=(p_1+p_2)^2$ and $t=(p_2+p_3)^2$. In the following equations, we drop the $\ord(\eps)$ terms.
The coproduct is evaluated order by order in the Laurent expansion in $\eps$.  At order $1/\eps^2$, it is trivial and there is clearly no discontinuity.  At order $1/\eps$, the coproduct is simply the function itself,
\bea
\Delta_{1} \left. B^{2mh}\right|_{1/\eps} &=& \frac{i}{st\eps}
\left[
\ln(-p_3^2) + \ln(-p_4^2)-\ln(-s)-2\ln(-t)
\right].
\eea
At order $\eps^0$, we are interested in 
the $\Delta_{1,1}$ term of the coproduct, which is given by
\beq\bsp
 \Delta_{1,1}& \left. B^{2mh}\right|_{\eps^0}  =
\frac{i}{st}\left[
\ln (-p_3^2)\otimes \ln (-p_4^2)
+\ln 
(-p_4^2)\otimes \ln (-p_3^2)-\ln 
(-p_3^2)\otimes \ln (-s)\right.
 \\ &
\left.-\ln (-s)\otimes \ln 
(-p_3^2)
-2 \ln (-p_3^2)\otimes \ln (-t) - 2 \ln (-t)
\otimes \ln (-p_3^2)\
 \right.
 \\ &
\left.- 2 \ln 
\left(\frac{t}{p_3^2}\right)\otimes \ln 
\left(1-\frac{t}{p_3^2}\right) + \ln (-p_3^2)\otimes 
\ln (-p_3^2)
- \ln (-p_4^2)\otimes \ln (-s) \right.
 \\ &
\left.- \ln 
(-s)\otimes \ln (-p_4^2) - 2 \ln (-p_4^2)\otimes 
\ln (-t) - 2 \ln (-t)\otimes \ln (-p_4^2)
\right.
 \\ &
\left.
- 2 \ln 
\left(\frac{t}{p_4^2}\right)\otimes \ln 
\left(1-\frac{t}{p_4^2}\right) + \ln (-p_4^2)\otimes 
\ln (-p_4^2) + 2 \ln (-s)\otimes \ln (-t)
\right.
 \\ &
\left. + 2 \ln 
(-t)\otimes \ln (-s)
 + \ln (-s)\otimes \ln (-s) + 4 \ln 
(-t)\otimes \ln (-t)
\right].
\label{eq:2mhDelta11}
\esp\eeq
Up to order $\mathcal{O}(\epsilon)$, the symbol alphabet can then be chosen to be
\begin{equation}
\mathcal{A}_{2mh}=\left\{p_3^2,p_4^2,t,s,t-p_3^2,t-p_4^2\right\} \, .
\end{equation}

\paragraph{Discontinuity in the $t$-channel.}
The discontinuity of $B^{2mh}$ in the $t$-channel, with $t>0$ and all other invariants negative, 
can be straightforwardly computed according to the definition (\ref{eq:def-disc}) starting with the expression for the function $B^{2mh}$ in (\ref{B2mh_result}), obtaining:
\beq\bsp
\textrm{Disc}_t B^{2mh}
=&\, 4\pi e^{\gamma_E \eps}r_\Gamma 
\frac{(-p_3^2)^\eps (-p_4^2)^\eps }{t^{1+2\eps}(-s)^{1+\eps}} \left[
\frac{1}{\eps}
  + \ln\left(1-\frac{t}{p_3^2}\right)+ \ln\left(1-\frac{t}{p_4^2}\right)
+\ord(\eps)\right]
\\ 
=&\,
-\frac{4\pi}{ s t} \left[
\frac{1}{\eps}
+
\ln \frac{ (-p_3^2) (-p_4^2)}{(-s) t^2} + \ln\left(1-\frac{t}{p_3^2}\right)+ \ln\left(1-\frac{t}{p_4^2}\right)
+\ord(\eps)\right].
\label{eq:2mhtdisc}
\esp\eeq
Considering instead the coproduct relation (\ref{eq:discequalsdelta}) and using the coproduct entry given in \refE{eq:2mhDelta11}, we find 
\beq\bsp
\delta_t B^{2mh}
\cong
\frac{i}{st}\Big[&\,-\frac{2}{\eps}
 - 2 \ln (-p_3^2)\
 - 2  \ln 
\left(1-\frac{t}{p_3^2}\right) 
 - 2 \ln (-p_4^2)
- 2  \ln 
\left(1-\frac{t}{p_4^2}\right)\\
&  + 2  \ln (-s)
 + 4  \ln (-t)
 + \ord(\eps)
\Big],
\label{eq:2mhtch11}
\esp\eeq
and thus $\Disc_t B^{2mh} \cong  -2\pi i\,\Theta\,\delta_t B^{2mh}$, as expected.

\paragraph{Sequential discontinuities.}
Since the two-mass-hard box has four momentum channels, there are six pairs to consider as sequential discontinuities.
\begin{figure}[!t]
\begin{subfigure}[]{0.24\linewidth}
\centering
\includegraphics[keepaspectratio=true, height=2.5cm]{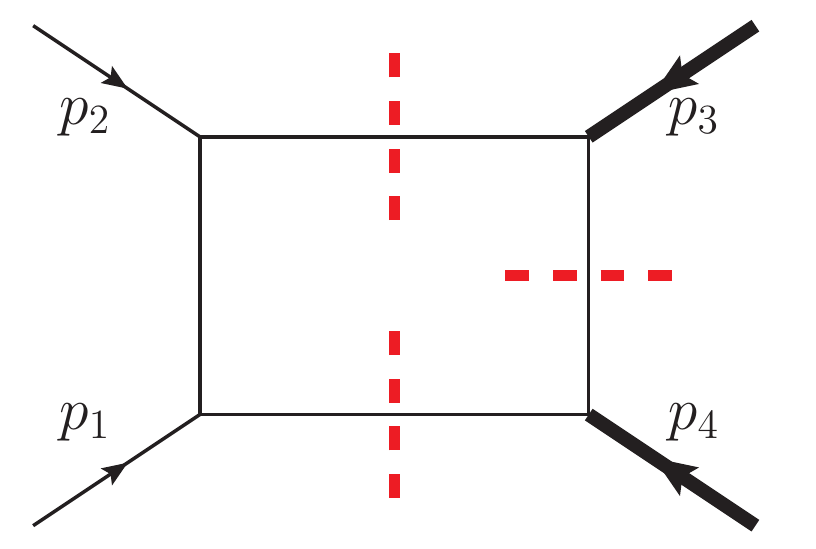}
\caption{}
\label{fig:2mhiterateda}
\end{subfigure}
\begin{subfigure}[]{0.24\linewidth}
\centering
\includegraphics[keepaspectratio=true, height=2.5cm]{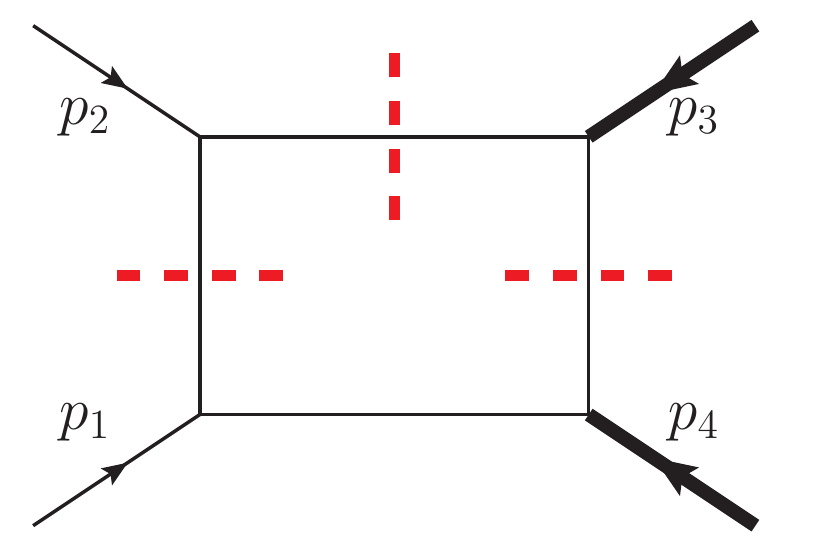}
\caption{}
\label{fig:2mhiteratedb}
\end{subfigure} 
\begin{subfigure}[]{0.24\linewidth}
\centering
\includegraphics[keepaspectratio=true, height=2.5cm]{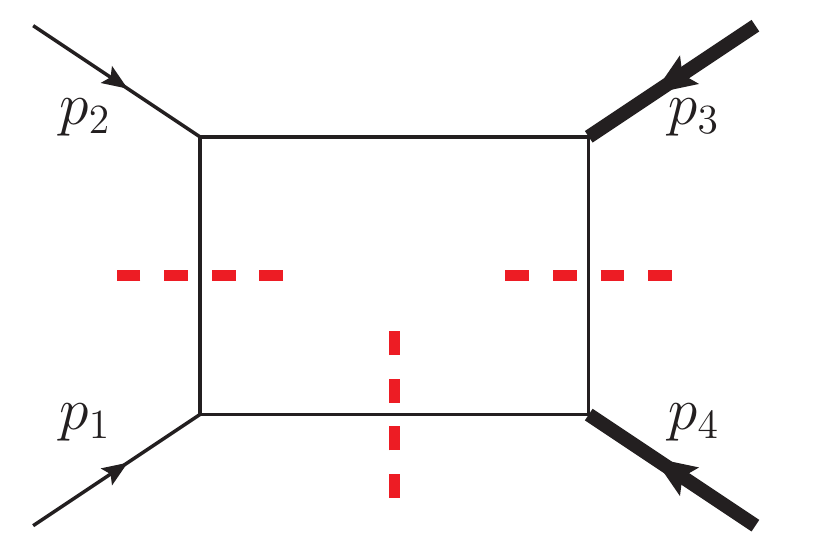}
\caption{}
\label{fig:2mhiteratedc}
\end{subfigure} 
\begin{subfigure}[]{0.24\linewidth}
\centering
\includegraphics[keepaspectratio=true, height=2.5cm]{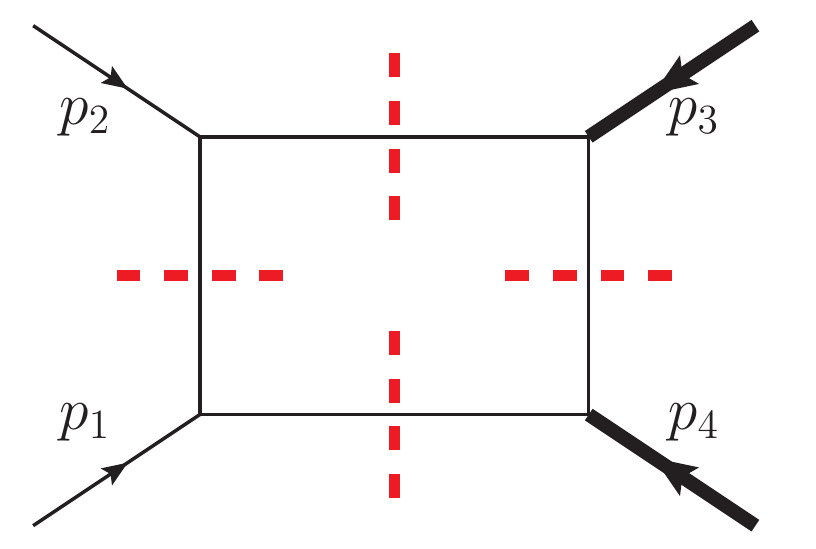}
\caption{}
\label{fig:2mhiteratedd}
\end{subfigure} 
\caption{Cut integral diagrams for sequential discontinuities of the two-mass-hard box, where legs 1 and 2 have null momenta. Here, we do not need the detailed information of physical cut channels or conjugated Feynman rules, since it makes no difference to the results. (a)  Channel pairs $(s, p_3^2), (s, p_4^2)$, or $(p_3^2, p_4^2)$.  (b) Channel pair $(t, p_3^2)$.  (c) Channel pair $(t, p_4^2)$. (d) Channel pair $(s, t)$.}
\label{fig:2mhiterated}
\end{figure}
Cutting any of the pairs of channels $(s, p_3^2), (s, p_4^2)$, or $(p_3^2, p_4^2)$ 
cuts the same set of three propagators, as shown in fig.~\ref{fig:2mhiterateda}, and gives the leading singularity.  The result of the integral (in the respective kinematic regions) is $-4\pi^2 i/(st)$, which matches the value computed from the coproduct, eq.~(\ref{eq:2mhtch11}), or the direct evaluation of discontinuities.

Let us now consider the sequential discontinuities on the channel pair $(t, p_i^2)$, where $i=3$ or $4$. For concreteness we focus on the case $i=4$. We first discuss the relation of the discontinuity to the coproduct as in eq.~(\ref{eq:discequalsdelta}); finally we will verify that the result is consistent with the iterated cut integral in the region where the latter is defined.

Specializing (\ref{eq:discequalsdelta}) to the case of interest, we have
\begin{equation}
\Disc_{t,p_4^2;R_2}B^{2mh}=\Theta\sum_{(x_1,x_2)\in\mathcal{A}_{2mh}^2}a_1(t,x_1)a_2(p_4^2,x_2)\delta_{x_1,x_2}B^{2mh}\,,
\end{equation}
where $R_1$ is the region where $t>0$ and $p_4^2<0$ and 
$R_2$ is the one where both $t>0$ and $p_4^2>0$. The relevant letters for $x_1$ can a priori be $t$ and $t-p_i^2$, however, by the first entry condition we know that $\delta_{t-p_i^2}B^{2mh}=0$ so we only need to consider $x_1=t$. We find: 
\begin{align}
a_1&(t,t)=\Disc_{t;t>0}\ln(-t-i0)=-2\pi i
\end{align}
For $x_2$ the relevant letters are $p_4^2$ and $t-p_4^2$, and both potentially contribute:
\begin{align}\bsp
a_2&(p_4^2,p_4^2)=\Disc_{p_4^2;R_2}\ln(-p_4^2+i0)=2\pi i \\
a_2&(p_4^2,t-p_4^2)=\Disc_{p_4^2;R_2}\ln(t-p_4^2+i0)=2\pi i\theta(p_4^2-t)\, .
\esp\end{align}
However using the coproduct of eq.~(\ref{eq:2mhDelta11}) we have $\delta_{t,p_4^2}B^{2mh}=0$, so we get
\begin{equation}\label{eq:disctp4b2m}
\Disc_{t,p_4^2;R_2}B^{2mh}=-(2\pi i)^2\theta(p_4^2-t)\delta_{t,t-p_4^2}B^{2mh}=-\frac{8\pi^2 i}{s t}\theta(p_4^2-t)\, .
\end{equation}

Next consider the sequential discontinuity in the reverse order: 
\begin{equation}
\Disc_{p_4^2,t;R_2}B^{2mh}=\Theta\sum_{(x_1,x_2)\in\mathcal{A}_{2mh}^2}a_1(p_4^2,x_1)a_2(t,x_2)\delta_{x_1,x_2}B^{2mh}\,,
\end{equation}
where $R_1$ is now the region where $p_4^2>0$ and $t<0$ and 
$R_2$ is the one where both $t>0$ and $p_4^2>0$.
Taking into account the first entry condition, there is only one relevant letter for the first discontinuity: $x_1=p_4^2$, and we find: 
\begin{align}
a_1&(p_4^2,p_4^2)=\Disc_{p_4^2;p_4^2>0}\ln(-p_4^2-i0)=-2\pi i \,.
\end{align}
For the second discontinuity $x_2$ can either be $t$ or $t-p_i^2$ for $i=3$ and $4$. 
\begin{align}\bsp
a_2&(t,t)=\Disc_{t;R_2}\ln(-t+i0)=2\pi i \\
a_2&(t,t-p_i^2)=\Disc_{t;R_2}\ln(p_i^2-t+i0)=2\pi i\,\theta(t-p_i^2)\, .
\esp\end{align}
Using the coproduct component in \refE{eq:2mhDelta11} we find that $\delta_{p_4^2,t-p_3^2}B^{2mh}$ vanishes and we finally obtain:
\begin{equation}\label{eq:discp4tb2m}
\Disc_{p_4^2,t;R_2}B^{2mh}=-(2\pi i)^2\left(\delta_{p_4^2,t}B^{2mh}+\delta_{p_4^2,t-p_4^2}B^{2mh}\right)=-\frac{8\pi^2 i}{s t}\theta(p_4^2-t)\, .
\end{equation}
We thus obtain the same result irrespectively of the order in which the two discontinuities are taken,
\begin{align}
\label{t_p4_commutative}
\Disc_{t,p_4^2}B^{2mh}=\Disc_{p_4^2,t}B^{2mh}\, .
\end{align}

Consider now the cut diagrams in the channel pair $(t, p_4^2)$. This cut  involves an on-shell massless three-point vertex, as shown in fig.~\ref{fig:2mhiterated}, diagrams (b) and (c). It is well known that this vertex, considered in real Minkowski space, requires collinear momenta.  Let us see how this property manifests itself in the computation of the cut integral.  Parametrize the loop momentum in fig.~\ref{fig:2mhiteratedb} by $\ell = x p_1 + y p_2 + w \vec{q}$, where  $q$ is integrated over all values satisfying $q^2 = -1$ and  $q \cdot p_i=0$ for $i=1,2$.  Then
\begin{align*}
\int d^D\ell~\delta(\ell^2)~\delta((\ell-p_2)^2) ~ f(\ell)
=& \int \frac{s}{2}~ dx~ dy~ w^{D-3} dw~d\Omega^{D-3}~\delta(xys-w^2)
\\&\delta((x(y-1)s-w^2)~f(\ell) 
\\=& \frac{1}{4} \int dy~ d\Omega^{D-3}~dw~\delta(w)~w^{D-4}~f(\ell). 
\end{align*}
The delta functions set $x=w=0$, so that $\ell = y p_2$, which is the familiar collinearity condition.  
For\footnote{We comment that for $D=4$ exactly one obtains a finite result for the double cut integral: it yields the leading singularity, $-4\pi^2 i/(st)$.} $D>4$, as needed to regularize the infrared divergences of the integral, the integral over $w$ vanishes
\begin{equation}
\cut_{p_3^2,t}B^{2mh}=\cut_{p_4^2,t}B^{2mh}=0\, .
\end{equation}
It should be emphasized that this result is in fact only valid for $t>p_4^2$, which is consistent with real external momenta (the complimentary region is unphysical; it can only be realized for complex external momenta).

Given that the iterated cut $\cut_{p_4^2,t}$ is only defined for $t>p_4^2$, where it was shown to vanish identically, and the discontinuities (\ref{eq:disctp4b2m}) and (\ref{eq:discp4tb2m}) also vanish in that region, we have verified that
\begin{equation}
\cut_{p_4^2,t}B^{2mh}=\Disc_{p_4^2,t}B^{2mh}=-(2\pi i)^2\Theta\left(\delta_{p_4^2,t}B^{2mh}+\delta_{p_4^2,t-p_4^2}B^{2mh}\right)=0\,.
\end{equation} 
Exactly the same conclusion holds for the double discontinuity on $p_3^2$ and $t$.

Finally, a comment is due concerning the channel pair $(s,t)$. This double discontinuity is 
excluded because the cuts cross in the sense described in the cutting rules of the previous section.
Indeed the relation between $\Cut$ and $\Disc$ does not apply for crossed cuts because the second $\Disc$ operation would not have an unambiguous $i\varepsilon$ prescription.
Note however that in the coproduct, \refE{eq:2mhDelta11}, there are terms proportional to $\log(-s)\otimes\log(-t)$ and $\log(-t)\otimes\log(-s)$.  If we were to compute the cut integral, it would be zero, not only because of the on-shell three-point vertices, but also because there is no real-valued momentum solution for any box with all four propagators on shell, even in $D=4$.

We have seen that iterated cut integrals, sequential discontinuities, and entries of coproducts agree in the appropriate kinematical region also in the case of dimensionally regularized infrared-divergent Feynman integrals. An important consequence is that on-shell three-point vertices force cut integrals to vanish. Having illustrated the relations of the previous section in a variety of one-loop examples we are now ready to proceed and study a more involved two-loop example.

\section{Unitarity cuts at two loops: the three-point ladder diagram}
\label{sec_UnitarityCut}

\begin{figure}[!t]
\begin{center}
\includegraphics[keepaspectratio=true, height=3.5cm]{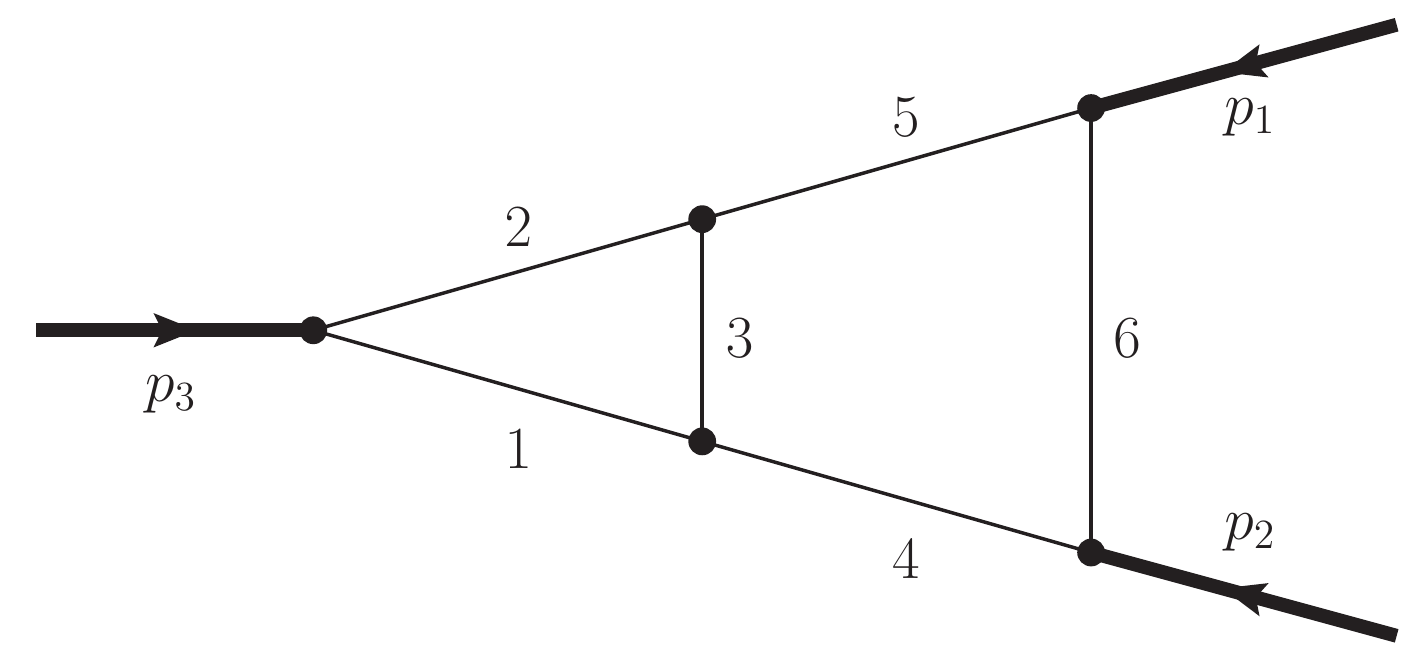}
\caption{Two-loop three-mass ladder.}
\label{twoLoop}
\end{center}
\end{figure}

The two-loop, three-point, three-mass ladder diagram with massless internal lines, \refF{twoLoop},
is finite in four dimensions \cite{Usyukina:1992jd}. In terms of the variables $z, \bz$  defined in \refE{eq:z}, it is given by a remarkably simple expression:
\beq\bsp
T_L(p_1^2,p_2^2,p_3^2)
&\,\equiv 
-\frac{i}{\pi^4}
\int d^{4}k_1\int d^{4}k_2
\,\frac{1}{k_1^2\, (p_3-k_1)^2(k_1+p_1)^2\, k_2^2 \,(p_3-k_2)^2 (k_1-k_2)^2}
\\
&\,=i\left(p_1^2\right)^{-2}
\frac{1}{(1-z)(1-\zbar)(z-\zbar)} F(z,\zbar)\label{ladderFunc}\,,
\esp\eeq
where we have defined the pure function
\begin{align}\bsp
F(z,\zbar)=&6\big[\Li_4\left(z\right)-\Li_4(\zbar)\big] 
-3\ln\left(z \zbar\right)\big[\Li_3\left(z\right)-\Li_3(\zbar)\big]\\
&+\frac{1}{2}\ln^2(z \zbar)\big[\Li_2(z)-\Li_2(\zbar)\big]\,.
\label{eq:ladder-defF}
\esp\end{align}
Because the two-loop three-point ladder in four dimensions is given by weight four functions, its coproduct structure is much richer than the one-loop cases of the preceding section. Since one of our goals is to match the entries in the coproduct to the cuts of the integral, we list below for later reference all the relevant components of the coproduct, of the form $\Delta_{\underbrace{1,\ldots,1}_{k\textrm{ times}},n-k}$.
We have 
\beq\bsp
\Delta_{1,3}(F(z,\zbar))&\,=\ln (z\zbar)\otimes[-3\,\Li_3(z)+3\,\Li_3(\zbar)+\ln (z\zbar)\, \left(\Li_2(z)-\Li_2(\zbar)\right)]\\
&\,+\ln ((1-z)(1-\zbar))\otimes\frac{1}{2}\ln z\ln\zbar\ln\frac{z}{\zbar}\,,\label{delta13}
\esp\eeq
\beq\bsp
\Delta_{1,1,2}(F(z,\zbar))
&=\ln ((1-z)(1-\zbar))\otimes\ln z\otimes\left(\ln z\ln\zbar-\frac{1}{2}\ln^2\zbar\right)\\
&-\ln ((1-z)(1-\zbar))\otimes\ln\zbar\otimes\left(\ln z\ln\zbar-\frac{1}{2}\ln^2z\right)\\
&-\ln (z\zbar)\otimes\ln(1-z)\otimes\left(\ln z\ln\zbar-\frac{1}{2}\ln^2z\right)\\
&+\ln (z\zbar)\otimes\ln(1-\zbar)\otimes\left(\ln z\ln\zbar-\frac{1}{2}\ln^2\zbar\right)\\
&+\ln (z\zbar)\otimes\ln(z \zbar)\otimes\left[\Li_2(z)-\Li_2(\zbar)\right]\,,
\label{delta112}
\esp\eeq
\beq\bsp
\Delta_{1,1,1,1}&(F(z,\zbar))=\ln ((1-z)(1-\zbar))\otimes\ln\frac{z}{\zbar}\otimes\left(\ln\zbar\otimes\ln z+\ln z\otimes\ln\zbar\right)\\
&-\ln ((1-z)(1-\zbar))\otimes\ln z \otimes\ln\zbar\otimes\ln\zbar\\
&+\ln ((1-z)(1-\zbar))\otimes\ln \zbar\otimes\ln z\otimes\ln z\\
&+\ln (z\zbar)\otimes\ln\frac{1-\zbar}{1-z}\otimes\ln z\otimes\ln\zbar+
\ln (z\zbar)\otimes\ln\frac{1-\zbar}{1-z}\otimes\ln\zbar\otimes\ln z\\
&-\ln (z\zbar)\otimes\ln(1-\zbar)\otimes\ln\zbar\otimes\ln\zbar
+\ln (z\zbar)\otimes\ln(1-z)\otimes\ln z\otimes\ln z\\
&-\ln (z\zbar)\otimes\ln(z \zbar)\otimes\ln(1-z)\otimes\ln z
+\ln (z\zbar)\otimes\ln(z \zbar)\otimes\ln(1-\zbar)\otimes\ln\zbar\,.
\label{eq_delta1111F}
\esp\eeq
Notice that the first entry of $\Delta_{1,1,1,1}$ is (the logarithm of) a Mandelstam invariant, in agreement with the first entry condition.


In the rest of this section we evaluate the standard unitarity cuts of the ladder graph of \refF{twoLoop}, which give the discontinuities across branch cuts of Mandelstam invariants in the time-like region.
Our goal is, first,  to relate these cuts to specific terms of $\Delta_{1,3}$ of $T_L(p_1^2,p_2^2,p_3^2)$, and, in the following section, to take cuts of these cuts and relate them to $\Delta_{1,1,2}$.

In contrast to the one-loop case, individual cut diagrams are infrared divergent. Again, we choose to use dimensional regularization. 
Even though $T_L(p_1^2,p_2^2,p_3^2)$ is finite in $D=4$ dimensions, its unitarity cuts need to be computed in $D=4-2\epsilon$ dimensions.
The finiteness of $T_L(p_1^2,p_2^2,p_3^2)$ for $\epsilon=0$ imposes cancellations between $\epsilon$-poles of individual cut diagrams.
These cancellations can be understood in the same way as the cancellation of infrared singularities between real and virtual
corrections in scattering cross sections.

The cut diagrams will be computed in the region $R^*_\triangle$, where $\zbar=z^*$ and all the Mandelstam invariants are timelike. 
This restriction is consistent with the physical
picture of amplitudes having branch cuts in the timelike region of their invariants. 
When comparing the results of cuts with $\delta$, but particularly with $\Disc$, we will be careful to analytically continue our result to the region where
only the cut invariant is positive, as this is where $\Cut$ is to be compared with $\Disc$.

Before we start computing the cut integrals, we briefly outline our approach to these calculations.
We will compute the cuts of this two-loop diagram by integrating first over a carefully chosen one-loop subdiagram,
with a carefully chosen parametrization of the internal propagators.
 We make our choices according to the following rules, which were designed
to simplify the calculations as much as possible:
\begin{itemize}
\item Always work in the center of mass frame of the cut channel $p_i^2$.  The momentum $p_i$ is taken to have positive energy.  
\item The routing of the loop momentum $k_1$ is such that $k_1$ is the momentum of a propagator, and there is either a propagator with momentum $(p_i-k_1)$ or a subdiagram with $(p_i-k_1)^2$ as one of its Mandelstam invariants.
\item The propagator with momentum $k_1$ is always cut.
\item Whenever possible, the propagator with momentum $(p_i-k_1)$ is cut.
\item Subdiagrams are chosen so to avoid the square root of the K\"all\'en function
as their leading singularity. This is always possible for this ladder diagram.
\end{itemize}
These rules, together with the parametrization of the momenta
\begin{align}\bsp
&p_i = \sqrt{p_i^2}(1,0,{\bf 0}_{D-2}), \qquad p_j = \sqrt{p_j^2}\left(\alpha,\sqrt{\alpha^2-1},{\bf 0}_{D-2}\right),
\\& k_1 =  (k_{1,0},|k_1| \cos \th,|k_1|\sin\th ~{\bf 1}_{D-2}),
\esp\label{paramMom}
\end{align}
where $\th \in [0,\pi]$, $|k_1|>0$, and ${\bf 1}_{D-2}$ ranges over unit vectors in the dimensions transverse to $p_i$ and $p_j$,
make the calculation of these cuts particularly simple. It is easy to show that
\begin{equation}
\alpha \sqrt{p_i^2} \sqrt{p_j^2}=\frac{p_l^2-p_i^2-p_j^2}{2} {\rm~~~~and~~~~} \sqrt{p_i^2} \sqrt{p_j^2}\sqrt{\alpha^2-1}=\frac{1}{2}\sqrt{\lambda(p_i^2,p_j^2,p_l^2)}\,.
\end{equation}
The changes of variables
\begin{equation}
\cos \th=2x-1 {\rm~~and~~} k_{1,0}=\frac{\sqrt{p_i^2}}{2}y\,,
\label{changeVarXY}
\end{equation}
are also useful (the $y$ variable is useful mainly when $(p_i-k_1)$ is not cut).

\subsection{Unitarity cut in the $p_3^2$ channel}
\label{sec:p3UnitarityCuts}

We present the computation of the cuts in the $p_3^2$ channel in some detail, in order to illustrate  our techniques for the evaluation of cut diagrams outlined above. We follow the conventions of appendix~\ref{app:conventions}. We then collect the different contributions and check the cancellation of divergent pieces and the agreement
with the term $\delta_{1-z}F(z,\zbar)$ in \refE{delta13} as predicted in \refE{C3}.

There are four cuts contributing to this channel,
\begin{equation}
\cut_{p^2_3}T_{L}(p_1^2,p_2^2,p_3^2)=\Big(\cut_{p^2_3,[45]}+\cut_{p^2_3,[12]}
+\cut_{p^2_3,[234]}+\cut_{p^2_3,[135]}\Big)T_{L}(p_1^2,p_2^2,p_3^2)\,,
\end{equation}
and our aim is to show that 
\beq\bsp
\cut_{p^2_3}T_{L}(p_1^2,p_2^2,p_3^2)&\,=-\Disc_{p_3^2}T_{L}(p_1^2,p_2^2,p_3^2)\\
&\, \cong 
-\,
\Theta
\,\frac{2\pi}{p_1^4}
\frac{1}{(1-z)(1-\zbar)(z-\zbar)}\delta_{1-z}F(z,\zbar).
\esp\eeq

\paragraph{Two-particle cuts.}

\begin{figure}[]
\begin{subfigure}[]{0.5\linewidth}
\centering
\includegraphics[keepaspectratio=true, height=3.5cm]{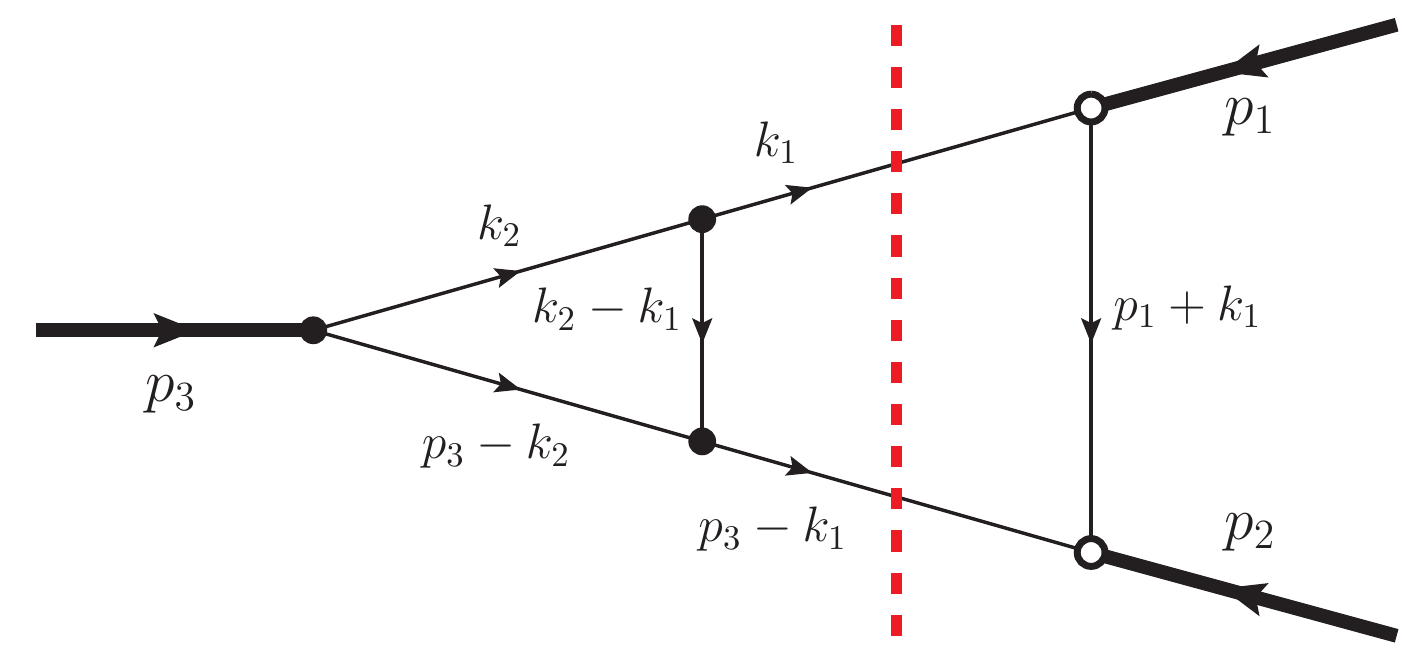}
\caption{Cut [45]}
\label{cut45Diag} 
\end{subfigure}
\begin{subfigure}[]{0.5\linewidth}
\centering
\includegraphics[keepaspectratio=true, height=3.5cm]{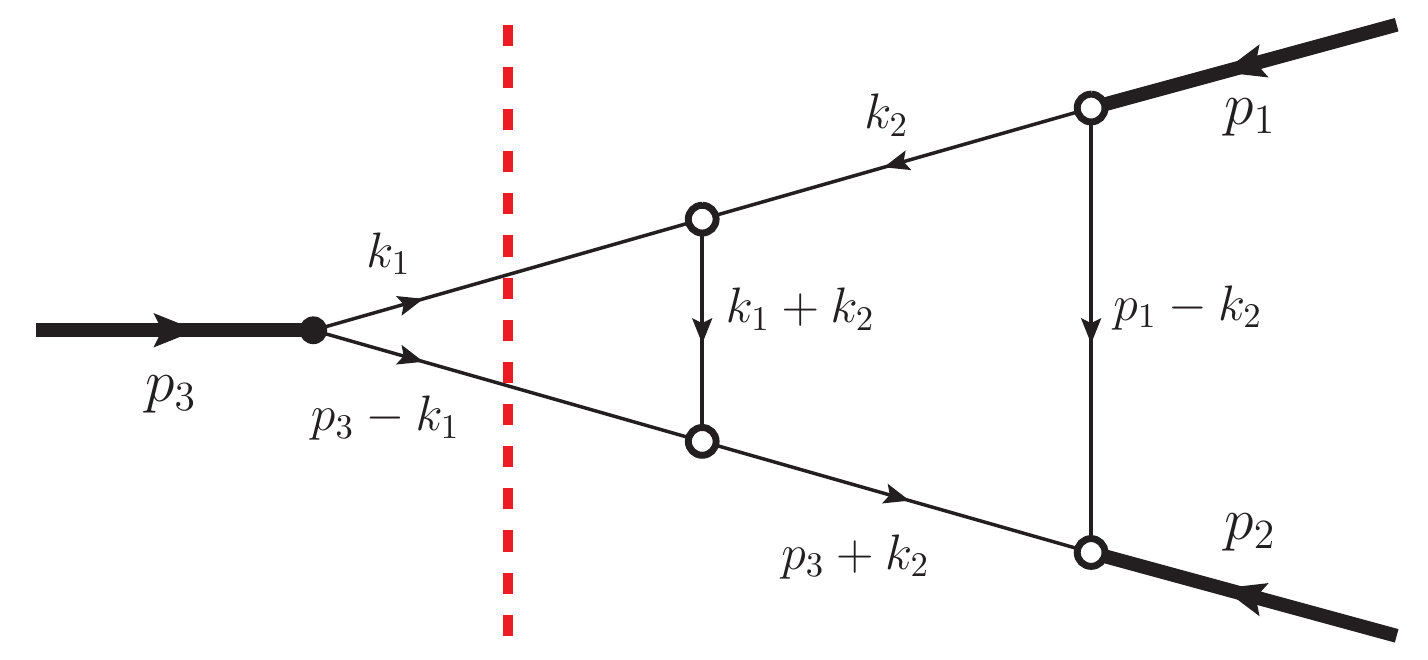}
\caption{Cut [12]}
\label{cut12Diag} 
\end{subfigure}  
\caption{Two-particle cuts in the $p^2_3$-channel.}
\label{twoPartCutsS}
\end{figure}

There are two two-particle cut diagrams contributing to the $p_3^2$-channel unitarity cut, $\cut_{p^2_3,[45]}T_{L}(p_1^2,p_2^2,p_3^2)$ and $\cut_{p^2_3,[12]}T_{L}(p_1^2,p_2^2,p_3^2)$, shown in
\refF{twoPartCutsS}.

We start by considering the diagram in \refF{cut45Diag}, which is very simple to compute because the cut completely factorizes the two loop momentum integrations into a one-mass triangle and the cut of a three-mass triangle:
\beq\bsp
\cut_{p^2_3,[45],R^*_\triangle}T_{L}(p_1^2,p_2^2,p_3^2)
=&-i\left[ \frac{e^{\gamma_E \epsilon}}{\pi^{2-\epsilon}} \int d^{4-2\epsilon}k_1(2\pi )^2\frac{\delta\left(k_1^2\right)\delta\left((p_3-k_1)^2\right)}{(k_1+p_1)^2-i\varepsilon}\right]\\
& \times\Bigg[\frac{e^{\gamma_E \epsilon}}{\pi^{2-\epsilon}}  \int d^{4-2\epsilon}k_2\frac{1}{k_2^2+i\varepsilon}\frac{1}{(p_3-k_2)^2+i\varepsilon}\frac{1}{(k_1-k_2)^2+i\varepsilon}\Bigg] \\
=& i\,T^{1m}(p_3^2)\, \cut_{p_3^2,R^*_\triangle}T(p_1^2,p_2^2,p_3^2)\,.
\esp\eeq

We substitute the following expressions for the one-loop integrals, which we have compiled in appendix~\ref{app_oneLoopRes},
\begin{align*}
\cut_{p_3^2,R^*_\triangle}T(p_1^2,p_2^2,p_3^2)=-2\pi \frac{e^{\gamma_E\epsilon}\Gamma(1-\epsilon)}{\Gamma(2-2\epsilon)}
(p_1^2)^{-1-\epsilon}\frac{u_3^{-\epsilon}}{z}\,_2F_1\left(1,1-\epsilon;2-2\epsilon;\frac{z-\zbar}{z}\right),
\end{align*}
\begin{align*}
T^{1m}(p_3^2)=ic_\Gamma\frac{1}{\epsilon^2}(-p_1^2)^{-1-\epsilon}\alpha^{-1-\epsilon}_{13}=ic_\Gamma\frac{1}{\epsilon^2}(e^{-i \pi}p_1^2)^{-1-\epsilon}\alpha^{-1-\epsilon}_{13}\,,
\label{oneLoopOneMass}
\end{align*}
where we have used $p_1^2=p_1^2+i\varepsilon$ to correctly identify the minus sign associated with $p_1^2$ in this region where $p_1^2>0$. 
As expected, the result is divergent for $\epsilon\to0$: the origin of the divergent terms is the one-loop
one-mass triangle subdiagram. Expanding up to $\mathcal{O}(\epsilon)$, we get
\beq\bsp
\cut_{p^2_3,[45],R^*_\triangle}&T_{L}(p_1^2,p_2^2,p_3^2)\\
&=i\frac{(p_1^2)^{-2-2\epsilon}}{(1-z)(1-\zbar)(z-\zbar)}\left\{\frac{1}{\epsilon^2}f_{[45]}^{(-2)}(z,\zbar)
+\frac{1}{\epsilon}f_{[45]}^{(-1)}(z,\zbar)+f_{[45]}^{(0)}(z,\zbar)\right\}
+\mathcal{O}(\epsilon)\,.
\esp\eeq
Expressions for the coefficients $f^{(i)}_{[45]}(z,\zbar)$ are given in appendix~\ref{singleCutRes}.

We now go on to \refF{cut12Diag}. 
We can see diagrammatically that the integration over $k_2$ is the (complex-conjugated) two-mass-hard box 
we have already studied in~\refS{sec_twoMassHard}, with
masses $p_1^2$ and $p_2^2$.
More precisely, we have
\beq\bsp
\cut_{p^2_3,[12],R^*_\triangle}&T_{L}(p_1^2,p_2^2,p_3^2)\\
&\,=\frac{e^{\gamma_E \epsilon}}{\pi^{2-\epsilon}}i\int d^{4-2\epsilon}k_1(2\pi)^2\delta\left(k_1^2\right)
\delta\left((p_3-k_1)^2\right)B^{2mh\dagger}(p_1^2,p_2^2;p_3^2,(p_1+k_1)^2)\,.
\label{temp1865}
\esp\eeq
 To proceed, we parametrize the momenta as in \refE{paramMom}, with $(i,j)=(3,1)$.
Then, we rewrite the momentum integration as
\beq\bsp\nonumber
 \frac{e^{\gamma_E \epsilon}}{\pi^{2-\epsilon}}\int d^{4-2\epsilon}k_1&(2\pi)^2\delta\left(k_1^2\right)
\delta\left((p_3-k_1)^2\right)=\\
=&\frac{4\pi }{\Gamma(1-\epsilon)}\int dk_{1,0}\int d\abs{\bold{k}_1}^2
\abs{\bold{k}_1}^{1-2\epsilon}\delta(k_{1,0}^2-\abs{\bold{k}_1}^2)\delta(p_3^2-2p_3\cdot k_1)\\
&\int_{-1}^1 d\cos\theta (1-\cos^2\theta)^{-\epsilon}\,.
\esp\eeq
The two delta functions allow us to trivially perform the $k_{1,0}$ and $\abs{\bold{k}_1}$ integrations. For the remaining integral, it is useful to change variables to $\cos\theta=2x-1$, as in \refE{changeVarXY}, and we get,
\begin{equation*}
 (p_1+k_1)^2=p_1^2\left(z-x(z-\zbar)\right)\,.
\end{equation*}
We finally have
\beq\bsp
\label{expCutHard}
 \cut_{p^2_3,[12],R^*_\triangle}&T_{L}(p_1^2,p_2^2,p_3^2)=2\pi \frac{c_\Gamma}{\Gamma(1-\epsilon)}
(p_1^2)^{-2-2\epsilon}e^{-i\pi\epsilon}u_2^\epsilon u_3^{-1-2\epsilon}
\int_0^1 dx\frac{x^{-\epsilon}(1-x)^{-\epsilon}}{(z-x(z-\zbar))^{1+2\epsilon}}\\
&\times\left[\frac{1}{\epsilon^2}+2\Li_2(1-z+x(z-\zbar))+2\Li_2\left(1-\frac{z-x(z-\zbar)}{z\zbar}\right)\right]
+\mathcal{O}(\epsilon)\,.
\esp\eeq
The factor $e^{-i\pi\epsilon}$ was determined according to the $i\varepsilon$ prescription of the invariants.
After expansion in $\epsilon$, all the integrals above are simple to evaluate in terms of multiple polylogarithms.
We write this expression as:
\beq\bsp
 \cut_{p^2_3,[12],R^*_\triangle}&T_{L}(p_1^2,p_2^2,p_3^2)\\
 &=i\frac{(p_1^2)^{-2-2\epsilon}}{(1-z)(1-\zbar)(z-\zbar)}\left\{\frac{1}{\epsilon^2}f_{[12]}^{(-2)}(z,\zbar)
+\frac{1}{\epsilon}f_{[12]}^{(-1)}(z,\zbar)+f_{[12]}^{(0)}(z,\zbar)\right\}+\mathcal{O}(\epsilon)\,,
\esp\eeq
and give the expressions for the coefficients $f_{[12]}^{(i)}(z,\zbar)$ in  appendix~\ref{singleCutRes}.

\paragraph{Three-particle cuts.}

\begin{figure}[]
\begin{subfigure}[]{0.5\linewidth}
\centering
\includegraphics[keepaspectratio=true, height=3.5cm]{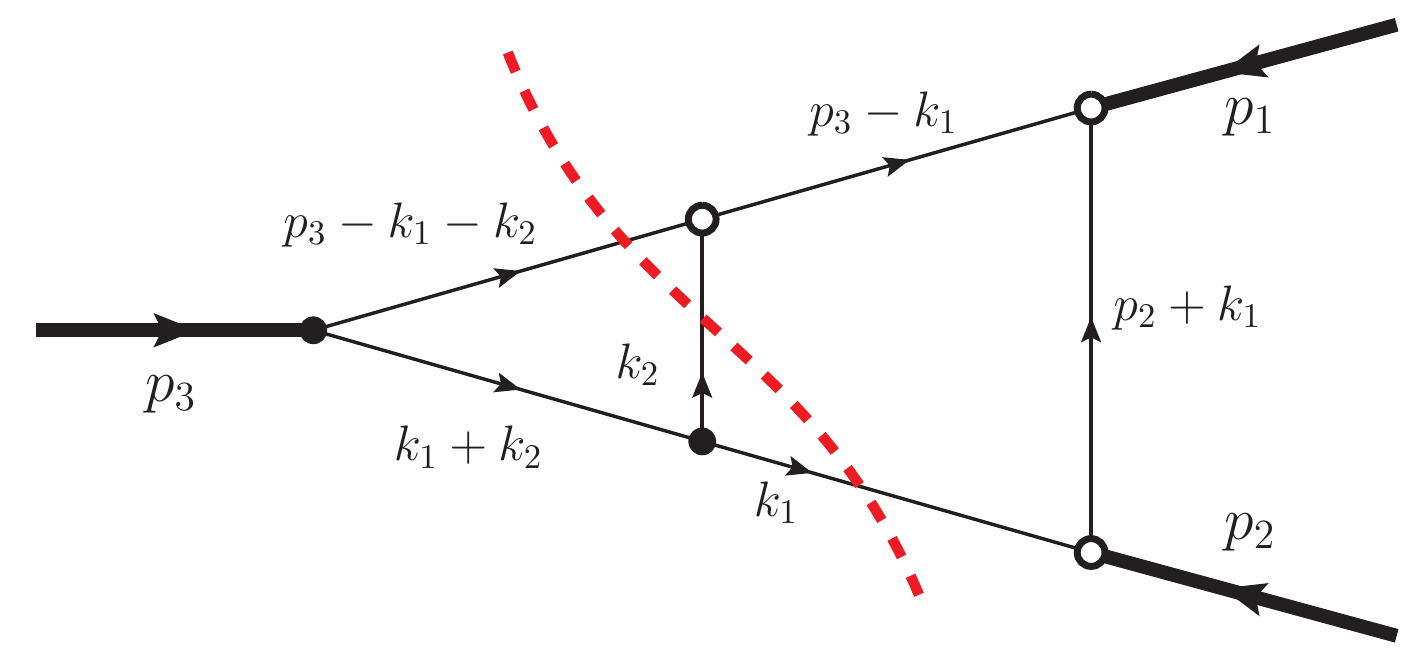}
\caption{Cut [234]}
\label{cut234Diag} 
\end{subfigure}
\begin{subfigure}[]{0.5\linewidth}
\centering
\includegraphics[keepaspectratio=true, height=3.5cm]{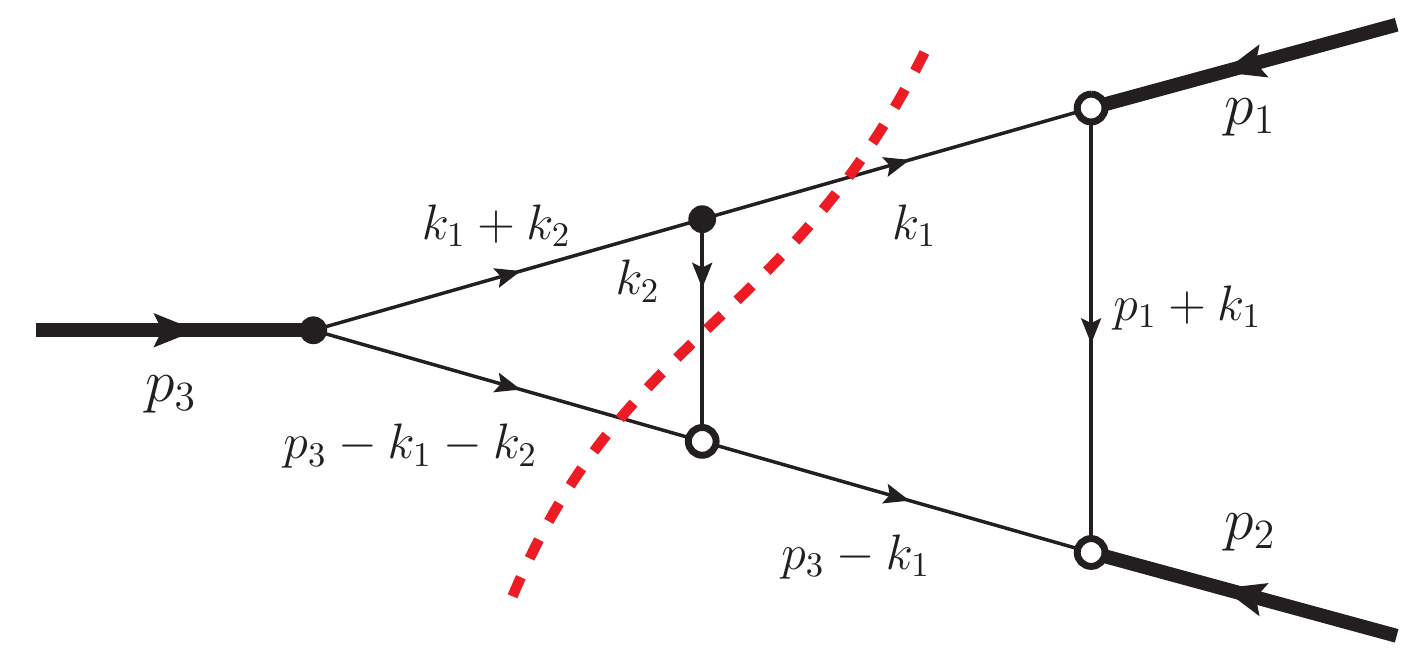}
\caption{Cut [135]}
\label{cut135Diag} 
\end{subfigure} 
\caption{Three-particle cuts in the $p^2_3$-channel.}
\label{threePartCutsS}
\end{figure}

There are two three-particle cut diagrams that contribute to the $p_3^2$-channel unitarity cut, $\cut_{p^2_3,[234]}T_{L}(p_1^2,p_2^2,p_3^2)$ and $\cut_{p^2_3,[135]}T_{L}(p_1^2,p_2^2,p_3^2)$, shown in \refF{threePartCutsS}. 
As these two cuts are very similar, we only present the details for the computation of the cut in \refF{cut234Diag}, and simply quote the result for \refF{cut135Diag}. In both cases, we note that the integration over $k_2$ is the cut in the $(p_3-k_1)^2$-channel of a two-mass one-loop triangle,
with masses $p_3^2$ and $(p_3-k_1)^2$. More precisely, for the cut in \refF{cut234Diag} we have
\beq\bsp
 \cut_{p^2_3,[234],R^*_\triangle}T_{L}(p_1^2,p_2^2,p_3^2)
=&\frac{e^{\gamma_E \epsilon}}{\pi^{2-\epsilon}}\int d^{4-2\epsilon}k_1(2\pi)\delta\left(k_1^2\right)
\frac{1}{(p_2+k_1)^2-i\varepsilon}\frac{1}{(p_3-k_1)^2-i\varepsilon}\\
\times&\cut_{(p_3-k_1)^2}T^{2m}(p_3^2,(p_3-k_1)^2) \label{temp99}\,.
\esp\eeq
We take the result for the cut of the two mass triangle given in appendix~\ref{app_oneLoopRes} and insert it into \refE{temp99},
\beq\bsp
\cut_{p^2_3,[234],R^*_\triangle}T_{L}(p_1^2,p_2^2,p_3^2)&\,=(2\pi)^2\frac{e^{2\gamma_E \epsilon}}{\pi^{2-\epsilon}}
\frac{\Gamma(1-\epsilon)}{\epsilon\Gamma(1-2\epsilon)}
\int d^{4-2\epsilon}k_1\,
\frac{\delta\left(k_1^2\right)}{(p_2+k_1)^2}\frac{\theta\left((p_3-k_1)^2\right)}{2p_3\cdot k_1}\\
&\,\times\left(\frac{1}{(p_3-k_1)^2}\right)^{1+\epsilon}\theta(k_{1,0})\,,
\esp\eeq
where we have used the $\delta$-function to set $k_1^2=0$, and we have dropped the $\pm i\varepsilon$. We have included the $\theta$-functions because the cut of the two-mass triangle is only nonzero when the $(p_3-k_1)^2$-channel
is positive. It is also important to recall that the positive energy flow across the cut requires $k_{1,0}>0$, so we have included this $\theta$-function explicitly.
We use the parametrization of \refE{paramMom}, with $(i,j)=(3,2)$ and both changes of variables in \refE{changeVarXY}, since the propagator with momentum $(p_3-k_1)$ is not cut.
The two conditions imposed by the $\theta$-functions imply that
\begin{equation}
0\leq y\leq 1\,.
\end{equation}
We then get
\beq\bsp
\cut_{p^2_3,[234],R^*_\triangle}T_{L}(p_1^2,p_2^2,p_3^2)&\,=-\frac{2\pi e^{2\gamma_E\epsilon}}{\epsilon\Gamma(1-2\epsilon)}(p_1^2)^{-2-2\epsilon}u_3^{-1-2\epsilon}
\int_0^1 dx\,x^{-\epsilon}(1-x)^{-1-\epsilon}\\
&\,\times\int_0^1 dy\frac{y^{-2\epsilon}(1-y)^{-1-\epsilon}}{u_2+y\left(z(1-\zbar)-x(z-\zbar)\right)}\\
&\,=-\frac{2\pi e^{2\gamma_E\epsilon}\Gamma(1-\epsilon)}{\epsilon^2\Gamma(1-3\epsilon)}(p_1^2)^{-2-2\epsilon}\frac{u_3^{-1-2\epsilon}}{u_2}
\int_0^1 dx\,x^{-\epsilon}(1-x)^{-1-\epsilon}\\
&\,\times\,_2F_1\left(1,1-2\epsilon;1-3\epsilon;1-\frac{z-x(z-\zbar)}{u_2}\right)\,.
\esp\eeq
We can now expand the hypergeometric function into a Laurent series in $\epsilon$ using standard techniques~\cite{Huber:2007dx}, and we then perform the remaining integration order by order. As usual, we write the result in the form
\beq\bsp
\cut_{p^2_3,[234],R^*_\triangle}&T_{L}(p_1^2,p_2^2,p_3^2)\\
&=i\frac{(p_1^2)^{-2-2\epsilon}}{(1-z)(1-\zbar)(z-\zbar)}\left\{\frac{1}{\epsilon^2}f_{[234]}^{(-2)}(z,\zbar)
+\frac{1}{\epsilon}f_{[234]}^{(-1)}(z,\zbar)+f_{[234]}^{(0)}(z,\zbar)\right\}+\mathcal{O}(\epsilon)\,.
\label{expCut234}
\esp\eeq

The diagram of \refF{cut135Diag} can be calculated following exactly the same steps, the only difference being that when using the parametrization of
\refE{paramMom} we have $(i,j)=(3,1)$.
The result is
\beq\bsp
\cut_{p^2_3,[135],R^*_\triangle}&T_{L}(p_1^2,p_2^2,p_3^2)=-\frac{2\pi e^{2\gamma_E\epsilon}\Gamma(1-\epsilon)}{\epsilon^2\Gamma(1-3\epsilon)}(p_1^2)^{-2-2\epsilon}u_3^{-1-2\epsilon}
\int_0^1 dx\,x^{-\epsilon}(1-x)^{-1-\epsilon}\\
&\,\times\,_2F_1\left(1,1-2\epsilon;1-3\epsilon;1-z+x(z-\zbar)\right)\\
&\,=i\frac{(p_1^2)^{-2-2\epsilon}}{(1-z)(1-\zbar)(z-\zbar)}\left\{\frac{1}{\epsilon^2}f_{[135]}^{(-2)}(z,\zbar)
+\frac{1}{\epsilon}f_{[135]}^{(-1)}(z,\zbar)+f_{[135]}^{(0)}(z,\zbar)\right\}+\mathcal{O}(\epsilon)\,.
\label{expCut135}
\esp\eeq
Explicit expressions for the $f_{[234]}^{(i)}(z,\zbar)$ and $f_{[135]}^{(i)}(z,\zbar)$ are given in appendix~\ref{singleCutRes}.

\paragraph{Summary and discussion.}

Let us now combine the results for each $p_3^2$-channel cut diagram and compare the total with Disc and the relevant terms in the coproduct. 
We observe the sum is very simple, compared to the expressions for each of the cuts.

Note that, as imposed by the fact that the two-loop ladder is finite in four dimensions, the sum of the divergent
terms of each diagram vanishes.
In fact, this cancellation happens in a very specific way: the sum of the two-particle cuts cancels with the sum of the three-particle cuts.
If we write
\begin{equation}
\frac{f_{[45]}^{(-2)}+f_{[12]}^{(-2)}}{\epsilon^2}+\frac{f_{[45]}^{(-1)}+f_{[12]}^{(-1)}}{\epsilon}\equiv\frac{f_\text{virt}^{(-2)}}{\epsilon^2}+\frac{f_\text{virt}^{(-1)}}{\epsilon^1}\,,
\end{equation}
\begin{equation}
\frac{f_{[234]}^{(-2)}+f_{[135]}^{(-2)}}{\epsilon^2}+\frac{f_{[234]}^{(-1)}+f_{[135]}^{(-1)}}{\epsilon}\equiv\frac{f_\text{real}^{(-2)}}{\epsilon^2}+\frac{f_\text{real}^{(-1)}}{\epsilon^1}\,,
\end{equation}
then this cancellation can be written as
\begin{equation}
f_\text{virt}^{(-2)}=-f_\text{real}^{(-2)}{\rm~~and~~} f_\text{virt}^{(-1)}=-f_\text{real}^{(-1)}\,.
\end{equation}
We call the divergent contribution of two particle cuts a \emph{virtual contribution} because it is associated with divergences of loop diagrams, whereas
the divergent contribution of three particle cuts, the \emph{real contribution}, comes from integrating over a three-particle phase space. This cancellation is similar to the cancellation of infrared divergences for inclusive cross sections, although in this case we are not directly dealing with a cross section, but merely with the unitarity cuts of a single finite Feynman integral. A better understanding of these cancellations might prove useful for the general study of the infrared properties of amplitudes, and
it would thus be interesting to understand how it generalizes to other cases.

As expected, the sum of the finite terms does not cancel. We get
\begin{equation}
f_{[45]}^{(0)}(z,\zbar)+f_{[12]}^{(0)}(z,\zbar)+f_{[234]}^{(0)}(z,\zbar)+f_{[135]}^{(0)}(z,\zbar)=i\pi\ln  z \ln \zbar \ln \frac{z}{\zbar}\,.
\end{equation}
Since all divergences have cancelled, we can set $\epsilon=0$ and write the cut-derived discontinuity of the integral as
\begin{align}
\label{cutP3Res}
\cut_{p^2_3,R^*_\triangle}T_{L}(p_1^2,p_2^2,p_3^2)&=-\frac{\pi(p_1^2)^{-2}}{(1-z)(1-\zbar)(z-\zbar)}\ln z\ln\zbar\ln\frac{z}{\zbar}\,.
\end{align}
For comparison with $\Disc$, we now analytically continue this result to the region $R^3_\triangle$ where only the cut invariant is positive: $p_3^2>0$ and 
$p_1^2,p_2^2<0$. In terms of the $z$ and $\bz$ variables, the region is: $z>1>\bz>0$.  None of the functions in \refE{cutP3Res}
has a branch cut in this region, and thus there is nothing to do for the analytic continuation and the result is valid in this region as it is given above,
\begin{equation*}
\cut_{p^2_3,R^3_\triangle}T_{L}(p_1^2,p_2^2,p_3^2)=\cut_{p^2_3,R^*_\triangle}T_{L}(p_1^2,p_2^2,p_3^2)\,.
\end{equation*}
This is consistent with the expectation that the discontinuity function would be real in the region where only the cut invariant is positive \cite{tHooft:1973pz,Veltman:1994wz}.

The relations with $\Disc$ and $\delta$ are now easy to verify. As expected, we find,
\beq\bsp\label{eq:Cutp3Res}
\cut_{p^2_3,R^3_\triangle}T_{L}(p_1^2,p_2^2,p_3^2)&=
-\Disc_{p_3^2}T_{L}(p_1^2,p_2^2,p_3^2)\\
&\cong- 2\pi \left(p_1^2\right)^{-2} \Theta\dfrac{1}{(1-z)(1-\zbar)(z-\zbar)}\delta_{1-z}F(z,\zbar)\,.\\
\esp\eeq
We can write this equation diagrammatically as
\begin{align}\label{eq:diagrams_delta_u3}
\delta_{1-z}T_L(p_1^2,p_2^2,p_3^2)
=&\frac{1}{2\pi i}\left(\raisebox{-5.2mm}{\includegraphics[keepaspectratio=true, height=1.15cm]{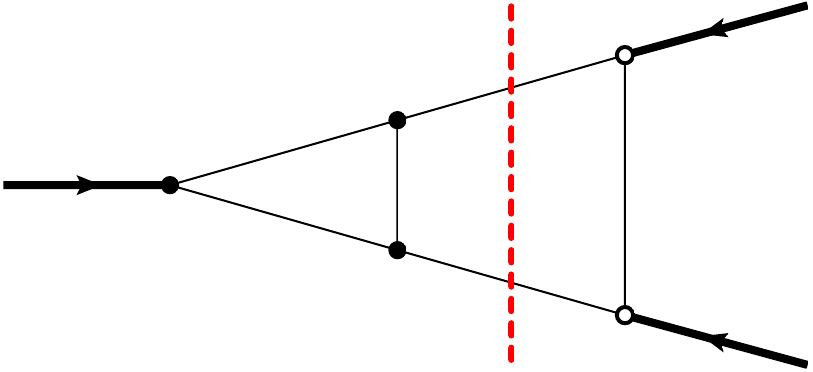}}
+\raisebox{-5.2mm}{\includegraphics[keepaspectratio=true, height=1.15cm]{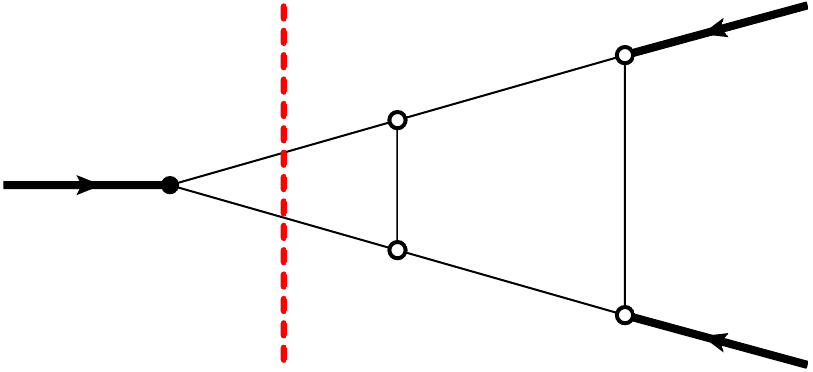}}\right.\nonumber\\
+&\left.\raisebox{-5.2mm}{\includegraphics[keepaspectratio=true, height=1.15cm]{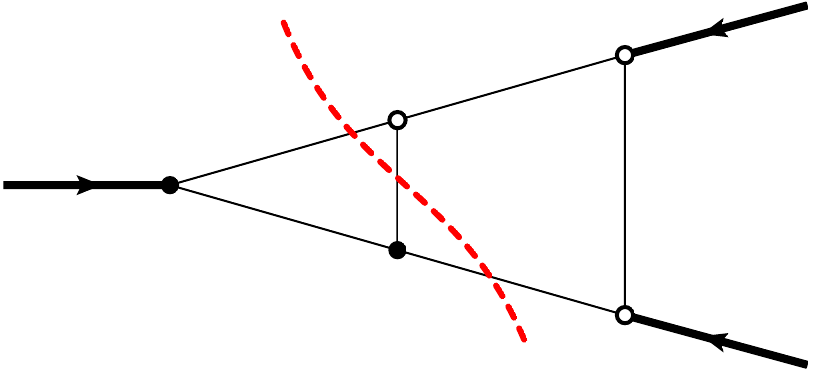}}
+\raisebox{-5.2mm}{\includegraphics[keepaspectratio=true, height=1.15cm]{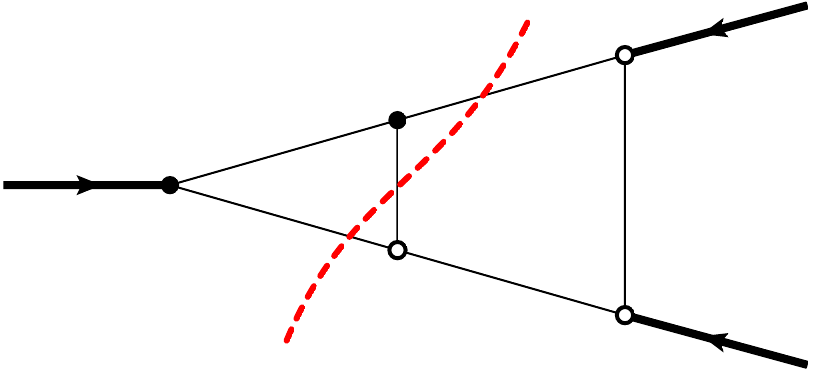}}\,\,\right)\,.\nonumber
\end{align}

\subsection{Unitarity cut in the $p_2^2$ channel}
\label{sec:p2UnitarityCuts}

\begin{figure}[!t]
\begin{subfigure}[]{0.5\linewidth}
\centering
\includegraphics[keepaspectratio=true, height=3.5cm]{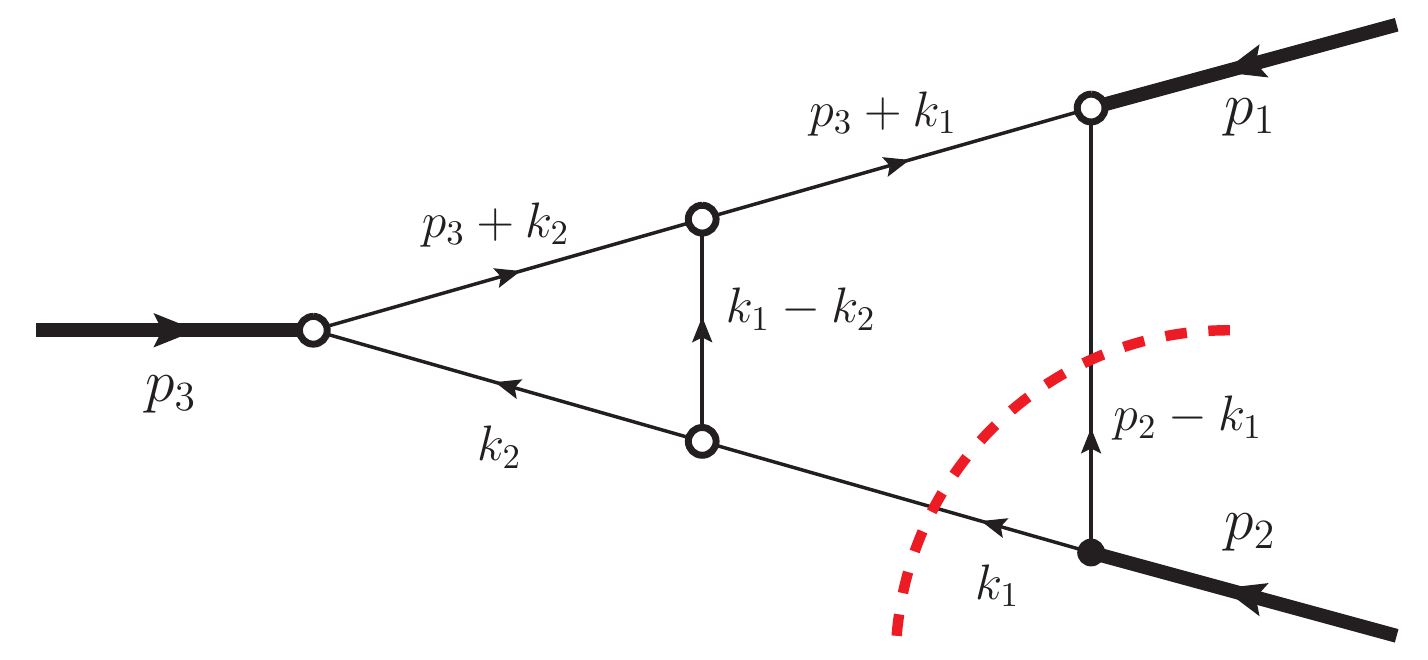}
\caption{Cut [46]}
\label{cut46Diag} 
\end{subfigure}
\begin{subfigure}[]{0.5\linewidth}
\centering
\includegraphics[keepaspectratio=true, height=3.5cm]{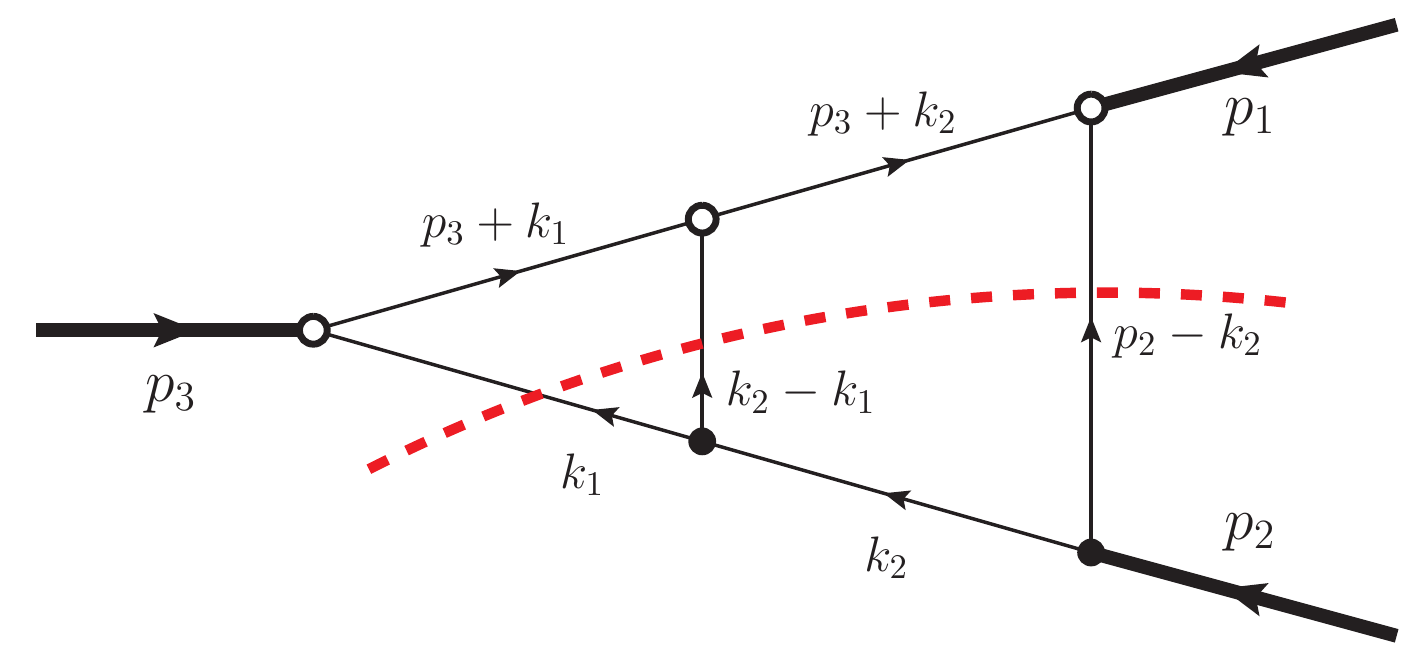}
\caption{Cut [136]}
\label{cut136Diag} 
\end{subfigure} 
\caption{Cuts in the $p^2_2$-channel}
\label{cutsP2}
\end{figure}

We now turn to the calculation of the cuts in the $p_2^2$ channel, in order to reproduce the $\delta_{z}T_L(p_1^2,p_2^2,p_3^2)$
entry of the coproduct in \refE{delta13} as in \refE{C2}. Only two cut diagrams contribute to this channel,
\begin{equation}
\cut_{p^2_2}T_{L}(p_1^2,p_2^2,p_3^2)=\Big(\cut_{p^2_2,[46]}+\cut_{p^2_2,[136]}\Big)T_{L}(p_1^2,p_2^2,p_3^2)\,.
\end{equation}
The computation of the two cuts diagrams follows the same strategy as before, i.e., we compute the cut of the two-loop diagram by integrating over a carefully chosen one-loop subdiagram.

\paragraph{Computation of the cut diagrams.}

We start by computing $\cut_{p^2_2,[46]}T_{L}(p_1^2,p_2^2,p_3^2)$. As suggested by the momentum routing in \refF{cut46Diag}, we identify the result of the $k_2$ integration with the complex
conjugate of an uncut two-mass triangle, with masses $(p_3+k_1)^2$ and $p_3^2$:
\beq\bsp
\cut_{p^2_2,[46],R^*_\triangle}&T_{L}(p_1^2,p_2^2,p_3^2)\\
&=-i\frac{e^{\gamma_E \epsilon}}{\pi^{2-\epsilon}}\int d^{4-2\epsilon}k_1 (2 \pi )^2\frac{\delta\left(k_1^2\right)\delta\left((p_2-k_1)^2\right)}{(p_3+k_1)^2-i\varepsilon}\,
T^{2m\dagger}(p_3^2,(p_3+k_1)^2)\,.
\esp\eeq
Using the result for the triangle given in appendix~\ref{app_oneLoopRes} and
proceeding in the same way as with the $p_3^2$-channel cuts, we get (setting $(i,j)=(2,3)$ in \refE{paramMom})
\begin{align}
\cut_{p^2_2,[46],R^*_\triangle}T_{L}(p_1^2,p_2^2,p_3^2)&=2\pi\frac{c_\Gamma e^{\gamma_E\epsilon}}{\epsilon^2\Gamma(1-\epsilon)}
u_2^{-\eps}e^{-i\pi\eps}(p_1^2)^{-2-2\epsilon}\int_0^1 dx\,(1-x)^{-\epsilon}x^{-\epsilon}\nonumber\\
&\times\frac{\left(u_3+z-u_2-x(z-\zbar)\right)^{-\epsilon}-u_3^{-\epsilon}}{\left(u_3+z-u_2-x(z-\zbar)\right)\left(z-u_2-x(z-\zbar)\right)}\\
&=i\frac{(p_1^2)^{-2-2\epsilon}}{(1-z)(1-\zbar)(z-\zbar)}\left\{\frac{1}{\epsilon}f_{[46]}^{(-1)}(z,\zbar)+f_{[46]}^{(0)}(z,\zbar)\right\}+\mathcal{O}(\epsilon)\nonumber\,.
\label{expCut46}
\end{align}\\

The cut integral $\cut_{p^2_2,[136]}T_{L}(p_1^2,p_2^2,p_3^2)$ is slightly more complicated.
Using the routing of loop momenta of  \refF{cut136Diag},
we look at it as the $k_1$-integration over the cut of a three-mass box,
\begin{align}
\cut_{p^2_2,[136],R^*_\triangle}T_{L}(p_1^2,p_2^2,p_3^2)
=& - \frac{e^{\gamma_E \epsilon}}{\pi^{2-\epsilon}}\int d^{4-2\epsilon}k_1\frac{2\pi\delta\left(k_1^2\right)}{(p_3+k_1)^2-i\varepsilon}\cut_tB^{3m}(l_2^2,l_3^2,l_4^2;s,t)\,,
\end{align}
where $\cut_tB^{3m}(l_2^2,l_3^2,l_4^2;s,t)$ is the $t$-channel cut of the three-mass box with masses $l_i^2$, for $i\in \{2,3,4\}$, $l_1^2=0$, $s=(l_1+l_2)^2$ and $t=(l_2+l_3)^2$.
In our case:
\begin{equation*}
l_2^2=(p_3+k_1)^2-i\varepsilon\,,\quad l_3^2=p_1^2-i\varepsilon\,,\quad l_4^2=p_2^2+i\varepsilon\,,\quad s=p_3^2-i\varepsilon\,,\quad t=(p_2-k_1)^2\,.
\end{equation*}
The result for the $t$-channel cut of the three-mass box is given in appendix~\ref{app_oneLoopRes} in the region where the uncut invariants are negative, and $t$ is positive.
Since we work in the region where all the $p_i^2$ are positive, some terms in the expression~\eqref{3mCut}
need to be analytically continued using the $\pm i\varepsilon$ prescriptions given above.
Using \refE{paramMom} with $(i,j)=(2,3)$ and introducing the variables $x$ and $y$ according to \refE{changeVarXY}, we have:\footnote{Strictly
speaking, this analytic continuation is valid for $\bz=z^*$, with $\Re(z)<1$. For the case of $\Re(z)>1$, the factors of $i\pi$ are distributed in other ways among the different terms, but the combination
of all terms is still the same.}
\beq\bsp
\ln(-s)&\,=\ln p_1^2+\ln u_3+i\pi\,,\\
(-l^2_2)^{-\epsilon}&\,=\left(e^{i\pi}p_1^2\right)^{-\epsilon}\left(u_3+y\left(z-u_2-x(z-\zbar)\right)\right)^{-\epsilon}\,,\\
(-l^2_3)^{-\epsilon}&\,=\left(e^{i\pi}p_1^2\right)^{-\epsilon}\,,\\
\ln\left(1-\frac{l_4^2}{t}\right)&\,=\ln y-\ln(1-y)-i\pi\,,\\
\ln\left(1-\frac{l_2^2l_4^2}{st}\right)&\,=\ln\left(u_3+z-u_2-x(z-\zbar)\right)+\ln y-\ln u_3-\ln(1-y)-i\pi\,.
\esp\eeq
Combining everything, $\cut_{p^2_2,[136],R^*_\triangle}T_{L}(p_1^2,p_2^2,p_3^2)$ is given by
\beq\bsp
\cut&_{p^2_2,[136],R^*_\triangle}T_{L}(p_1^2,p_2^2,p_3^2)\\
&=2\pi\frac{e^{2\gamma_{E}\epsilon}}{\Gamma(1-2\epsilon)}u_2^{-\epsilon}(p_1^2)^{-2-2\epsilon}\int_0^1 dx\,(1-x)^{-\epsilon}x^{-\epsilon}
\int_0^1 dy\,y^{-2\epsilon}\frac{1}{u_3+z-u_2-x(z-\zbar)}\\
&\times\frac{1}{u_3+y\left(z-u_2-x(z-\zbar)\right)}\Bigg[-\frac{u_2^{\epsilon}}{\epsilon}(1-y)^{\epsilon}\left(u_3+y\left(z-u_2-x(z-\zbar)\right)\right)^{-\epsilon}\Big.\\
&\Big.+\frac{2}{\epsilon}u_2^{-\epsilon}(1-y)^{-\epsilon}-2\ln\left(u_3+z-u_2-x(z-\zbar)\right)+2\ln u_2+2\ln(1-y)\Bigg]+\mathcal{O}(\epsilon)\\
&=i\frac{(p_1^2)^{-2-2\epsilon}}{(1-z)(1-\zbar)(z-\zbar)}\left\{\frac{1}{\epsilon}f_{[136]}^{(-1)}(z,\zbar)+f_{[136]}^{(0)}(z,\zbar)\right\}+\mathcal{O}(\epsilon)\,.
\esp\eeq
Explicit results for $f_{[46]}^{(i)}(z,\zbar)$ and $f_{[136]}^{(i)}(z,\zbar)$ are given in appendix~\ref{singleCutRes}.

\paragraph{Summary and discussion.}

Similarly to the $p_3^2$-channel cuts, we first analyze the cancellation of the singularities in the sum of the two cuts contributing to the $p_2^2$ channel,
and check the agreement with $\delta_{z}T_L(p_1^2,p_2^2,p_3^2)$ given in \refE{delta13}. In this case we only have single poles, and we see that the poles cancel, as expected:
\begin{equation}
f_{[46]}^{(-1)}(z,\zbar)+f_{[136]}^{(-1)}(z,\zbar)=0\,.
\end{equation}
This cancellation can again be understood as the cancellation between virtual (from cut [46]) and real contributions (from cut [136]).

Adding the finite contributions, we find
\begin{equation}
f_{[46]}^{(0)}(z,\zbar)+f_{[136]}^{(0)}(z,\zbar)=2\pi i\bigg\{3\big[\Li_3(\zbar)-\Li_3(z)\big]+\big(\ln (z\zbar)-i\pi\big)\big[\Li_2(z)-\Li_2(\zbar)\big]\bigg\}\,.
\end{equation}
Hence, the cut of the two-loop ladder in the $p_2^2$ channel is
\beq\bsp
\cut_{p^2_2,R^*_\triangle}&T_{L}(p_1^2,p_2^2,p_3^2)\\
&= -\frac{2\pi(p_1^2)^{-2}}{(1-z)(1-\zbar)(z-\zbar)}\bigg\{3\big[\Li_3(\zbar)-\Li_3(z)\big]+\big(\ln (z\zbar)-i\pi\big)\big[\Li_2(z)-\Li_2(\zbar)\big]\bigg\}\,.
\label{p2CutRes}
\esp\eeq
Since this result was computed in the region where all invariants are positive, we now analytically continue to the region $R^2_\triangle$ where $p_2^2>0$ and $p_1^2,p_3^2<0$.
For the $z$ and $\bz$ variables, this corresponds to $1>z>0>\bz$. 
The analytic continuation of the $\Li_2$ and $\Li_3$ functions is trivial, because
their branch cuts lie in the $[1,\infty)$ region of their arguments. However, the continuation of $\ln (z\zbar)$ needs to be done with some care, since $(z\zbar)$ becomes
negative. We can determine the sign of the $i\varepsilon$ associated with $(z\zbar)$ by noticing that
\begin{equation*}
\ln\left(-\frac{p_2^2}{p_1^2-i\varepsilon}\right)=\ln\left(-z\zbar-i\varepsilon\right)\,,
\end{equation*}
where we associate a $-i\varepsilon$ to $p_1^2$ because it is in the complex-conjugated region of the cut diagrams. We thus see that the
$-i \pi$ term in \refE{p2CutRes} is what we get from the analytic continuation of $\ln\left(-z\zbar-i\varepsilon\right)$ to positive $(z\zbar)$. In region $R^2_\triangle$, we thus have
\beq\bsp
\cut_{p^2_2,R^2_\triangle}&T_{L}(p_1^2,p_2^2,p_3^2)\\
&= -\frac{2\pi(p_1^2)^{-2}}{(1-z)(1-\zbar)(z-\zbar)}\bigg\{3\big[\Li_3(\zbar)-\Li_3(z)\big]+\ln(-z\zbar
-i\varepsilon
)\big[\Li_2(z)-\Li_2(\zbar)\big]\bigg\}\,.
\esp\eeq
This agrees with the expectation that the discontinuity function should be real in the region where only the cut invariant is positive \cite{tHooft:1973pz,Veltman:1994wz}.
Furthermore, we again observe the expected relations with $\Disc$ and $\delta$,
\beq\bsp\label{eq:Cutp2Res}
\cut_{p^2_2,R^2_\triangle}T_{L}(p_1^2,p_2^2,p_3^2)&=-\Disc_{p_2^2}T_{L}(p_1^2,p_2^2,p_3^2)\\
&\cong- 2 \pi \, \Theta \left(p_1^2\right)^{-2}\dfrac{1}{(1-z)(1-\zbar)(z-\zbar)}\delta_{z}F(z,\zbar)\,.
\esp\eeq
Diagrammatically, the relation can be written as follows:
\begin{align}\label{eq:diagrams_delta_u2}
\delta_{z}T_L(p_1^2,p_2^2,p_3^2)=&\frac{1}{2\pi i}
\left(\raisebox{-5.2mm}{\includegraphics[keepaspectratio=true, height=1.2cm]{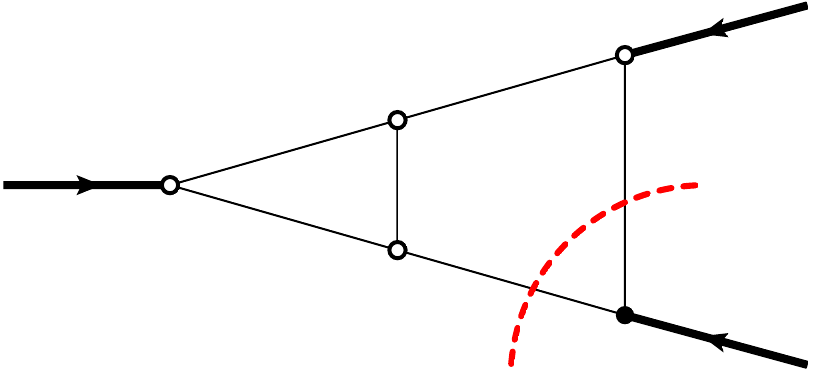}}
+\raisebox{-5.2mm}{\includegraphics[keepaspectratio=true, height=1.2cm]{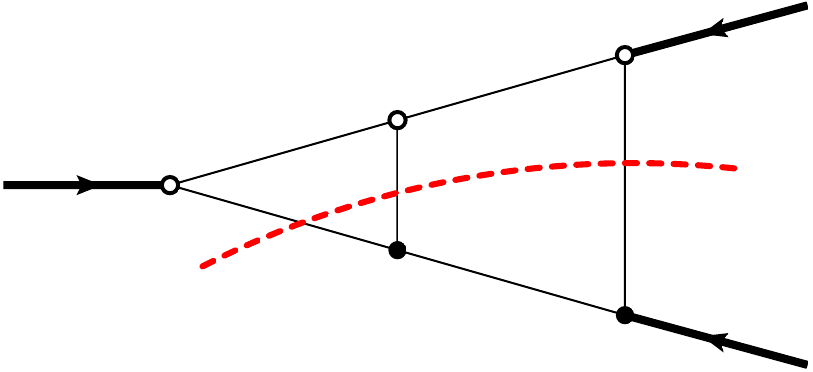}}
\,\,\right)\,.
\end{align}

\subsection{Unitarity cut in the $p_1^2$ channel}

\begin{figure}[!t]
\begin{subfigure}[]{0.5\linewidth}
\centering
\includegraphics[keepaspectratio=true, height=3.5cm]{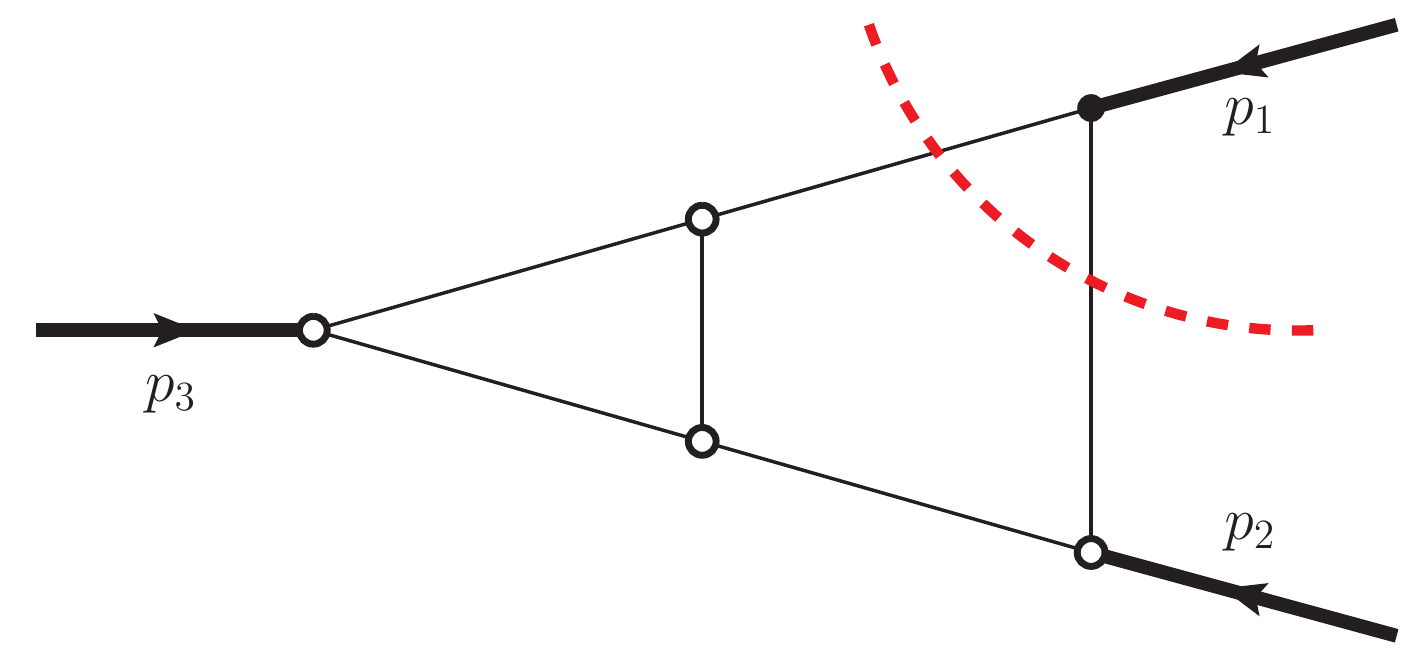}
\caption{Cut [56]}
\label{cut56Diag} 
\end{subfigure}
\begin{subfigure}[]{0.5\linewidth}
\centering
\includegraphics[keepaspectratio=true, height=3.5cm]{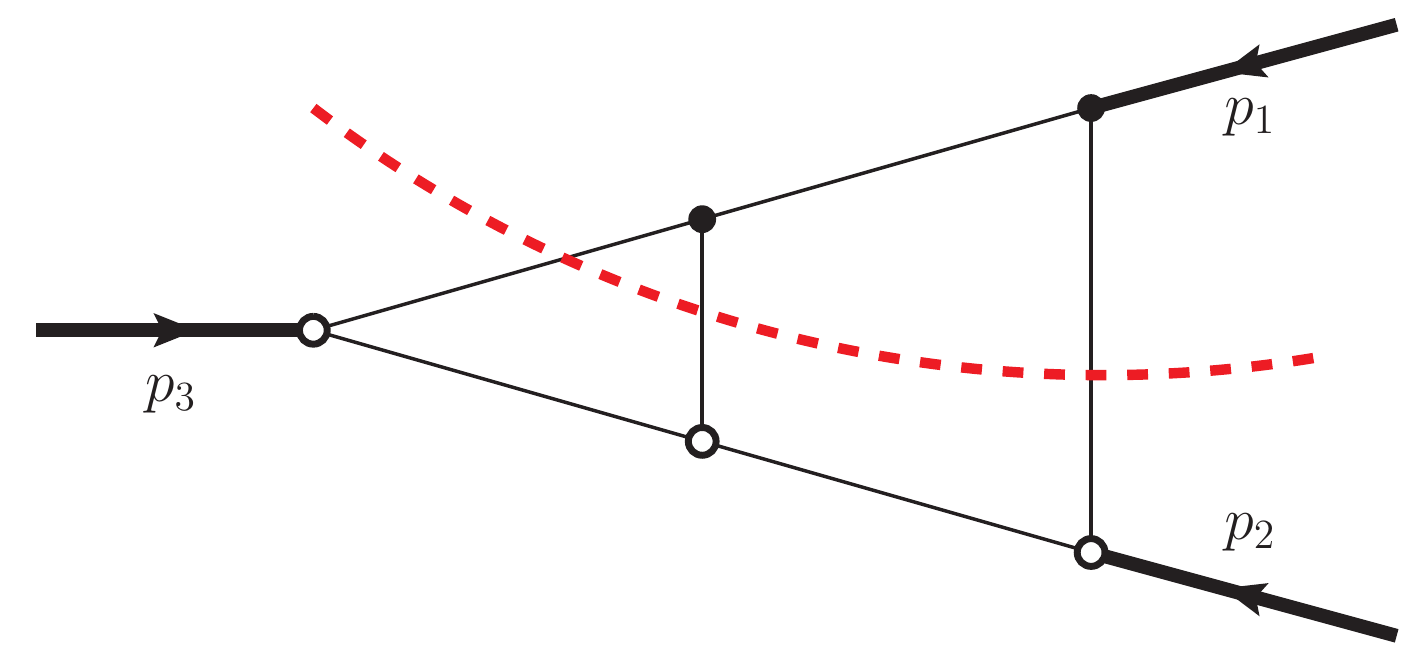}
\caption{Cut [236]}
\label{cut236Diag} 
\end{subfigure} 
\caption{Cuts in the $p^2_1$-channel}
\label{cutsP1}
\end{figure}

Given the symmetry of the three-point ladder, 
the cut in the $p_1^2$ channel shown in \refF{cutsP1} can be done in exactly the same way as the $p_2^2$ channel, so we will be brief in listing the results for completeness.

For the sum of the two cut integrals, the reflection symmetry can be implemented by exchanging $p_1$ and $p_2$ in \refE{p2CutRes}, along with transforming $z \to 1/\bz$ and $\bz \to 1/z$.  The total cut integral is then
\beq\bsp
 \cut_{p^2_1,R^*_\triangle}T_{L}(p_1^2,p_2^2,p_3^2)=& -\frac{2\pi(p_1^2)^{-2}}{(1-z)(1-\bz) (z-\bz)}\bigg\{3\big[\Li_3\left(\frac{1}{z}\right)-\Li_3\left(\frac{1}{\bz}\right)\big]\\
&-\big(\ln (z\zbar) + i\pi\big)\big[\Li_2\left(\frac{1}{\bz}\right)-\Li_2\left(\frac{1}{z}\right)\big]\bigg\} \, .
\esp\eeq
We now analytically continue $p_2^2$ and $p_3^2$ to the region $R^1_\triangle$ where we should match Disc.  In this region, we have $\bz<0$ and $z>1$.  Similarly to the previous case, we take $p_2^2-i\varepsilon$ to find that $\ln(z\zbar-i\varepsilon)  \to \ln(-z\zbar)-i\pi$, and thus
\beq\bsp
& \cut_{p^2_1,R^1_\triangle}T_{L}(p_1^2,p_2^2,p_3^2)\\
&= -\frac{2\pi(p_1^2)^{-2}}{(1-z)(1-\bz) (z-\bz)}\bigg\{3\left[\Li_3\left(\frac{1}{z}\right)-\Li_3\left(\frac{1}{\bz}\right)\right]-\ln (-z\zbar) \left[\Li_2\left(\frac{1}{\bz}\right)-\Li_2\left(\frac{1}{z}\right)\right]\bigg\} 
\label{eq:p1CutRes}
\\
&= -\Disc_{p^2_1}T_{L}(p_1^2,p_2^2,p_3^2).
\esp\eeq
In the last line, we have confirmed that the cut result agrees with a direct evaluation of the discontinuity of $T_{L}(p_1^2,p_2^2,p_3^2)$ in the region $R^1_\triangle$.

The $\delta$ discontinuity evaluated from the coproduct is simply related to the discontinuities in the $p_2^2$ and $p_3^2$ channels. 
 Indeed, we
can rewrite \refE{delta13} as
\begin{align*}
\Delta_{1,3}(F(z,\zbar))
=&\ln\left(-p_2^2\right)\otimes\delta_{p^2_2}F(z,\zbar)+\ln\left(-p_3^2\right)\otimes\delta_{p^2_3}F(z,\zbar)+\ln\left(-p_1^2\right)\otimes\delta_{p^2_1}F(z,\zbar)\,,
\end{align*}
where
\begin{equation}
\delta_{p^2_2}F(z,\zbar)=\delta_{z}F(z,\zbar)\,,\,\,\delta_{p^2_3}F(z,\zbar)=\delta_{1-z}F(z,\zbar)\,,\,\,\delta_{p^2_1}F(z,\zbar)=-\delta_{z}F(z,\zbar)-\delta_{1-z}F(z,\zbar)\,.
\end{equation}
Explicitly, 
\begin{align}\bsp
 (-2\pi i)\delta_{p_1^2}T_L &=-\frac{2\pi(p_1^2)^{-2}}{(1-z)(1-\zbar)(z-\zbar)}\bigg\{3\big[\Li_3(\zbar)-\Li_3(z)\big]
 \\&+\ln(-z\zbar)\big[\Li_2(z)-\Li_2(\zbar)\big]
+ \frac{1}{2}\ln z\ln \zbar\ln\frac{z}{\zbar}
\bigg\}\,,
\esp\end{align}
which agrees with $\Disc_{p^2_1}T_{L}$ from \refE{eq:p1CutRes} modulo $\pi^2$.

\section{Sequence of unitarity cuts}

In the previous section we gave a diagrammatic interpretation of the $\delta_{z}F(z,\zbar)$ and $\delta_{1-z}F(z,\zbar)$ terms of
\refE{delta13} as unitarity cuts in $p_2^2$ and $p_3^2$ respectively. In this section we will take sequences of two unitarity cuts as defined in section~\ref{sec:cut_def} and match the result to entries of the coproduct as predicted in eqs.~(\ref{eq:tableeqns}).

 Unlike the single unitarity cuts, which could be computed in the kinematic
region $R^*_\triangle$ where $\sqrt{\lambda}$ is imaginary and thus $\zbar=z^*$, and then analytically continued back to the region in which $\Disc$ is evaluated, the calculation of double unitarity cuts (in real kinematics) has to be done
in the region where $z,\zbar$ and $\sqrt{\lambda}$ are real in order to get a nonzero result. 
Moreover, we must work in the specific region in terms of $z$ and $\zbar$ corresponding to positive cut invariants and negative uncut invariant.

We focus on the cases of
$\Cut_{p_3^2,p_1^2}$ and $\Cut_{p_2^2,p_1^2}$. We present our method to evaluate the necessary cuts.  We check that we indeed reproduce
the expected terms of the coproduct and satisfy the relations \eqref{eq:tableeqns} that we expect, so that the relations \eqref{eq:cutequalsdisc} and \eqref{eq:discequalsdelta} among $\Disc$, $\cut$, and the coproduct components hold.  We stress that the fact that we reproduce the expected relations between $\Disc$, $\cut$ and the coproduct components is a highly nontrivial check on the consistency of the extended cutting rules of section~\ref{sec:cut_def}. In particular, we see that the restriction to real kinematics is justified. We observe that, unlike for the case of single unitarity cuts, it is insufficient to define cut diagrams only through the set of propagators that go on shell, as the results for the integrals strongly depend on the phase space boundaries which are specified by the correct choice of kinematic region.
Finally, we comment on the triple discontinuity of the ladder.

\subsection{Double unitarity cuts}
\label{sec:doubleCutCalculation}

\begin{figure}[!t]
\begin{subfigure}[]{0.5\linewidth}
\centering
\includegraphics[keepaspectratio=true, height=3.3cm]{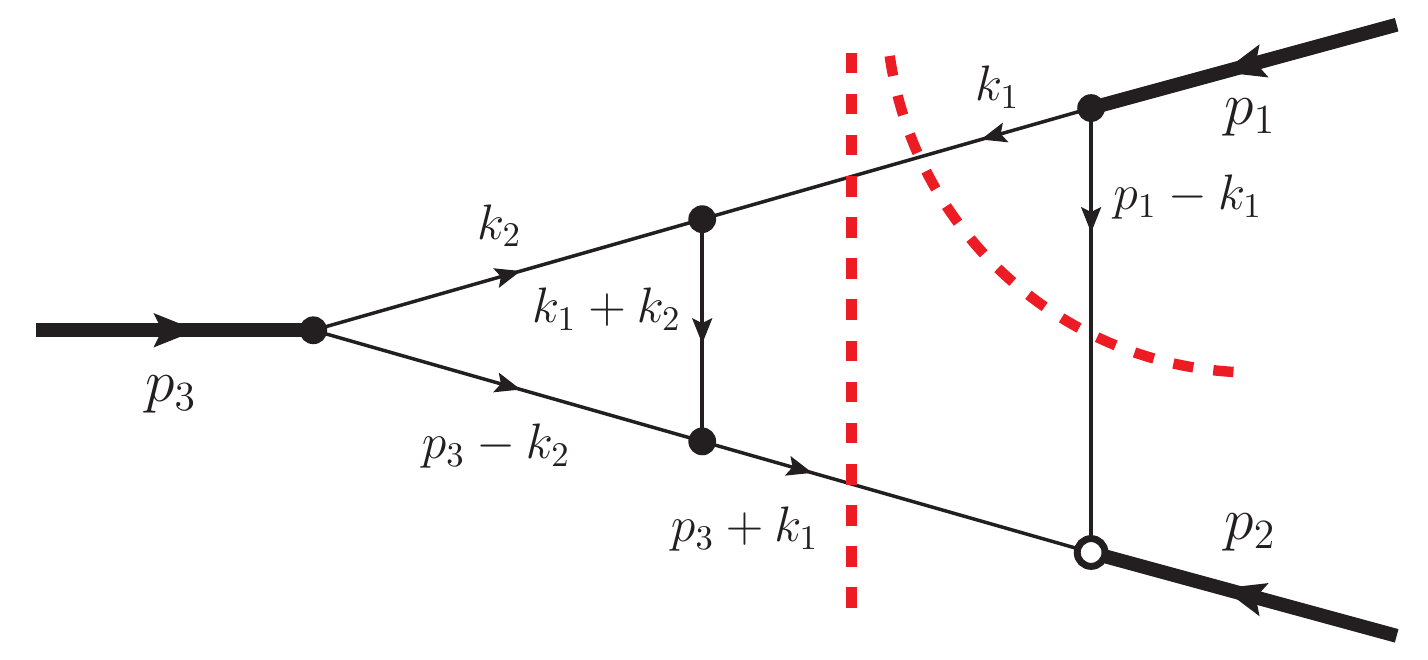}
\caption{Cut [456]}
\label{cut456Diag} 
\end{subfigure}
\begin{subfigure}[]{0.5\linewidth}
\centering
\includegraphics[keepaspectratio=true, height=3.3cm]{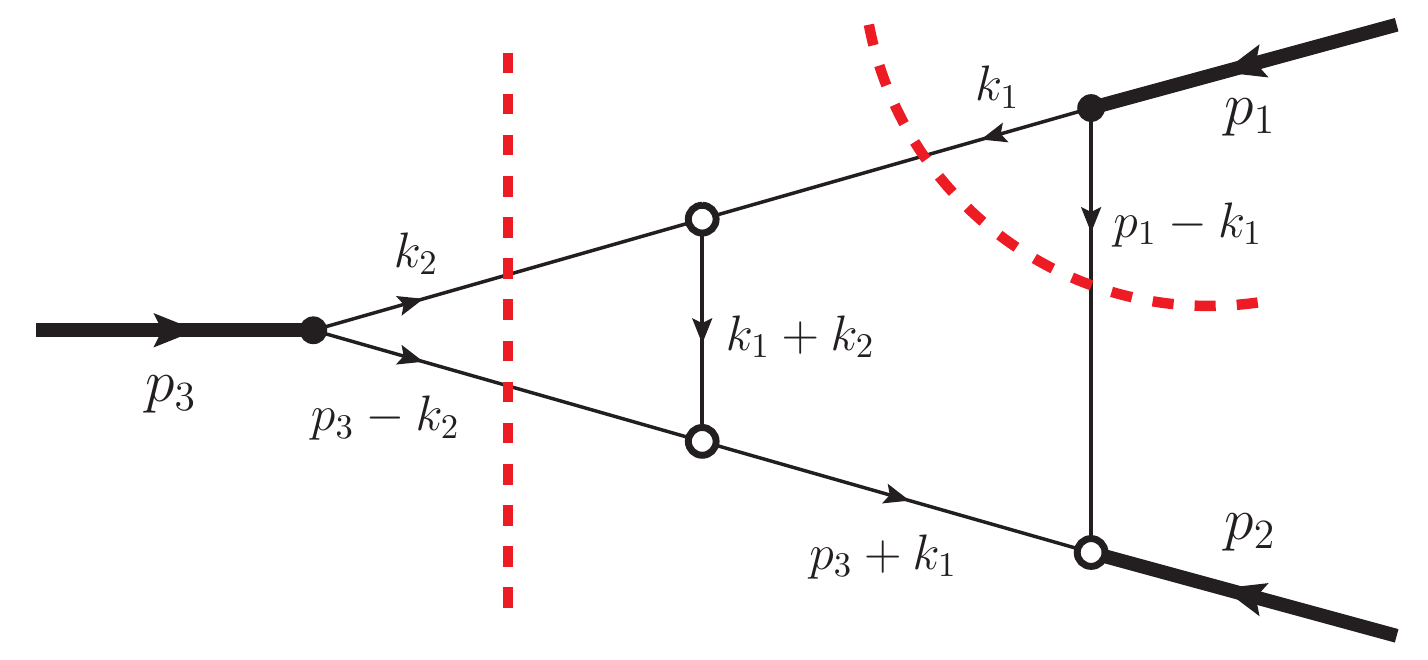}
\caption{Cut [1256]}
\label{cut1256Diag} 
\end{subfigure}
\begin{subfigure}[]{0.5\linewidth}
\centering
\includegraphics[keepaspectratio=true, height=3.3cm]{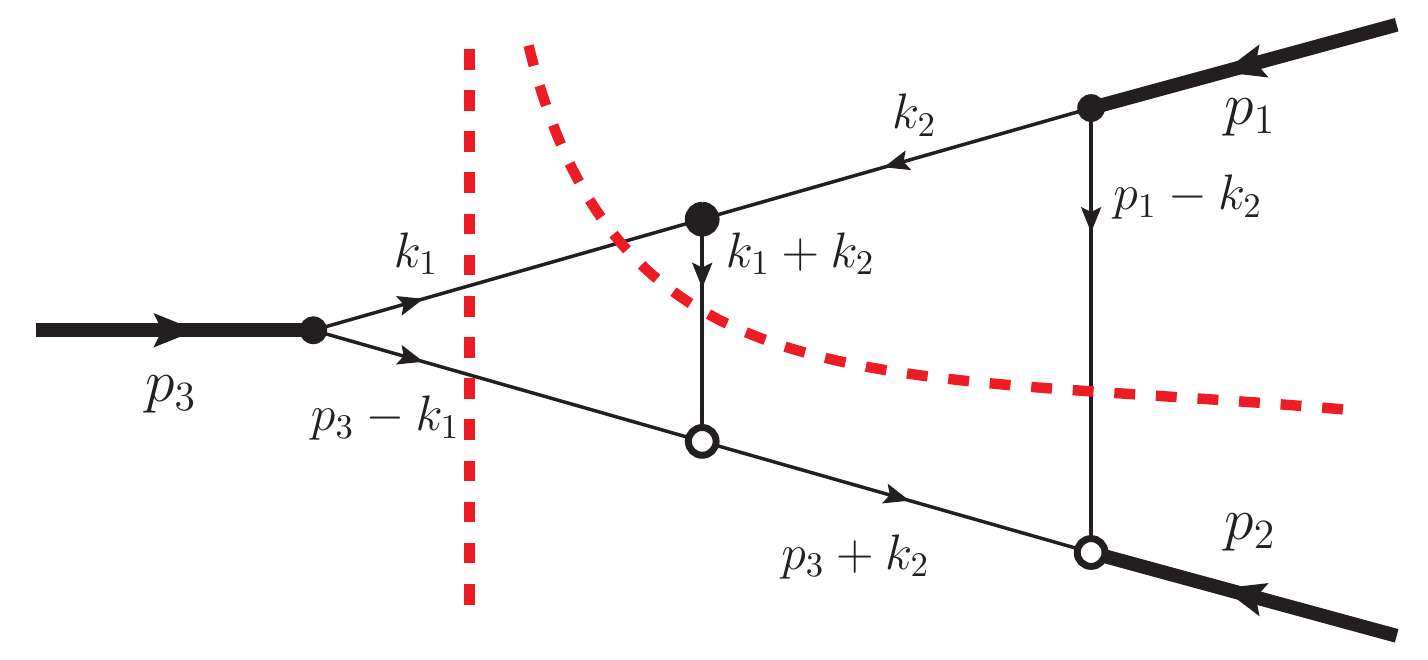}
\caption{Cut [1236]}
\label{cut1236Diag} 
\end{subfigure} 
\begin{subfigure}[]{0.5\linewidth}
\centering
\includegraphics[keepaspectratio=true, height=3.3cm]{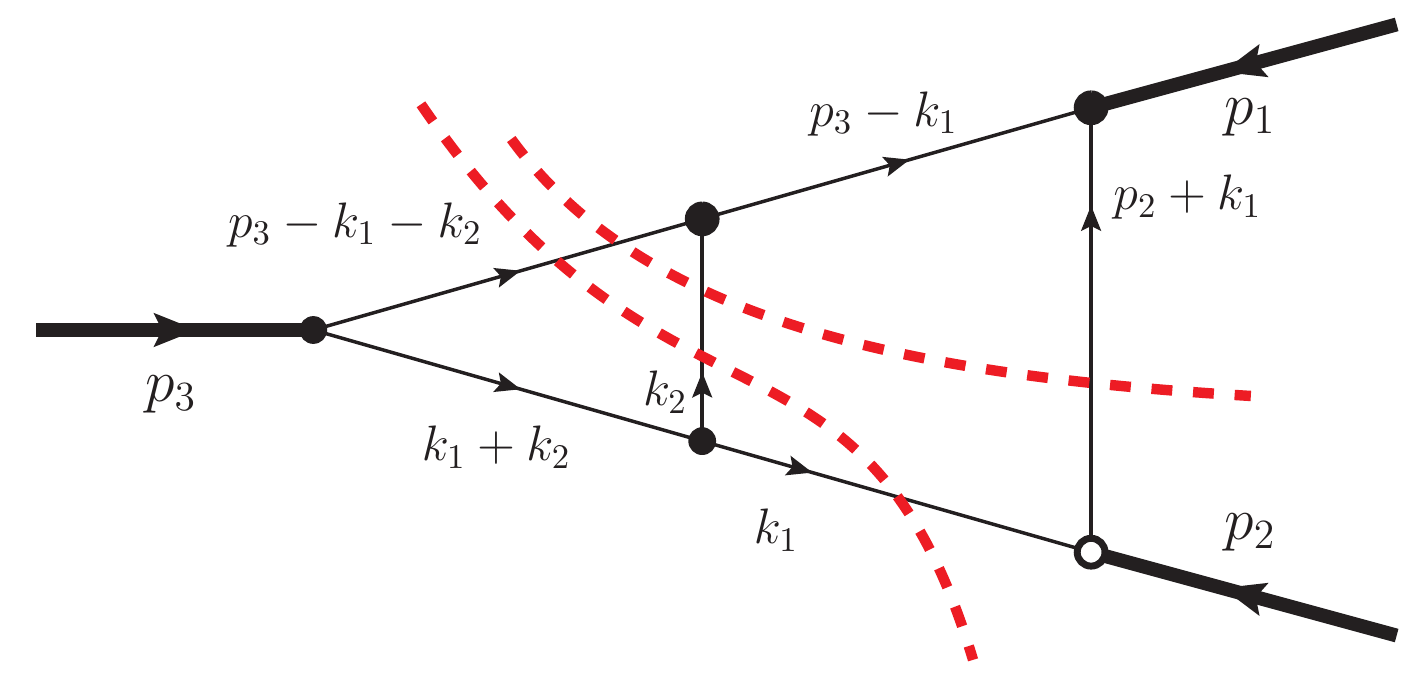}
\caption{Cut [2346]}
\label{cut2346Diag} 
\end{subfigure}
\begin{subfigure}[]{0.5\linewidth}
\centering
\includegraphics[keepaspectratio=true, height=3.3cm]{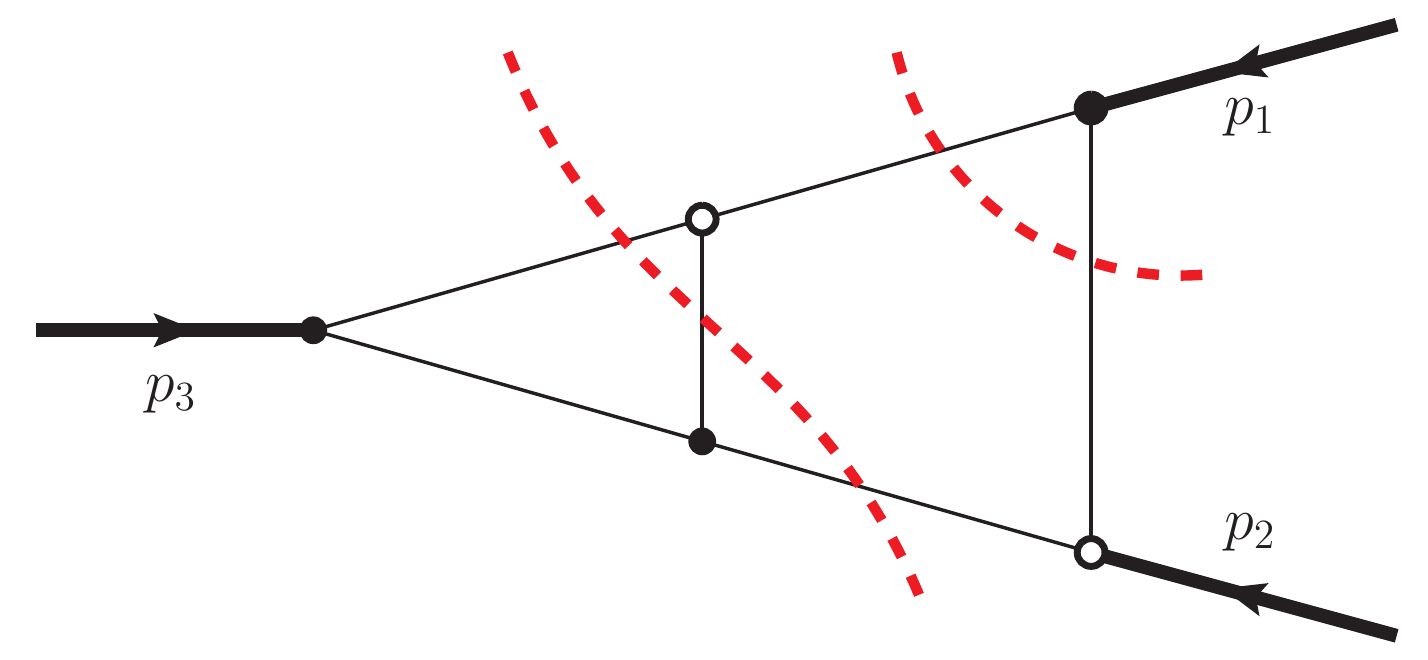}
\caption{Cut [23456]}
\label{cut23456BisDiag} 
\end{subfigure} 
\begin{subfigure}[]{0.5\linewidth}
\centering
\includegraphics[keepaspectratio=true, height=3.3cm]{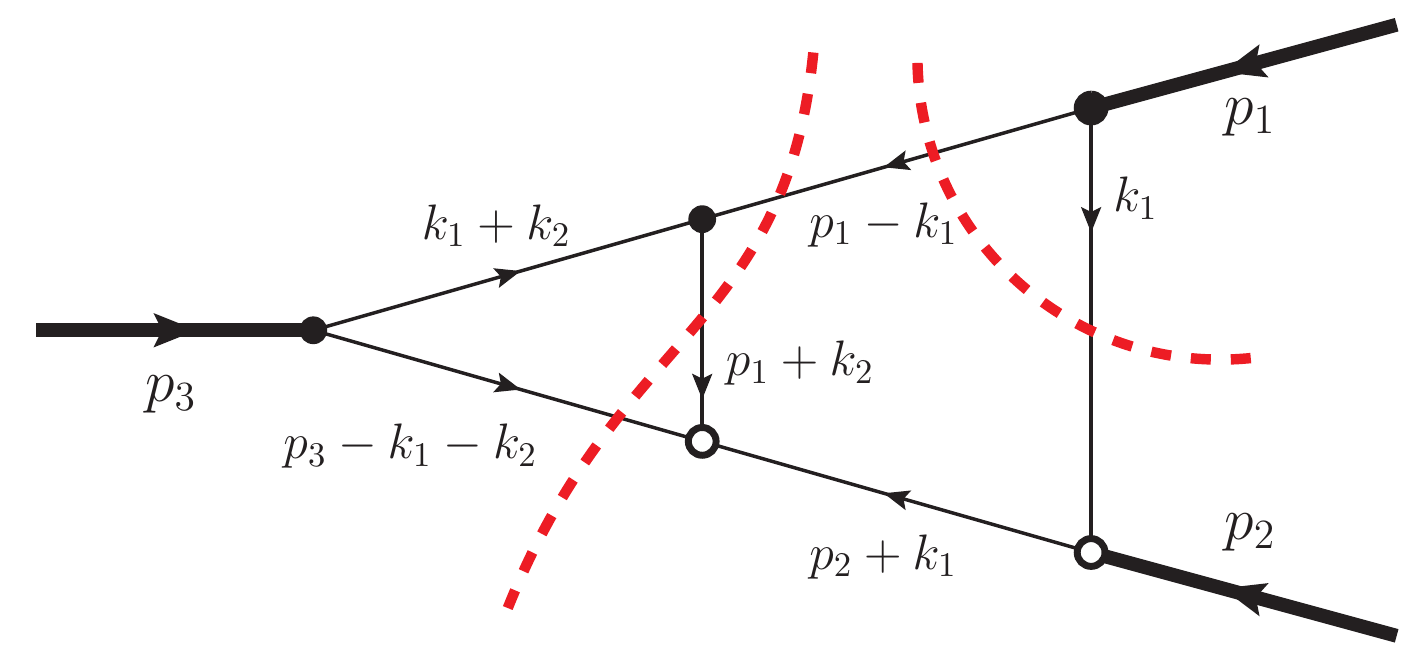}
\caption{Cut [1356]}
\label{cut1356Diag} 
\end{subfigure}
\caption{Cut diagrams contributing to the $\cut_{p_1^2}\circ \cut_{p_3^2}$ sequence of
unitarity cuts.}
\label{doubleCutsp1p3}
\end{figure}

\begin{figure}[!t]
\begin{subfigure}[]{0.5\linewidth}
\centering
\includegraphics[keepaspectratio=true, height=3.3cm]{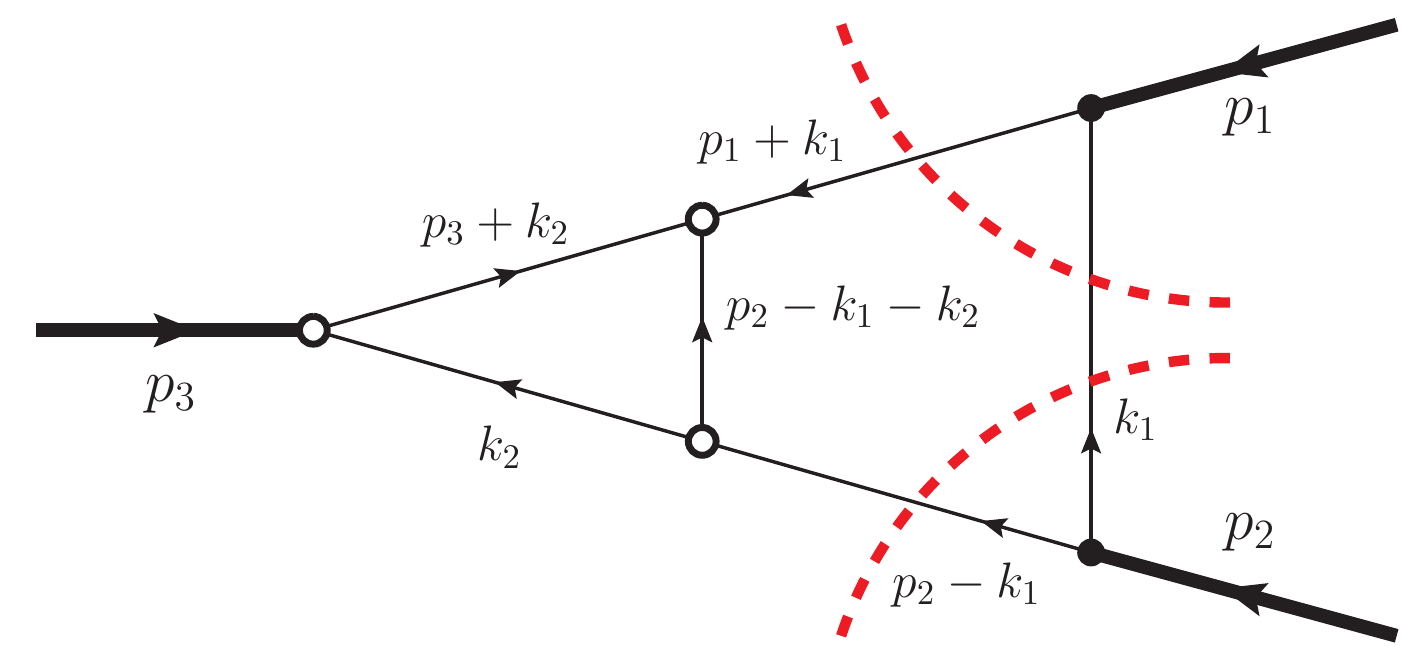}
\caption{Cut [456]}
\label{cut456R3Diag} 
\end{subfigure}
\begin{subfigure}[]{0.5\linewidth}
\centering
\includegraphics[keepaspectratio=true, height=3.3cm]{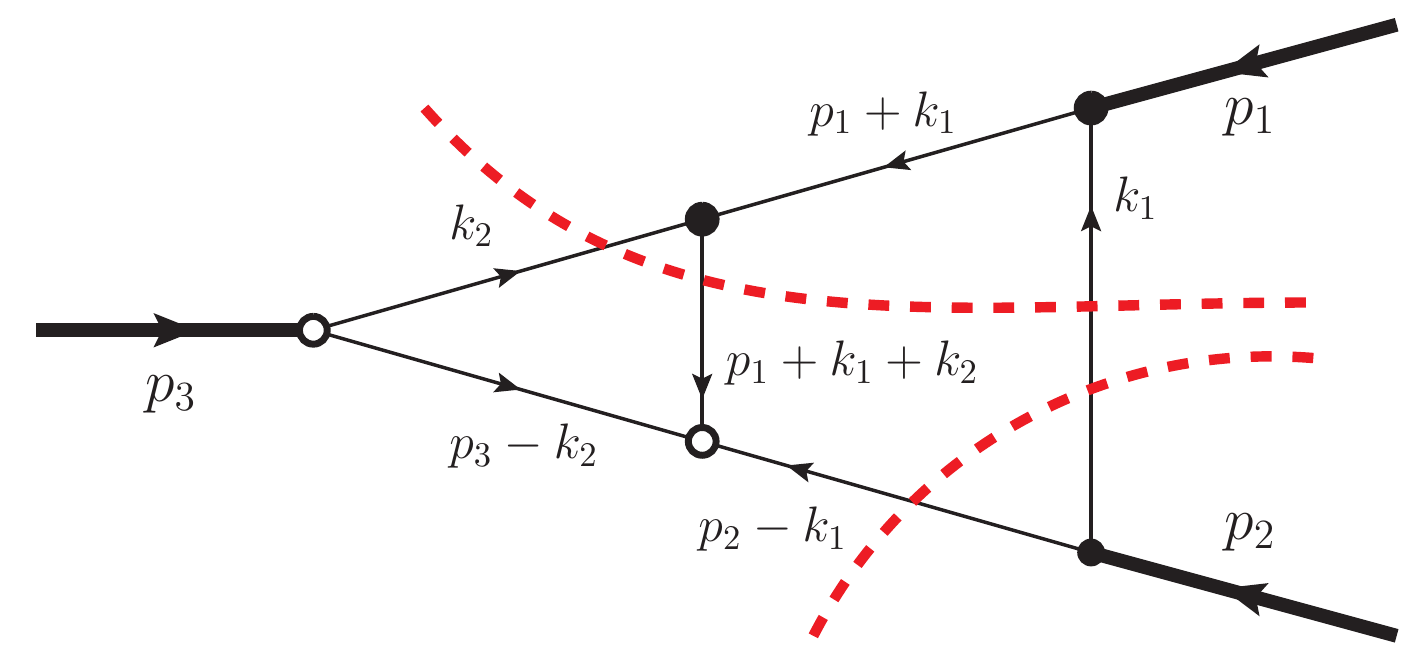}
\caption{Cut [2346]}
\label{cut2346R3Diag} 
\end{subfigure} 
\begin{subfigure}[]{0.5\linewidth}
\centering
\includegraphics[keepaspectratio=true, height=3.3cm]{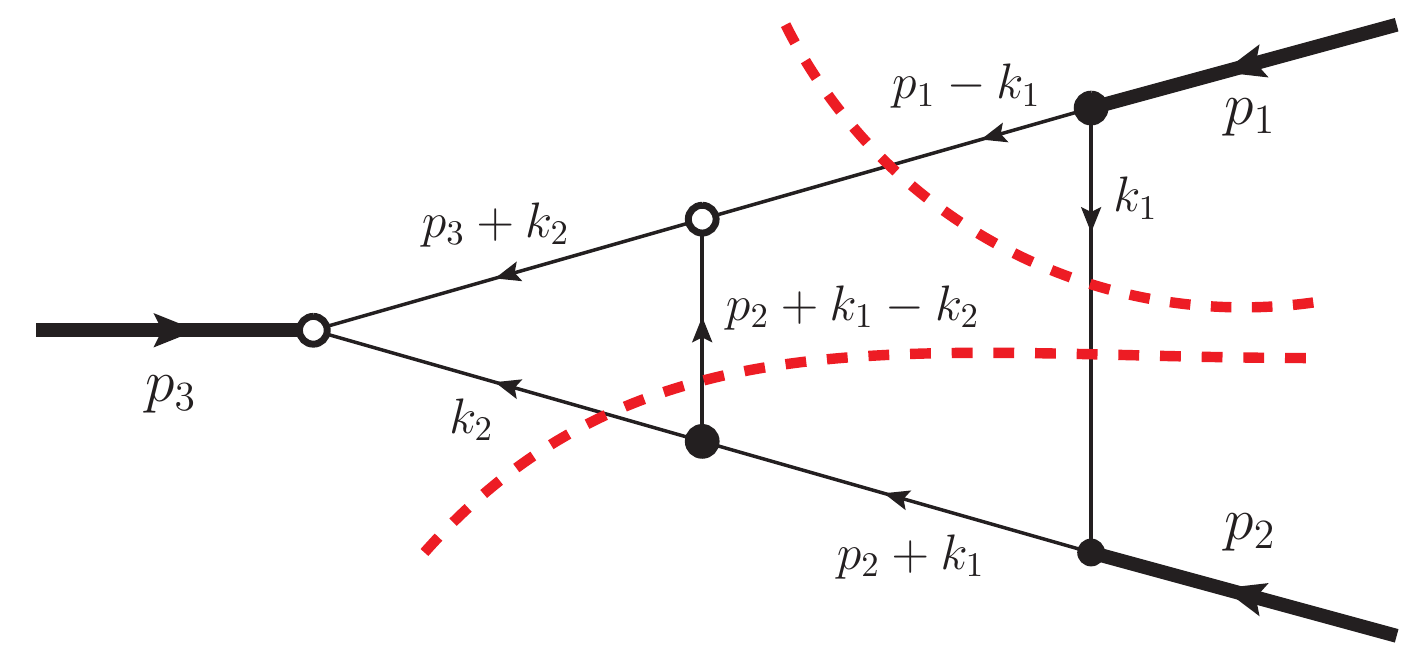}
\caption{Cut [1356]}
\label{cut1356R3Diag} 
\end{subfigure}
\begin{subfigure}[]{0.5\linewidth}
\centering
\includegraphics[keepaspectratio=true, height=3.3cm]{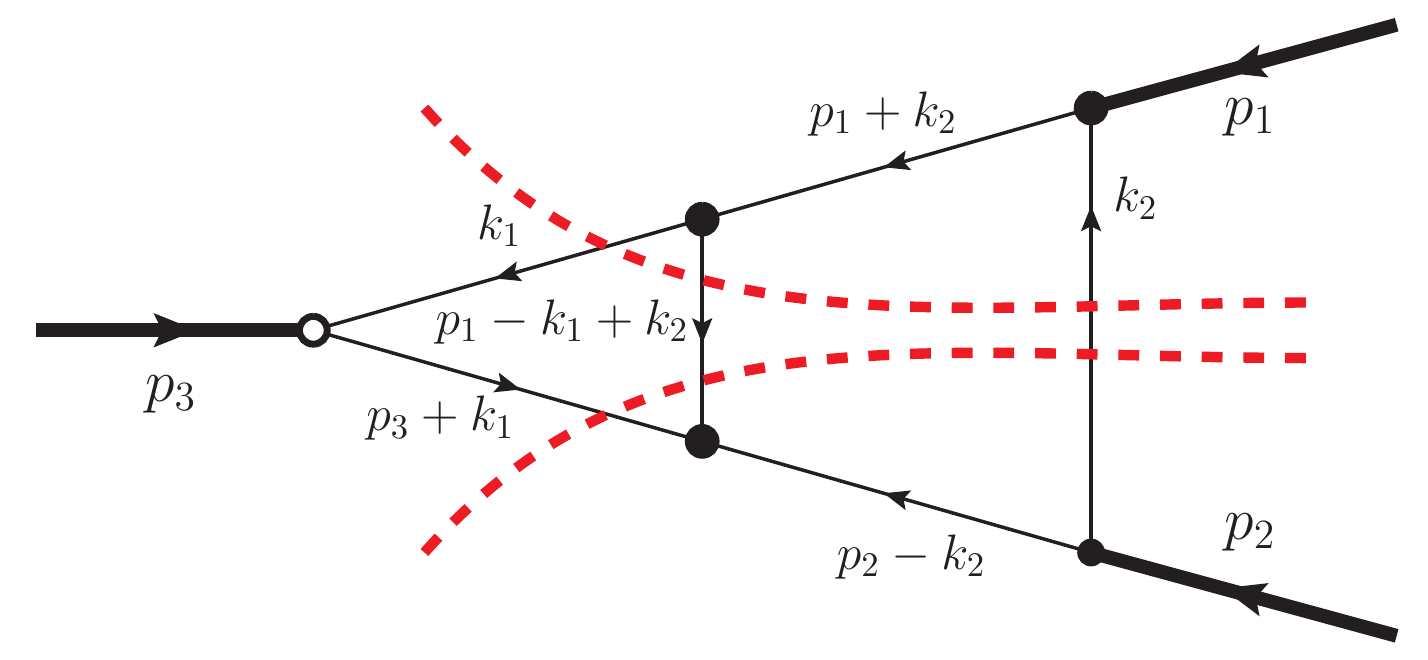}
\caption{Cut [1236]}
\label{cut1236R3Diag} 
\end{subfigure} 
\caption{Cut diagrams contributing to the $\cut_{p_1^2}\circ \cut_{p_2^2}$ sequence of
unitarity cuts.}
\label{doubleCutsp1p2}
\end{figure}

In this section we describe the computation of the sequences of two unitarity cuts corresponding to {$\cut_{p_1^2}\circ \cut_{p_3^2}$} and $\cut_{p_1^2}\circ \cut_{p_2^2}$; see \refF{doubleCutsp1p3} and \refF{doubleCutsp1p2}.
All the cut integrals can be computed following similar techniques as the ones outlined in \refS{sec_UnitarityCut}, so we will be brief and only comment on some special features of the computation.
Details on how to compute the integrals can be found in appendix~\ref{calculationDoubleCuts}, and the explicit results for all the cuts in \refF{doubleCutsp1p3} and \refF{doubleCutsp1p2} are given in appendices \ref{resultsDoubleCutsR2} and \ref{resultsDoubleCutsR3} respectively.

First, we note that, since we are dealing with sequences of unitarity cuts, the cut diagrams correspond to the extended cutting rules introduced in section~\ref{sec:cut_def}. In particular, in section~\ref{sec:cut_def} we argued that cut diagrams with crossed cuts should be discarded, and such diagrams are therefore not taken into account in our computation.  (In this example, all possible crossed cut diagrams would vanish anyway, for the reason given next.)

Second, some of the cut integrals vanish because of energy-momentum constraints. Indeed the cut in \refF{cut23456BisDiag}  vanishes in real kinematics
because it contains a three-point vertex where all the connected legs are massless and on shell. Hence, the cut diagram  cannot satisfy energy momentum conservation in real kinematics with $D>4$ (recall the example of the two-mass-hard box).
We will set this diagram to zero, and we observe a posteriori that this is consistent with the other results, again supporting our approach of working in real kinematics.

We make one further remark on kinematic restrictions.  Recall that the generalized cutting rules allow for all possible directions of energy flow across each cut (as illustrated in \refF{fig:cutcolors1} for the triangle).  In this example of the ladder cut in channels $p_1^2$ and $p_3^2$, all diagrams except \refF{cut1256Diag} would vanish if the energy components of $p_1$ and $p_3$ had the same sign.  However, it follows from the conditions of the cut region ($p_1^2, p_3^2>0,~ p_2^2<0$), in real kinematics, that the energy components of $p_1$ and $p_3$ must have opposite signs.  Thus we will find that we always have nonvanishing contributions from all diagrams except \refF{cut23456BisDiag}.  It is important to be aware of these types of restrictions on the existence of the cut region, since they do not necessarily show up explicitly in the cut integrals.

Let us now focus on the cuts that do not vanish. As we mentioned previously, the cuts
are computed by integrating over carefully chosen one-loop subdiagrams. In particular,
for simplicity we avoid integrating over three-mass triangles, cut or uncut, because
the leading singularity of this diagram is the square root of the K\"all\'en function,
which leads to integrands that are not directly integrable using the tools developed for multiple
polylogarithms. In Tables \ref{tableP1P3} and \ref{tableP1P2} we summarize the preferred choices of subdiagrams for the first loop integration.
We observe that it is insufficient to define a cut integral by the subset of propagators that are cut. Indeed, some cut integrals in the two tables have the same cut propagators, but are computed in different kinematic regions due to the rules of \refS{sec:disccutdelta}, leading to very different results.\footnote{One might think that the results would be related by analytic continuation, but this is not generally true.}

\begin{table}[!t]
\centering
\begin{tabular}{l l}
\hline\hline
Cut (computed in $R^{1,3}_\triangle$) & One-loop subdiagram \\
\hline
$\cut_{p^2_1,[56]}\circ\cut_{p^2_3,[45]}=\cut_{[456],R^{1,3}_\triangle}$ &
 One-mass triangle, mass $p_3^2$, \refF{cut456Diag} \\
& (this cut completely factorizes).\\
$\cut_{p^2_1,[56]}\circ\cut_{p^2_3,[12]}=\cut_{[1256],R^{1,3}_\triangle}$ &
Cut two-mass triangle, masses $p_3^2$ and \\
& $(p_3+k_1)^2$, in $p_3^2$ channel, \refF{cut1256Diag}.  \\
$\cut_{p^2_1,[236]}\circ\cut_{p^2_3,[12]}=\cut_{[1236],R^{1,3}_\triangle}$ &
Cut two-mass-hard box, masses $p_1^2$\\
&and $p_2^2$, in $t=(p_1-k_1)^2$ channel, \refF{cut1236Diag}. \\
$\cut_{p^2_1,[236]}\circ\cut_{p^2_3,[234]}=\cut_{[2346],R^{1,3}_\triangle}$ &
Cut two-mass triangle, masses $p_3^2$ and\\
&  $(p_3-k_1)^2$, in $(p_3-k_1)^2$ channel, \refF{cut2346Diag}. \\
$\cut_{p^2_1,[56]}\circ\cut_{p^2_3,[135]}=\cut_{[1356],R^{1,3}_\triangle}$ &
Cut two-mass triangle, masses $p_3^2$ and\\
&  $(p_3-k_1)^2$, in $(p_3-k_1)^2$ channel, \refF{cut1356Diag}. \\
\hline\hline
\end{tabular}
\caption{Nonvanishing cuts contributing to the $\cut_{p_1^2}\circ \cut_{p_3^2}$ sequence of
unitarity cuts.}
\label{tableP1P3}
\end{table}

Finally, depending on the cut integral and the kinematic region where the cut is computed, the integrands
might become divergent at specific points, and we need to make sense of these divergences
to perform the integrals. In the case where the integral develops an end-point singularity, we explicitly subtract the divergence before expanding in $\eps$, using the technique known as the plus prescription.
  For example, if $g(y,\epsilon)$ is regular for all $y\in[0,1]$,
then, for $\epsilon<0$, we have:
\begin{align}
\int_0^1 dy\,\frac{g(y,\epsilon)}{(1-y)^{1+\epsilon}}=\frac{g(1,\epsilon)}{\epsilon}+\int_0^1 dy\,\frac{g(y,\epsilon)-g(1,\epsilon)}{(1-y)^{1+\epsilon}} \, .
\label{divSubtraction}
\end{align}
The remaining integral is manifestly finite, and we can thus expand in $\eps$ under the integration sign. However, we also encounter integrands which, at first glance, develop simple poles inside the integration region. A careful analysis however reveals that the singularities are shifted into the complex plane due to the Feynman $i\varepsilon$ prescription for the propagators. As a consequence, the integral develops an imaginary part, which can be extracted by the usual principal value prescription,
\begin{equation}
\lim_{\varepsilon\to0}\frac{1}{a\pm i\varepsilon}=\text{PV}\frac{1}{a}\mp i\pi\, \delta(a),
\label{PVReI}
\end{equation}
where PV denotes the Cauchy principal value, defined by
\beq
\textrm{PV}\int_0^1 dy\,\frac{g(y)}{y-y_0} = \lim_{\eta\to0}\left[\int_{0}^{y_0-\eta} dy\,\frac{g(y)}{y-y_0} + \int_{y_0+\eta}^1 dy\,\frac{g(y)}{y-y_0}\right]\,,
\eeq
where $g(y)$ is regular on $[0,1]$ and $y_0\in[0,1]$. Note that the consistency throughout the calculation of the signs of the $i\varepsilon$ of uncut propagators and subdiagram invariants, as derived from the conventions of the extended cutting rules of section~\ref{sec:cut_def} (see also appendix \ref{app:conventions}), is a nontrivial consistency check of these cutting rules.

\begin{table}[!t]
\centering
\begin{tabular}{l l}
\hline\hline
Cut  (computed in $R^{1,2}_\triangle$) & One-loop subdiagram \\
\hline
$\cut_{p^2_1,[56]}\circ\cut_{p^2_2,[46]}=\cut_{[456],R^{1,2}_\triangle}$ &
 One-mass triangle, mass $p_3^2$, \refF{cut456R3Diag}. \\
& (this cut completely factorizes)\\
$\cut_{p^2_1,[236]}\circ\cut_{p^2_2,[46]}=\cut_{[2346],R^{1,2}_\triangle}$ &
Cut two-mass triangle, masses $p_3^2$ and\\
& $(p_1+k_1)^2$, in $(p_1+k_1)^2$ channel, \refF{cut2346R3Diag}.  \\
$\cut_{p^2_1,[56]}\circ\cut_{p^2_2,[136]}=\cut_{[1356],R^{1,2}_\triangle}$ &
Cut two-mass triangle, masses $p_3^2$ and\\
& $(p_2+k_1)^2$, in $(p_2+k_1)^2$ channel, \refF{cut1356R3Diag}. \\
$\cut_{p^2_1,[236]}\circ\cut_{p^2_2,[136]}=\cut_{[1236],R^{1,2}_\triangle}$ &
Cut two-mass-hard box, masses $p_1^2$\\
& and $p_2^2$, in $t=(p_1-k_1)^2$ channel, \refF{cut1236R3Diag}. \\
\hline\hline
\end{tabular}
\caption{Cuts contributing to the $\cut_{p_1^2}\circ \cut_{p_2^2}$ sequence of
unitarity cuts.}
\label{tableP1P2}
\end{table}

\subsection{Summary and discussion}

Let us now look at explicit results for $\Cut_{p_3^2,p_1^2}T_L$ and $\Cut_{p_2^2,p_1^2}T_L$. From the explicit calculations collected in Appendix \ref{app:DoubleCutRes}, we get
\begin{equation}
\Cut_{p_3^2,p_1^2}T_L\left(p_1^2,p_2^2,p_3^2\right)=\frac{4\pi^2 i(p_1^2)^{-2}}{(1-z)(1-\bz)(z-\bz)}\left(\ln(z)\ln(\zbar)-\frac{1}{2}\ln^2(z)\right)\, ,
\end{equation}
and
\begin{align}\bsp
\Cut_{p_2^2,p_1^2}T_L\left(p_1^2,p_2^2,p_3^2\right)=\frac{4\pi^2 i(p_1^2)^{-2}}{(1-z)(1-\bz)(z-\bz)}\bigg(\ln(z)\ln&(\zbar)-\frac{1}{2}\ln^2(z)\\
&\qquad-\Li_2(z)+\Li_2(\zbar)\bigg)\, .
\esp\end{align}

Comparing with the coproduct in \refE{delta112}, we verify from these results that  the relations \eqref{eq:cutequalsdisc} and \eqref{eq:discequalsdelta} between $\cut$ and $\delta$, as written in eqs.~(\ref{eq:tableeqns}), are satisfied.  We have confirmed by direct calculation from the original ladder function \eqref{ladderFunc} that the $\Disc$ operation gives the expect results as well.

Diagrammatically, for the specific cuts considered above, we have
\beq\bsp
\left[\delta_{1-z,\zbar}+\delta_{1-z,1-z}\right]&T_L(p_1^2,p_2^2,p_3^2)\\
&=\frac{1}{(2\pi i)^2}
\left(\raisebox{-5.2mm}{\includegraphics[keepaspectratio=true, height=1.2cm]{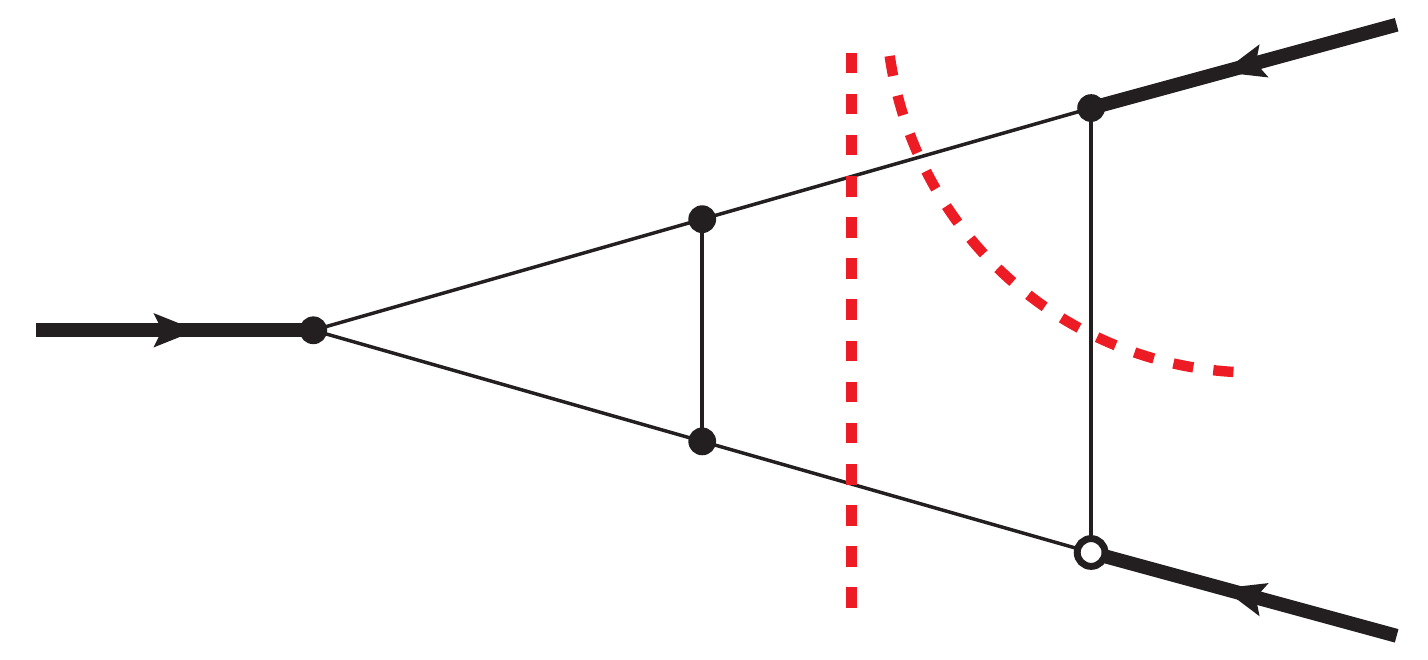}}
+\raisebox{-5.2mm}{\includegraphics[keepaspectratio=true, height=1.2cm]{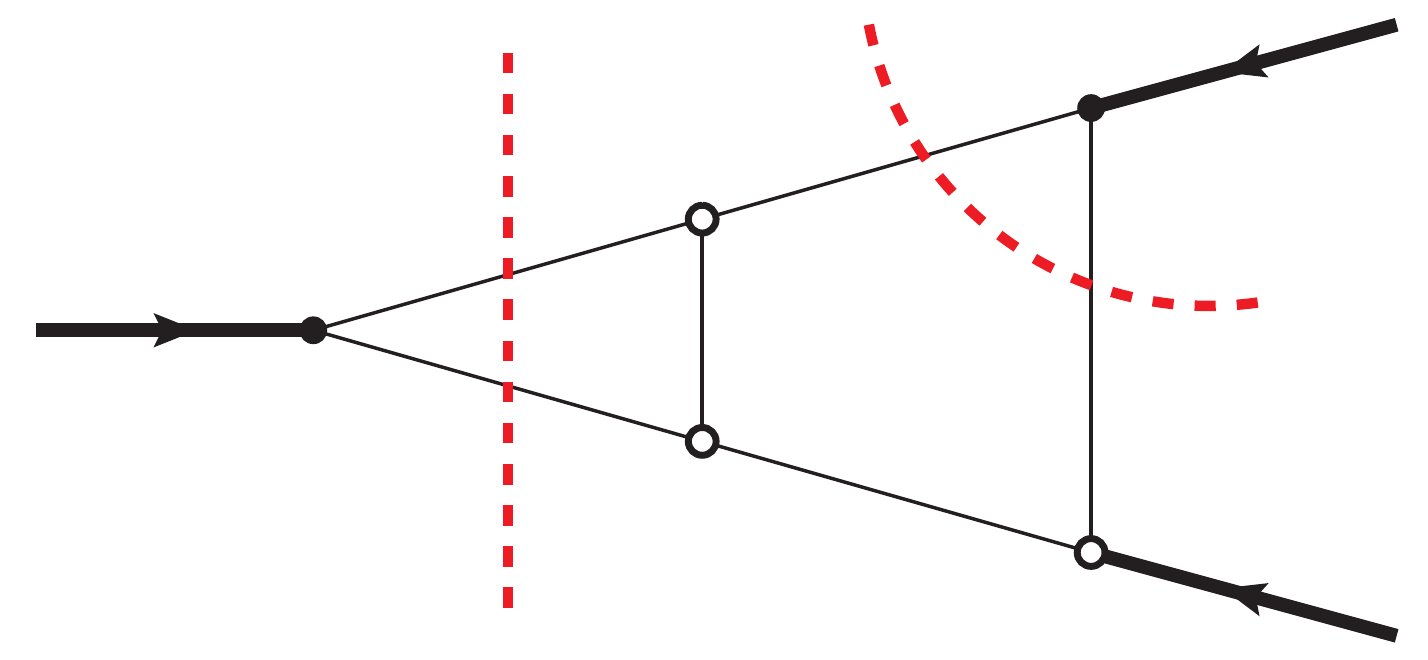}}
+\raisebox{-5.2mm}{\includegraphics[keepaspectratio=true, height=1.2cm]{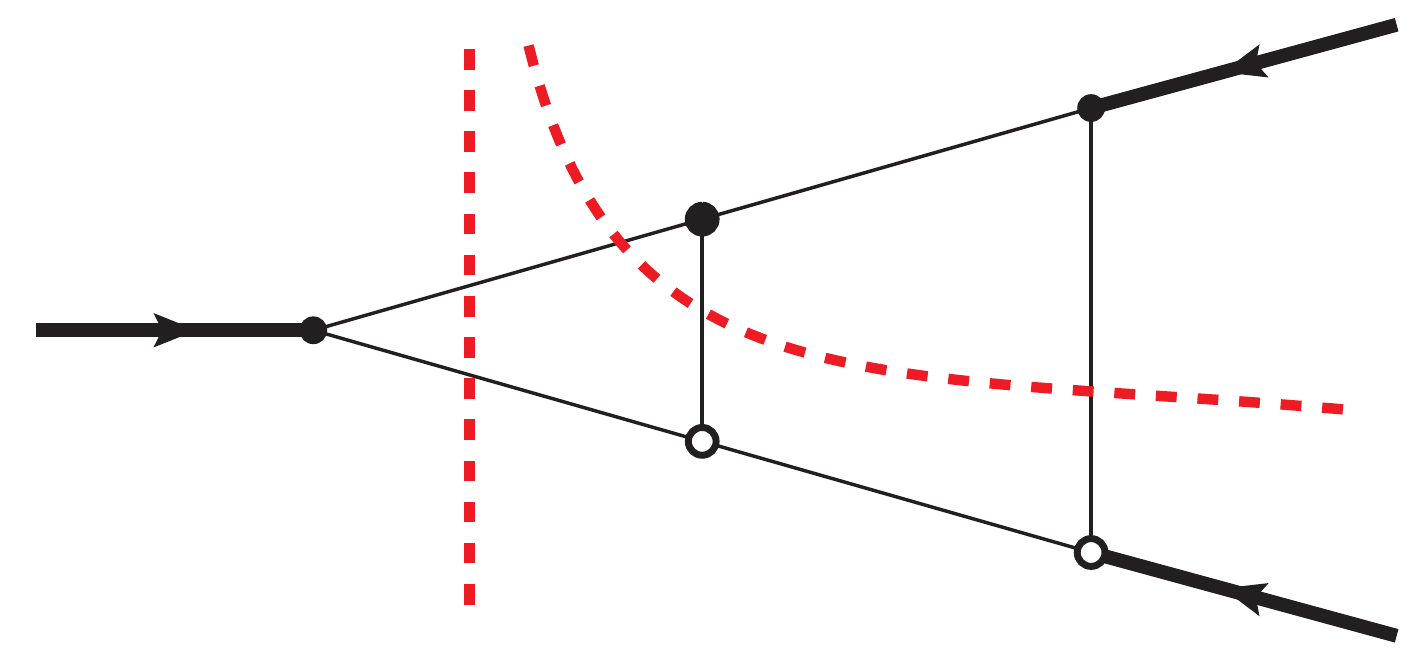}}\,\,\right.\\
&
\left.\phantom{=\frac{1}{(2\pi i)^2}(}+\raisebox{-5.2mm}{\includegraphics[keepaspectratio=true, height=1.2cm]{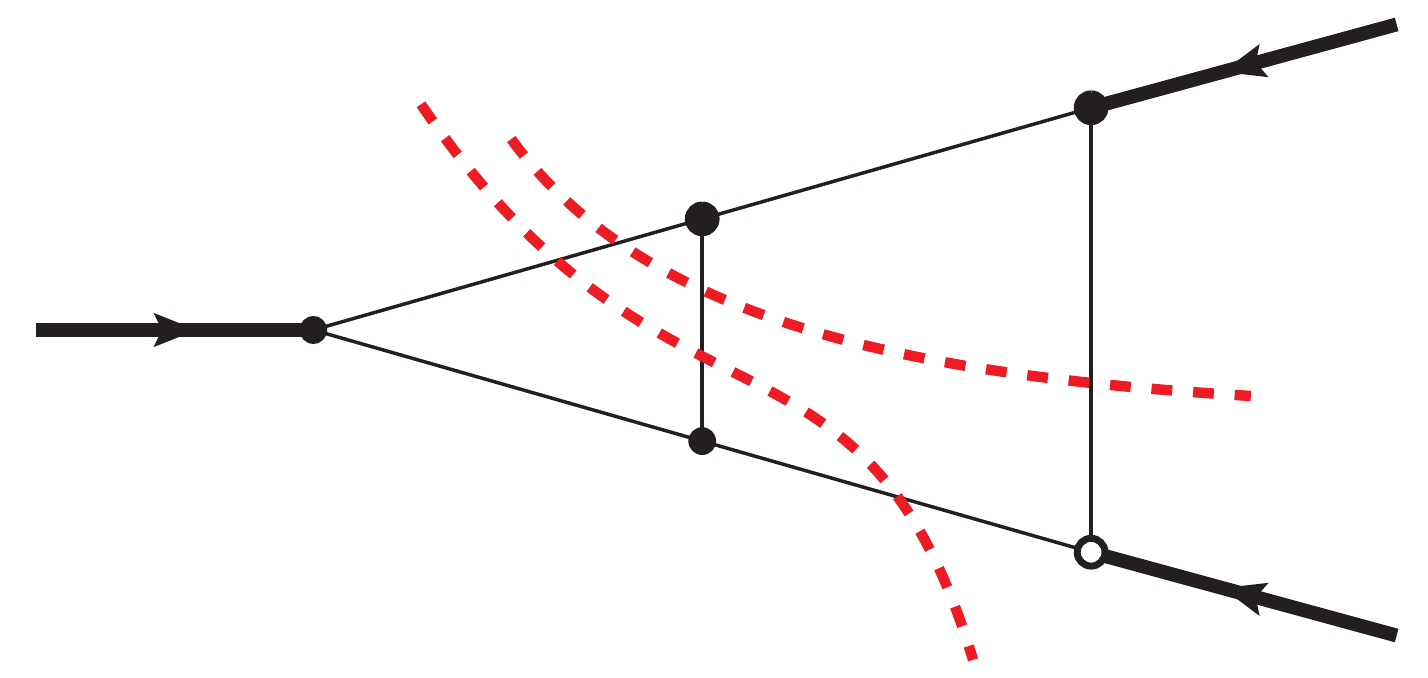}}
+\raisebox{-5.2mm}{\includegraphics[keepaspectratio=true, height=1.2cm]{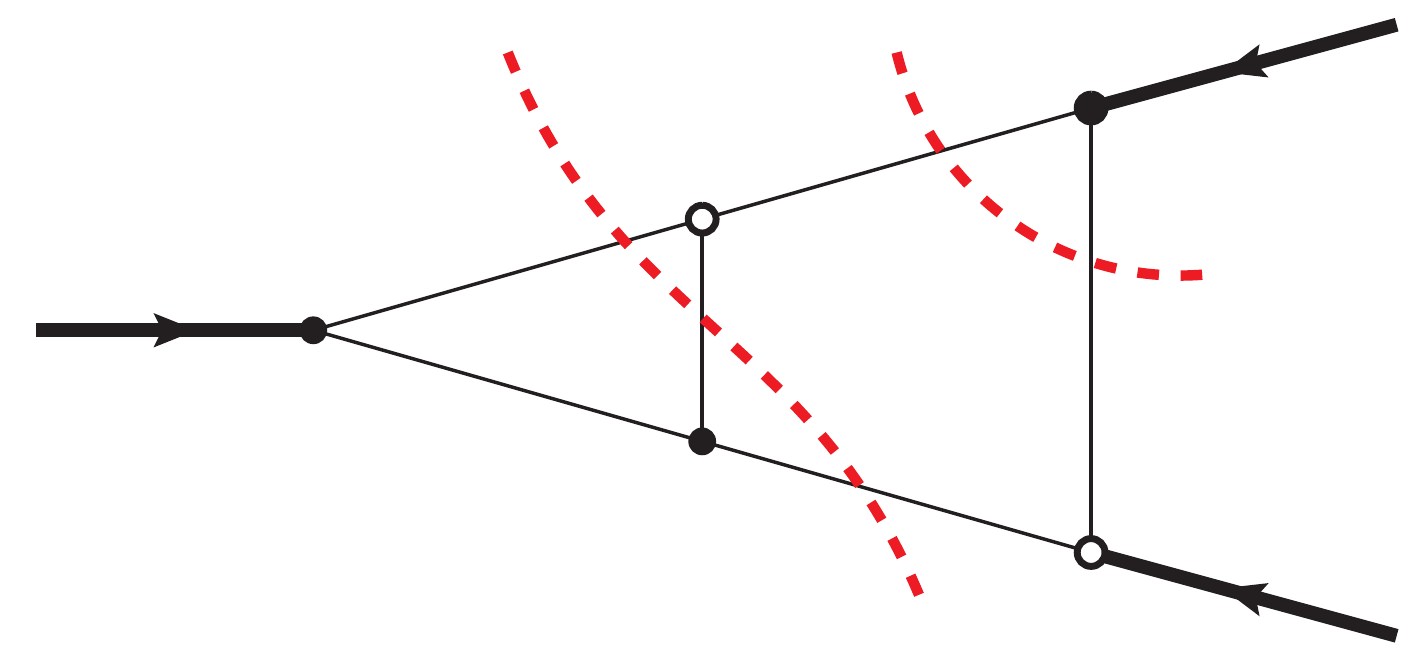}}
+\raisebox{-5.2mm}{\includegraphics[keepaspectratio=true, height=1.2cm]{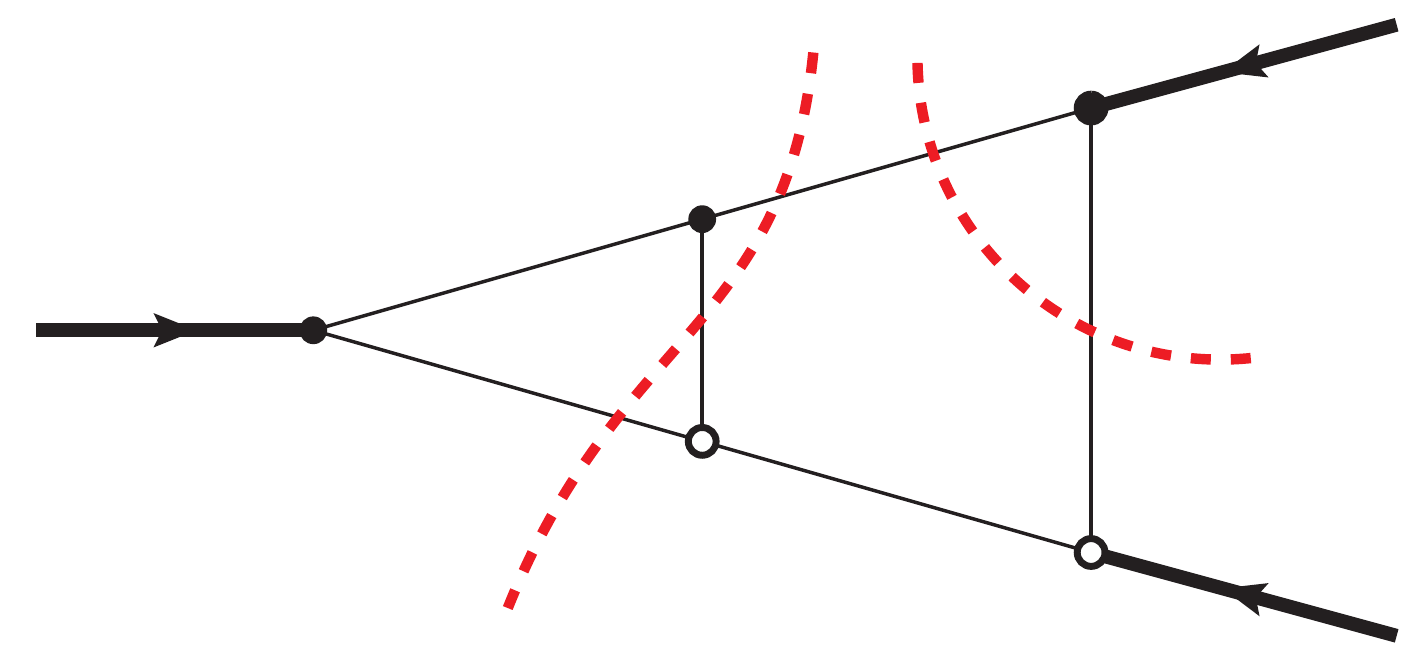}}
\,\,\right)_{R^{1,3}_\triangle}\,,\nonumber
\esp\eeq
and,
\begin{align}
\left[\delta_{z,\zbar}+\delta_{z,1-z}\right]T_L(p_1^2,p_2^2,p_3^2)=\frac{1}{(2\pi i)^2}
&\left(\raisebox{-5.2mm}{\includegraphics[keepaspectratio=true, height=1.2cm]{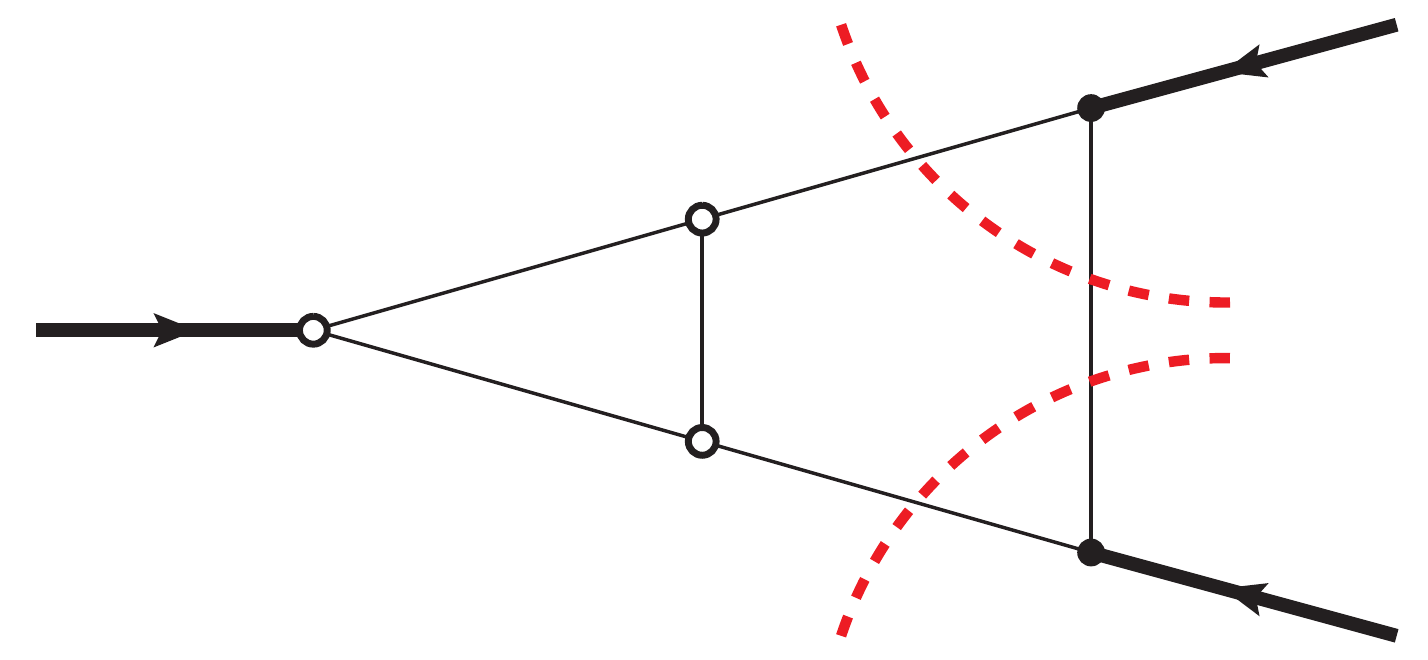}}
+\raisebox{-5.2mm}{\includegraphics[keepaspectratio=true, height=1.2cm]{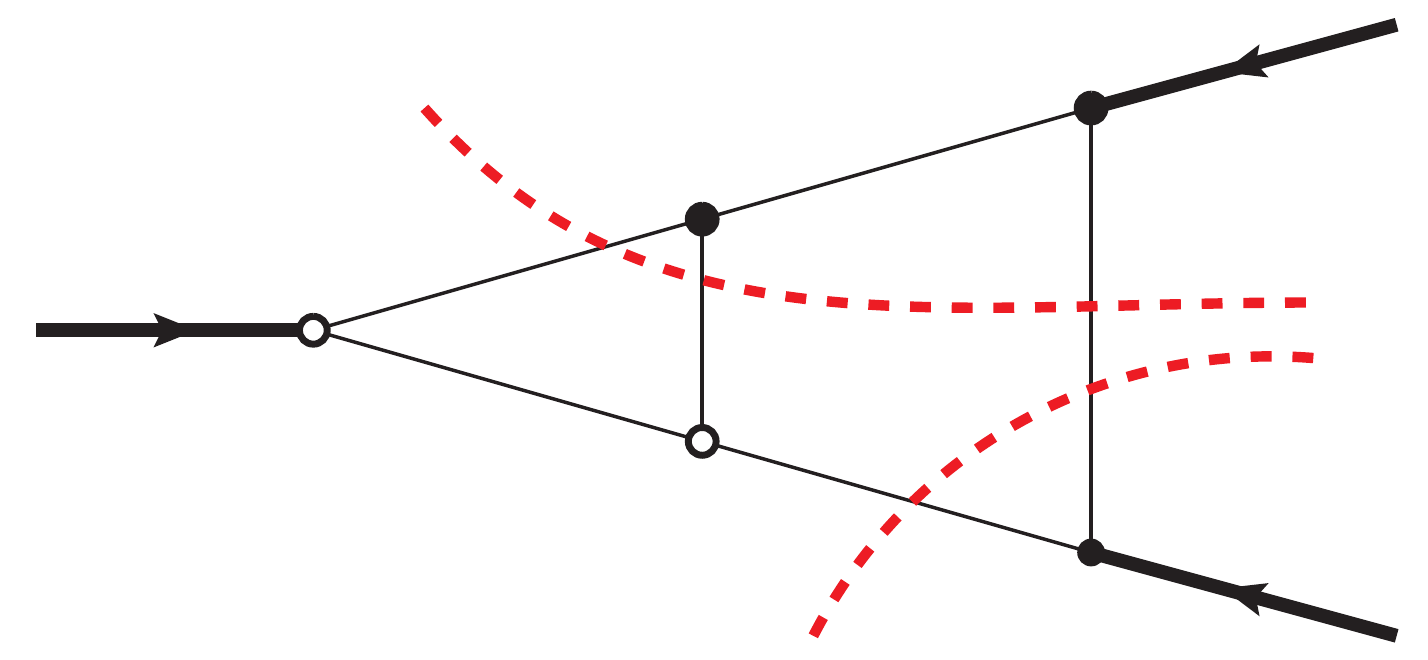}}\,\,\right.\nonumber\\
&\left.
+\raisebox{-5.2mm}{\includegraphics[keepaspectratio=true, height=1.2cm]{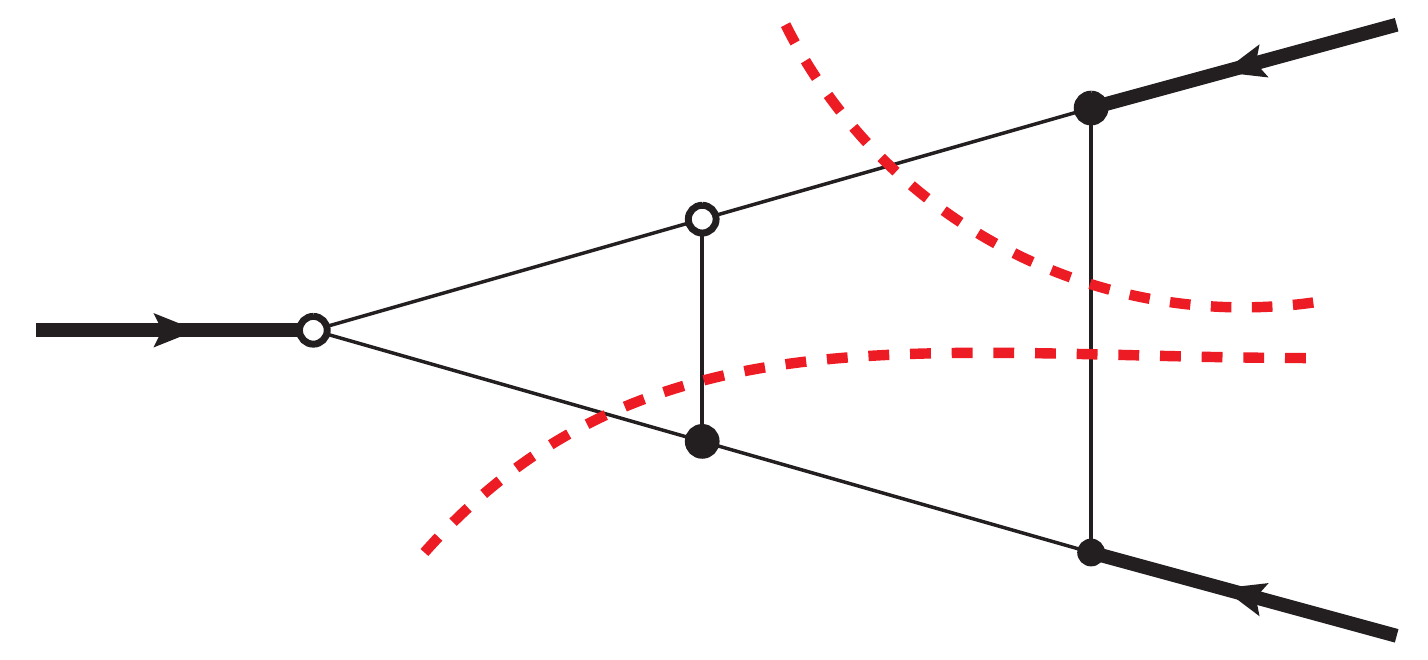}}
+\raisebox{-5.2mm}{\includegraphics[keepaspectratio=true, height=1.2cm]{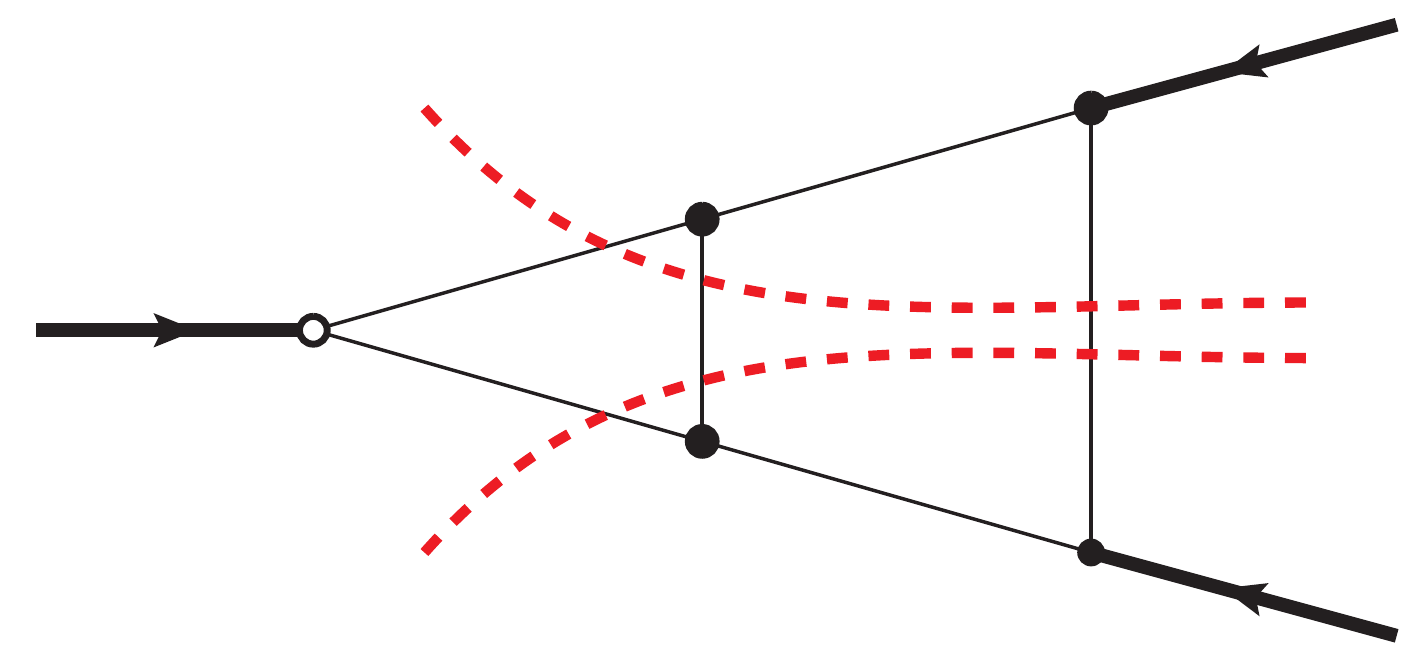}}
\,\,\right)_{R^{1,2}_\triangle}.\nonumber
\end{align}

One could wonder about a sequence of three unitarity cuts in the three distinct channels of the ladder. As argued in Section \ref{sec:ex-relations}, the region where one would hope to compute this triple cut has all $p_i^2>0$. Since $F$ is only a function of the ratios of the $p_i^2$, this region is indistinguishable from the Euclidean region,
so the triple cut must vanish and contains no nontrivial information on the analytic structure of the function.

\section{From cuts to dispersion relations and coproducts}
\label{sec:integrable}

In previous sections we introduced computational tools to compute cut integrals, and we showed that extended cutting rules  in real kinematics lead to consistent results. Furthermore, we argued that the entries in the coproduct of a Feynman integral can be related to its discontinuities and cut integrals. While these results are interesting in their own right, in this section we take a step further and put them to use: we present several ways of using the knowledge of (sequences of) cut integrals to reconstruct the original Feynman integral based on the knowledge of its cuts.

It is obvious from the first entry condition that if all cuts are known, we can immediately write down the coproduct component of weight $(1,n-1)$ of a pure integral of weight $n$. In particular, for the one- and two-loop triangle integrals investigated in previous sections, we  immediately obtain
\beq\bsp\label{eq:(1,n-1)_for_triangle}
\Delta_{1,1}(\cT(z,\zbar)) &\,= \log (z\zbar) \otimes \delta_{z}\cT(z,\zbar) + \log ((1-z)(1-\zbar)) \otimes \delta_{1-z}\cT(z,\zbar)\,,\\
\Delta_{1,3}(F(z,\zbar)) &\,= \log (z\zbar) \otimes \delta_{z}F(z,\zbar) + \log ((1-z)(1-\zbar)) \otimes \delta_{1-z}F(z,\zbar)\,,
\esp\eeq
and the quantities $\delta_{z}\cT(z,\zbar)$, $\delta_{1-z}\cT(z,\zbar)$, $\delta_{z}F(z,\zbar)$ and $\delta_{1-z}F(z,\zbar)$ are directly related to the discontinuities of the integral through eqs.~\eqref{eq:tablecut1}. 
These components of the  coproduct in term determine the functions $\cT(z,\zbar)$ and $F(z,\zbar)$ up to terms that vanish when acting with $\Delta_{1,1}$ and $\Delta_{1,3}$. We will see how this information can be recovered in the following.

Similarly, in \refE{eq:tableeqns} we have shown how the double discontinuities of the two-loop ladder triangle are related to the entries in the coproduct. We can then immediately write
\beq\label{eq:(1,1,2)_ladder}
\Delta_{1,1,2}(F(z,\zbar))  = \sum_{(x_1,x_2)\in\mathcal{A}^2_\triangle}\log x_1\otimes \log x_2\otimes\delta_{x_1,x_2}F(z,\zbar)\,,
\eeq
and the values of $\delta_{x_1,x_2}F(z,\zbar)$ can be read off from \refE{eq:tableeqns}.\footnote{As written, \refE{eq:tableeqns} gives solutions for four of the sixteen  functions, $\delta_{x_1,x_2}F(z,\zbar)$.  The remaining ones can be obtained trivially by imposing the first entry condition, so that $\delta_z F(z,\zbar)=\delta_{\bz} F(z,\zbar)$ and $\delta_{1-z} F(z,\zbar)=\delta_{1-\bz} F(z,\zbar)$, and by extending the kinematic analysis to regions in which $\zbar>z$, thus restoring the symmetry of the full function under exchange of $z$ and $\zbar$.} 
Thus, we see that the knowledge of \emph{all} double discontinuities enables us to immediately write down the answer for the (1,1,2) component of the two-loop ladder triangle. Just as in the case of a single unitarity cut, this component of the coproduct determines $F(z,\zbar)$ up to terms that vanish when acting with $\Delta_{1,1,2}$. In the following, we show how this ambiguity can be lifted.

While the previous application is trivial and follows immediately from the first entry condition and the knowledge of the set of variables that can enter the symbol in these particular examples, it is less obvious that we should be able to reconstruct information about the full function by looking at 
a single unitarity cut, or at a specific sequence of two unitarity cuts.
In the rest of this section we give evidence that this is true nevertheless.

The classic tool
for determining a Feynman integral from its cuts is the  \emph{dispersion relation}, which expresses a given Feynman integral as the integral of its discontinuity across a certain branch cut. Traditionally used in the context of the study of strongly interacting theories, dispersion relations appear more generally
as a consequence of the unitarity of the $S$-matrix, and of the analytic structure of
amplitudes \cite{SMatrix}. These relations are valid in perturbation theory, order by order in an expansion of the coupling constant.
It was shown in refs.~\cite{Landau:1959fi,Cutkosky:1960sp,tHooft:1973pz,Remiddi:1981hn,Veltman:1994wz} that individual Feynman integrals can also be written as dispersive integrals.
The fundamental ingredient in the proof of the existence of this representation is the largest time equation \cite{tHooft:1973pz}, which is also the basis of the cutting rules.
In the first part of this section we review dispersion relations for Feynman integrals, illustrating them with the examples of the one-loop three-mass triangle integral and the two-loop three-point three-mass ladder integral.

We then show that, at least in the case of the integrals considered in this paper, we can use the modern Hopf algebraic language to determine the symbol of the integrals from either a single unitarity cut or a single sequence of unitarity cuts. We note however that the reconstructibility procedure presented here works for the full integral, and not for individual terms in the Laurent expansion in $\epsilon$.\footnote{An example of an IR divergent integral where the reconstructibility of individual terms in the Laurent expansion would fail is the two-mass-hard box:  it is clear from eq.~\eqref{eq:2mhtdisc} that a cut in a single channel does not capture all terms of the symbol.}  We therefore focus on examples which are finite in four dimensions, so that we can set $\epsilon=0$.

\subsection{Dispersion relations}
\label{Sec_dispersion_relation}

Dispersion relations are a prescription for computing an integral from its discontinuity across a branch cut, taking the form
\begin{equation}
\label{disp_rel} 
F(p_1^2,p_2^2,\ldots)=\frac{1}{2\pi i}\int_{C} \frac{ds}{s-(p_2^2+i \varepsilon)}\,\rho(p_1^2,s,\ldots)\,,
\end{equation}
where
\begin{equation*}
\rho(p_1^2,s,\ldots)=\Disc_{p_2^2}F(p_1^2,p_2^2,\ldots)\big\vert_{p_2^2=s}\,,
\end{equation*}
as computed with eq.~(\ref{eq:def-disc}), 
and the integration contour $C$ goes along that same branch cut. The above relation can be checked using eqs.~(\ref{eq:def-disc}) and (\ref{PVReI}).

We start with a simple generalisation of the above expression. Let
\begin{equation*}
G(p_i^2)=r(p_i^2)F(p_i^2)\,,
\end{equation*}
where $r(p_i^2)$ is a rational function of the Mandelstam invariants $p_i^2$. Then, because $G(p_i^2)$
and $F(p_i^2)$ have the same branch point and branch cut structure, $G(p_i^2)$ itself has a dispersive representation of the form \refE{disp_rel}.
This in turn provides an alternative representation for $F(p_i^2)$. Indeed, using that
\begin{equation*}
\Disc_{p_2^2}G(p_1^2,p_2^2,\ldots)=r(p_1^2,p_2^2,\ldots)\Disc_{p_2^2}F(p_1^2,p_2^2,\ldots)\,,
\end{equation*}
one gets:
\begin{equation}
\label{disp_rel_subDisc}
F(p_1^2,p_2^2,\ldots)=\frac{1}{2\pi i}\frac{1}{r(p_1^2,p_2^2,\ldots)}\int_{C} \frac{ds}{s-(p_2^2+i \varepsilon)}\,r(p_1^2,s,\ldots)\rho(p_1^2,s,\ldots)\,,
\end{equation}
provided the integral on the right hand side is well defined, and where for simplicity we assumed $\rho(p_1^2,s,\ldots)$ has no poles in the integration region (if this is not the case, we need
to add the contribution of the residues at those poles, as dictated by the residue theorem).
Eq.~(\ref{disp_rel}) can be seen as a particular case of \refE{disp_rel_subDisc}, with $r(p_i^2)=1$.
If the integral in \refE{disp_rel} is not well defined, typically by becoming divergent at some end point of the integration region, a judicious choice of $r(p_i^2)$ can in fact be made
to find a dispersive representation for $F(p_i^2)$. These are called subtracted dispersion relations (see e.g. Appendix B in \cite{vanNeerven:1985xr} for an example in the context of dispersive representations of Feynman integrals). 

In light of the relation between discontinuities and cuts presented in the paper, if $F(p_1^2,p_2^2,\ldots)$ is a Feynman integral \refE{disp_rel_subDisc} can also be written as:
\begin{equation}
\label{disp_rel_sub}
F(p_1^2,p_2^2,\ldots)=-\frac{1}{2\pi i}\frac{1}{r(p_1^2,p_2^2,\ldots)}\int_{C} \frac{ds}{s-(p_2^2+i \varepsilon)}\,r(p_1^2,s,\ldots)\left(\Cut_{p_2^2}F(p_1^2,p_2^2,\ldots)\right)\Big\vert_{p_2^2=s}\,.
\end{equation}

In order to illustrate the use of dispersion relations, we first look at the case of the scalar three-mass triangle.
Its $p_2^2$-channel cut was computed in eq.~\eqref{3mass_triangle_cut_result}, and we recall it here expressed in terms of Mandelstam invariants,
\begin{align}
\label{cut_3mass_tri}
\begin{split}
\Cut_{p_2^2}T(p_1^2,p_2^2,p_3^2)
&= \frac{2\pi}{\sqrt{\lambda(p_1^2,p_2^2,p_3^2)}} \,\ln \frac{p_1^2-p_2^2+p_3^2-\sqrt{\lambda(p_1^2,p_2^2,p_3^2)}}{p_1^2-p_2^2+p_3^2+\sqrt{\lambda(p_1^2,p_2^2,p_3^2)}}\,+\,{\cal O}(\epsilon)\,.
\end{split}
\end{align}
This leads to a dispersive representation for the three-mass triangle of the form ($r(p_i^2)=1$)
\begin{equation}
T(p_1^2,p_2^2,p_3^2)=-\frac{1}{2\pi i}\int_0^\infty \frac{ds}{s-(p_2^2+i \varepsilon)} 
\frac{2\pi}{\sqrt{\lambda(p_1^2,s,p_3^2)}} \,\ln \frac{p_1^2-s+p_3^2-\sqrt{\lambda(p_1^2,s,p_3^2)}}{p_1^2-s+p_3^2+\sqrt{\lambda(p_1^2,s,p_3^2)}}\,.
\label{dispRepTriangle}
\end{equation}
Note that the integration contour runs along the real positive axis: it corresponds to the branch cut for timelike invariants of Feynman integrals with massless internal propagators.
Already for this not too complicated diagram we see that the dispersive representation involves a rather complicated integration.

The main difficulty in performing the integral above comes from the square root of the K\"all\'en function, whose arguments depend on the integration variable. However, defining $x=s/p_1^2$, and introducing variables $w$ and $\bar{w}$ similar to eq.~\eqref{eq:zalpha}, which are are a particular case of the more general \refE{eq:s_to_xi}, defined as
\begin{equation}
w \bar{w}= x {\rm~~and~~} (1-w)(1-\bar{w})=u_3\,,
\end{equation}
or equivalently,
\begin{equation}
w=\frac{1+x-u_3+\sqrt{\lambda(1,x,u_3)}}{2} {\rm~~and~~} \bar{w}=\frac{1+x-u_3-\sqrt{\lambda(1,x,u_3)}}{2}\,,
\label{wwbarDef}
\end{equation}
we can rewrite the dispersive integral as, 
\begin{align}
T(p_1^2,p_2^2,p_3^2)&=\frac{i}{p_1^2}\int du_3\int\frac{dx}{x-u_2}\frac{1}{w-\bar{w}}\log\frac{1-w}{1-\bar{w}}\,\delta\big(u_3-(1-w)(1-\bar{w})\big)\,\theta(-x)\,\theta(u_3)\nonumber\\
&=\frac{-i}{p_1^2}\int_0^1dw\int_{-\infty}^{0}d\bar{w}\frac{1}{w\bar{w}-u_2}\delta\big(u_3-(1-w)(1-\bar{w})\big)\log\frac{1-w}{1-\bar{w}}\label{dispRepTri}\\
&=\frac{-i}{p_1^2}\frac{1}{z-\bz}\int_0^1 dw\left(\frac{1}{w-\bz}-\frac{1}{w-z}\right)\Big[2\ln(1-w)-\ln u_3\Big]\nonumber\,,
\end{align}
where the integration region for $w$ and $\bar{w}$ is deduced from the region where the discontinuity is computed (see, e.g., table \ref{table:regionstriangle}).
Written in this form, the remaining integration is trivial to perform in terms of polylogarithms, and we indeed
recover the result of the three-mass triangle, eq.~\eqref{eq:triangle}.

For the three-mass triangle, we can in fact take a second discontinuity and reconstruct the result through a double dispersion relation because
the discontinuity function, \refE{cut_3mass_tri}, has a dispersive representation itself \cite{Cutkosky:1960sp,Ball:1991bs}.
Note that this representation falls outside of what is discussed in ref.~\cite{Remiddi:1981hn}, and we are not aware of a proof of its existence from first principles.
The double discontinuity is simply given, up to overall numerical and scale factors, by the inverse of the square root of the K\"all\'en function, see eq.~\eqref{eq:cuttrip3p2}. We obtain
\begin{align}\bsp
T(p_1^2,p_2^2,p_3^2)&=-\frac{1}{(2\pi i)^2}\int\frac{dx}{x-u_2}\int\frac{dy}{y-u_3}\left(\Cut_{p_3^2,p_2^2}T(p_1^2,p_2^2,p_3^2)\right)\Big\vert_{u_2=x,u_3=y}\\
&=\frac{1}{(2\pi i)^2}\frac{4\pi^2i}{p_1^2}\int\frac{dx}{x-u_2}\int\frac{dy}{y-u_3}\frac{1}{\sqrt{\lambda(1,x,y)}}\,\theta(-x)\,\theta(-y)\\
&=\frac{-i}{p_1^2}\int_1^\infty dw\int_{-\infty}^{0}d\bar{w}\frac{1}{w \bar{w}-z\bz}\frac{1}{(1-w)(1-\bar{w})-(1-z)(1-\bz)}\,.
\label{doubleDispRepTri}
\esp\end{align}
The integral is trivial to perform\footnote{We have redefined $w$ and $\bar{w}$ by replacing $u_3$ by $y$ in \refE{wwbarDef}. Just as for the single dispersion integral, the integration region is deduced from the region where the double discontinuity is computed, $R_{2,3}$ in this case. Changing variables to $\beta=\frac{1}{w}$ and $\gamma=\frac{1}{1-\bar{w}}$ makes the integral particularly simple to evaluate.} and leads to the correct result.

We now turn to the case of the two-loop ladder.
As long as we are using suitable variables, from the point of view of dispersion relations it is trivial to go from the three-mass triangle to the two-loop ladder. 
The only new feature we need to deal with is a more complicated rational prefactor: instead of just having the inverse of the square root of the K\"all\'en function, it appears multiplied by $1/u_3$.
This makes the dispersive integral over $p_3^2$ as written in \refE{disp_rel} non convergent. However, we can easily overcome this difficulty by setting $r(p_i^2)=p_3^2/p_1^2=u_3$ in \refE{disp_rel_sub}.
When considering a dispersive integral over $p_2^2$ this is not necessary for the convergence of the integral, but the same choice of $r(p_i^2)$ still simplifies the integrand and makes de calculation simpler. Having made this choice, and proceeding as with the 
three-mass triangle, the remaining integral is trivial to perform in terms of polylogarithms.

As an example, we consider  the dispersive integral over $p_3^2$:
\begin{align}\bsp
T_L(p_1^2,p_2^2,p_3^2)&=-\frac{1}{2\pi i}\frac{1}{u_3}\int_{-\infty}^0\frac{dy}{y-u_3}\,y\int_0^\infty du_2\,\delta(u_2-w\bar{w})\left(\Cut_{p_3^2}T_L(p_1^2,p_2^2,p_3^2)\right)\Big\vert_{u_3=y}\\
&=\frac{i (p_1^2)^{-2}}{(z-\zbar)(1-z)(1-\zbar)}\int_0^1d\bar{w}\left(\frac{1}{\bar{w}-\bar{z}}-\frac{1}{\bar{w}-z}\right)\frac{1}{2}\log\bar{w}\log\frac{u_2}{\bar{w}}\log\frac{u_2}{\bar{w}^2}\,,
\esp\end{align}
where the variables $w$ and $\bar{w}$ are similar to the ones defined in \refE{wwbarDef} but with $x$ replaced by $u_2$ and $u_3$ by $y$, and we used $\Cut_{p^2_3}T_L(p_1^2,p_2^2,p_3^2)$ as obtained from eq.~(\ref{cutP3Res}). This integral does indeed reproduce the expected result, \refE{ladderFunc}.

Similarly to the one-loop three-mass triangle, the two-loop three-mass ladder also has a representation as a double dispersive integral. Given the variables we chose to work with it is more convenient to consider the double unitarity cut on $p_2^2$ and $p_3^2$. Using \refE{eq:tableeqns}, with the necessary prefactors as in \refE{ladderFunc},
\begin{align*}
\Cut_{p^2_3,p_2^2}T_L(p_1^2,p_2^2,p_3^2)
&=-\frac{4\pi^2 i\left(p_1^2\right)^{-2}}{(1-z)(1-\zbar)(z-\zbar)}\left(\ln z\ln\zbar-\frac{1}{2}\ln^2z\right)\,,
\end{align*}
from which we get:
\begin{align}\bsp
T_L(p_1^2,p_2^2,p_3^2)&=-\frac{1}{(2\pi i)^2}\frac{1}{u_3}\int^0_{-\infty}\frac{dx}{x-u_2}\int^0_{-\infty}\frac{dy}{y-u_3}\,y\left(\Cut_{p^2_3,p_2^2}T_L(p_1^2,p_2^2,p_3^2)\right)\Big\vert_{u_2=x,u_3=y}\\
&=-i\frac{(p_1^2)^{-2}}{u_3}\int_1^\infty dw\int_{-\infty}^0d\bar{w}\frac{\ln w\ln \bar{w}-\frac{\ln^2 w}{2}}{(w\bar{w}-u_2)((1-w)(1-\bar{w})-u_3)}\,,
\esp\end{align}
where we again used $r(p_i^2)=u_3$ and exactly the same comments as the ones accompanying \refE{doubleDispRepTri} apply. The remaining
integrals are trivial to perform, and we indeed recover the correct result, \refE{ladderFunc}. As far as we are aware, this is the first
time such a representation of the two-loop three-mass ladder has been given.

We see that we can obtain the full result for the one-loop three-mass triangle and the two-loop three-point three-mass ladder from the knowledge of either its single or double cuts. A fundamental ingredient necessary to perform the dispersive integral was the choice of variables in which to write the dispersive integral. While for the one-loop example we studied one might still consider performing the integration in terms of the Mandelstam invariants, for the two-loop ladder this does not seem feasible anymore given the complexity of the expression for the discontinuity in any of the channels when written in terms of the Mandelstam invariants. Choosing the variables of the type given in \refE{wwbarDef}, which we showed are naturally found by computing cuts, the increase in complexity in going from the one-loop diagram to the two-loop is not as great as one might naively expect.

We finish with a comment: we believe the dispersive representation for the three-mass triangle provides one of the simplest ways to compute the diagram to any order in the expansion of the dimensional regularization parameter $\epsilon$. While we only considered the leading order in \refE{dispRepTri}, following the same arguments we could as easily have written
\begin{align}\bsp
T(p_1^2,p_2^2,p_3^2)=&-i\frac{(p_1^2)^{-1-\epsilon}}{z-\zbar}\frac{e^{\gamma_E\epsilon}\Gamma(1-\epsilon)}{\Gamma(2-2\epsilon)}\int_0^1 dw\left(\frac{1}{w-\bz}-\frac{1}{w-z}\right)\frac{u_3-(1-w)^2}{u_3}\\
&\left(\frac{w(1-w-u_3)}{1-w}\right)^{-\epsilon}\,_2F_1\left(1,1-\epsilon;2-2\epsilon;\frac{u_3-(1-w)^2}{u_3}\right)
\esp\end{align}
where we used the $D$-dimensional result for the cut given in \refE{cutP2TriDDim}.
Integration over $w$ and expansion in $\epsilon$ on the right hand side commute, and we are left at any order with one integration to perform. The expansion in $\epsilon$ of the hypergeometric
function, although not trivial, has been automatised \cite{Huber:2007dx}. Aside from one overall rational prefactor that cancels the one remaining in the integrand, it will only produce polylogarithms and thus the remaining integration is trivial to perform in terms of multiple polylogarithms. The result will already be expressed as a function of the variables in terms of which this diagram is known to be most simply written.

We believe a deeper understanding of the connection between multiple cuts and sequential discontinuities as defined in this paper can provide a way to prove the existence of multiple dispersive representations. We expect they will in turn be useful in the actual calculation of Feynman integrals in cases where more traditional techniques fail.

\subsection{Reconstructing the coproduct from a single unitarity cut}
\label{sec_singleCutRec}
As discussed above, Feynman diagrams can be fully recovered from unitarity cuts in a given channel through dispersion relations.
These relations rely on two ingredients: the discontinuity of a function across a specific branch cut, and the position of that particular branch cut. Given the relations between the
$(1,n-1)$ entries of the coproduct, discontinuities, and single unitarity cuts established in previous sections, it is clear that the full information about the Feynman integral is encoded in 
any one of these entries of the coproduct,
since it contains the same information about the function as a dispersive representation.
We should thus be able to reconstruct information about the full function by looking at a single cut in a given channel.

For simplicity, we work mainly at the level of the symbol in the rest of this section, keeping in mind that we lose information about terms proportional to $\pi$ and zeta values in doing so. We will find that this information can easily be recovered in our examples.  In a nutshell, we observe that if we combine the first entry condition and the results for the discontinuities with the integrability condition~\eqref{eq:integrability}, we immediately obtain the symbol of the full function.
In the following, we illustrate this procedure in the examples of the one-loop triangle and two-loop three-point ladder. Starting from the result for the unitarity cut in a single channel, the procedure to obtain the symbol of the full function can be formulated in terms of a simple algorithm, which involves two steps:
\begin{enumerate}
\item[(i)] Check if the tensor satisfies the integrability condition, and if not, add the relevant terms required to make the tensor integrable. 
\item[(ii)] Check if the symbol obtained from the previous step satisfies the first entry condition, and if not, add the relevant terms. Then return to step (i).
\end{enumerate}

We start by illustrating this procedure on the rather simple example of the three-mass triangle of \refS{sec_threeMassTriangle}. 
From eq.~\eqref{3mass_triangle_cut_result}, the symbol of the cut on the $p_2^2$ channel is
\begin{equation*}
 \frac{1}{2}\,\frac{1-\zz}{1-\bz}\,,
 \end{equation*}
 where we emphasize that the rational function is to be interpreted as the symbol of a logarithm. 
 Since we considered a cut in the $p_2^2$ channel, the first entry condition implies that we need to prepend $u_2=\zz\bz$  to the symbol of the discontinuity. Thus we begin with the tensor
\begin{equation*}
 \frac{1}{2}\,(\zz\bz)\otimes\frac{1-\zz}{1-\bz}\,.
\end{equation*}
 We then proceed as follows.
\begin{itemize}
\item Step $(i)$:
This tensor is not the symbol of a function, as it violates the integrability condition.
To satisfy the integrability condition, we need to add the two terms
\begin{equation*}
  \frac{1}{2}\,(1-\zz)\otimes \bz-  \frac{1}{2}\,(1-\bz)\otimes \zz\,.
\end{equation*}
The full tensor is not the symbol of a Feynman diagram, since the two new terms do not satisfy the first entry condition.
\item Step $(ii)$:
To satisfy the first entry condition, we add two new terms:
\begin{equation*}
  \frac{1}{2}\,(1-\bz)\otimes \bz-  \frac{1}{2}\,(1-\zz)\otimes \zz\,.
\end{equation*}
At this stage, the sum of terms obeys the first entry condition and the symbol obeys the integrability condition, so we stop our process.
\end{itemize}
Putting all the terms together, we obtain
\begin{equation}
  \cS(\cT(z,\zbar)) = \frac{1}{2}\,z \zbar\otimes \frac{ 1-\zz}{ 1-\bz}+  \frac{1}{2}\,(1-z)(1-\zbar)\otimes \frac{\bz}{\zz},
\end{equation}
which agrees with the symbol of the one-loop three mass triangle in $D=4$ dimensions, eq.~\eqref{eq:trianglecoproduct}. 

Note that we can easily integrate this symbol to the full function. Indeed, the cut computation has allowed us to determine the symbol, and hence also the symbol alphabet $\cA_\triangle=\{z,\zbar,1-z,1-\zbar\}$. It is well known that the most general class of functions giving rise to this symbol alphabet and satisfying the first entry condition are the single-valued harmonic polylogarithms~\cite{BrownSVHPLs}. Up to overall normalization, there is a unique single-valued harmonic polylogarithm of weight two that is odd under the exchange of $z$ and $\zbar$, namely the function $\cP_2(z)$ defined in eq.~\eqref{eq:p2asmpl}. We therefore immediately recover the analytic expression for $\cT(z,\zbar)$ given in Section~\ref{sec:OneLoop}.

While the previous example might seem too simple to be representative, we show next that the same conclusion still holds for the two-loop ladder.
In the following we use our knowledge of the cut in the $p_3^2$ channel, eq.~\eqref{cutP3Res}, and show that we can again reconstruct the symbol of the full integral $F(\zz,\zbar)$. Combining eq.~\eqref{cutP3Res} with the first entry condition, we conclude that $\cS(F(\zz,\zbar))$ must contain the following terms:
\bean
(1-z)(1-\zbar)\otimes\left[\zz\otimes\zz\otimes\bz+\zz\otimes\bz\otimes\zz+\bz\otimes\zz\otimes\zz-\zz\otimes\bz\otimes\bz-\bz\otimes\zz\otimes\bz-\bz\otimes\bz\otimes\zz\right]\,.
\eean
If we follow the same steps as in the one-loop case, we can again reconstruct the symbol of the full function from the knowledge of the symbol of the cut in the $p_3^2$ channel alone. More precisely, we perform the following operations:
\begin{itemize}
\item Step $(i)$:
To obey the integrability condition, we must add to the expression above the following eight terms:
\begin{align*}
&+\zz\otimes(1-\bz)\otimes\zz\otimes\bz+\zz\otimes\zz\otimes(1-\bz)\otimes\bz+\zz\otimes(1-\bz)\otimes\bz\otimes\zz\\
&+\bz\otimes(1-\zz)\otimes\zz\otimes\zz-\zz\otimes(1-\bz)\otimes\bz\otimes\bz-\bz\otimes(1-\zz)\otimes\zz\otimes\bz\\
&-\bz\otimes(1-\zz)\otimes\bz\otimes\zz-\bz\otimes\bz\otimes(1-\zz)\otimes\zz\,.
\end{align*}
\item Step $(ii)$:
The terms we just added violate the first entry condition. To restore it we must add eight more terms that combine
with the ones above to have Mandelstam invariants in the first entry,
\begin{align*}
&+\bz\otimes(1-\bz)\otimes\zz\otimes\bz+\bz\otimes\zz\otimes(1-\bz)\otimes\bz+\bz\otimes(1-\bz)\otimes\bz\otimes\zz\\
&+\zz\otimes(1-\zz)\otimes\zz\otimes\zz-\bz\otimes(1-\bz)\otimes\bz\otimes\bz-\zz\otimes(1-\zz)\otimes\zz\otimes\bz\\
&-\zz\otimes(1-\zz)\otimes\bz\otimes\zz-\zz\otimes\bz\otimes(1-\zz)\otimes\zz\,.
\end{align*}
\item Step $(i)$:
The newly added terms violate the integrability condition. To correct it, we must add two new terms,
\bean
\zz\otimes\bz\otimes(1-\bz)\otimes\bz-\bz\otimes\zz\otimes(1-z)\otimes\zz\,.
\eean
\item Step $(ii)$:
We again need to add terms that combine with the two above to have invariants in the first entry,
\bean
\bz\otimes\bz\otimes(1-\bz)\otimes\bz-\zz\otimes\zz\otimes(1-z)\otimes\zz\,.
\eean
\end{itemize}
At this point the symbol satisfies both the first entry and integrability conditions, and we obtain a tensor which agrees with the symbol for $F(\zz,\zbar)$ in \refE{eq_delta1111F}. 

Note that we can again easily promote the symbol to the full function. Indeed, the symbol alphabet $\cA_\triangle=\{z,\zbar,1-z,1-\zbar\}$ combined with the first entry condition again implies that $F(\zz,\zbar)$ can be expressed in terms of single-valued harmonic polylogarithms. Taking into account the antisymmetry under exchange of $z$ and $\zbar$ we find that there is a one-parameter family of functions with the correct symbol,
\beq\bsp\label{eq:gen_sol}
F(\zz,\zbar) =&\, 6\big[\Li_4\left(z\right)-\Li_4(\zbar)\big] 
-3\ln\left(z \zbar\right)\big[\Li_3\left(z\right)-\Li_3(\zbar)\big]\\
&\,+\frac{1}{2}\ln^2(z \zbar)\big[\Li_2(z)-\Li_2(\zbar)\big]+ c\,\cP_2(z)\,,
\esp\eeq
where $c$ is a real constant (of weight two). This constant can be fixed by explicitly computing the discontinuity of the function $F(\zz,\zbar)$ in the variable $p_3^2$, and imposing that the discontinuity agrees with the result for the cut integral~\eqref{cutP3Res}, i.e., by requiring that (cf. eq.~\eqref{eq:Cutp3Res}),
\beq
\Disc_{p_3^2}F(z,\zbar) = i\,(p_1^2)^2\,(1-z)(1-\zbar)(z-\zbar)\,\cut_{p^2_3,R^3_\triangle}T_{L}(p_1^2,p_2^2,p_3^2)\,.
\eeq
It is easy to check that we must have $c=0$. Note that if the free parameters in the solution multiply functions that vanish when a discontinuity in a given channel is taken, we can supplement this procedure by considering cuts in other channels. In this way we can fix the initial condition up to a polylogarithmic function that does not have any discontinuities, and must thus be a constant. This constant can easily be fixed by computing the value of the original Feynman integral numerically in a single point.

We note that for both examples considered above, the same exercise could have been done using the results for cuts in other channels.

\subsection{Reconstructing the coproduct from double unitarity cuts}

While the possibility of reconstructing the function from a single cut in a given channel might not be surprising, due to the fact that Feynman integrals can be written as dispersive integrals over the discontinuity in a given channel, we show in this section that in this particular case we are able to reconstruct the full answer for  $\Delta_{1,1,2}F$ from the knowledge of  just \emph{one} sequential double cut, along with the symbol alphabet. Note that $\Delta_{1,1,2}F$ is completely equivalent to the symbol $\cS(F)$. Indeed, the weight two part of $\Delta_{1,1,2}F$ is defined only modulo $\pi$, which is precisely the amount of information contained in the symbol.

Assuming that the symbol letters are drawn from the set $\mathcal{A}_\triangle=\{z,\zbar,1-z,1-\zbar\}$, we can write $\Delta_{1,1,2}F$ in the following general form:
\begin{equation*}
\Delta_{1,1,2}F = \sum_{(x_1,x_2)\in\mathcal{A}_{\triangle}^2}\log x_1\otimes \log x_2\otimes f_{x_1,x_2}\, ,
\end{equation*}
where the $f_{x_1,x_2}$ denote 16 a priori unknown functions of weight two (defined only modulo $\pi^2$).
Imposing the first entry condition and the integrability condition in the first two entries of the coproduct gives the following constraints among the $f_{x_1,x_2}$:
\begin{align}\bsp
\label{integrability_constraints}
&f_{z,z}=f_{\zbar,z}=f_{z,\zbar}=f_{\zbar,\zbar}\,,\\
&f_{1-z,z}=f_{1-\zbar,z}=f_{z,1-\zbar}=f_{\zbar,1-\zbar}\,,\\
&f_{z,1-z}=f_{\zbar,1-z}=f_{1-z,\zbar}=f_{1-\zbar,\zbar}\,,\\
&f_{1-z,1-z}=f_{1-\zbar,1-z}=f_{1-z,1-\zbar}=f_{1-\zbar,1-\zbar}\,,
\esp\end{align}
which reduces the number of unknown functions to 4. Defining $\tilde F(z,\zbar)=F(\zbar,z)$, we must require in addition that $\tilde F(z,\zbar)=-F(z,\zbar)$ (because its leading singularity is likewise odd under this exchange), which gives further constraints. For instance,
\begin{equation}
f_{1-\zbar,z}=-\tilde f_{1-z,\zbar}\, .
\end{equation}
We can thus write
  \begin{align}\bsp
 \Delta_{1,1,2}F =& \log (z\zbar)\otimes\log (z\zbar)\otimes f_{z,z} + \log ((1-z)(1-\zbar))\otimes\log ((1-z)(1-\zbar))\otimes f_{1-z,1-z}  \\ 
&+ \left[\log (z\zbar)\otimes \log(1-z) + \log ((1-z)(1-\zbar))\otimes\log \zbar\right]\otimes f_{1-z,\zbar}\\
&- \left[\log (z\zbar)\otimes \log(1-\zbar) + \log ((1-z)(1-\zbar))\otimes\log z \right]\otimes  \tilde f_{1-z,\zbar}\,.
 \label{eq:112_ansatz}
\esp\end{align}

Notice that up to this stage all the steps are generic: we have not used our knowledge of the functional form of any of the double cuts which determine the $f_{x_1,x_2}$, but only the knowledge of the set of variables entering its symbol
and the antisymmetry of the leading singularity under the exchange of $z$ and $\bz$.

We now assume that we know the value of $\cut_{p_3^2,p_2^2}F$, and thus by \refE{C32} we have determined that 
\beq\label{eq:Fu2zb}
\delta_{1-z,\bz}F=-\log z \log \bz + \half \log^2 z\,.
\eeq

Next, we have to require that eq.~\eqref{eq:112_ansatz} be integrable in the second and third component. Assuming again that we only consider symbols with letters drawn from the set $\mathcal{A}_\triangle$, we use eq.~\eqref{eq:Fu2zb} and impose the integrability condition eq.~\eqref{eq:integrability}, and we see that the symbols of the two unknown functions in eq.~\eqref{eq:112_ansatz} are uniquely fixed,
\begin{equation*}\bsp
\cS(f_{z,z}) &\,= -z\otimes(1-z)+\zbar\otimes(1-\zbar) =\cS(\Li_2(z)-\Li_2(\bz))\,,\\
\cS(f_{1-z,1-z}) &\,= 0\,,
\esp\end{equation*}
in agreement with eq.~\eqref{delta112}.

Note that once again we can easily integrate the symbol to the full function by an argument similar to the one presented in Section~\ref{sec_singleCutRec}: the most general function having the correct symbol is again given by eq.~\eqref{eq:gen_sol}, and the constant $c$ can easily be shown to vanish by requiring the function to have the correct double discontinuity, i.e., by imposing that
\beq
\Disc_{p_3^2,p_2^2}F(z,\zbar) = -i\,(p_1^2)^2\,(1-z)(1-\zbar)(z-\zbar)\,\cut_{p_3^2,p_2^2}T_{L}(p_1^2,p_2^2,p_3^2)\,.
\eeq

We stress that the fact that we can reconstruct $\Delta_{1,1,2}F $ from a single sequence of cuts is not related to the specific sequence we chose.  For example, if we had computed only $\cut_{p_1^2,p_2^2}F$ and thus determined that $-f_{z,\bz}-f_{1-z,\bz}=-\Li_2(z)+\Li_2(\bz)+\log z \log \bz - \half \log^2 z$, the integrability condition would fix the remaining two free coefficients in a similar way.
Finally, we could consider $\cut_{p_3^2,p_1^2}F$, but since this cut is obtained by a simple change of variables from $\cut_{p_3^2,p_2^2}F$ through the reflection symmetry of the ladder, it is clear that integrability fixes the full symbol once again.  

Let us briefly consider the analogous construction for the one-loop triangle, where the $f_{x_1,x_2}$ are simply constant functions.  The analog of \refE{eq:112_ansatz} above is
  \begin{align}\bsp
 \Delta_{1,1}\mathcal{T} =& f_{z,z}\left(\log (z\zbar)\otimes\log (z\zbar)\right) +f_{1-z,1-z}  \left(\log ((1-z)(1-\zbar))\otimes\log ((1-z)(1-\zbar))\right)  \\ 
&+f_{1-z,\zbar} \left[\log (z\zbar)\otimes \log(1-z) + \log ((1-z)(1-\zbar))\otimes\log \zbar\right] \\
&+ f_{1-\bz,z} \left[\log (z\zbar)\otimes \log(1-\zbar) + \log ((1-z)(1-\zbar))\otimes\log z \right]  \,.
 \label{eq:11_ansatz}
\esp\end{align}
A specific double cut, without loss of generality say $\Cut_{p_3^2,p_2^2}$, gives a constant value for $f_{1-z,\zbar}$, as seen from \refE{eq:cuttrip3p2} and \refE{C32}.
We have a consistent solution with $f_{1-z,\zbar}=- f_{1-\zbar,z}=1/2$ and $f_{z,z}=f_{1-z,1-z}=0$, which is indeed the $\Delta_{1,1}$ of the triangle, obtained by a consistent completion algorithm as in the previous subsection.

While it is quite clear that the reason why the algorithm of section \ref{sec_singleCutRec} converged was the existence of a dispersive representation of Feynman integrals, we do not know whether the existence of a double dispersive representation is a necessary condition for the reconstruction based on the knowledge of $\Delta_{1,1,2}$ done in this section to work, although it does seem reasonable that it would be the case.

In closing, we notice that in this example, the integrability condition \refE{integrability_constraints} implies that $\Cut_{p_i^2,p_j^2}=\Cut_{p_j^2,p_i^2}$, through the relations listed in \refE{eq:tableeqns}.  It would be interesting to see whether there is a general link between the integrability of the symbol and the permutation invariance of a sequence of cuts.

\section{Discussion}

In this paper we have studied the analytic structure of Feynman integrals revealed by their unitarity cuts.  The final objective of our investigation is the reconstruction of Feynman integrals through the knowledge of their cuts.  For the class of Feynman integrals with massless propagators that may be expressed in terms of the iterated integrals known as multiple polylogarithms, we have formulated precise relations between discontinuities across their physical branch cuts, their unitarity cuts, and their coproduct.  We have proposed that the structure of iterated integrals and their Hopf algebra form a natural framework for constructing Feynman integrals from their cuts.  Furthermore, we have presented techniques to be used in the analytic evaluation of cut integrals.

The ultimate advantage of a cut-based computation is that multi-loop cut diagrams reduce to integrals over products of simpler lower-loop integrals with on-shell external legs.   Techniques for direct computation of cut integrals in $D$ spacetime dimensions are far less developed than those for ordinary (uncut) loop integrals. A well established technique for the calculation of multi-loop diagrams is the integration over an off-shell subdiagram. This was seen here for cut integrals at the two-loop level, where different cuts were computed using one-loop triangle and box integrals with massless or a limited number of massive external legs. This method has the potential to be applied to more complicated multi-loop and multi-leg cut integrals.

In this paper we took $D=4-2\epsilon$-dimensional cuts. This is a necessity when dealing with infrared-divergent cut integrals: notably, individual cuts of (multi-loop) integrals that are themselves finite in four dimensions may be divergent when the internal propagators that are put on shell are massless. The sum of all cuts in a given channel corresponds, according to the largest time equation~\cite{tHooft:1973pz,Veltman:1994wz}, to the discontinuity of the uncut integral; given that the latter is finite, one expects complete cancellation of the singularities among the different cuts. This situation was encountered here upon taking unitarity cuts of the two-loop ladder graph, where we saw that the pattern of cancellation is similar to the familiar real-virtual cancellation mechanism in cross sections, although this example does not correspond to a cross section. Understanding this pattern of cancellation is useful for the general program of developing efficient subtraction procedures for infrared singularities, and it would be interesting to explore how this generalizes for other multi-loop integrals. 

Taking a step beyond the familiar case of a single unitarity cut, we developed the concept of a sequence of unitarity cuts. To define this notion consistently, we extended the cutting rules of refs.~\cite{tHooft:1973pz,Veltman:1994wz} to accommodate multiple cuts in different channels in an appropriately chosen kinematic region.
The cutting rules specify a unique prescription for complex conjugation of certain vertices and propagators, which is dictated by the channels on which cuts are taken. Importantly, the result does not depend on the order in which the cuts are applied.
The kinematic region is chosen such that the Mandelstam invariants corresponding to the cut channels are positive, corresponding to timelike kinematics. In its center-of-mass Lorentz frame, this invariant defines the energy flowing through the set of on-shell propagators. The energy flow through all these propagators has a consistent direction that is dictated by the external kinematics; for any given propagator, this direction must be consistent with the direction of energy flow assigned to it by any other cut in the sequence. We further exclude crossed cuts, as well as iterated cuts in the same channel since they are not related to discontinuities as computed in this paper. Finally, we restrict ourselves to real kinematics.
These cutting rules pass numerous consistency checks and they form a central result of the present paper.  In the future, it will of course be interesting to study what information is contained in crossed cuts and in iterated cuts in the same channel, as well as what can be obtained by allowing for complex kinematics.  It will also be important to study nonplanar examples, in which it is less obvious how to identify suitable kinematic regions.

Having specified the definition of a sequence of unitarity cuts, we find the following correspondence, which we conjecture to be general, among 
\begin{itemize}
\item[(a)]{} the sum of all cut diagrams in the channels $s_1,\ldots s_k$, which we denote by $\Cut_{s_1,\ldots,s_k}$;
\item[(b)]{}a sequence of discontinuity operations, which we denote by $\Disc_{s_1,\ldots,s_k}$;
\item[(c)]{} the weight $n-k$ cofactors of the terms in the coproduct of the form $\Delta_{1,1,\ldots, 1, n-k}$, where each of the $k$ weight one entries of a specific term in $\Delta_{1,1,\ldots, 1, n-k}$ is associated with the $s_i$ in a well defined manner.
\end{itemize}
The correspondence is formulated in eqs.~(\ref{eq:cutequalsdisc}) and (\ref{eq:discequalsdelta}). 
We illustrated it using the three-mass triangle and the two-loop three-point ladder examples where one may take up to two sequential cuts with any combination of channels, obtaining nontrivial results; the relations are summarized by eqs.~(\ref{eq:tableeqns}). 
In examples with more loops and legs, we expect that a deeper sequence of unitarity cuts may be attainable.

The entries of the symbol (or equivalently of the $\Delta_{1,\ldots, 1, n-k}$ terms in the coproduct) are drawn from a list $\{x_i\}$ of algebraic functions of the Mandelstam invariants, which we have referred to as the symbol alphabet $\cA$. These are also the natural variables appearing as arguments of logarithms and polylogarithms in both cut diagrams and the original uncut one. 
For example, in the two-loop ladder triangle considered through ${\cal O}(\epsilon^0)$, the alphabet consists of four letters $\{z,\zbar,1-z,1-\zbar\}$ defined in eq.~(\ref{eq:z}). 
The letters in the symbol  are the solutions of quadratic equations which emerge upon solving the simultaneous on-shell conditions imposed by cuts. Consequently, cuts may be used to identify the relevant variables in terms of which the uncut integral can be most naturally expressed.

Because the arguments of polylogarithms, and equivalently the entries of the coproduct $\Delta_{1,1,\ldots, 1, n-k}$ terms, are not the Mandelstam invariants themselves, 
while any unitarity cut is defined by a channel that does correspond to a Mandelstam invariant $s_i$, 
it is not immediately obvious how to formulate the relation to the coproduct, \refE{eq:discequalsdelta}.  Here, we have given a precise correspondence in terms of discontinuities of ordinary logarithms, analytically continued between two kinematic regions corresponding to those before and after the cut in each invariant $s_i$.  The $i\varepsilon$ prescription of $x_i$ is inherited from that of $s_i$, and the relation \refE{eq:discequalsdelta} is thus precise.

We verified that the expected relations between sequences of cuts, sequences of discontinuities and the relevant terms of the coproduct hold in the cases of the double cut of the one-loop triangle, the four-mass box and the two-mass-hard box. We then explored in detail the much less trivial two-loop three-mass ladder diagram, and also there we found complete agreement with the expected relations.

Given that cut diagrams may be simpler to compute (because they reduce to integrals over products of simpler lower-loop amplitudes) and may identify the most convenient variables, it is natural to ponder whether the result of a cut diagram can be uplifted to obtain the uncut function. In the case of a single unitarity cut, this can always be done through a dispersion relation~\cite{Landau:1959fi,Cutkosky:1960sp,tHooft:1973pz,Remiddi:1981hn,Veltman:1994wz}.   
In the case of a sequence of unitarity cuts, this requires a multiple dispersion relation, and the general conditions for these to exist are not known. 

In section~\ref{sec:integrable}, we made progress in developing methods for the reconstruction of a Feynman integral from its cuts. 
Our first observation, considering the reconstruction of the one-loop three-mass triangle from either its single or double cut, was that while dispersion relations may appear as complicated integrals, they become simple when expressed in terms of the symbol alphabet $\cA$. The dispersion integral then falls into the class of iterated integrals amenable to Hopf algebra techniques. This is of course consistent with the fact that each dispersion integral is expected to raise the transcendental weight of the function by one: it is the opposite operation to taking the discontinuity of the function across its branch cut.
It is clearly important to study this connection between dispersion integrals and iterated polylogarithmic integrals for other examples.

We next presented algebraic ways to reconstruct information about the full function from the knowledge of a single set of cuts, along with the symbol alphabet. This was achieved by using two main constraints: the integrability of the symbol and the first entry condition.  More precisely, we showed how to reconstruct the symbol of the full integral from the knowledge of a single unitarity cut in one of the channels.   We believe that our approach to reconstruction is valid generally, provided the existence of a dispersive representation of Feynman integrals.
We also showed that in the case of the two-loop ladder (and the much simpler one-loop triangle) it is possible to reconstruct all the terms of the $\Delta_{1,1,2}$ component of the coproduct, and then the full function, of the uncut integral from the knowledge of a single sequence of double cuts. How general this procedure is is less obvious to us, and it is certainly worth investigating.

Another very intriguing observation based on the examples at hand concerns the connection between the integrability condition of the symbol and the equality of sequences of unitarity cuts in which the order is permuted. 
As mentioned above, the result of a sequence of unitarity cuts does not depend on the order in which the cuts are applied. Therefore the double cut relations summarized in eqs.~(\ref{eq:tableeqns}) must satisfy $\Cut_{p_i^2, p_j^2} = \Cut_{p_j^2, p_i^2}$. This in turn implies highly nontrivial relations among different $\Delta_{1,1,2}$ components; for example the r.h.s.\ of eq.~(\ref{C12}) must be the same as the r.h.s.\ of eq.~(\ref{C21}), and similarly for the other pairs. The crucial observation is that these relations indeed hold owing to the integrability constraints as summarized in eq.~(\ref{integrability_constraints}). Note that the latter  are based solely on the symbol alphabet and the integrability condition of eq.~(\ref{eq:integrability}). We leave it for future study to determine how general the connection is between integrability and permutation invariance of a sequence of cuts. 

In conclusion, we developed new techniques to evaluate cut Feynman integrals and relate these to the original uncut ones. In dealing with complicated multi-loop and multi-leg Feynman integrals there is a marked advantage to computing cuts, where lower-loop information can be systematically put to use. While cut integrals are simpler than uncut ones, they depend on the kinematics through the same variables, the symbol alphabet $\cA$,
 which characterize the analytic structure of the integral. Identifying this alphabet is crucial in relating cuts to terms in the coproduct, and then  
either integrating the dispersion relation or reconstructing the symbol of the uncut integral algebraically. We have demonstrated that the language of the Hopf algebra of polylogarithms is highly suited for understanding the analytic structure of Feynman integrals and their cuts. Finally, we have shown that there is great potential for computing Feynman integrals by using multiple unitarity cuts, and further work in this direction is in progress.

\acknowledgments

We have benefitted from discussions with Leandro Almeida, Vladimir Braun, Hanna Gr\"onqvist, Gregory Korchemsky, and Edoardo Mirabella.
S.A. and C.D. were supported by the Research Executive Agency (REA) of the European Union under the Grant Agreement number PITN-GA-2010-264564 (LHCPhenoNet).
S.A. acknowledges support from Funda\c{c}\~ao para a Ci\^encia e a Tecnologia, Portugal, through a doctoral degree fellowship (SFRH/BD/69342/2010).
R.B. was supported in part by the Agence Nationale de la Recherche under grant ANR-09-CEXC-009-01 and is grateful to the BCTP of Bonn University for extensive hospitality in the course of this project.  R.B. and C.D. thank the Higgs Centre for Theoretical Physics at the University of Edinburgh for its hospitality.
E.G. was supported in part by the STFC grant ``Particle Physics at the Tait Institute'' and thanks the IPhT of CEA-Saclay for its hospitality.




\appendix

\section{Notation and conventions}
\label{app:conventions}

\paragraph{Feynman rules.}

Here we summarize the Feynman rules for cut diagrams in massless scalar theory.  For a discussion of their origin, as well as the rules for determining whether a propagator is cut or uncut, see section \ref{sec:disccutdelta}.

\begin{itemize}

\item Vertex:
\begin{align}
\raisebox{-0.1mm}{\includegraphics[keepaspectratio=true, height=0.2cm]{./diagrams/blackVertex.pdf}}=i
\end{align}

\item Complex conjugated vertex:
\begin{align}
\raisebox{-0.1mm}{\includegraphics[keepaspectratio=true, height=0.2cm]{./diagrams/whiteVertex.pdf}}=-i
\end{align}

\item Propagator:
\begin{align}
\raisebox{-0.1mm}{\includegraphics[keepaspectratio=true, height=0.55cm]{./diagrams/propagator.pdf}}=\frac{i}{p^2+i\varepsilon}
\end{align}

\item Complex conjugated propagator:
\begin{align}
\raisebox{-0.1mm}{\includegraphics[keepaspectratio=true, height=0.55cm]{./diagrams/whitePropagator.pdf}}=\frac{-i}{p^2-i\varepsilon}
\end{align}

\item Cut propagator:
\begin{align}
\raisebox{-4.7mm}{\includegraphics[keepaspectratio=true, height=1.2cm]{./diagrams/CutBBPropagator.pdf}}
= 
\raisebox{-4.7mm}{\includegraphics[keepaspectratio=true, height=1.2cm]{./diagrams/CutBWPropagator.pdf}}
=
\raisebox{-4.7mm}{\includegraphics[keepaspectratio=true, height=1.2cm]{./diagrams/CutWBPropagator.pdf}}
= 
\raisebox{-4.7mm}{\includegraphics[keepaspectratio=true, height=1.2cm]{./diagrams/CutWWPropagator.pdf}}
=
2\pi\, \delta\left(p^2\right)
\end{align}
There can be multiple dashed lines, indicating cuts, on the same propagator, without changing its value.  
There is a theta function restricting the direction of energy flow on a cut propagator, whose origin is detailed in Section \ref{sec:disccutdelta}.  In the examples, we omit writing the theta function, as there is always at most one nonvanishing configuration.

\item Loop factor (for loop momentum $k$):
\begin{equation}
\left(\frac{e^{\gamma_E\epsilon}}{\pi^{2-\epsilon}}\right)\int d^{4-2\epsilon}k\,.
\end{equation}

\end{itemize}

\paragraph{Kinematic regions.}

For the three-point triangle and ladder, we use the following shorthand for different kinematic regions, with variables $z,\bz$ as defined in eq.~\eqref{eq:z},
\bea
&& R^*_\triangle: \quad p_1^2, p_2^2, p_3^2>0, \quad \bz=z^*\,, \\
&& R^{i}_\triangle: \quad p_i^2>0, ~\textrm{and}~ p_k^2<0 ~\textrm{for all}~k \neq i\,, \\
&& R^{i,j}_\triangle: \quad p_i^2, p_j^2>0, ~\textrm{and}~ p_k^2<0 ~\textrm{for all}~k \neq i,j .
\eea

\section{Some explicit results of one-loop diagrams and one-loop cut diagrams}
\label{app_oneLoopRes}

All the results are computed according to the conventions of appendix \ref{app:conventions}. Unless indicated otherwise, expressions are given for spacelike invariants.

\subsection{Three-point functions}

\paragraph{One mass}
\begin{align}
T^{1m}(p_1^2)&=-\left(\frac{e^{\gamma_E\epsilon}}{\pi^{2-\epsilon}}\right)\int  dk^{4-2\epsilon}\frac{1}{k^2+i\varepsilon}\frac{1}{(p_1-k)^2+i\varepsilon}\frac{1}{(p_3+k)^2+i\varepsilon}\nonumber\\
&=i\frac{c_\Gamma}{\epsilon^2}(-p_1^2)^{-1-\epsilon}\,.
\end{align}
\begin{align}
\cut_{p_1^2}T^{1m}(p_1^2)&=-(2\pi)^2\left(\frac{e^{\gamma_E\epsilon}}{\pi^{2-\epsilon}}\right)\int  dk^{4-2\epsilon}\frac{\delta(k^2)\delta\left((p_1-k)^2\right)}{(p_3+k)^2-i\varepsilon}\nonumber\\
&=-2\pi\frac{e^{\gamma_E\epsilon}\Gamma(1-\epsilon)}{\epsilon\Gamma(1-2\epsilon)}(p_1^2)^{-1-\epsilon}\theta(p_1^2)\,.
\end{align}

\paragraph{Two masses}

\begin{align}
T^{2m}(p_1^2,p_2^2)&=-\left(\frac{e^{\gamma_E\epsilon}}{\pi^{2-\epsilon}}\right)\int  dk^{4-2\epsilon}\frac{1}{k^2+i\varepsilon}\frac{1}{(p_1-k)^2+i\varepsilon}\frac{1}{(p_3+k)^2+i\varepsilon}\nonumber\\
&=-i\frac{c_\Gamma}{\epsilon^2}\frac{(-p_1^2)^{-\epsilon}-(-p_2^2)^{-\epsilon}}{p_1^2-p_2^2}\,.
\end{align}

\begin{align}
\cut_{p_1^2} T^{2m}(p_1^2,p_2^2) &=-(2\pi)^2\left(\frac{e^{\gamma_E\epsilon}}{\pi^{2-\epsilon}}\right)\int  dk^{4-2\epsilon}\frac{\delta(k^2)\delta\left((p_1-k)^2\right)}{(p_3+k)^2-i\varepsilon}\nonumber\\
&=-2\pi\frac{e^{\gamma_E\epsilon}\Gamma(1-\epsilon)}{\epsilon\Gamma(1-2\epsilon)}\frac{(p_1^2)^{-\epsilon}}{p_1^2-p_2^2}\theta(p_1^2)\,.
\end{align}

\paragraph{Three masses}

\beq\bsp
\cut_{p_1^2}&T(p_1^2,p_2^2,p_3^2)=-(2\pi)^2\frac{e^{\gamma_E \epsilon}}{\pi^{2-\epsilon}}\int d^{4-2\epsilon}k\frac{\delta(k^2)\delta((p_1-k)^2)}{(p_3+k)^2-i\varepsilon}\\
&=-2\pi\frac{e^{\gamma_E \epsilon}\Gamma(1-\epsilon)}{\Gamma(2-2\epsilon)}(p_1^2)^{-1-\epsilon}\frac{1}{z-1+u_3}\,_2F_1\left(1,1-\epsilon;2-2\epsilon;\frac{z-\zbar}{z-1+u_3}\right)\theta(p_1^2)\\
&=\frac{2\pi}{p_1^2(z-\zbar)}\log\left(\frac{z(1-\zbar)}{\zbar(1-z)}\right)\theta(p_1^2) +\mathcal{O}(\epsilon) \,.
\label{cutP1TriDDim}
\esp\eeq

\beq\bsp
\cut_{p_2^2}&T(p_1^2,p_2^2,p_3^2)=-(2\pi)^2\frac{e^{\gamma_E \epsilon}}{\pi^{2-\epsilon}}\int d^{4-2\epsilon}k\frac{\delta(k^2)\delta((p_2-k)^2)}{(p_1+k)^2-i\varepsilon}\\
&=-2\pi\frac{e^{\gamma_E \epsilon}\Gamma(1-\epsilon)}{\Gamma(2-2\epsilon)}u_2^{-\epsilon}(p_1^2)^{-1-\epsilon}\frac{1}{1-\zbar}\,_2F_1\left(1,1-\epsilon;2-2\epsilon;\frac{z-\zbar}{1-\zbar}\right)\theta(p_2^2)\\
&=\frac{2\pi}{p_1^2(z-\zbar)}\log\left(\frac{1-z}{1-\zbar}\right)\theta(p_2^2) +\mathcal{O}(\epsilon) \,.
\label{cutP2TriDDim}
\esp\eeq

\beq\bsp
\cut_{p_3^2}&T(p_1^2,p_2^2,p_3^2)=-(2\pi)^2\frac{e^{\gamma_E \epsilon}}{\pi^{2-\epsilon}}\int d^{4-2\epsilon}k\frac{\delta(k^2)\delta((p_3-k)^2)}{(p_1+k)^2-i\varepsilon}\\
&=-2\pi\frac{e^{\gamma_E \epsilon}\Gamma(1-\epsilon)}{\Gamma(2-2\epsilon)}u_3^{-\epsilon}(p_1^2)^{-1-\epsilon}\frac{1}{z}\,_2F_1\left(1,1-\epsilon;2-2\epsilon;\frac{z-\zbar}{z}\right)\theta(p_3^2)\\
&=\frac{2\pi}{p_1^2(z-\zbar)}\log\left(\frac{\zbar}{z}\right)\theta(p_3^2) +\mathcal{O}(\epsilon) \,.
\label{cutP3TriDDim}
\esp\eeq

\beq\bsp
\left(\cut_{p_1^2}\right.&\left.\circ\cut_{p_3^2}\right)_{R^{1,3}_\triangle}T(p_1^2,p_2^2,p_3^2)=i(2\pi)^3\frac{e^{\gamma_E \epsilon}}{\pi^{2-\epsilon}}\int d^{4-2\epsilon}k\delta(k^2)\delta((p_3-k)^2)\delta((p_1+k)^2)\\
&=4\pi^2i\frac{e^{\gamma_E \epsilon}}{\Gamma(1-\epsilon)}u_3^{-\epsilon}(p_1^2)^{-1-\epsilon}(z-\zbar)^{-1+2\epsilon}z^{-\epsilon}(-\bar{z})^{-\epsilon}\theta(p_1^2)\theta(p_3^2)\,.
\esp\eeq

\subsection{Four-point functions}

The results for the uncut diagrams are taken from ref.~\cite{Bern:1993kr}. Cuts are computed using $\cut_s=-\Disc_s$.

\paragraph{Two masses, hard ($p^2_3$, $p_4^2\neq0$)}
\beq\bsp
&B^{2mh}(p_3^2,p_4^2;s,t)
\\
&=\left(\frac{e^{\gamma_E\epsilon}}{\pi^{2-\epsilon}}\right)\int  dk^{4-2\epsilon}\frac{1}{k^2+i\varepsilon}\frac{1}{(p_1+k)^2+i\varepsilon}\frac{1}{(p_1+p_2+k)^2+i\varepsilon}
\frac{1}{(k-p_4)^2+i\varepsilon}\\
&=ic_\Gamma\frac{(-p_3^2)^\epsilon(-p_4^2)^{\epsilon}}{(-t)^{1+2\epsilon}(-s)^{1+\epsilon}}\left[\frac{1}{\epsilon^2}+2\Li_2\left(1-\frac{t}{p_3^2}\right)+2\Li_2\left(1-\frac{t}{p_4^2}\right)\right]+\mathcal{O}(\epsilon)\,.
\esp\eeq

\beq\bsp
\cut_{t}&B^{2mh}(p_3^2,p_4^2;s,t)=(2\pi)^2\frac{e^{\gamma_E \epsilon}}{\pi^{2-\epsilon}}\int d^{4-2\epsilon}k\frac{\delta\left((k+p_1)^2\right)\delta((k-p_4)^2)}{(k^2+i\varepsilon)((k+p_1+p_2)^2-i\varepsilon)}\\
&=-4\pi c_\Gamma\frac{(-p_3^2)^\epsilon(-p_4^2)^{\epsilon}}{t^{1+2\epsilon}(-s)^{1+\epsilon}}\left[\frac{1}{\epsilon}+\ln\left(1-\frac{t}{p_3^2}\right)+\ln\left(1-\frac{t}{p_4^2}\right)\right]\theta(t)+\mathcal{O}(\epsilon)\,.
\esp\eeq

\paragraph{Three masses ($p_2^2$, $p^2_3$, $p_4^2\neq0$)}
\beq\bsp
& B^{3m}(p_2^2,p_3^2,p_4^2;s,t)
\\
&=\left(\frac{e^{\gamma_E\epsilon}}{\pi^{2-\epsilon}}\right)\int  dk^{4-2\epsilon}\frac{1}{k^2+i\varepsilon}\frac{1}{(p_1+k)^2+i\varepsilon}\frac{1}{(p_1+p_2+k)^2+i\varepsilon}
\frac{1}{(k-p_4)^2+i\varepsilon}\\
&=i\frac{c_\Gamma}{st-p_2^2p_4^2}\left\{\frac{2}{\epsilon^2}\left[(-s)^{-\epsilon}+(-t)^{-\epsilon}-(-p_2^2)^{-\epsilon}-(-p_3^2)^{-\epsilon}-(-p_4^2)^{-\epsilon}\right]\right.\\
&~~+\frac{1}{\epsilon^2}\frac{(-p_2^2)^{-\epsilon}(-p_3^2)^{-\epsilon}}{(-t)^{-\epsilon}}+\frac{1}{\epsilon^2}\frac{(-p_3^2)^{-\epsilon}(-p_4^2)^{-\epsilon}}{(-s)^{-\epsilon}}-2\Li_2\left(1-\frac{p_2^2}{s}\right)\\
&~~\left.-2\Li_2\left(1-\frac{p_4^2}{t}\right)+2\Li_2\left(1-\frac{p_2^2p_4^2}{st}\right)-\ln^2\frac{s}{t}\right\}+\mathcal{O}(\epsilon)\,.
\esp\eeq

\beq\bsp
\cut_{t}&B^{3m}(p_2^2,p_3^2,p_4^2;s,t)=(2\pi)^2\frac{e^{\gamma_E \epsilon}}{\pi^{2-\epsilon}}\int d^{4-2\epsilon}k
\frac{\delta\left((k+p_1)^2\right)\delta((k-p_4)^2)}{(k^2+i\varepsilon)((k+p_1+p_2)^2-i\varepsilon)}\\
=&\frac{e^{\gamma_E\epsilon}\Gamma(1-\epsilon)}{\Gamma(1-2\epsilon)}\frac{2\pi}{st-p_2^2p_4^2}\left[\frac{2}{\epsilon}t^{-\epsilon}-\frac{1}{\epsilon}t^{\epsilon}(-p_2^2)^{-\epsilon}(-p_3^2)^{-\epsilon}
+2\ln\left(1-\frac{p_4^2}{t}\right)\right.\\
&\left.-2\ln\left(1-\frac{p_2^2p_4^2}{st}\right)-2\ln(-s)+2\ln t\right]\theta(t)+\mathcal{O}(\epsilon)\,.
\label{3mCut}
\esp\eeq

\section{Explicit results for the single unitarity cuts}
\label{singleCutRes}

We present the results we obtained for the single unitarity cuts. These results were computed and numerically
checked in the region where $\zbar=z^*$. For cut [45] the hypergeometric function
was expanded using {\tt HypExp}~\cite{Huber:2007dx}. We write everything in terms of multiple polylogarithms as defined in \refS{sec:Hopf}
to simplify the comparison between different terms.

\subsection{Unitarity cuts in the $p_3^2$ channel}

\beq\bsp
&\mathrm{Cut}_{p_3^2,[12],R^*_\triangle}T_L(p_1^2,p_2^2,p_3^2) \\
&= \frac{i\,c_\Gamma^2\,(p_1^2)^{-2-2\eps}}{(1-z)(1-\bar{z})(z-\bar{z})}\,\sum_{k=-2}^\infty{\eps^k}\,\left[(-2\pi i)\,f_{[12]}^{(k,1)}(z,\bz)+(-2\pi i)^2\,f_{[12]}^{(k,2)}(z,\bz)\right]\,,
\esp\eeq

\beq\bsp
f_{[12]}^{(-2,1)}(z,\bz) &\,=\log\frac{z}{\bz}\,,\\
f_{[12]}^{(-2,2)}(z,\bz) &\,=0\,,\\
f_{[12]}^{(-1,1)}(z,\bz) &\,=-G\left(0,\frac{1}{z};\frac{1}{\bz}\right)+G\left(0,\frac{1}{\bz};\frac{1}{z}\right)+\log\frac{z}{\bz}\left[G\left(\frac{1}{z};\frac{1}{\bz}\right)+G\left(\frac{1}{\bz};\frac{1}{z}\right)\right]\\
&\,-2\log[(1-z)(1-\bz)]\log\frac{z}{\bz}\,,\\
f_{[12]}^{(-1,2)}(z,\bz) &\,=\frac{1}{2}\log\frac{z}{\bz}\,,\\
f_{[12]}^{(0,1)}(z,\bz) &\,=-2\left[G\left(0,\frac{1}{z},\frac{1}{\bz};\frac{1}{\bz}\right)-G\left(0,\frac{1}{\bz},\frac{1}{z};\frac{1}{z}\right)\right]+\frac{1}{2}\log\frac{z}{\bz}\left[G\left(\frac{1}{z};\frac{1}{\bz}\right)+G\left(\frac{1}{\bz};\frac{1}{z}\right)\right]^2\\
&\,-\frac{1}{12}\log^3\frac{z}{\bz}-2\log[(1-z)(1-\bz)]\log\frac{z}{\bz}\left[G\left(\frac{1}{z};\frac{1}{\bz}\right)+G\left(\frac{1}{\bz};\frac{1}{z}\right)\right]\\
&\,+2\log^2[(1-z)(1-\bz)]\log\frac{z}{\bz}-\frac{1}{4}\log^2(z\bz)\log\frac{z}{\bz}\\
&\,+2\log[(1-z)(1-\bz)]\left[G\left(0,\frac{1}{z};\frac{1}{\bz}\right)-G\left(0,\frac{1}{\bz};\frac{1}{z}\right)\right]+\frac{\pi^2}{6}\log\frac{z}{\bz}\,,\\
f_{[12]}^{(0,2)}(z,\bz) &\,=-\frac{1}{2}\left[G\left(0,\frac{1}{z};\frac{1}{\bz}\right)-G\left(0,\frac{1}{\bz};\frac{1}{z}\right)\right]-\frac{1}{2}\log^2\frac{z}{\bz}\\
&\,+\frac{1}{2}\log\frac{z}{\bz}\left[G\left(\frac{1}{z};\frac{1}{\bz}\right)+G\left(\frac{1}{\bz};\frac{1}{z}\right)\right]-\log[(1-z)(1-\bz)]\log\frac{z}{\bz}\,.
\esp\eeq


\beq\bsp
&\mathrm{Cut}_{p_3^2,[45],R^*_\triangle}T_L(p_1^2,p_2^2,p_3^2) \\
&= \frac{i\,c_\Gamma^2\,(p_1^2)^{-2-2\eps}}{(1-z)(1-\bar{z})(z-\bar{z})}\,\sum_{k=-2}^\infty{\eps^k}\,\left[(-2\pi i)\,f_{[45]}^{(k,1)}(z,\bz)+(-2\pi i)^2\,f_{[45]}^{(k,2)}(z,\bz)\right]\,,
\esp\eeq

\beq\bsp 
f_{[45]}^{(-2,1)}(z,\bz) &\,=\log\frac{z}{\bz}\,,\\
f_{[45]}^{(-2,2)}(z,\bz) &\,=0\,,\\
f_{[45]}^{(-1,1)}(z,\bz) &\,=-G\left(0,\frac{1}{z};\frac{1}{\bz}\right)+G\left(0,\frac{1}{\bz};\frac{1}{z}\right)+\log\frac{z}{\bz}\left[G\left(\frac{1}{z};\frac{1}{\bz}\right)+G\left(\frac{1}{\bz};\frac{1}{z}\right)\right]\\
&\,-2\log[(1-z)(1-\bz)]\log\frac{z}{\bz}\,,\\
f_{[45]}^{(-1,2)}(z,\bz) &\,=-\frac{1}{2}\log\frac{z}{\bz}\,,\\
f_{[45]}^{(0,1)}(z,\bz)&\,=
-2\left[G\left(0,\frac{1}{z},\frac{1}{\bz};\frac{1}{\bz}\right)-G\left(0,\frac{1}{\bz},\frac{1}{z};\frac{1}{z}\right)\right]+2\log^2[(1-z)(1-\bz)]\log\frac{z}{\bz}\\
&\,-2\log[(1-z)(1-\bz)]\log\frac{z}{\bz}\left[G\left(\frac{1}{z};\frac{1}{\bz}\right)+G\left(\frac{1}{\bz};\frac{1}{z}\right)\right]\\
&\,+2\log[(1-z)(1-\bz)]\left[G\left(0,\frac{1}{z};\frac{1}{\bz}\right)-G\left(0,\frac{1}{\bz};\frac{1}{z}\right)\right]-\frac{1}{3}\log^3\frac{z}{\bz}\\
&\,+\frac{1}{2}\log\frac{z}{\bz}\left[G\left(\frac{1}{z};\frac{1}{\bz}\right)+G\left(\frac{1}{\bz};\frac{1}{z}\right)\right]^2+\frac{\pi^2}{6}\log\frac{z}{\bz}\,,\\
f_{[45]}^{(0,2)}(z,\bz)&\,=
-\frac{1}{2}\left[G\left(0,\frac{1}{\bz};\frac{1}{z}\right)-G\left(0,\frac{1}{z};\frac{1}{\bz}\right)\right]-\frac{1}{2}\log\frac{z}{\bz}\left[G\left(\frac{1}{z};\frac{1}{\bz}\right)+G\left(\frac{1}{\bz};\frac{1}{z}\right)\right]\\
&\,-\frac{1}{2}\log^2\frac{z}{\bz}+\log[(1-z)(1-\bz)]\log\frac{z}{\bz}\,.
\esp\eeq


\beq\bsp
&\mathrm{Cut}_{p_3^2,[135],R^*_\triangle}T_L(p_1^2,p_2^2,p_3^2)\\
& = \frac{i\,c_\Gamma^2\,(p_1^2)^{-2-2\eps}}{(1-z)(1-\bar{z})(z-\bar{z})}\,\sum_{k=-1}^\infty{\eps^k}\,\left[(-2\pi i)\,f_{[135]}^{(k,1)}(z,\bz)+(-2\pi i)^2\,f_{[135]}^{(k,2)}(z,\bz)\right]\,,
\esp\eeq

\beq\bsp
f_{[135]}^{(-2,1)}(z,\bz) &\,=-\log\frac{z}{\bz}\,,\\
f_{[135]}^{(-2,2)}(z,\bz) &\,=0\,,\\
f_{[135]}^{(-1,1)}(z,\bz) &\,=G\left(0,\frac{1}{z};\frac{1}{\bz}\right)-G\left(0,\frac{1}{\bz};\frac{1}{z}\right)-\log\frac{z}{\bz}\left[G\left(\frac{1}{z};\frac{1}{\bz}\right)+G\left(\frac{1}{\bz};\frac{1}{z}\right)\right]\\
&\,+2\log[(1-z)(1-\bz)]\log\frac{z}{\bz}+\frac{1}{2}\log\frac{z}{\bz}\log(z\bz)\,,\\
f_{[135]}^{(-1,2)}(z,\bz) &\,=0\,,\\
f_{[135]}^{(0,1)}(z,\bz) &\,=2\left[G\left(0,\frac{1}{z},\frac{1}{\bz};\frac{1}{\bz}\right)-G\left(0,\frac{1}{\bz},\frac{1}{z};\frac{1}{z}\right)\right]+6[\text{Li}_3(z)-\text{Li}_3(\bz)]\\
&\,+2\log[(1-z)(1-\bz)]\log\frac{z}{\bz}\left[G\left(\frac{1}{z};\frac{1}{\bz}\right)+G\left(\frac{1}{\bz};\frac{1}{z}\right)\right]\\
&\,-\frac{1}{2}\log\frac{z}{\bz}\left[G\left(\frac{1}{z};\frac{1}{\bz}\right)+G\left(\frac{1}{\bz};\frac{1}{z}\right)\right]^2-\frac{3}{2}[\text{Li}_2(z)-\text{Li}_2(\bz)]\log(z\bz)\\
&\,+\frac{1}{2}\log(z\bz)\log\frac{z}{\bz}\left[G\left(\frac{1}{z};\frac{1}{\bz}\right)+G\left(\frac{1}{\bz};\frac{1}{z}\right)\right]-\frac{3}{2}[\text{Li}_2(z)+\text{Li}_2(\bz)]\log\frac{z}{\bz}\\
&\,-2\log[(1-z)(1-\bz)]\left[G\left(0,\frac{1}{z};\frac{1}{\bz}\right)-G\left(0,\frac{1}{\bz};\frac{1}{z}\right)\right]\\
&\,-\frac{1}{2}\log(z\bz)\left[G\left(0,\frac{1}{z};\frac{1}{\bz}\right)-G\left(0,\frac{1}{\bz};\frac{1}{z}\right)\right]+\frac{5}{24}\log^3\frac{z}{\bz}\\
&\,-2\log^2[(1-z)(1-\bz)]\log\frac{z}{\bz}-\log[(1-z)(1-\bz)]\log(z\bz)\log\frac{z}{\bz}\\
&\,-\frac{1}{8}\log^2(z\bz)\log\frac{z}{\bz}-\frac{2}{3}\pi^2\log\frac{z}{\bz}\,,\\
f_{[135]}^{(0,2)}(z,\bz) &\,=\frac{1}{2}\log^2\frac{z}{\bz}\,.
\esp\eeq


\beq\bsp
&\mathrm{Cut}_{p_3^2,[234],R^*_\triangle}T_L(p_1^2,p_2^2,p_3^2)\\
&= \frac{i\,c_\Gamma^2\,(p_1^2)^{-2-2\eps}}{(1-z)(1-\bar{z})(z-\bar{z})}\,\sum_{k=-1}^\infty{\eps^k}\,\left[(-2\pi i)\,f_{[234]}^{(k,1)}(z,\bz)+(-2\pi i)^2\,f_{[234]}^{(k,2)}(z,\bz)\right]\,,
\esp\eeq

\beq\bsp
f_{[234]}^{(-2,1)}(z,\bz) &\,=-\log\frac{z}{\bz}\,,\\
f_{[234]}^{(-2,2)}(z,\bz) &\,=0\,,\\
f_{[234]}^{(-1,1)}(z,\bz) &\,=G\left(0,\frac{1}{z};\frac{1}{\bz}\right)-G\left(0,\frac{1}{\bz};\frac{1}{z}\right)-\log\frac{z}{\bz}\left[G\left(\frac{1}{z};\frac{1}{\bz}\right)+G\left(\frac{1}{\bz};\frac{1}{z}\right)\right]\\
&\,+2\log[(1-z)(1-\bz)]\log\frac{z}{\bz}-\frac{1}{2}\log\frac{z}{\bz}\log(z\bz)\,,\\
f_{[234]}^{(-1,2)}(z,\bz) &\,=0\,,\\
f_{[234]}^{(0,1)}(z,\bz) &\,=2\left[G\left(0,\frac{1}{z},\frac{1}{\bz};\frac{1}{\bz}\right)-G\left(0,\frac{1}{\bz},\frac{1}{z};\frac{1}{z}\right)\right]-6[\text{Li}_3(z)-\text{Li}_3(\bz)]\\
&\,-\frac{1}{2}\log\frac{z}{\bz}\left[G\left(\frac{1}{z};\frac{1}{\bz}\right)+G\left(\frac{1}{\bz};\frac{1}{z}\right)\right]^2+\frac{3}{2}[\text{Li}_2(z)+\text{Li}_2(\bz)]\log\frac{z}{\bz}\\
&\,+2\log[(1-z)(1-\bz)]\log\frac{z}{\bz}\left[G\left(\frac{1}{z};\frac{1}{\bz}\right)+G\left(\frac{1}{\bz};\frac{1}{z}\right)\right]\\
&\,-\frac{1}{2}\log(z\bz)\log\frac{z}{\bz}\left[G\left(\frac{1}{z};\frac{1}{\bz}\right)+G\left(\frac{1}{\bz};\frac{1}{z}\right)\right]\\
&\,-2\log[(1-z)(1-\bz)]\left[G\left(0,\frac{1}{z};\frac{1}{\bz}\right)-G\left(0,\frac{1}{\bz};\frac{1}{z}\right)\right]\\
&\,+\frac{1}{2}\log(z\bz)\left[G\left(0,\frac{1}{z};\frac{1}{\bz}\right)-G\left(0,\frac{1}{\bz};\frac{1}{z}\right)\right]+\frac{1}{3}\log^3\frac{z}{\bz}\\
&\,+\frac{3}{2}[\text{Li}_2(z)-\text{Li}_2(\bz)]\log(z\bz)-2\log^2[(1-z)(1-\bz)]\log\frac{z}{\bz}\\
&\,+\frac{1}{4}\log^2(z\bz)\log\frac{z}{\bz}+\log[(1-z)(1-\bz)]\log(z\bz)\log\frac{z}{\bz}+\frac{\pi^2}{3}\log\frac{z}{\bz}\,,\\
f_{[234]}^{(0,2)}(z,\bz) &\,=\frac{1}{2}\log^2\frac{z}{\bz}\,.
\esp\eeq


\subsection{Unitarity cuts in the $p_2^2$ channel}

\beq\bsp
&\mathrm{Cut}_{p_2^2,[46],R^*_\triangle}T_L(p_1^2,p_2^2,p_3^2)\\
& = \frac{i\,c_\Gamma^2\,(p_1^2)^{-2-2\eps}}{(1-z)(1-\bar{z})(z-\bar{z})}\,\sum_{k=-1}^\infty{\eps^k}\,\left[(-2\pi i)\,f_{[46]}^{(k,1)}(z,\bz)+(-2\pi i)^2\,f_{[46]}^{(k,2)}(z,\bz)\right]\,,
\esp\eeq

\beq\bsp
f_{[46]}^{(-1,1)}(z,\bz) &\,=-\text{Li}_2(z)+\text{Li}_2(\bz)\,,\\
f_{[46]}^{(-1,2)}(z,\bz) &\,=0\,,\\
f_{[46]}^{(0,1)}(z,\bz) &\,=G\left(0,\frac{1}{z},\frac{1}{\bz};1\right)-G\left(0,\frac{1}{\bz},\frac{1}{z};1\right)-4[\text{Li}_3(1-z)-\text{Li}_3(1-\bz)]\\
&\,-2[\text{Li}_3(z)-\text{Li}_3(\bz)]-2[\text{Li}_2(z)+\text{Li}_2(\bz)]\log\frac{1-z}{1-\bz}+[\text{Li}_2(z)-\text{Li}_2(\bz)]\log(z\bz)\\
&\,-\frac{1}{2}\log\frac{z}{\bz}\log^2\frac{1-z}{1-\bz}-\frac{1}{2}\log^2[(1-z)(1-\bz)]\log\frac{z}{\bz}\\
&\,-\log[(1-z)(1-\bz)]\log(z\bz)\log\frac{1-z}{1-\bz}+\frac{2\pi^2}{3}\log\frac{1-z}{1-\bz}\,,\\
f_{[46]}^{(0,2)}(z,\bz) &\,=\frac{1}{2}[\text{Li}_2(z)-\text{Li}_2(\bz)]\,.
\esp\eeq


\beq\bsp
&\mathrm{Cut}_{p_2^2,[136],R^*_\triangle}T_L(p_1^2,p_2^2,p_3^2)\\
& = \frac{i\,c_\Gamma^2\,(p_1^2)^{-2-2\eps}}{(1-z)(1-\bar{z})(z-\bar{z})}\,\sum_{k=-1}^\infty{\eps^k}\,\left[(-2\pi i)\,f_{[136]}^{(k,1)}(z,\bz)+(-2\pi i)^2\,f_{[136]}^{(k,2)}(z,\bz)\right]\,,
\esp\eeq

\beq\bsp
f_{[136]}^{(-1,1)}(z,\bz) &\,=\text{Li}_2(z)-\text{Li}_2(\bz)\,,\\
f_{[136]}^{(-1,2)}(z,\bz) &\,=0\,,\\
f_{[136]}^{(0,1)}(z,\bz) &\,=-G\left(0,\frac{1}{z},\frac{1}{\bz};1\right)+G\left(0,\frac{1}{\bz},\frac{1}{z};1\right)+4[\text{Li}_3(1-z)-\text{Li}_3(1-\bz)]\\
&\,+5[\text{Li}_3(z)-\text{Li}_3(\bz)]+2[\text{Li}_2(z)+\text{Li}_2(\bz)]\log\frac{1-z}{1-\bz}-2[\text{Li}_2(z)-\text{Li}_2(\bz)]\log(z\bz)\\
&\,+\frac{1}{2}\log\frac{z}{\bz}\log^2\frac{1-z}{1-\bz}+\frac{1}{2}\log^2[(1-z)(1-\bz)]\log\frac{z}{\bz}\\
&\,+\log[(1-z)(1-\bz)]\log(z\bz)\log\frac{1-z}{1-\bz}-\frac{2}{3}\pi^2\log\frac{1-z}{1-\bz}\,,\\
f_{[136]}^{(0,2)}(z,\bz) &\,=0\,.
\esp\eeq

\section{Computation and explicit results for double unitarity cuts}
\label{app:DoubleCutRes}

We briefly outline our approach to the calculation of the double unitarity cuts of \refF{doubleCutsp1p3} and \refF{doubleCutsp1p2}. We then give explicit results for these integrals, written in terms of multiple polylogarithms to simplify the comparison between different terms.

\subsection{Calculation of double unitarity cuts}
\label{calculationDoubleCuts}

\paragraph{Cut [456], $R^{1,3}_\triangle$, \refF{cut456Diag}:\\}
Because cut $[45]$ factorizes the two loop integrations, this cut is just the product of an uncut one-loop triangle with one mass ($p_3^2$) and the double cut of a three-mass triangle, with masses
$p_1^2$, $p_2^2$ and $p_3^2$, in the channels $p_1^2$ and $p_3^2$.
\begin{align}\bsp
\cut_{[456],R^{1,3}_\triangle}T_{L}(p_1^2,p_2^2,p_3^2)
=&-i\frac{e^{\gamma_E \epsilon}}{\pi^{2-\epsilon}}(2\pi)^3\int\text{d}^{4-2\epsilon}k_1\delta\left(k_1^2\right)\delta\left((p_3-k_1)^2\right)\\
&\quad\delta\left((p_1+k_1)^2\right)T^{1m}(p_3^2)\\
=&-4\pi^2 i\frac{c_\Gamma e^{\gamma_E\epsilon}}{\epsilon^2\Gamma(1-\epsilon)}(p_1^2)^{-2-2\epsilon}u_3^{-1-2\epsilon}e^{i\pi\epsilon}\frac{z^{-\epsilon}(-\zbar)^{-\epsilon}}{(z-\zbar)^{1-2\epsilon}}\,.
\esp\end{align}

\paragraph{Cut [1256], $R^{1,3}_\triangle$, \refF{cut1256Diag}:\\}
The integrand has a simple pole inside the integration region. We can still make sense of the integral by keeping
track of the $i\varepsilon$ prescription associated to the propagators and the invariants, and we obtain
\beq\bsp
\cut&_{[1256],R^{1,3}_\triangle}T_{L}(p_1^2,p_2^2,p_3^2)\\
=&-i\frac{e^{\gamma_E \epsilon}}{\pi^{2-\epsilon}}(2\pi)^2\int\text{d}^{4-2\epsilon}k_1\frac{\delta\left(k_1^2\right)\delta\left((p_1-k_1)^2\right)}{(p_3+k_1)^2-i\varepsilon}
\cut_{p_3^2}T^{2m}\left(p_3^2,\left(p_3+k\right)^2\right)\\
=&i\frac{e^{2\gamma_E \epsilon}}{\pi^{2-\epsilon}}\frac{\Gamma(1-\epsilon)}{\epsilon\Gamma(1-2\epsilon)}(2\pi)^3\int\text{d}^{4-2\epsilon}k\frac{\delta\left(k^2\right)\delta\left((p_1-k)^2\right)}{(p_3+k)^2-i\varepsilon}
\frac{(p_3^2+i\varepsilon)^{-\epsilon}}{(p_3^2+i\varepsilon)-((p_3+k)^2-i\varepsilon)}\\
=&-4\pi^2i\frac{e^{2\gamma_E\epsilon}}{\epsilon\Gamma(1-2\epsilon)}(p_1^2)^{-2-2\epsilon}u_3^{-\epsilon}\\
&\int_0^1\text{d}x\frac{x^{-\epsilon}(1-x)^{-\epsilon}}{(u_3+z-1-x(z-\zbar)-i\varepsilon)(z-1-x(z-\zbar)-i\varepsilon)}\,,
\esp\eeq
where in each line we were careful to keep the $\pm i\varepsilon$ prescription associated with propagators and invariants. For $\varepsilon=0$, the integrand in the last line has poles at
\begin{equation*}
 0<x_p\equiv\frac{(1-z)(-\zbar)}{z-\zbar}<1 \qquad \text{and} \qquad x=\frac{z-1}{z-\zbar}<0\,.
\end{equation*}
While the location of the second pole lies outside the integration region, the first singularity lies inside, and we must hence split the integral into its principle value and imaginary part,
\begin{equation*}
\lim_{\varepsilon\to0}\frac{1}{a\pm i\varepsilon}=\text{PV}\left(\frac{1}{a}\right)\mp i\pi \delta(a)\,.
\end{equation*}
which is valid in a distribution sense. We then obtain
\beq\bsp
\int_0^1&\text{d}x\frac{x^{-\epsilon}(1-x)^{-\epsilon}}{(u_3+z-1-x(z-\zbar)-i\varepsilon)(z-1-x(z-\zbar))}\\
&=\text{PV}\int_0^1\text{d}x\frac{x^{-\epsilon}(1-x)^{-\epsilon}}{(u_3+z-1-x(z-\zbar))(z-1-x(z-\zbar))}\\
&+i\pi\int_0^1\text{d}x\frac{x^{-\epsilon}(1-x)^{-\epsilon}}{(z-1-x(z-\zbar))}\delta(u_3+z-1-x(z-\zbar))\,.
\esp\eeq
Both integrals are finite and can easily be performed order by order in $\eps$ in terms of polylogarithms.

\paragraph{Cut [1236], $R^{1,3}_\triangle$, \refF{cut1236Diag}:\\}
Using the strategy outlined in \refS{sec:doubleCutCalculation}, we immediately obtain
\beq\bsp
\cut&_{[1236],R^{1,3}_\triangle}T_{L}(p_1^2,p_2^2,p_3^2)\\
&=i\frac{e^{\gamma_E \epsilon}}{\pi^{2-\epsilon}}(2\pi)^2\int\text{d}^{4-2\epsilon}k\delta\left(k^2\right)\delta\left((p_3-k)^2\right)\cut_{(p_1-k)^2}B^{2mh}(p_1^2,p_2^2;p_3^2,(p_1-k)^2)\,.
\esp\eeq
Inserting the analytic expression for the cut box (see appendix~\ref{app_oneLoopRes}) and parametrizing the remaining cut integration, we obtain an integral with an endpoint singularity. After subtraction of the singularity, all the integrals are finite and can be expanded under the integration sign. We obtain
\beq\bsp
\cut&_{[1236],R^{1,3}_\triangle}T_{L}(p_1^2,p_2^2,p_3^2)\\
&=8\pi^2i\frac{e^{\gamma_E\epsilon}c_\Gamma}{\Gamma(1-\epsilon)}(p_1^2)^{-2-2\epsilon}u_3^{-1-2\epsilon}z^\epsilon(-\zbar)^{\epsilon}
\int_0^{\frac{z}{z-\zbar}}\text{d}x\frac{x^{-\epsilon}(1-x)^{-\epsilon}}{(z-x(z-\zbar))^{1+2\epsilon}}\\
&\times\left[\frac{1}{\epsilon}+\ln\left(1-\frac{z-x(z-\zbar)}{z\zbar}\right)+\ln\left(1-z+x(z-\zbar)\right)\right]\,.
\esp\eeq
The remaining integral is easy to perform.

\paragraph{Cut [2346], $R^{1,3}_\triangle$, \refF{cut2346Diag}:\\}

Using the routing of the loop momenta shown in \refF{cut2346Diag},
we compute this cut by integrating over the cut of a two-mass triangle. However, when using the result for the cut triangle, we need
to correct for the fact that the vertex attached to propagators 2, 3 and 5 has a different color, compared to the usual cut triangle. Note also that it is convenient to introduce the variable $y$ defined in \refE{changeVarXY}. We obtain
\beq\bsp
\cut&_{[2346],R^{1,3}_\triangle}T_{L}(p_1^2,p_2^2,p_3^2)\\
&=i\left(\frac{e^{\gamma_E \epsilon}}{\pi^{2-\epsilon}}\right)
(2\pi)^2\int\text{d}^{4-2\epsilon}k_1
\frac{\delta\left(k_1^2\right)\delta\left((p_2+k_1)^2\right)}{(p_3-k_1)^2}i^2\cut_{(p_3-k_1)^2}T^{2m}\left(p_3^2,\left(p_3-k_1\right)^2\right)\\
&=-4\pi^2i\frac{e^{2\gamma_E\epsilon}}{\epsilon\Gamma(1-2\epsilon)}(p_1^2)^{-2-2\epsilon}u_3^{-1-2\epsilon}
\int_0^1\text{d}x\, x^{-\epsilon} (1-x)^{-\epsilon}\\
&\times\int_0^1\text{d}y\,y^{-2\epsilon} (1-y)^{-1-\epsilon}\delta\left(u_2+y(z(1-\bz)-x(z-\bz))\right)\\
&=-4\pi^2i\frac{e^{2\gamma_E\epsilon}}{\epsilon\Gamma(1-2\epsilon)}(p_1^2)^{-2-2\epsilon}u_3^{-1-2\epsilon}z^{-2\epsilon}(-\zbar)^{-2\epsilon}\\
&\times\int_0^{\frac{z}{z-\zbar}}\text{d}x\frac{x^{-\epsilon}(1-x)^{-\epsilon}}{(z-x(z-\zbar))^{1+\epsilon}}\frac{1}{(z-z\zbar-x(z-\zbar))^{-3\epsilon}}\,.
\esp\eeq
The integral has an endpoint singularity that needs to be subtracted before expansion in $\eps$ under the integration sign.
The $y$ variable is restricted to the interval $[0,1]$ because of the $\theta$-function of the cut triangle subdiagram.
We find it simpler to use the $\delta$-function associated with the cut on $(p_2+k_1)^2$
to perform the $y$ integration, which in turn imposes some limits on the range of integration of $x$.

\paragraph{Cut [1356], $R^{1,3}_\triangle$, \refF{cut1356Diag}:\\}

The integral is
\beq\bsp
\cut&_{[1356],R^{1,3}_\triangle}T_{L}(p_1^2,p_2^2,p_3^2)\\
&=-i\left(\frac{e^{\gamma_E \epsilon}}{\pi^{2-\epsilon}}\right)
(2\pi)^2\int\text{d}^{4-2\epsilon}k_1\frac{\delta\left(k_1^2\right)\delta\left((p_1-k_1)^2\right)}{(p_2+k_1)^2}\cut_{(p_2+k_1)^2}T^{2m}\left(p_3^2,\left(p_2+k_1\right)^2\right)\\
&=-4\pi^2i\frac{e^{2\gamma_E\epsilon}}{\epsilon\Gamma(1-2\epsilon)}(p_1^2)^{-2-2\epsilon}\int_0^{\frac{-\bz(1-z)}{z-\bz}}\text{d}x\frac{x^{-\epsilon}(1-x)^{-\epsilon}(-\bz(1-z)-x(z-\bz))^{-1-\epsilon}}{(1-z+x(z-\bz))}\,.
\esp\eeq
The restriction on the integration range of $x$ is imposed by the $\theta$-function of the cut triangle subdiagram. 
After subtracting the singularity, the integral can be performed oder by order in $\eps$.

\paragraph{Cut [456], $R^{1,2}_\triangle$, \refF{cut456R3Diag}:\\}

The calculation of this cut in region $R^{1,2}_\triangle$ is done in exactly the same way as in region $R^{1,3}_\triangle$. However, we write the result differently so that
we are away from the branch cuts in this region:
\begin{align}\bsp
\cut_{[456],R^{1,2}_\triangle}&T_{L}(p_1^2,p_2^2,p_3^2)\\
=&-i\frac{e^{\gamma_E \epsilon}}{\pi^{2-\epsilon}}(2\pi)^3\int\text{d}^{4-2\epsilon}k_1\delta\left(k_1^2\right)\delta\left((p_3-k_1)^2\right)\delta\left((p_1+k_1)^2\right)T^{1m}(p_3^2)\\
=&-4\pi^2 i\frac{c_\Gamma e^{\gamma_E\epsilon}}{\epsilon^2\Gamma(1-\epsilon)}(p_1^2)^{-2-2\epsilon}e^{i\pi\epsilon}((z-1)(1-\bz))^{-1-2\epsilon}\frac{(z\bz)^{-\epsilon}}{(z-\zbar)^{1-2\epsilon}}\,.
\esp\end{align}

\paragraph{Cut [2346], $R^{1,2}_\triangle$, \refF{cut2346R3Diag}:\\}

The calculation of this cut in $R^{1,2}_\triangle$ is simpler than in region $R^{1,3}_\triangle$. We get
\beq\bsp
\cut&_{[2346],R^{1,2}_\triangle}T_{L}(p_1^2,p_2^2,p_3^2)\\
&=-i\left(\frac{e^{\gamma_E \epsilon}}{\pi^{2-\epsilon}}\right)
(2\pi)^2\int\text{d}^{4-2\epsilon}k_1
\frac{\delta\left(k_1^2\right)\delta\left((p_2-k_1)^2\right)}{(p_1+k_1)^2}\cut_{(p_1+k_1)^2}T^{2m}\left(p_3^2,\left(p_1+k_1\right)^2\right)\\
&=4\pi^2i\frac{e^{2\gamma_E\epsilon}}{\epsilon\Gamma(1-2\epsilon)}(p_1^2)^{-2-2\epsilon}u_2^{-\epsilon}\int_0^{\frac{1-\bz}{z-\bz}}
\text{d}x\,x^{-\epsilon}(1-x)^{-\epsilon}\frac{(1-\bz-x(z-\bz))^{-1-\epsilon}}{z(1-\bz)-x(z-\bz)}\,.
\esp\eeq
After subtraction of the singularity, the integral is easy to perform.

\paragraph{Cut [1356], $R^{1,2}_\triangle$, \refF{cut1356R3Diag}:\\}

The computation of this cut is very similar to the previous one. We have
\beq\bsp
\cut&_{[1356],R^{1,2}_\triangle}T_{L}(p_1^2,p_2^2,p_3^2)\\
&=-i\left(\frac{e^{\gamma_E \epsilon}}{\pi^{2-\epsilon}}\right)
(2\pi)^2\int\text{d}^{4-2\epsilon}k_1
\frac{\delta\left(k_1^2\right)\delta\left((p_1-k_1)^2\right)}{(p_2+k_1)^2}\cut_{(p_2+k_1)^2}T^{2m}\left(p_3^2,\left(p_2+k_1\right)^2\right)\\
&=4\pi^2i\frac{e^{2\gamma_E\epsilon}}{\epsilon\Gamma(1-2\epsilon)}(p_1^2)^{-2-2\epsilon}\int_0^{\frac{\bz(z-1)}{z-\bz}}
\text{d}x\,x^{-\epsilon}(1-x)^{-\epsilon}\frac{(\bz(z-1)-x(z-\bz))^{-1-\epsilon}}{z-1-x(z-\bz)}\,.
\esp\eeq
The restriction on the integration range of $x$ is imposed by the $\theta$-function of the cut triangle subdiagram. The endpoint singularity is dealt with as before.

\paragraph{Cut [1236], $R^{1,2}_\triangle$, \refF{cut1236R3Diag}:\\}

This cut is slightly harder to compute in region $R^{1,2}_\triangle$ than in region $R^{1,3}_\triangle$. We follow the same technique of integrating over the cut
of a two-mass hard box, although we have to be careful to correct for the different factors of $\pm i$ between the subdiagram entering in \refF{cut1236R3Diag}
and a standard cut box that would have black vertices on one side of the cut and white vertices on the other side. It is also useful to introduce the
$y$ variable defined in \refE{changeVarXY}, and to integrate over it with the $\delta$-function on propagator $(p_3+k)$. The $y$ variable is 
restricted to the interval $[0,1]$ because of the $\theta$-function on $(p_1-k)^2$:
\beq\bsp
\cut&_{[1236],R^{1,2}_\triangle}T_{L}(p_1^2,p_2^2,p_3^2)\\
&=-i\frac{e^{\gamma_E \epsilon}}{\pi^{2-\epsilon}}(2\pi)^2\int\text{d}^{4-2\epsilon}k\delta\left(k^2\right)\delta\left((p_3+k)^2\right)i^6\cut_{(p_1-k)^2}B^{2mh}(p_1^2,p_2^2;p_3^2,(p_1-k)^2)\\
&=-8\pi^2i\frac{e^{\gamma_E\epsilon}c_\Gamma}{\Gamma(1-\epsilon)}(p_1^2)^{-2-2\epsilon}\frac{u_2^{\epsilon}}{((z-1)(1-\bz))^{1+\epsilon}}\int_0^1\text{d}x\, x^{-\epsilon} (1-x)^{-\epsilon}\\
&\times\int_0^1\text{d}y\,y^{1-2\epsilon} (1-y)^{-1-2\epsilon}\delta\big(u_3+y(z-1-x(z-\bz))\big)\\
&\times\left[\frac{1}{\epsilon}+\ln y+\ln\big(u_2-(1-y)\big)-\ln u_2 \right]\\
&=-8\pi^2i\frac{e^{\gamma_E\epsilon}c_\Gamma}{\Gamma(1-\epsilon)}(p_1^2)^{-2-2\epsilon}\frac{u_2^{\epsilon}}{((z-1)(1-\bz))^{3\epsilon}}
\int_0^{\frac{\bz(z-1)}{z-\zbar}}\text{d}x\frac{x^{-\epsilon}(1-x)^{-\epsilon}}{(z-1-x(z-\zbar))^{1-4\epsilon}}\\
&\times\big(\bz(z-1)-x(z-\bz)\big)^{-1-2\epsilon}\left[\frac{1}{\epsilon}+\ln\big((z-1)(1-\bz)\big)-\ln\big(z-1-x(z-\bz)\big)\right.\\
&\left.-\ln(z \bz)+\ln\left(z \bz-\frac{\bz(z-1)-x(z-\bz)}{z-1-x(z-\bz)}\right)\right]
\esp\eeq
The restriction on the integration range of $x$ is imposed when integrating over $y$. The endpoint singularity is dealt with as before.

\subsection{Double unitarity cuts in the $p_1^2$ and $p_3^2$ channels in region $R^{1,3}_\triangle$}
\label{resultsDoubleCutsR2}
In this section we present the analytic results for all the nonvanishing cuts in the $p_1^2$ and $p_3^2$ channels in region $R^{1,3}_\triangle$, where $\zbar<0<z<1$.

\beq\bsp
&\mathrm{Cut}_{[456], R^{1,3}_\triangle}T_L(p_1^2,p_2^2,p_3^2) \\
&= \frac{i\,c_\Gamma^2\,(p_1^2)^{-2-2\eps}}{(1-z)(1-\bar{z})(z-\bar{z})}\,\sum_{k=-2}^\infty{\eps^k}\,\left[(-2\pi i)^2\,f_{[456], R^{1,3}_\triangle}^{(k,2)}(z,\bz)+(-2\pi i)^3\,f_{[456], R^{1,3}_\triangle}^{(k,3)}(z,\bz)\right]\,,
\esp\eeq

\beq\bsp
f_{[456],R^{1,3}_\triangle}^{(-2,2)}&\,=1\,,\\
f_{[456],R^{1,3}_\triangle}^{(-2,3)}&\,=0\,,\\
f_{[456],R^{1,3}_\triangle}^{(-1,2)}&\,=2 \log\left(z-\bar{z}\right)-2 \log \left[(1-z) \left(1-\bar{z}\right)\right]-\log \left(-z\bar{z}\right)\,,\\
f_{[456],R^{1,3}_\triangle}^{(-1,3)}&\,=-\frac{1}{2}\,,\\
f_{[456],R^{1,3}_\triangle}^{(0,2)}&\,=\frac{1}{2} \left[-2 \log \left(z-\bar{z}\right)+2 \log \left[(1-z) \left(1-\bar{z}\right)\right]+\log \left(-z \bar{z}\right)\right]^2-\frac{\pi ^2}{2}\,,\\
f_{[456],R^{1,3}_\triangle}^{(0,3)}&\,=\log \left[(1-z) \left(1-\bar{z}\right)\right]-\log \left(z-\bar{z}\right)+\frac{1}{2} \log \left(-z\bar{z}\right)\,.
\esp\eeq

\beq\bsp
&\mathrm{Cut}_{[1236], R^{1,3}_\triangle}T_L(p_1^2,p_2^2,p_3^2) \\
&= \frac{i\,c_\Gamma^2\,(p_1^2)^{-2-2\eps}}{(1-z)(1-\bar{z})(z-\bar{z})}\,\sum_{k=-2}^\infty{\eps^k}\,\left[(-2\pi i)^2\,f_{[1236], R^{1,3}_\triangle}^{(k,2)}(z,\bz)+(-2\pi i)^3\,f_{[1236], R^{1,3}_\triangle}^{(k,3)}(z,\bz)\right]\,,
\esp\eeq

\beq\bsp
f_{[1236],R^{1,3}_\triangle}^{(-2,2)}&\,=1\,,\\
f_{[1236],R^{1,3}_\triangle}^{(-2,3)}&\,=0\,,\\
f_{[1236],R^{1,3}_\triangle}^{(-1,2)}&\,=2 \log \left(z-\bar{z}\right)-2 \log \left[(1-z)(1-\bar{z})\right]-2 \log z\,,\\
f_{[1236],R^{1,3}_\triangle}^{(-1,3)}&\,=0\,,\\
f_{[1236],R^{1,3}_\triangle}^{(0,2)}&\,=2 \text{Li}_2(z)-2 \text{Li}_2\left(\frac{z}{\bar{z}}\right)-2 \text{Li}_2\left(\bar{z}\right)+2 \log ^2\left(1-\bar{z}\right)+2 \log ^2\left(z-\bar{z}\right)-\log ^2\left(-\bar{z}\right)\\
&\,+4 \log (1-z) \log \left(1-\bar{z}\right)-4 \log (1-z) \log \left(z-\bar{z}\right)+4 \log z \log \left(1-\bar{z}\right)\\
&\,-4 \log z \log \left(z-\bar{z}\right)-4 \log \left(1-\bar{z}\right) \log \left(z-\bar{z}\right)+2 \log ^2(1-z)+2 \log ^2z\\
&\,+4 \log z \log (1-z)-\frac{2 \pi ^2}{3}\,,\\
f_{[1236],R^{1,3}_\triangle}^{(0,3)}&\,=0\,.
\esp\eeq

\beq\bsp
&\mathrm{Cut}_{[1256], R^{1,3}_\triangle}T_L(p_1^2,p_2^2,p_3^2)\\
& = \frac{i\,c_\Gamma^2\,(p_1^2)^{-2-2\eps}}{(1-z)(1-\bar{z})(z-\bar{z})}\,\sum_{k=-1}^\infty{\eps^k}\,\left[(-2\pi i)^2\,f_{[1256], R^{1,3}_\triangle}^{(k,2)}(z,\bz)+(-2\pi i)^3\,f_{[1256], R^{1,3}_\triangle}^{(k,3)}(z,\bz)\right]\,,
\esp\eeq

\beq\bsp
f_{[1256],R^{1,3}_\triangle}^{(-1,2)}&\,=\log z-\log \left(-\bar{z}\right)\,,\\
f_{[1256],R^{1,3}_\triangle}^{(-1,3)}&\,=\frac{1}{2}\,,\\
f_{[1256],R^{1,3}_\triangle}^{(0,2)}&\,=2 \text{Li}_2\left(\frac{z}{\bar{z}}\right)+2 \text{Li}_2\left(\bar{z}\right)-2 \text{Li}_2(z)+\frac{3}{2} \log ^2\left(-\bar{z}\right)-2 \log z \log \left(1-\bar{z}\right)\\
&\,+2 \log z \log \left(z-\bar{z}\right)-\log z \log \left(-\bar{z}\right)+2 \log (1-z) \log \left(-\bar{z}\right)\\
&\,+2 \log \left(1-\bar{z}\right) \log \left(-\bar{z}\right)-2 \log \left(z-\bar{z}\right) \log \left(-\bar{z}\right)-\frac{1}{2} \log^2z\\
&\,-2 \log (1-z) \log z+\frac{\pi ^2}{6}\,,\\
f_{[1256],R^{1,3}_\triangle}^{(0,3)}&\,=\log \left(z-\bar{z}\right) - \log \left[(1-z) \left(1-\bar{z}\right)\right]-\frac{1}{2}\log \left(-z \bar{z}\right)\,.
\esp\eeq

\beq\bsp
&\mathrm{Cut}_{[1356], R^{1,3}_\triangle}T_L(p_1^2,p_2^2,p_3^2)\\
& = \frac{i\,c_\Gamma^2\,(p_1^2)^{-2-2\eps}}{(1-z)(1-\bar{z})(z-\bar{z})}\,\sum_{k=-2}^\infty{\eps^k}\,\left[(-2\pi i)^2\,f_{[1356], R^{1,3}_\triangle}^{(k,2)}(z,\bz)+(-2\pi i)^3\,f_{[1356], R^{1,3}_\triangle}^{(k,3)}(z,\bz)\right]\,,
\esp\eeq

\beq\bsp
f_{[1356],R^{1,3}_\triangle}^{(-2,2)}&\,=-1\,,\\
f_{[1356],R^{1,3}_\triangle}^{(-2,3)}&\,=0\,,\\
f_{[1356],R^{1,3}_\triangle}^{(-1,2)}&\,=2 \log \left[(1-z)(1-\bar{z})\right]-2 \log \left(z-\bar{z}\right)+2 \log \left(-\bar{z}\right)+\log (z)\,,\\
f_{[1356],R^{1,3}_\triangle}^{(-1,3)}&\,=0\,,\\
f_{[1356],R^{1,3}_\triangle}^{(0,2)}&\,=-\text{Li}_2\left(\frac{z}{\bar{z}}\right)-3 \text{Li}_2\left(\bar{z}\right)-2 \log ^2\left(1-\bar{z}\right)-2 \log ^2\left(z-\bar{z}\right)-\frac{5}{2} \log ^2\left(-\bar{z}\right)\\
&\,-4 \log (1-z) \log \left(1-\bar{z}\right)+4 \log (1-z) \log \left(z-\bar{z}\right)-4 \log (1-z) \log \left(-\bar{z}\right)\\
&\,-2 \log z \log \left(1-\bar{z}\right)+2 \log z \log \left(z-\bar{z}\right)+4 \log \left(1-\bar{z}\right) \log \left(z-\bar{z}\right)\\
&\,-\log z \log \left(-\bar{z}\right)-4 \log \left(1-\bar{z}\right) \log \left(-\bar{z}\right)+4 \log \left(z-\bar{z}\right) \log \left(-\bar{z}\right)-2 \log ^2(1-z)\\
&\,-\log ^2z-2 \log z \log (1-z)+\frac{\pi ^2}{6}\,,\\
f_{[1356],R^{1,3}_\triangle}^{(0,3)}&\,=0\,.
\esp\eeq

\beq\bsp
&\mathrm{Cut}_{[2346], R^{1,3}_\triangle}T_L(p_1^2,p_2^2,p_3^2)\\
& = \frac{i\,c_\Gamma^2\,(p_1^2)^{-2-2\eps}}{(1-z)(1-\bar{z})(z-\bar{z})}\,\sum_{k=-2}^\infty{\eps^k}\,\left[(-2\pi i)^2\,f_{[2346], R^{1,3}_\triangle}^{(k,2)}(z,\bz)+(-2\pi i)^3\,f_{[2346], R^{1,3}_\triangle}^{(k,3)}(z,\bz)\right]\,,
\esp\eeq

\beq\bsp
f_{[2346],R^{1,3}_\triangle}^{(-2,2)}&\,=-1\,,\\
f_{[2346],R^{1,3}_\triangle}^{(-2,3)}&\,=0\,,\\
f_{[2346],R^{1,3}_\triangle}^{(-1,2)}&\,=2 \log \left[(1-z)(1-\bar{z})\right]-2 \log \left(z-\bar{z}\right)+\log z\,,\\
f_{[2346],R^{1,3}_\triangle}^{(-1,3)}&\,=0\,,\\
f_{[2346],R^{1,3}_\triangle}^{(0,2)}&\,=\text{Li}_2\left(\frac{z}{\bar{z}}\right)+3 \text{Li}_2\left(\bar{z}\right)-2 \log ^2\left(1-\bar{z}\right)-2 \log ^2\left(z-\bar{z}\right)+\frac{3}{2} \log ^2\left(-\bar{z}\right)\\
&\,-4 \log (1-z) \log \left(1-\bar{z}\right)+4 \log (1-z) \log \left(z-\bar{z}\right)-2 \log z \log \left(1-\bar{z}\right)\\
&\,+2 \log z \log \left(z-\bar{z}\right)+4 \log \left(1-\bar{z}\right) \log \left(z-\bar{z}\right)-2 \log ^2(1-z)-\frac{1}{2}\log ^2z\\
&\,-2 \log z \log (1-z)+\frac{5 \pi ^2}{6}\,,\\
f_{[2346],R^{1,3}_\triangle}^{(0,3)}&\,=0\,.
\esp\eeq


\subsection{Double unitarity cuts in the $p_1^2$ and $p_2^2$ channels in region $R^{1,2}_\triangle$}
\label{resultsDoubleCutsR3}
In this section we present the analytic results for all the nonvanishing cuts in the $p_1^2$ and $p_2^2$ channels in region $R^{1,2}_\triangle$, where $0<\zbar<1<z$.

\beq\bsp
&\mathrm{Cut}_{[456], R^{1,2}_\triangle}T_L(p_1^2,p_2^2,p_3^2)\\
& = \frac{i\,c_\Gamma^2\,(p_1^2)^{-2-2\eps}}{(1-z)(1-\bar{z})(z-\bar{z})}\,\sum_{k=-2}^\infty{\eps^k}\,\left[(-2\pi i)^2\,f_{[456], R^{1,2}_\triangle}^{(k,2)}(z,\bz)+(-2\pi i)^3\,f_{[456], R^{1,2}_\triangle}^{(k,3)}(z,\bz)\right]\,,
\esp\eeq

\beq\bsp
f_{[456],R^{1,2}_\triangle}^{(-2,2)}&\,=1\,,\\
f_{[456],R^{1,2}_\triangle}^{(-2,3)}&\,=0\,,\\
f_{[456],R^{1,2}_\triangle}^{(-1,2)}&\,=2 \log \left(z-\bar{z}\right)-2 \log \left[(z-1) \left(1-\bar{z}\right)\right]-\log \left(z \bar{z}\right)\,,\\
f_{[456],R^{1,2}_\triangle}^{(-1,3)}&\,=0\,,\\
f_{[456],R^{1,2}_\triangle}^{(0,2)}&\,=2 \log ^2\left(1-\bar{z}\right)+2 \log ^2\left(z-\bar{z}\right)+\frac{1}{2} \log ^2\bar{z}+4 \log (z-1) \log \left(1-\bar{z}\right)\\
&\,-4 \log (z-1) \log \left(z-\bar{z}\right)+2 \log (z-1) \log \bar{z}+2 \log z \log \left(1-\bar{z}\right)\\
&\,-2 \log z \log \left(z-\bar{z}\right)-4 \log \left(1-\bar{z}\right) \log \left(z-\bar{z}\right)+\log z \log \bar{z}+2 \log \left(1-\bar{z}\right) \log \bar{z}\\
&\,-2 \log \left(z-\bar{z}\right) \log \bar{z}+2 \log ^2(z-1)+\frac{1}{2}\log ^2z+2 \log z \log (z-1)\,,\\
f_{[456],R^{1,2}_\triangle}^{(0,3)}&\,=0\,.
\esp\eeq

\beq\bsp
&\mathrm{Cut}_{[1236], R^{1,2}_\triangle}T_L(p_1^2,p_2^2,p_3^2) \\
&= \frac{i\,c_\Gamma^2\,(p_1^2)^{-2-2\eps}}{(1-z)(1-\bar{z})(z-\bar{z})}\,\sum_{k=-2}^\infty{\eps^k}\,\left[(-2\pi i)^2\,f_{[1236], R^{1,2}_\triangle}^{(k,2)}(z,\bz)+(-2\pi i)^3\,f_{[1236], R^{1,2}_\triangle}^{(k,3)}(z,\bz)\right]\,,
\esp\eeq

\beq\bsp
f_{[1236],R^{1,2}_\triangle}^{(-2,2)}&\,=1\,,\\
f_{[1236],R^{1,2}_\triangle}^{(-2,3)}&\,=0\,,\\
f_{[1236],R^{1,2}_\triangle}^{(-1,2)}&\,=2 \log \left(z-\bar{z}\right)-2 \log \left[(z-1) \left(1-\bar{z}\right)\right]-2 \log \bar{z}\,,\\
f_{[1236],R^{1,2}_\triangle}^{(-1,3)}&\,=0\,,\\
f_{[1236],R^{1,2}_\triangle}^{(0,2)}&\,=2 \text{Li}_2(1-z)-2 \text{Li}_2\left(\bar{z}\right)-2 \text{Li}_2\left(\frac{\bar{z}}{z}\right)+2 \log ^2\left(1-\bar{z}\right)+2 \log ^2\left(z-\bar{z}\right)+2 \log ^2\bar{z}\\
&\,+4 \log (z-1) \log \left(1-\bar{z}\right)-4 \log (z-1) \log \left(z-\bar{z}\right)+4 \log (z-1) \log \bar{z}\\
&\,-4 \log \left(1-\bar{z}\right) \log \left(z-\bar{z}\right)+4 \log \left(1-\bar{z}\right) \log \bar{z}-4 \log \left(z-\bar{z}\right) \log \bar{z}+2 \log ^2(z-1)\\
&\,-\log ^2z+2 \log z \log (z-1)\,,\\
f_{[1236],R^{1,2}_\triangle}^{(0,3)}&\,=0\,.
\esp\eeq

\beq\bsp
&\mathrm{Cut}_{[1356], R^{1,2}_\triangle}T_L(p_1^2,p_2^2,p_3^2)\\
& = \frac{i\,c_\Gamma^2\,(p_1^2)^{-2-2\eps}}{(1-z)(1-\bar{z})(z-\bar{z})}\,\sum_{k=-2}^\infty{\eps^k}\,\left[(-2\pi i)^2\,f_{[1356], R^{1,2}_\triangle}^{(k,2)}(z,\bz)+(-2\pi i)^3\,f_{[1356], R^{1,2}_\triangle}^{(k,3)}(z,\bz)\right]\,,
\esp\eeq

\beq\bsp
f_{[1356],R^{1,2}_\triangle}^{(-2,2)}&\,=-1\,,\\
f_{[1356],R^{1,2}_\triangle}^{(-2,3)}&\,=0\,,\\
f_{[1356],R^{1,2}_\triangle}^{(-1,2)}&\,=2 \log \left[(z-1) \left(1-\bar{z}\right)\right]-2 \log \left(z-\bar{z}\right)+2 \log \bar{z}+\log z\,,\\
f_{[1356],R^{1,2}_\triangle}^{(-1,3)}&\,=0\,,\\
f_{[1356],R^{1,2}_\triangle}^{(0,2)}&\,=-3 \text{Li}_2\left(\bar{z}\right)+\text{Li}_2\left(\frac{\bar{z}}{z}\right)-2 \log ^2\left(1-\bar{z}\right)-2 \log ^2\left(z-\bar{z}\right)-2 \log ^2\bar{z}\\
&\,-4 \log (z-1) \log \left(1-\bar{z}\right)+4 \log (z-1) \log \left(z-\bar{z}\right)-4 \log (z-1) \log \bar{z}\\
&\,-2 \log z \log \left(1-\bar{z}\right)+2 \log z \log \left(z-\bar{z}\right)+4 \log \left(1-\bar{z}\right) \log \left(z-\bar{z}\right)-2 \log z \log \bar{z}\\
&\,-4 \log \left(1-\bar{z}\right) \log\bar{z}+4 \log \left(z-\bar{z}\right) \log \bar{z}-2 \log ^2(z-1)-\frac{\log ^2(z)}{2}\\
&\,-2 \log z \log (z-1)+\frac{\pi ^2}{3}\,,\\
f_{[1356],R^{1,2}_\triangle}^{(0,3)}&\,=0\,.
\esp\eeq

\beq\bsp
&\mathrm{Cut}_{[1356], R^{1,2}_\triangle}T_L(p_1^2,p_2^2,p_3^2)\\
& = \frac{i\,c_\Gamma^2\,(p_1^2)^{-2-2\eps}}{(1-z)(1-\bar{z})(z-\bar{z})}\,\sum_{k=-2}^\infty{\eps^k}\,\left[(-2\pi i)^2\,f_{[1356], R^{1,2}_\triangle}^{(k,2)}(z,\bz)+(-2\pi i)^3\,f_{[1356], R^{1,2}_\triangle}^{(k,3)}(z,\bz)\right]\,,
\esp\eeq

\beq\bsp
f_{[1356],R^{1,2}_\triangle}^{(-2,2)}&\,=-1\,,\\
f_{[1356],R^{1,2}_\triangle}^{(-2,3)}&\,=0\,,\\
f_{[1356],R^{1,2}_\triangle}^{(-1,2)}&\,=2 \log \left[(z-1) \left(1-\bar{z}\right)\right]-2 \log \left(z-\bar{z}\right)+\log \bar{z}\,,\\
f_{[1356],R^{1,2}_\triangle}^{(-1,3)}&\,=0\,,\\
f_{[1356],R^{1,2}_\triangle}^{(0,2)}&\,=\text{Li}_2\left(\frac{\bar{z}}{z}\right)-3 \text{Li}_2(1-z)-2 \log ^2\left(1-\bar{z}\right)-2 \log ^2\left(z-\bar{z}\right)-\frac{1}{2} \log ^2\bar{z}\\
&\,-4 \log (z-1) \log \left(1-\bar{z}\right)+4 \log (z-1) \log \left(z-\bar{z}\right)-2 \log (z-1) \log \bar{z}\\
&\,+4 \log \left(1-\bar{z}\right) \log \left(z-\bar{z}\right)-2 \log \left(1-\bar{z}\right) \log \bar{z}+2 \log \left(z-\bar{z}\right) \log \bar{z}-2 \log ^2(z-1)\\
&\,+\frac{3 }{2}\log ^2z-3 \log z \log (z-1)-\frac{\pi ^2}{6}\,,\\
f_{[1356],R^{1,2}_\triangle}^{(0,3)}&\,=0\,.
\esp\eeq

\bibliographystyle{JHEP}
\bibliography{bibFullDraft.bib}

\providecommand{\href}[2]{#2}\begingroup\raggedright\begin{thebibliography}{10}

\bibitem{Cutkosky:1960sp}
R.~Cutkosky, {\it {Singularities and discontinuities of Feynman amplitudes}},
  {\em J.Math.Phys.} {\bf 1} (1960) 429--433.

\bibitem{tHooft:1973pz}
G.~'t~Hooft and M.~Veltman, {\it {DIAGRAMMAR}},  {\em NATO Adv.Study Inst.Ser.B
  Phys.} {\bf 4} (1974) 177--322.

\bibitem{Veltman:1994wz}
M.~Veltman, {\it {Diagrammatica: The Path to Feynman rules}},  {\em Cambridge
  Lect.Notes Phys.} {\bf 4} (1994) 1--284.

\bibitem{Landau:1959fi}
L.~Landau, {\it {On analytic properties of vertex parts in quantum field
  theory}},  {\em Nucl.Phys.} {\bf 13} (1959) 181--192.

\bibitem{Remiddi:1981hn}
E.~Remiddi, {\it {Dispersion Relations for Feynman Graphs}},  {\em
  Helv.Phys.Acta} {\bf 54} (1982) 364.

\bibitem{Bern:1994zx}
Z.~Bern, L.~J. Dixon, D.~C. Dunbar, and D.~A. Kosower, {\it {One loop n point
  gauge theory amplitudes, unitarity and collinear limits}},  {\em Nucl.Phys.}
  {\bf B425} (1994) 217--260,
  [\href{http://xxx.lanl.gov/abs/hep-ph/9403226}{{\tt hep-ph/9403226}}].

\bibitem{Bern:1994cg}
Z.~Bern, L.~J. Dixon, D.~C. Dunbar, and D.~A. Kosower, {\it {Fusing gauge
  theory tree amplitudes into loop amplitudes}},  {\em Nucl.Phys.} {\bf B435}
  (1995) 59--101, [\href{http://xxx.lanl.gov/abs/hep-ph/9409265}{{\tt
  hep-ph/9409265}}].

\bibitem{Britto:2004nc}
R.~Britto, F.~Cachazo, and B.~Feng, {\it {Generalized unitarity and one-loop
  amplitudes in N=4 super-Yang-Mills}},  {\em Nucl.Phys.} {\bf B725} (2005)
  275--305, [\href{http://xxx.lanl.gov/abs/hep-th/0412103}{{\tt
  hep-th/0412103}}].

\bibitem{Britto:2005ha}
R.~Britto, E.~Buchbinder, F.~Cachazo, and B.~Feng, {\it {One-loop amplitudes of
  gluons in SQCD}},  {\em Phys.Rev.} {\bf D72} (2005) 065012,
  [\href{http://xxx.lanl.gov/abs/hep-ph/0503132}{{\tt hep-ph/0503132}}].

\bibitem{Buchbinder:2005wp}
E.~I. Buchbinder and F.~Cachazo, {\it {Two-loop amplitudes of gluons and
  octa-cuts in N=4 super Yang-Mills}},  {\em JHEP} {\bf 0511} (2005) 036,
  [\href{http://xxx.lanl.gov/abs/hep-th/0506126}{{\tt hep-th/0506126}}].

\bibitem{Forde:2007mi}
D.~Forde, {\it {Direct extraction of one-loop integral coefficients}},  {\em
  Phys.Rev.} {\bf D75} (2007) 125019,
  [\href{http://xxx.lanl.gov/abs/0704.1835}{{\tt 0704.1835}}].

\bibitem{Britto:2007tt}
R.~Britto and B.~Feng, {\it {Integral coefficients for one-loop amplitudes}},
  {\em JHEP} {\bf 0802} (2008) 095,
  [\href{http://xxx.lanl.gov/abs/0711.4284}{{\tt 0711.4284}}].

\bibitem{Mastrolia:2009dr}
P.~Mastrolia, {\it {Double-Cut of Scattering Amplitudes and Stokes' Theorem}},
  {\em Phys.Lett.} {\bf B678} (2009) 246--249,
  [\href{http://xxx.lanl.gov/abs/0905.2909}{{\tt 0905.2909}}].

\bibitem{Kosower:2011ty}
D.~A. Kosower and K.~J. Larsen, {\it {Maximal Unitarity at Two Loops}},  {\em
  Phys.Rev.} {\bf D85} (2012) 045017,
  [\href{http://xxx.lanl.gov/abs/1108.1180}{{\tt 1108.1180}}].

\bibitem{Goncharov:2005sla}
A.~Goncharov, {\it {Galois symmetries of fundamental groupoids and
  noncommutative geometry}},  {\em Duke Math.J.} {\bf 128} (2005) 209,
  [\href{http://xxx.lanl.gov/abs/math/0208144}{{\tt math/0208144}}].

\bibitem{GoncharovMixedTate}
A.~B. Goncharov, {\it {Multiple polylogarithms and mixed Tate motives}},
  \href{http://xxx.lanl.gov/abs/math/0103059}{{\tt math/0103059}}.

\bibitem{Gaiotto:2011dt}
D.~Gaiotto, J.~Maldacena, A.~Sever, and P.~Vieira, {\it {Pulling the straps of
  polygons}},  {\em JHEP} {\bf 1112} (2011) 011,
  [\href{http://xxx.lanl.gov/abs/1102.0062}{{\tt 1102.0062}}].

\bibitem{Caffo:1998du}
M.~Caffo, H.~Czyz, S.~Laporta, and E.~Remiddi, {\it {The Master differential
  equations for the two loop sunrise selfmass amplitudes}},  {\em Nuovo Cim.}
  {\bf A111} (1998) 365--389,
  [\href{http://xxx.lanl.gov/abs/hep-th/9805118}{{\tt hep-th/9805118}}].

\bibitem{MullerStach:2011ru}
S.~M{\"u}ller-Stach, S.~Weinzierl, and R.~Zayadeh, {\it {A Second-Order
  Differential Equation for the Two-Loop Sunrise Graph with Arbitrary Masses}},
   {\em Commun.Num.Theor.Phys.} {\bf 6} (2012) 203--222,
  [\href{http://xxx.lanl.gov/abs/1112.4360}{{\tt 1112.4360}}].

\bibitem{CaronHuot:2012ab}
S.~Caron-Huot and K.~J. Larsen, {\it {Uniqueness of two-loop master contours}},
   {\em JHEP} {\bf 1210} (2012) 026,
  [\href{http://xxx.lanl.gov/abs/1205.0801}{{\tt 1205.0801}}].

\bibitem{Adams:2013nia}
L.~Adams, C.~Bogner, and S.~Weinzierl, {\it {The two-loop sunrise graph with
  arbitrary masses}},  \href{http://xxx.lanl.gov/abs/1302.7004}{{\tt
  1302.7004}}.

\bibitem{Bloch:2013tra}
S.~Bloch and P.~Vanhove, {\it {The elliptic dilogarithm for the sunset graph}},
   \href{http://xxx.lanl.gov/abs/1309.5865}{{\tt 1309.5865}}.

\bibitem{Remiddi:2013joa}
E.~Remiddi and L.~Tancredi, {\it {Schouten identities for Feynman graph
  amplitudes; the Master Integrals for the two-loop massive sunrise graph}},
  \href{http://xxx.lanl.gov/abs/1311.3342}{{\tt 1311.3342}}.

\bibitem{Remiddi:1999ew}
E.~Remiddi and J.~Vermaseren, {\it {Harmonic polylogarithms}},  {\em
  Int.J.Mod.Phys.} {\bf A15} (2000) 725--754,
  [\href{http://xxx.lanl.gov/abs/hep-ph/9905237}{{\tt hep-ph/9905237}}].

\bibitem{Gehrmann:2000zt}
T.~Gehrmann and E.~Remiddi, {\it {Two loop master integrals for $\gamma^*\to 3$
  jets: The Planar topologies}},  {\em Nucl.Phys.} {\bf B601} (2001) 248--286,
  [\href{http://xxx.lanl.gov/abs/hep-ph/0008287}{{\tt hep-ph/0008287}}].

\bibitem{Kalmykov:2000qe}
M.~Y. Kalmykov and O.~Veretin, {\it {Single scale diagrams and multiple
  binomial sums}},  {\em Phys.Lett.} {\bf B483} (2000) 315--323,
  [\href{http://xxx.lanl.gov/abs/hep-th/0004010}{{\tt hep-th/0004010}}].

\bibitem{Kalmykov:2007dk}
M.~Y. Kalmykov, B.~Ward, and S.~Yost, {\it {Multiple (inverse) binomial sums of
  arbitrary weight and depth and the all-order epsilon-expansion of generalized
  hypergeometric functions with one half-integer value of parameter}},  {\em
  JHEP} {\bf 0710} (2007) 048, [\href{http://xxx.lanl.gov/abs/0707.3654}{{\tt
  0707.3654}}].

\bibitem{Bonciani:2010ms}
R.~Bonciani, G.~Degrassi, and A.~Vicini, {\it {On the Generalized Harmonic
  Polylogarithms of One Complex Variable}},  {\em Comput.Phys.Commun.} {\bf
  182} (2011) 1253--1264, [\href{http://xxx.lanl.gov/abs/1007.1891}{{\tt
  1007.1891}}].

\bibitem{Ablinger:2011te}
J.~Ablinger, J.~Blumlein, and C.~Schneider, {\it {Harmonic Sums and
  Polylogarithms Generated by Cyclotomic Polynomials}},  {\em J.Math.Phys.}
  {\bf 52} (2011) 102301, [\href{http://xxx.lanl.gov/abs/1105.6063}{{\tt
  1105.6063}}].

\bibitem{Ablinger:2013cf}
J.~Ablinger, J.~Bl{\"u}mlein, and C.~Schneider, {\it {Analytic and Algorithmic
  Aspects of Generalized Harmonic Sums and Polylogarithms}},  {\em
  J.Math.Phys.} {\bf 54} (2013) 082301,
  [\href{http://xxx.lanl.gov/abs/1302.0378}{{\tt 1302.0378}}].

\bibitem{Goncharov:1998kja}
A.~B. Goncharov, {\it {Multiple polylogarithms, cyclotomy and modular
  complexes}},  {\em Math.Res.Lett.} {\bf 5} (1998) 497--516,
  [\href{http://xxx.lanl.gov/abs/1105.2076}{{\tt 1105.2076}}].

\bibitem{ChenSymbol}
K.~T. Chen, {\it {Iterated path integrals}},  {\em Bull.\ Amer.\ Math.\ Soc.}
  {\bf 83} (1977) 831.

\bibitem{Goncharov:2010jf}
A.~B. Goncharov, M.~Spradlin, C.~Vergu, and A.~Volovich, {\it {Classical
  Polylogarithms for Amplitudes and Wilson Loops}},  {\em Phys.Rev.Lett.} {\bf
  105} (2010) 151605, [\href{http://xxx.lanl.gov/abs/1006.5703}{{\tt
  1006.5703}}].

\bibitem{Goncharov:2009tja}
A.~Goncharov, {\it {A simple construction of Grassmannian polylogarithms}},
  \href{http://xxx.lanl.gov/abs/0908.2238}{{\tt 0908.2238}}.

\bibitem{Brown:2009qja}
F.~C. Brown, {\it {Multiple zeta values and periods of moduli spaces
  $\mathfrak{M}_{0,n}$}},  {\em Annales Sci.Ecole Norm.Sup.} {\bf 42} (2009)
  371, [\href{http://xxx.lanl.gov/abs/math/0606419}{{\tt math/0606419}}].

\bibitem{Duhr:2011zq}
C.~Duhr, H.~Gangl, and J.~R. Rhodes, {\it {From polygons and symbols to
  polylogarithmic functions}},  {\em JHEP} {\bf 1210} (2012) 075,
  [\href{http://xxx.lanl.gov/abs/1110.0458}{{\tt 1110.0458}}].

\bibitem{Brown:2011ik}
F.~Brown, {\it {On the decomposition of motivic multiple zeta values}},
  \href{http://xxx.lanl.gov/abs/1102.1310}{{\tt 1102.1310}}.

\bibitem{Duhr:2012fh}
C.~Duhr, {\it {Hopf algebras, coproducts and symbols: an application to Higgs
  boson amplitudes}},  {\em JHEP} {\bf 1208} (2012) 043,
  [\href{http://xxx.lanl.gov/abs/1203.0454}{{\tt 1203.0454}}].

\bibitem{ArkaniHamed:2010gh}
N.~Arkani-Hamed, J.~L. Bourjaily, F.~Cachazo, and J.~Trnka, {\it {Local
  Integrals for Planar Scattering Amplitudes}},  {\em JHEP} {\bf 1206} (2012)
  125, [\href{http://xxx.lanl.gov/abs/1012.6032}{{\tt 1012.6032}}].

\bibitem{Cachazo:2008vp}
F.~Cachazo, {\it {Sharpening The Leading Singularity}},
  \href{http://xxx.lanl.gov/abs/0803.1988}{{\tt 0803.1988}}.

\bibitem{Chetyrkin:1981qh}
K.~Chetyrkin and F.~Tkachov, {\it {Integration by Parts: The Algorithm to
  Calculate beta Functions in 4 Loops}},  {\em Nucl.Phys.} {\bf B192} (1981)
  159--204.

\bibitem{Henn:2013pwa}
J.~M. Henn, {\it {Multiloop integrals in dimensional regularization made
  simple}},  {\em Phys.Rev.Lett.} {\bf 110} (2013) 251601,
  [\href{http://xxx.lanl.gov/abs/1304.1806}{{\tt 1304.1806}}].

\bibitem{Henn:2013tua}
J.~M. Henn, A.~V. Smirnov, and V.~A. Smirnov, {\it {Analytic results for planar
  three-loop four-point integrals from a Knizhnik-Zamolodchikov equation}},
  {\em JHEP} {\bf 1307} (2013) 128,
  [\href{http://xxx.lanl.gov/abs/1306.2799}{{\tt 1306.2799}}].

\bibitem{Henn:2013woa}
J.~M. Henn and V.~A. Smirnov, {\it {Analytic results for two-loop master
  integrals for Bhabha scattering I}},  {\em JHEP} {\bf 1311} (2013) 041,
  [\href{http://xxx.lanl.gov/abs/1307.4083}{{\tt 1307.4083}}].

\bibitem{Henn:2013nsa}
J.~M. Henn, A.~V. Smirnov, and V.~A. Smirnov, {\it {Evaluating single-scale
  and/or non-planar diagrams by differential equations}},
  \href{http://xxx.lanl.gov/abs/1312.2588}{{\tt 1312.2588}}.

\bibitem{Argeri:2014qva}
M.~Argeri, S.~Di~Vita, P.~Mastrolia, E.~Mirabella, J.~Schlenk, {\em et~al.},
  {\it {Magnus and Dyson Series for Master Integrals}},
  \href{http://xxx.lanl.gov/abs/1401.2979}{{\tt 1401.2979}}.

\bibitem{Bogner:2007mn}
C.~Bogner and S.~Weinzierl, {\it {Periods and Feynman integrals}},  {\em
  J.Math.Phys.} {\bf 50} (2009) 042302,
  [\href{http://xxx.lanl.gov/abs/0711.4863}{{\tt 0711.4863}}].

\bibitem{Dixon:2011pw}
L.~J. Dixon, J.~M. Drummond, and J.~M. Henn, {\it {Bootstrapping the three-loop
  hexagon}},  {\em JHEP} {\bf 1111} (2011) 023,
  [\href{http://xxx.lanl.gov/abs/1108.4461}{{\tt 1108.4461}}].

\bibitem{Dixon:2011nj}
L.~J. Dixon, J.~M. Drummond, and J.~M. Henn, {\it {Analytic result for the
  two-loop six-point NMHV amplitude in N=4 super Yang-Mills theory}},  {\em
  JHEP} {\bf 1201} (2012) 024, [\href{http://xxx.lanl.gov/abs/1111.1704}{{\tt
  1111.1704}}].

\bibitem{Drummond:2012bg}
J.~Drummond, {\it {Generalised ladders and single-valued polylogarithms}},
  {\em JHEP} {\bf 1302} (2013) 092,
  [\href{http://xxx.lanl.gov/abs/1207.3824}{{\tt 1207.3824}}].

\bibitem{Chavez:2012kn}
F.~Chavez and C.~Duhr, {\it {Three-mass triangle integrals and single-valued
  polylogarithms}},  {\em JHEP} {\bf 1211} (2012) 114,
  [\href{http://xxx.lanl.gov/abs/1209.2722}{{\tt 1209.2722}}].

\bibitem{Dixon:2012yy}
L.~J. Dixon, C.~Duhr, and J.~Pennington, {\it {Single-valued harmonic
  polylogarithms and the multi-Regge limit}},  {\em JHEP} {\bf 1210} (2012)
  074, [\href{http://xxx.lanl.gov/abs/1207.0186}{{\tt 1207.0186}}].

\bibitem{Drummond:2013nda}
J.~Drummond, C.~Duhr, B.~Eden, P.~Heslop, J.~Pennington, {\em et~al.}, {\it
  {Leading singularities and off-shell conformal integrals}},
  \href{http://xxx.lanl.gov/abs/1303.6909}{{\tt 1303.6909}}.

\bibitem{Dixon:2013eka}
L.~J. Dixon, J.~M. Drummond, M.~von Hippel, and J.~Pennington, {\it {Hexagon
  functions and the three-loop remainder function}},  {\em JHEP} {\bf 1312}
  (2013) 049, [\href{http://xxx.lanl.gov/abs/1308.2276}{{\tt 1308.2276}}].

\bibitem{Dixon:2014voa}
L.~J. Dixon, J.~M. Drummond, C.~Duhr, and J.~Pennington, {\it {The four-loop
  remainder function and multi-Regge behavior at NNLLA in planar N=4
  super-Yang-Mills theory}},  \href{http://xxx.lanl.gov/abs/1402.3300}{{\tt
  1402.3300}}.

\bibitem{Kotikov:1990kg}
A.~V. Kotikov, {\it {Differential equations method: New technique for massive
  Feynman diagrams calculation}},  {\em Phys. Lett.} {\bf B254} (1991)
  158--164.

\bibitem{Kotikov:1991hm}
A.~V. Kotikov, {\it {Differential equations method: The Calculation of vertex
  type Feynman diagrams}},  {\em Phys. Lett.} {\bf B259} (1991) 314--322.

\bibitem{Kotikov:1991pm}
A.~V. Kotikov, {\it {Differential equation method: The Calculation of N point
  Feynman diagrams}},  {\em Phys. Lett.} {\bf B267} (1991) 123--127.

\bibitem{Gehrmann:1999as}
T.~Gehrmann and E.~Remiddi, {\it {Differential equations for two loop four
  point functions}},  {\em Nucl.Phys.} {\bf B580} (2000) 485--518,
  [\href{http://xxx.lanl.gov/abs/hep-ph/9912329}{{\tt hep-ph/9912329}}].

\bibitem{steinmann1}
O.~Steinmann, {\it {\"Uber den Zusammenhang zwischen den Wightmanfunktionen und
  der retardierten Kommutatoren}},  {\em Helv.Phys.Acta.} {\bf 33} (1960) 257.

\bibitem{steinmann2}
O.~Steinmann, {\it Wightman-funktionen und retardierten kommutatoren},  {\em
  Helv.Phys.Acta.} {\bf 33} (1960) 347.

\bibitem{stapp1}
H.~Stapp, {\it {Inclusive cross-sections are discontinuities}},  {\em
  Phys.Rev.} {\bf D3} (1971) 3177.

\bibitem{Usyukina:1993ch}
N.~Usyukina and A.~I. Davydychev, {\it {Exact results for three and four point
  ladder diagrams with an arbitrary number of rungs}},  {\em Phys.Lett.} {\bf
  B305} (1993) 136--143.

\bibitem{Bern:1993kr}
Z.~Bern, L.~J. Dixon, and D.~A. Kosower, {\it {Dimensionally regulated pentagon
  integrals}},  {\em Nucl.Phys.} {\bf B412} (1994) 751--816,
  [\href{http://xxx.lanl.gov/abs/hep-ph/9306240}{{\tt hep-ph/9306240}}].

\bibitem{Anastasiou:1999ui}
C.~Anastasiou, E.~N. Glover, and C.~Oleari, {\it {Scalar one loop integrals
  using the negative dimension approach}},  {\em Nucl.Phys.} {\bf B572} (2000)
  307--360, [\href{http://xxx.lanl.gov/abs/hep-ph/9907494}{{\tt
  hep-ph/9907494}}].

\bibitem{Usyukina:1992jd}
N.~Usyukina and A.~I. Davydychev, {\it {An Approach to the evaluation of three
  and four point ladder diagrams}},  {\em Phys.Lett.} {\bf B298} (1993)
  363--370.

\bibitem{Davydychev:2000na}
A.~I. Davydychev and M.~Y. Kalmykov, {\it {New results for the epsilon
  expansion of certain one, two and three loop Feynman diagrams}},  {\em
  Nucl.Phys.} {\bf B605} (2001) 266--318,
  [\href{http://xxx.lanl.gov/abs/hep-th/0012189}{{\tt hep-th/0012189}}].

\bibitem{Birthwright:2004kk}
T.~Birthwright, E.~N. Glover, and P.~Marquard, {\it {Master integrals for
  massless two-loop vertex diagrams with three offshell legs}},  {\em JHEP}
  {\bf 0409} (2004) 042, [\href{http://xxx.lanl.gov/abs/hep-ph/0407343}{{\tt
  hep-ph/0407343}}].

\bibitem{Ball:1991bs}
P.~Ball, V.~M. Braun, and H.~G. Dosch, {\it {Form-factors of semileptonic D
  decays from QCD sum rules}},  {\em Phys.Rev.} {\bf D44} (1991) 3567--3581.

\bibitem{Mastrolia:2006ki}
P.~Mastrolia, {\it {On Triple-cut of scattering amplitudes}},  {\em Phys.Lett.}
  {\bf B644} (2007) 272--283,
  [\href{http://xxx.lanl.gov/abs/hep-th/0611091}{{\tt hep-th/0611091}}].

\bibitem{Feynman:1963ax}
R.~Feynman, {\it {Quantum theory of gravitation}},  {\em Acta Phys.Polon.} {\bf
  24} (1963) 697--722.

\bibitem{Feynman:tree1}
R.~Feynman, {\it Problems in quantizing the gravitational field, and the
  massless yang-mills field},  in {\em Magic without magic} (J.~Klauder, ed.),
  pp.~355--375.
\newblock San Francisco, 1972.

\bibitem{Feynman:tree2}
R.~Feynman, {\it Closed loop and tree diagrams},  in {\em Magic without magic}
  (J.~Klauder, ed.), pp.~377--408.
\newblock San Francisco, 1972.

\bibitem{Huber:2007dx}
T.~Huber and D.~Maitre, {\it {HypExp 2, Expanding Hypergeometric Functions
  about Half-Integer Parameters}},  {\em Comput.Phys.Commun.} {\bf 178} (2008)
  755--776, [\href{http://xxx.lanl.gov/abs/0708.2443}{{\tt 0708.2443}}].

\bibitem{SMatrix}
R.~Eden, P.~Landshoff, D.~Olive, and J.~Polkinghorne, {\em The Analytic
  S-Matrix}.
\newblock Cambridge at the University Press, 1966.

\bibitem{vanNeerven:1985xr}
W.~van Neerven, {\it {Dimensional Regularization Of Mass And Infrared
  Singularities In Two Loop On-Shell Vertex Functions}},  {\em Nucl.Phys.} {\bf
  B268} (1986) 453.

\bibitem{BrownSVHPLs}
F.~C. Brown, {\it {Single-valued multiple polylogarithms in one variable}},
  {\em C. R. Acad. Sci. Paris, Ser. I} {\bf 338} (2004) 527.

\end{thebibliography}\endgroup

\end{document}